\documentclass[12pt]{article}

\usepackage{newtxtext,newtxmath}
\usepackage{hyperref}
\usepackage{graphicx}
\graphicspath{{./}}
\usepackage[letterpaper,margin=1in]{geometry}
\renewenvironment{abstract}
	{\quotation}
	{\endquotation}

\date{}

\makeatletter
\renewcommand{\fnum@figure}{\textbf{Figure \thefigure}}
\renewcommand{\fnum@table}{\textbf{Table \thetable}}
\makeatother

\usepackage{scicite}
\usepackage{url}

\usepackage{appendix}
\usepackage{titletoc}
\usepackage{bibunits}
\usepackage{longtable}
\usepackage{booktabs}
\usepackage{array}
\usepackage{float}
\usepackage{comment}
\usepackage{ragged2e}
\usepackage{multirow}
\usepackage{xcolor}
\usepackage[most]{tcolorbox}

\def\scititle{
	Digital Twins are Funhouse Mirrors: Five Systematic Distortions
}
\title{\bfseries \boldmath \scititle}

\author{
  Tianyi\ Peng,$^{1}$
  George\ Gui,$^{1}$
  Melanie\ Brucks,$^{1}$
  Daniel\ J.\ Merlau,$^{1}$
  Grace\ Jiarui\ Fan,$^{1}$\and
  Malek\ Ben\ Sliman,$^{1}$
  Eric\ J.\ Johnson,$^{1}$
  Abdullah\ Althenayyan,$^{1}$
  Silvia\ Bellezza,$^{1}$\and
  Dante\ Donati,$^{1}$
  Hortense\ Fong,$^{1}$
  Elizabeth\ Friedman,$^{1}$
  Ariana\ Guevara,$^{1}$\and
  Mohamed\ Hussein,$^{1}$
  Kinshuk\ Jerath,$^{1}$
  Bruce\ Kogut,$^{1}$
  Akshit\ Kumar,$^{2}$
  Kristen\ Lane,$^{1}$\and
  Hannah\ Li,$^{1}$
  Vicki\ Morwitz,$^{1}$
  Oded\ Netzer,$^{1}$
  Patryk\ Perkowski,$^{3}$
  Olivier\ Toubia$^{1\ast}$\and
  \small$^{1}$Columbia University, New York, NY 10027, USA\and
  \small$^{2}$Yale University, New Haven, CT 06511, USA\and
  \small$^{3}$Yeshiva University, New York, NY 10033, USA\and
  \small$^{\ast}$Corresponding author.\ Email: ot2107@columbia.edu
}

\begin{document}

\maketitle
\thispagestyle{plain}

\begin{abstract} \bfseries \boldmath
Scientists and practitioners are aggressively moving to deploy digital twins---LLM-based models of real individuals---across social science and policy research. We conducted 19 pre-registered studies with 164 diverse outcomes (e.g., attitudes towards hiring algorithms, intention to share misinformation) and compared human responses to those of their digital twins (trained on each person's previous answers to over 500 questions). We establish an empirical benchmark for digital twin performance: digital twins' answers are only modestly more accurate than those from the (homogeneous) base LLM and correlate weakly with human responses (average r = 0.20). To guide future twin development, we further document five ways in which digital twins distort human behavior: (i) insufficient individuation, (ii) stereotyping, (iii) representation bias, (iv) ideological biases, (v) hyper-rationality. Finally, we make our full dataset and code public as a standardized testbed for novel digital twin methodologies. Together, our results caution against premature deployment while laying the groundwork for the transparent, replicable, and iterative science necessary for responsible deployment of twins.
\end{abstract}

\textbf{One-sentence summary: Despite bold industry promises for digital twins, a large-scale mega-study finds they perform only marginally better than generic AI personas and documents five key distortions that serve as a diagnostic roadmap for improvement}.

\noindent
Large language models (LLMs) hold the capacity to fundamentally transform psychological measurement by overcoming the long-standing cognitive, ethical, and practical barriers inherent to human subjects research \cite{argyle2023out,dillion2023can,horton2023large,hewitt2024predicting}. Through the scalable simulation of human responses, LLMs offer not only faster and cheaper data collection, but also potentially ``better'' data for researchers and managers: LLMs do not tire, can be repeatedly surveyed, and can simulate high-stakes or potentially harmful scenarios without ethical risk. Although LLMs can show human-like behavior \cite{jones2025large,westwood2025potential} and directionally replicate experimental treatment effects \cite{hewitt2024predicting, cui2025large}, many researchers have expressed skepticism, both on theoretical and empirical grounds \cite{broska2025mixed,schroder2025large,gao2025take, bisbee2024synthetic,lin2025six}. Studies have shown that LLMs often struggle to reproduce human opinions \cite{santurkar2023whose,motoki2024more,li2025llm}, tend to lack heterogeneity \cite{wang2025large}, and overestimate experimental effect sizes \cite{cui2025large}.

To address these concerns, experts and practitioners alike have proposed using digital twins: base LLM models (e.g., GPT, Llama) augmented with rich personal information to simulate specific individual behavior \cite{park2024generative}. At first glance, digital twins present a remarkably elegant solution, offering what appears to be a panacea to the foundational limitations documented in prior synthetic data research. First, distinct, tailored models offer the potential to infuse human variance into responses, overcoming the homogenization that often plagues general models \cite{wang2025large}. Second, anchoring the model in rich and real human data may provide the detail necessary to prevent hallucinations or confabulations and address default sycophancy and normativity that undermines base model accuracy \cite{cui2025large,gao2025take,bisbee2024synthetic,santurkar2023whose}. Together, this would allow for ``simulations that better reflect the myriad, often idiosyncratic, factors that influence individuals' attitudes, beliefs, and behaviors'' \cite{park2024generative}. The allure of digital twins is further amplified by the unique applications they unlock, from agentic AI (e.g., negotiating, working, or dating on one's behalf) to self-reflection, where individuals interact with digital copies of themselves.

Given these promises, digital twin applications are being frantically built, deployed and sold across academia and industry. For example, the startup simile.ai, based on \cite{park2024generative}, recently raised \$100 M,\footnote{\href{https://techfundingnews.com/100m-for-stanford-spinout-simile-ai-that-simulates-human-decisions/}{https://techfundingnews.com/100m-for-stanford-spinout-simile-ai-that-simulates-human-decisions/}, \href{https://www.wsj.com/cio-journal/can-ai-replace-humans-for-market-research-4f818890?st=MF12bP}{https://www.wsj.com/cio-journal/can-ai-replace-humans-for-market-research-4f818890?st=MF12bP}.} joining a growing list of synthetic data startups with valuations in the order of \$1B.\footnote{e.g., \href{https://techcrunch.com/2025/12/05/ai-synthetic-research-startup-aaru-raised-a-series-a-at-a-1b-headline-valuation/?utm_campaign=daily_weekend}{https://techcrunch.com/2025/12/05/ai-synthetic-research-startup-aaru-raised-a-series-a-at-a-1b-headline-valuation/?utm\_campaign=daily\_weekend}.}

Despite fervent interest, the empirical foundation of digital twins remains remarkably thin. To date, only two initial studies have directly examined whether twins can serve as human surrogates. One study created digital twins of over 1,000 individuals based on transcripts from two-hour interviews \cite{park2024generative} and reported encouraging results (relative accuracy of 85\% on the General Social Survey, based on the ratio of digital twin accuracy to test-retest accuracy). However, the data are not publicly available, presenting a significant hurdle to the scientific understanding of digital twins. And without knowing the degree of separation between the original transcripts and validation questions, it is hard to fully evaluate performance.

The second study developed Twin-2K-500, a public dataset permitting the creation of digital twins of over 2,000 individuals based on their answers to over 500 questions \cite{toubia2025database} and reported relatively high accuracy of digital twins on holdout data (average accuracy of 72\%, relative accuracy of 88\% based on the ratio of digital twin accuracy to test-retest accuracy). As a first test, however, it was necessarily limited in scope: twins were not compared against benchmarks such as demographic-only personas. Furthermore, twins only replicated about half the experimental effects in humans. And in both studies \cite{park2024generative,toubia2025database}, validation was based on a relatively narrow set of well-known experiments and surveys, which may overlap with the training data of the base model, raising concerns of leakage \cite{ludwig2025large}. Consequently, the field is rapidly adopting a technology that lacks essential empirical validation, leaving open the risk of making consequential decisions based on an illusion of personalization rather than genuine individual-level insight.

To address this urgent empirical gap, we conducted a large-scale, pre-registered mega-study designed in the spirit of true scientific inquiry. We assembled 23 co-authors from various backgrounds within social science, each curious to test the validity of digital twins in their domain of interest. Together, we jointly designed and ran the most comprehensive empirical test of digital twins to date, with 19 new pre-registered (\href{https://researchbox.org/4145}{https://researchbox.org/4145}) sub-studies covering 164 diverse outcomes (cumulative sample size of N=13,506 human participants across substudies, 1,784 unique participants). We matched each human answer to their digital twin's answer to the same question. To give digital twins their best shot, we conducted our study on the validated Twin-2K-500 human sample and their corresponding digital twins. We invited the original Twin-2K-500 participants to complete entirely new studies (including novel stimuli) and then compared their answers against their twins'. This mega-study, unprecedented in scale and scope, presents an earnest, transparent, and replicable empirical benchmark of digital twin performance.

Our evaluation demonstrated underwhelming performance overall. While digital twins showed revealing variance across domains and individuals, they largely failed to faithfully mimic human responses. We also gathered human expert predictions on twin performance to gauge relative understanding of this nascent technology. We find that these experts generally failed to predict the behavior of digital twins (see SI), underscoring the value of empirical tests.

In light of these disappointing results, we conducted exploratory analyses to identify how digital twins distort human behavior and therefore how they may be improved. We document five distortions: (i) insufficient individuation, (ii) stereotyping, (iii) representation bias, (iv) ideological biases, (v) hyper-rationality. These distortions are generally consistent with previous literature on synthetic data, suggesting that state-of-the-art digital twins are not yet able to fully resolve known limitations of synthetic data. Without actively addressing these distortions, the premature deployment of twins risks systematically misrepresenting human cognition in ways that could undermine both scientific understanding and practical applications.

Together, this work advances the scientific foundation of digital twins on three fronts. First, we establish an empirical baseline for where the technology stands today, grounding a field thus far driven more by promise than evidence. Second, by consolidating disparate prior findings into a unified evaluative framework, we offer a roadmap for the improvement of digital twins and a clear set of indicators on which to evaluate them. This enables more targeted improvements in future digital twin research and direct comparisons across methodologies. Third, by releasing our full dataset and code, we provide a standardized testbed for the improvement of digital twin pipelines. By quantifying current performance, establishing clear, standardized metrics and benchmarks for safety and accuracy, and engaging in open science practices, we enable transparent, replicable, and iterative science necessary to move this technology forward to the point where it can be deployed responsibly in both academia and industry.

\section*{Empirical Context}

Our studies cover a wide range of domains and topics including creativity, politics, privacy preferences, storytelling, fairness perceptions, interactions with technology platforms, luxury consumption, news consumption, and labor market preferences, among others. Unlike prior work that primarily examined digital twins' performance in published studies \cite{park2024generative, toubia2025database}, our sub-studies combined established paradigms from published research, new paradigms from unpublished research, and newly designed studies. Our contexts range from behavioral economics paradigms to personality scale development and include correlational, within-subject, and between-subject designs. Together, they span a considerably broader and richer scope of use cases than any paper to date, providing an ecologically valid test of digital twins as they are available today.

To create each digital twin, we used in-context learning based on answers to over 500 questions (approximately 128K characters) provided by the twin's human counterpart. These questions covered demographics (14 questions), personality traits (279 questions from 19 personality tests measuring 26 constructs), cognitive abilities (85 questions covering 11 measures), economic preferences (34 questions covering 10 measures), as well as heuristics and biases experiments (48 questions from 16 experiments), and a pricing study (40 questions), collected during a four-wave, longitudinal study (average completion time of 145 min). These (human) data have high test-retest accuracy, good face validity, and largely replicate known effects, suggesting they provide high-quality training data for digital twins \cite{toubia2025database}. While no single study can evaluate all possible instantiations of digital twins, our implementation represents the most replicable and scientifically rigorous approach currently available. Unlike twins built on proprietary data, our twins are built using the most comprehensive (and widely used, with approximately 15,000 downloads) dataset publicly available. We complement this public resource by contributing an even more comprehensive dataset (including responses to novel stimuli and paradigms), which can be merged with the Twin-2K-500 dataset by matching participants across the two datasets.\footnote{Researchers may combine our data with the original Twin-2K-500 dataset, with twins matched using a Twin ID. For privacy reasons we are unable to share the Prolific ID associated with each Twin ID; therefore, other researchers are not able to run new studies on the human sample as we did.}

\begin{figure}[h!]
    \centering
    \includegraphics[width=\textwidth]{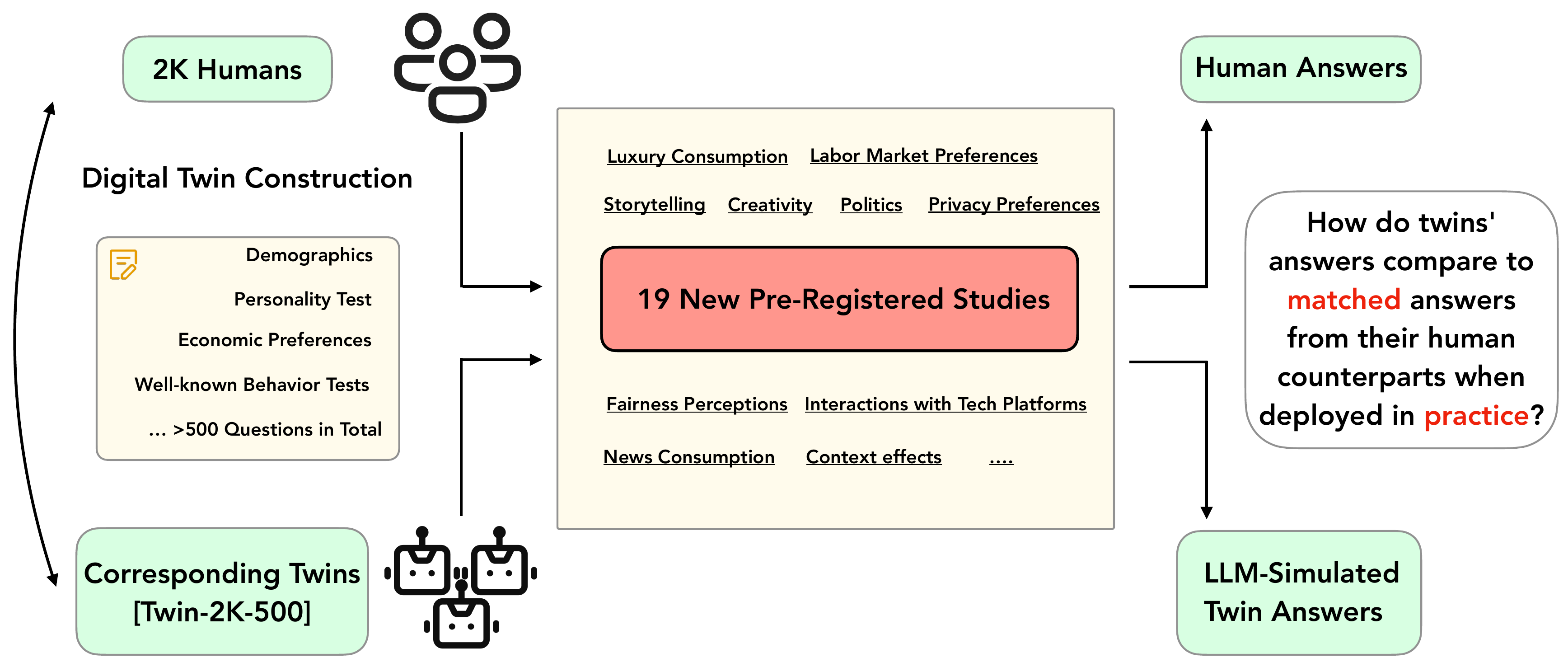}
    \caption{\textbf{Mega-Study Overview.} We ran 19 pre-registered studies on digital twins from the Twin-2K-500 dataset and their human counterparts. The studies were proposed by a diverse group of scholars and cover a wide range of behaviors and domains. As a set, they represent how digital twins may be leveraged today by social scientists. We match the answer of each digital twin to each question with the answer from their human counterpart, allowing us to explore the performance of digital twins at both the individual and population levels.}
    \label{fig:overview}
\end{figure}

In each sub-study, both human participants and their corresponding digital twins answered the exact same questions. By matching each digital twin's answers to those given by its human counterpart to the same questions, we are able to test, across a wide range of domains, the individual-level accuracy of digital twins as well as the correlation between responses from the digital twins and their human counterparts. Further, because our digital twins are based on a representative sample of the U.S. population, we can explore not only how performance varies across types of domains and questions, but also which groups are well represented by digital twins and which are not. We share the detailed analyses of each sub-study for researchers interested in using digital twins in specific domains (see SI). We make our data (\href{https://huggingface.co/datasets/LLM-Digital-Twin/Twin-2K-500-Mega-Study}{\url{https://huggingface.co/datasets/LLM-Digital-Twin/Twin-2K-500-Mega-Study}}) and code (\href{https://github.com/TianyiPeng/Twin-2K-500-Mega-Study}{\url{https://github.com/TianyiPeng/Twin-2K-500-Mega-Study}}) publicly available.\footnote{Researchers may combine our data with the original Twin-2K-500 dataset, with twins matched using a Twin ID. For privacy reasons we are unable to share the Prolific ID associated with each Twin ID; therefore, other researchers are not able to run new studies on the human sample as we did.} See Fig.~\ref{fig:overview} for an overview and SI for details of our mega-study. We pre-registered all our studies and associated outcomes, as well as our basic performance evaluation. In our pre-registration we also indicated that, beyond these metrics, this mega-study is exploratory in nature. We did not pre-register any specific hypotheses as to how well digital twins would perform, and we allowed ourselves to explore systematic ways in which digital twins differed from humans.

\section*{Results}

\subsection*{Digital Twin Performance}

\subsubsection*{Performance metrics}

We pre-registered 164 outcomes (e.g., willingness to share a particular article on social media, likelihood of applying for a particular job) across our 19 sub-studies.
We focus on performance measures that directly match the response of each digital twin to that of its human counterpart. Specifically, for each outcome, we match each twin's response to that of its corresponding human, and compute two metrics:

\begin{itemize}
    \item \textbf{Individual-level accuracy}: Following previous research \cite{toubia2025database,kolluri2025finetuning}, we measure individual-level accuracy as 1-(MAD/range), where MAD is the mean absolute deviation between a human's response and that of their twin, and range is the natural range of the outcome.\footnote{We exclude outcomes without a defined range, such as open-ended price estimates, resulting in 161 outcomes for this metric.} We compute the average individual-level accuracy across participants for each outcome, where higher values indicate greater accuracy (between 0 and 1, where 1 denotes perfect accuracy). For example, on a 1 to 7 scale, if the human answered 5 and the twin answered 6, accuracy would be 0.833.
    \item \textbf{Correlation}: We calculate the correlation between human and twin responses across participants, for each outcome. This metric reflects the degree to which twins capture individual-level heterogeneity in human responses. Following previous research \cite{park2024generative}, we compute an overall average correlation across outcomes by first applying a Fisher z-transformation to individual correlations, averaging these values, and then converting back using an inverse transformation.
\end{itemize}

In addition to the metrics above, which are based on matched human-twin data, we also compute two distribution-level performance metrics that compare the aggregate distributions of responses from twins and humans for each outcome. These metrics are more commonly available (no individual-level matching required), but arguably less informative:
\begin{itemize}
    \item \textbf{Comparison of averages}: We compare the average response of humans to that of their twins for each outcome using the absolute value of Glass's $\Delta$.\footnote{Glass's $\Delta$ is an effect size measure calculated as: $\Delta = (\hat{X}_{twin} - \hat{X}_{human}) / s_{human}$, where $\hat{X}_{twin}$ and $\hat{X}_{human}$ are the sample means of the twin and human groups respectively, and $s_{human}$ is the standard deviation of the human group. It is similar to Cohen's d, but it relies on the standard deviation from the human sample only, as the assumption of equal variances between the two samples is not valid in our setting.} Higher values of this metric indicate larger differences between group means (expressed in standard deviations).
    \item \textbf{Comparison of standard deviations}: We compute the ratio of standard deviations between humans and twins for each outcome. Values below (above) 1 suggest lower (higher) variance in twins relative to humans.
\end{itemize}

\subsubsection*{Performance comparisons}

\begin{figure}[h!]
    \centering
    \includegraphics[width=\textwidth]{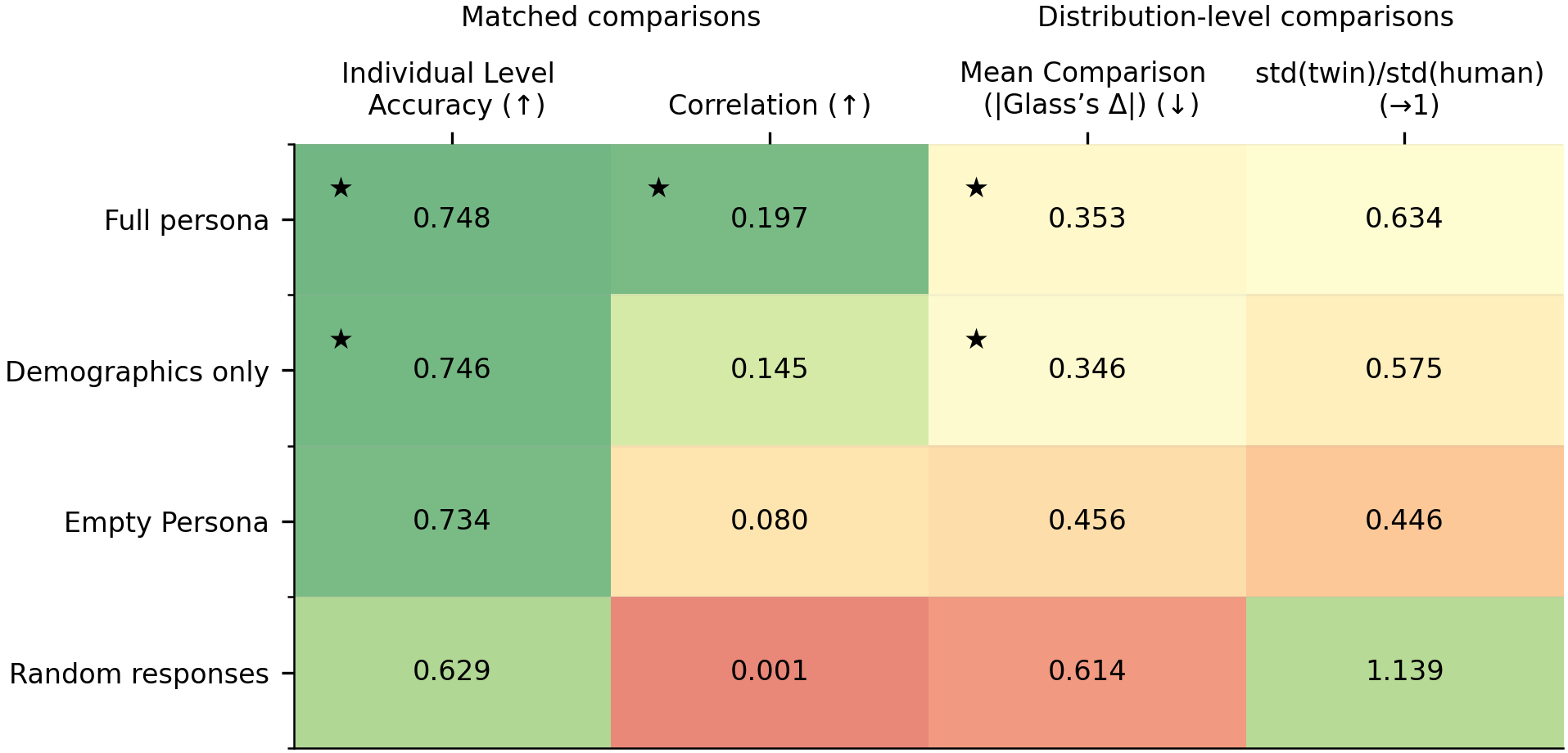}
    \caption{\textbf{Gains from Leveraging Individual-Level Data.} *: best performing benchmark, or not significantly different from best at p$<$0.05 (not applicable to ratio of standard deviations), after applying Bonferroni correcting for multiple comparisons in each column. Creating digital twins using rich individual-level data leads to a modest improvement in accuracy over predictions made without any individual-level data (empty persona) and no significant improvement over personas based on demographics only. Full personas improve the correlation between twin and human responses compared to these benchmarks, although the correlation is modest and digital twins remain under-dispersed.}
    \label{fig:summary stat}
\end{figure}

By default (and unless otherwise specified), we build each twin using a ``full persona'' approach which augments the LLM's context window with that human's complete data from the Twin-2K-500 dataset.
Across all outcomes, the average individual-level accuracy of the ``full persona'' approach is 0.748. See Fig.~\ref{fig:summary stat}, and the SI for histograms of this and other performance metrics across outcomes for the ``full persona'' approach.

To contextualize this level of accuracy and quantify gains from leveraging individual-level data for out-of-distribution predictions, we compare this performance to several benchmarks. First, we compare performance to that of random responses, where responses to each outcome are drawn from a uniform random distribution with support equal to the outcome's range (i.e., no LLM is required for this benchmark). The random benchmark already achieves an individual accuracy level of 0.629, underscoring that absolute levels of individual-level accuracy should be interpreted relative to this baseline. As an illustration, if the true responses were uniformly distributed on [0,1], always predicting 0 would yield an individual-level accuracy of 0.5.

Next, we compare the ``full persona'' digital twin approach with a simple ``empty persona'' benchmark, which uses an identical prompt for all respondents (i.e., no individual information) and thus only reflects the base model's behavior (with serving status and characteristics of the base model such as temperature being the only source of variation). The empty persona benchmark achieves an average accuracy of 0.734, which is only modestly lower than full personas (a difference of 0.014, albeit statistically significant).\footnote{We perform paired t-tests to compare each pair of benchmarks on each metric, with appropriate corrections for multiple comparisons. We drop one outcome in these comparisons, the creativity rating of ideas generated by humans and their twins in one of the sub-studies, as it was provided by humans and was only available for ideas from twins based on the full persona approach.}

We lastly introduce a ``demographics only'' benchmark, which uses a persona description that only includes the 14 demographic variables (region, sex, age, education, race, citizenship, marital status, religion, religious attendance, political party, household income, political ideology, household size, employment status) which are part of the Twin-2K-500 dataset (and therefore also included in the full persona description). This benchmark captures the performance that would be obtained from generic personas based on demographic characteristics only. We find that individual-level accuracy of full-persona twins is not significantly better than personas based only on demographic information ($Accuracy_{full \ persona}=0.748$, $Accuracy_{demographics \ only}=0.746$, $p=0.37$). This provides initial evidence that, when predicting an individual's response, digital twins may mostly rely on stereotypical demographic-based tendencies rather than modeling an individual's distinct cognition. We explore this phenomenon and the comparison between full-persona and demographics-only twins more thoroughly in a subsequent section.

At the aggregate level (comparing the mean twin response to the mean human response for each outcome), twin responses differ from human responses by 0.352 standard deviations on average;\footnote{This means the MSE of the digital twin's mean estimate is about $(0.352\sigma)^2 \approx 0.12\sigma^2$, which is approximately equal to the MSE of the sample mean from 10 humans, $0.1\sigma^2$.} a paired t-test (using Bonferroni correction for multiple comparisons) indicates a statistically significant difference ($p<0.05/164$) between humans and digital twins in 105 out of 164 outcomes (64.0\%).

We next turn to our correlation metric, which captures how well twins reflect relative human-level differences in individual responses across participants. The correlation between twin and human responses is positive in 157 of the 164 outcomes (95.7\%), and significantly positive ($p<0.05/165$ --- applying Bonferroni correction) in 97 of these cases (59.1\%). The overall average correlation is 0.197. On that metric, full personas perform markedly better than demographics-only personas (r=0.145), empty personas (r=0.080) and random responses (r=0.001).\footnote{For correlation, we apply a z-transformation before conducting t-tests. We set correlation to 0 for cases where there is no variation among twin answers, which may occur with the ``empty persona'' or ``demographics only'' benchmarks. For the three outcomes that do not have a natural range, we use the empirical range in humans' answers as support for the random distribution. The correlation achieved by empty personas is not 0 due to variations in stimuli within an outcome across participants, e.g., different respondents may see different versions of the same question or may be exposed to various conditions. This creates meaningful variations across respondents, even if predictions are not customized based on individual characteristics.} Nevertheless, the correlation achieved by digital twins is modest by most social science standards, and comparable to the correlation found between height and intelligence \cite{beauchamp2011sources}. Note that \cite{park2024generative} report a much higher correlation in their digital twin study. However, they compute the correlation across questions for each participant. In contrast, we compute the correlation across participants for each outcome. This is more often the measure of interest, as social scientists often ask questions such as who is more likely to vote for a candidate or who is more likely to purchase a specific product.

In sum, enriching digital twins with detailed individual-level information significantly improves individual-level prediction, that is, the ability to reproduce the exact responses given by specific participants. However, the improvement over an empty persona is arguably negligible. Second, digital twins significantly improve our ability to capture heterogeneity across participants and predict relative differences between them, that is, it enhances our ability to distinguish one participant from another. However, the correlation between answers from digital twins and their human counterparts is modest. Together, these findings establish an empirical benchmark for where digital twin technology stands today, revealing that although they show promise, digital twins fall well short of their promise when placed in a rigorous and ecologically valid test.

\subsubsection*{Digital Twin Performance Across Implementation Methods}

All results reported so far use GPT4.1 (dated 2025-04-14) with a default temperature of 0.7. One might wonder whether poor digital twin performance was in part due the specific configurations we employed rather than fundamental limitations of current digital twin technology. To explore the role of these factors, we systematically varied the temperature and base LLM (GPT-5, Deepseek, Gemini --- including the latest Gemini-3 Pro model --- and a version of GPT4.1 fine-tuned on the Twin-2K-500 dataset). We find the best results overall using GPT4.1 with temperature = 0 (e.g., highest correlation of 0.232 across all implementation options, tied for best on individual-level accuracy at 0.752). We also experimented with replacing the full persona information with a concise (approximately 13K characters), statement-based summary of the questions and responses with distributional information (e.g., Big 5 personality scores with percentile ranks rather than 44 detailed questions and answers on which the scores are based). We find that this simpler version performs very similarly to the full persona, offering a viable lower-cost alternative. See SI for details.

We also explored using an LLM fine-tuned specifically for human behavioral prediction as the base model for the creation of digital twins \cite{binz2025foundation, kolluri2025finetuning}. These fine-tuned models, trained on large cross-sectional datasets, may indeed serve as alternative base LLMs, i.e., they can be combined with additional panel data in which participants are tracked across multiple experiments or studies, allowing the creation of twins associated with specific individuals. We explored using Centaur \cite{binz2025foundation} as a base model. Performance was lower than that achieved using GPT as the base model. The relatively low performance may be due to limitations of the Llama base model on which Centaur is built, or to our formatting and out-of-distribution testing, among other reasons. We leave a thorough investigation of the training of digital twins on such enhanced LLMs for future work. See SI for details.

Overall, twin performance across temperatures, base models, and persona formatting was, at best, modest.

\subsubsection*{Digital Twin Performance Across Domains}

The breadth of our study allows exploring the performance of digital twins across domains, using a meta-analytic approach. We conduct regressions with each performance metric as the dependent variable. Given that correlation is the performance metric most sensitive to the individual-level information captured by digital twins, we focus on the regression with the (z-transformed) correlation for each outcome as the main dependent variable.\footnote{We z-transform the correlation due to better statistical properties, e.g., being approximately normally distributed.} Our independent variables include a set of categorical labels that characterize each outcome, related to the following domains: social, preferences/attitudes, cognitive skills/rationality, content evaluation, human-tech interactions. We also include mechanical features that might influence performance (e.g., sample size).

We estimate a mixed linear model, including random intercepts for each sub-study (see SI for details). We find that correlation was on average higher in the cognitive domain, on outcomes related to human-technology interactions, and on outcomes that use response scales. Digital twins also performed relatively better overall in the social domain. The correlation was higher when outcomes related to conflict, pro-social issues, social cognition, or personality. However, the correlation was significantly lower for outcomes where social desirability was salient, suggesting twins are less capable of mimicking human responses in socially-sensitive contexts (descriptive evidence suggests twins are more likely to provide socially-desirable responses). Correlations were also lower in the political domain (consistent with previous findings such as \cite{santurkar2023whose} or \cite{motoki2024more}), on outcomes with valenced (positive versus negative) evaluations, and when questions varied across participants (consistent with prior findings that LLMs struggle to treat prompt-level variation as exogenous \cite{gui2023challenge}).

\subsection*{Five Key Distortions}

A natural objection to benchmarking digital twin performance today is that the technology will undoubtedly improve as LLMs continue to evolve, making current performance largely beside the point. However, meaningful improvement requires both an understanding of \emph{how} digital twins deviate from humans and a clear rubric of indicators against which progress can be measured. To that end, we next characterize systematic distortions in digital twins' representations of human behavior that may underlie their poor performance. These distortions are consistent with previous literature on synthetic data, suggesting that digital twins in their current form are not able to solve known limitations of synthetic data. Of course, we cannot claim that our list is exhaustive, but addressing these distortions is a necessary, if not sufficient, condition for improving the validity of digital twins.

\subsubsection*{1. Insufficient Individuation}

Adopting a Bayesian framework, one may think of the base LLM as reflecting a \emph{prior} distribution over how an individual may respond to a particular question. The additional information fed to the LLM to build that person's digital twin leads to an updated, \emph{posterior} distribution of answers tailored to that individual. Previous work on other forms of synthetic data found that LLM-generated responses often suffer from excessive homogeneity \cite{wang2025large}. \cite{santurkar2023whose} argue that LLMs are steerable through customized prompting, but that steerability does not guarantee alignment with human responses.

By augmenting the base LLM with extensive individual-level data, digital twins offer the potential to address this issue. To empirically assess whether this is the case, we first compare the amount of variation in answers from twins vs. humans. The standard deviation of the twin responses is lower than that of human responses in 154 of 164 cases (93.9\%), indicating under-dispersion in twin responses. This difference is statistically significant ($p<0.05/164$ --- applying Bonferroni correction) in 140 of those 154 cases.

The under-dispersion of digital twin answers relative to their human counterparts suggests that while the additional data is able to steer the distribution in an appropriate direction (i.e., the variations across digital twins better mirror the variations across people), the base model (i.e., the prior) still carries significant weight and influence on the digital twins' responses. In other words, the answers are overly ``shrunk'' towards a base model. This suggests that providing extensive individualized data does not necessarily address the homogeneity bias in LLMs.

To further quantify this, we compare the MAD between answers from full-persona twins vs. humans to the MAD between answers from full-persona twins vs. empty-persona twins (which only reflect the base LLM). Ideally, full-persona twins would behave more similarly to humans than to empty-persona twins. We find the opposite: full-persona twins are closer to empty-persona twins than they are to humans ($MAD_{full \ vs. \ empty}=0.175$, $MAD_{full \ vs. \ humans}=0.252$, $p<0.01$).

In sum, we find that digital twins distort human behavior by insufficiently deviating from the base model. This yields the first benchmark for progress: comparing standard deviation of twin responses to human responses. If full-persona twins approximate the standard deviation of human responses, this suggests that twins have overcome the homogenization that plagues LLMs.

\subsubsection*{2. Stereotyping}

We have shown that digital twins do not deviate sufficiently from the base LLM. To the extent they deviate, do they do so in a way that over-relies on generic characteristics such as demographics? Previous research has documented that LLMs prompted to mimic specific demographic groups tend to provide answers that are stereotypical of these groups \cite{gupta2023bias,wang2025large}. Here again, we explore whether augmenting persona description with rich individual-level data is enough to correct such phenomenon.

The fact that full-persona twins achieve only slightly higher individual-level accuracy than those based only on demographic information provides initial evidence for demographic-based stereotyping. However, to test this more directly, we compute the mean absolute deviation (MAD) between the answers (normalized between 0 and 1) given by full-persona twins vs. demographics-only twins. The average MAD across outcomes is 0.132, which is significantly lower than the average MAD we reported above between full-persona twins and empty-persona twins ($MAD_{full \ vs. \ demographics}=0.132$, $MAD_{full \ vs. \ empty}=0.175$, $p<0.01$) or the actual human answers ($MAD_{full \ vs. \ demographics}=0.132$, $MAD_{full \ vs. \ humans}=0.252$, $p<0.01$). That is, answers from full-persona twins are closer to those from demographics-only twins than to those from empty-persona twins or real humans. This suggests that digital twins rely heavily on demographic characteristics rather than the wealth of additional individual-level information they were provided.

As discussed earlier, correlation on the other hand improves more substantially when using full vs. demographics-only personas. This suggests that, to the extent full-persona twins deviate from demographics-only twins, they do so in a way that is somewhat consistent with variations across humans.

\begin{figure}[h!]
    \centering
    \includegraphics[width=0.48\textwidth]{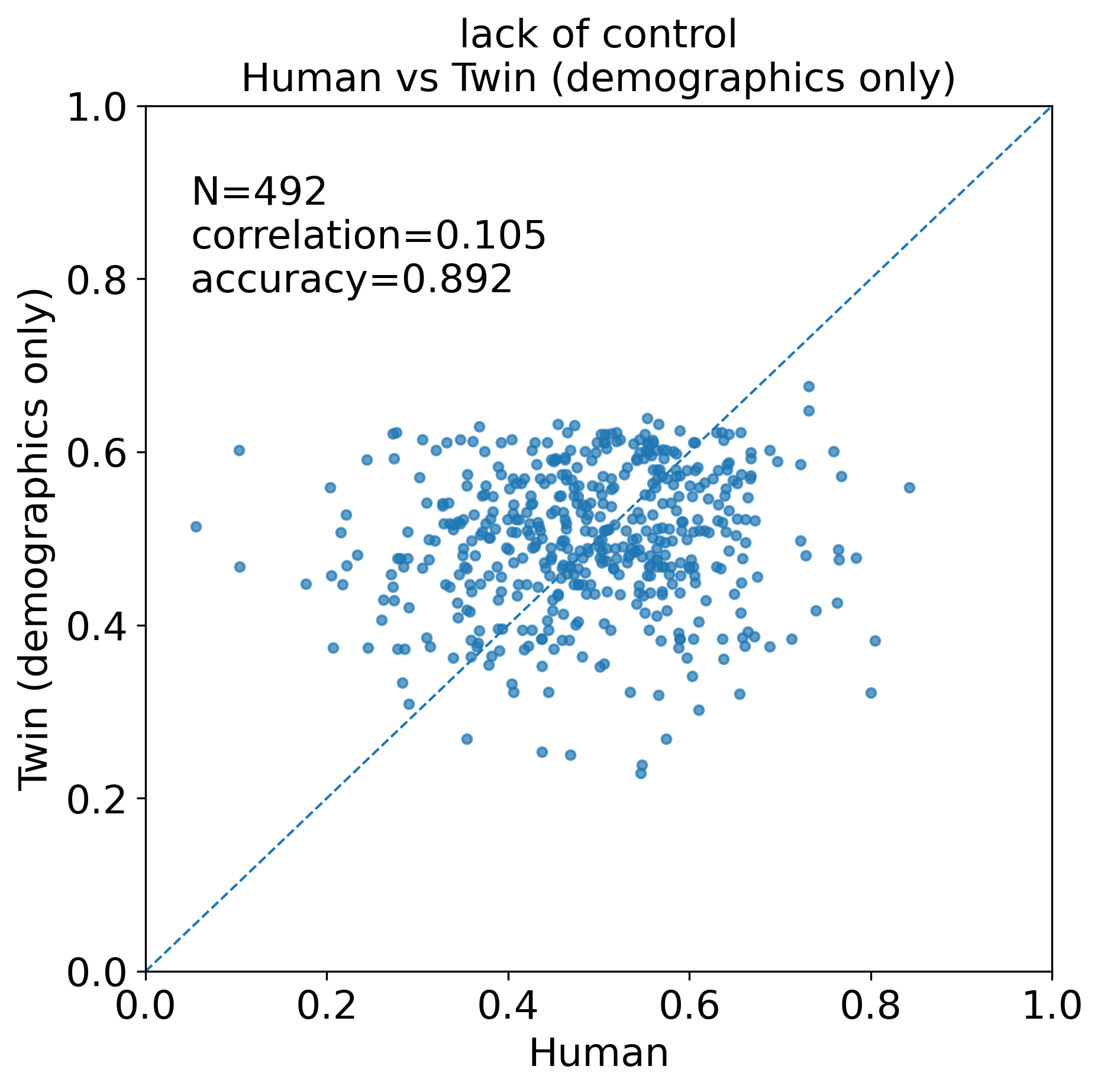}
    \includegraphics[width=0.48\textwidth]{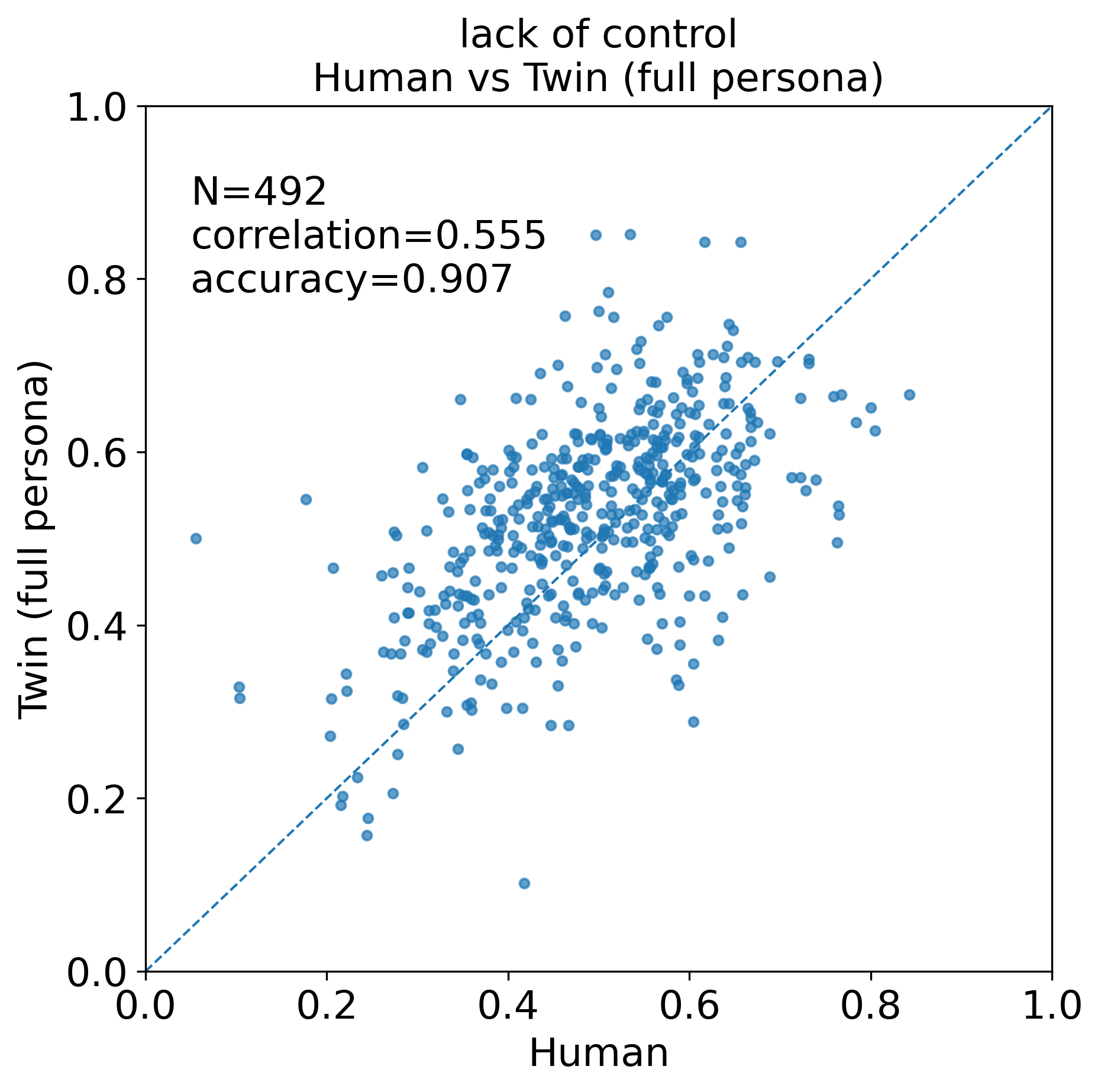}
    \caption{\textbf{Human Responses versus Digital Twin Responses Based on Demographics Only (Left) and Full Persona (Right), for One Particular Outcome.} The 45\textdegree \ line is included. This scatter plot illustrates how correlation may be improved when full personas are used, without significant change to individual-level accuracy (captured by the average distance of each point to the 45\textdegree \ line).}
    \label{fig:compare}
\end{figure}

It may be counter-intuitive that full personas significantly improve correlation compared to demographics-based personas, without improving accuracy. However, accuracy and correlation are distinct constructs. To illustrate the distinction between improving correlation and improving individual-level accuracy, we select one outcome (Lack of Control) from one sub-study (Substudy 2, Affective Primes), for which the distinction between individual-level accuracy and correlation is particularly sharp. Fig.~\ref{fig:compare} reports scatter plots of the predictions from synthetic personas created using demographics only versus human responses (left panel), and predictions from digital twins created using full personas versus human responses (right panel). These plots clearly illustrate how the correlation is much improved when full personas are used ($r_{demo}=0.105$ versus $r_{full}=.555$). Yet, individual-level accuracy (captured by the average distance of each point to the 45\textdegree \ line) is virtually unchanged ($Accuracy_{demo}=0.892$ versus $Accuracy_{full}=0.907$).

To illustrate this comparison more concretely, consider two people with similar demographics. One person rates themselves a 2 out of 7 and the other rates themselves a 4 out of 7 on feeling a lack of control. Using demographics only, the twins might report 3 for each, 1 point away from the human's actual response on average. Using the full persona, the twins might now report 3 and 5 respectively, still 1 point off on average, failing to capture the individual's subjective experience to a similar extent, but now capturing the relative difference across people. This suggests that personalized twins can sort people more accurately (who feels more or less control) but still fail to capture individuals' actual cognition and subjective experiences. Note that individual-level accuracy should not be confused with other \emph{aggregate} measures of accuracy, such as the accuracy of the average difference between groups of participants (e.g., average treatment effect). Such measures may be more strongly impacted by improvements in correlation.

In sum, one way digital twins distort human behavior is by over-relying on demographic information rather than modeling the complexity of individual cognition. This yields the second benchmark for progress: comparing responses of full-persona twins to those of (1) demographic personas, (2) empty personas, and (3) humans. If full-persona twins are closer to demographic personas than humans, that suggests an over-reliance on demographics at the expense of true individuation.

\subsubsection*{3. Representation Bias}

We have so far documented two distortions that can explain inaccuracy. But is this inaccuracy uniform across the humans that digital twins are attempting to represent? As training data are typically based on corpora that systematically over-represent certain demographics \cite{dodge2021documenting}, previous research has shown that LLMs struggle to reflect the opinions of certain demographic groups \cite{zewail2026moral,santurkar2023whose}. Does creating digital twins using demographic data combined with extensive individual-level data address this issue? To explore this, we compute accuracy (across all outcomes) between each participant's normalized responses and those of their twin.\footnote{We use accuracy for this analysis as a more natural way to compare the answers between a particular human and their twin across outcomes.} We then analyze how this accuracy relates to participants' demographic characteristics. All demographic variables are dummy-coded, yielding 61 demographic dummy variables coming from the 14 demographic questions. To identify which participant characteristics are most predictive of digital twin performance, we train an XGBoost (Extreme Gradient Boosting) model to predict twin accuracy using these features \cite{chen2016xgboost}.

\begin{figure}[h!]
    \centering
    \includegraphics[width=\textwidth]{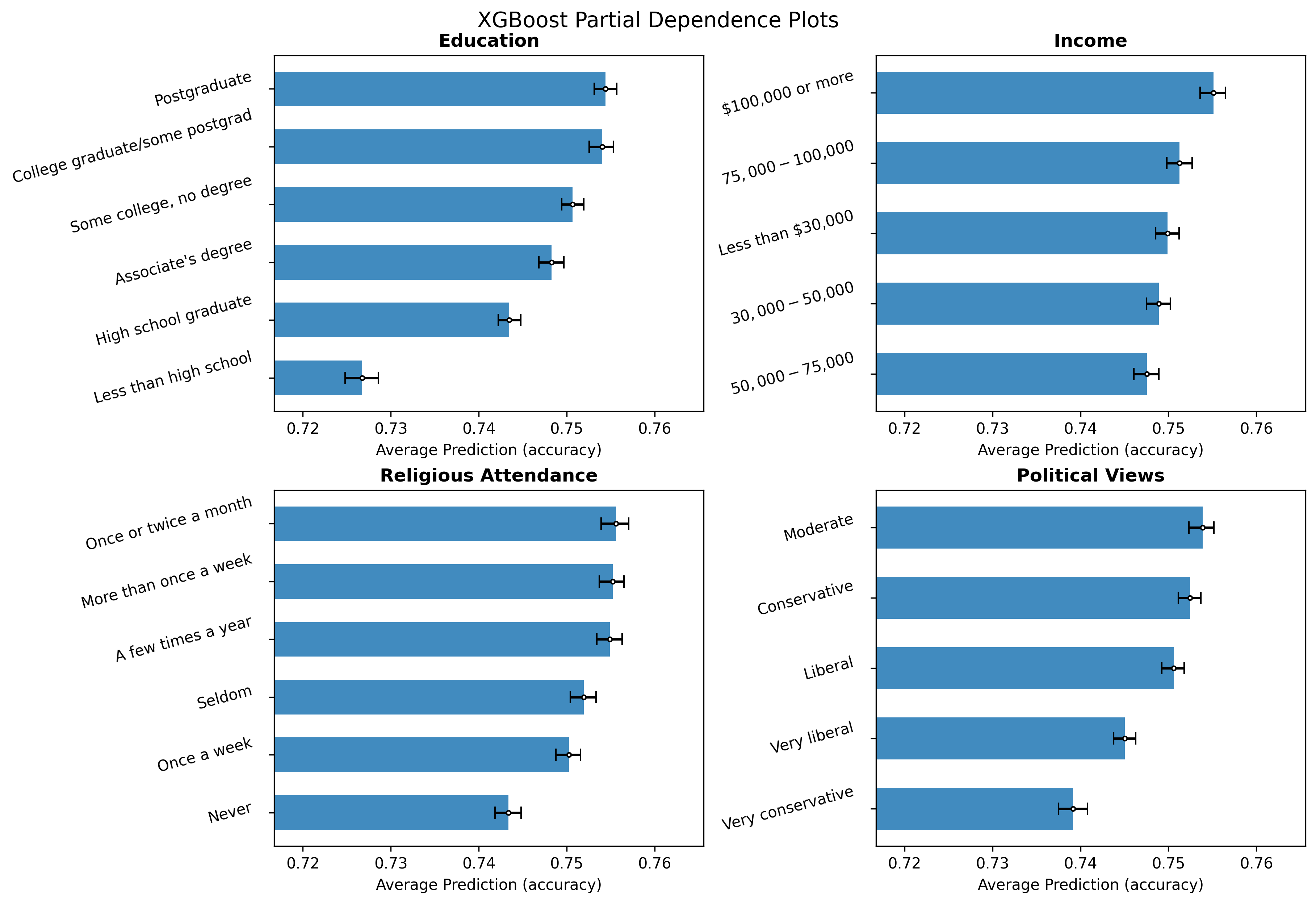}
    \caption{\textbf{Partial Dependence Plots for Understanding Heterogeneity in Digital Twin Performance.} Twins tend to be more accurate for participants with higher education levels and higher income, as well as those with moderate political views and religious attendance habits.}
    \label{fig:charac}
\end{figure}

To interpret the model's findings, we generate grouped partial dependence plots (PDPs) for each categorical variable. PDPs visualize the average predicted accuracy for each level of a feature while holding all other variables constant. For example, the PDP for political ideology shows predicted accuracy for participants identifying as ``Very Conservative,'' ``Liberal,'' ``Moderate,'' ``Liberal,'' and ``Very Liberal'' on a single chart, enabling direct comparison across different demographic groups. Plots for education, income, religious attendance, and political views are presented in Fig.~\ref{fig:charac}, while the remaining plots are in the SI.

Our findings suggest representation bias, i.e., systematic differences in twin accuracy across demographic groups \cite{shahbazi2023representation}. Twins tend to be more accurate for participants with higher education levels and higher income, indicating a potential bias in favor of more socioeconomically advantaged individuals, which could exacerbate a focus on WEIRD (Western, Educated, Industrialized, Rich, Democratic) populations in the social sciences \cite{zewail2026moral,henrich2010weirdest}. Accuracy also tends to be higher for participants with moderate political views and religious attendance habits.

In sum, this differential accuracy constitutes another distortion: digital twins systematically misrepresent certain groups more than others. This yields the third benchmark for progress: visualizing partial dependence plots across segments (such as demographic or psychographic variables). Flat plots suggest that these different groups of people are represented equally well by their digital twins.

\subsubsection*{4. Ideological Biases}

The first three distortions discussed above represent failures to capture individual cognition by over-relying on demographic information, insufficiently adjusting away from the base model, or insufficiently capturing the specific behaviors of certain groups. These can be seen as errors of omission. However, another possible avenue of distortion is the introduction of bias in responses. Our data provide a unique opportunity to contribute to the growing literature on LLM biases by directly comparing opinions expressed by humans to those expressed by an LLM instructed to mimic those humans, across a wide range of substantive areas.

Specifically, in addition to 164 detailed outcomes (e.g., willingness to share a particular article on social media, likelihood of applying for a particular job), we also pre-registered 31 higher-level comparisons between digital twins and humans (1-4 comparisons per sub-study). These comparisons evaluated whether twins responded similarly to humans on average to a particular question or set of questions (e.g., ``How does the knowledge of digital twins regarding fees and surcharges in the marketplace compare to that of their human counterparts?''), whether they replicated known treatment effects (e.g., ``Can digital twins demonstrate default effects?''), etc. A complete list of these comparisons and their results is available in the SI; we only highlight a subset here, which reveal systematic ways in which the content of digital twin answers differ from that of humans. In particular, our diverse mega-study allowed us to uncover subtle ideological biases embedded in the twins' responses.

Specifically, we found that twins tended to express views that were often more ``pro-human'' than those expressed by humans. For example, compared to their human counterparts, twins were more likely to believe that people are fair and can be trusted (Substudy 15), that people should take care of themselves (Substudy 15), and were relatively more favorable to people who donate to both political parties (Substudy 16). They were also more favorable toward government regulation of fees and surcharges in the marketplace (Substudy 7), and more willing to pay taxes to improve healthcare for all people (Substudy 15).

Interestingly, at the same time, twins also tended to express views that could be described as ``pro-technology.'' In particular, twins tended to provide answers that were consistent with a view of technology as a safe tool under the control of humans (e.g., Substudies 6, 9, 14, 19). For example, twins were relatively more accepting of algorithmic hiring compared to their human counterparts (Substudy 9), perceived online targeting as less intrusive on their privacy (Substudy 14), and under-reported usage of platforms like Netflix and TikTok (Substudy 19). While it is possible to be simultaneously ``pro-human'' and ``pro-technology,'' do digital twins show any \emph{preference} in favor of humans versus technology? When asked to rate the creativity of ideas (Substudy 10), twins showed human aversion (rating ideas from humans lower due to their source).

These biases may come from various inputs into the base model. In particular, they may be reflected in the training corpus of the base model, or they may have been acquired through Reinforcement Learning with Human Feedback \cite{santurkar2023whose} or through the base model's system prompt which provides overarching instructions on how the model should behave. Irrespective of their exact sources, they imply that blindly deploying digital twins might lead to distorted conclusions about human behavior and opinions, which may, sometimes in very subtle ways, push specific ideological agendas.

In sum, we find that digital twins distort human behavior by exerting ideological sway. This yields the fourth benchmark for progress: checking for systematic ideological biases (including pro-technology, pro-social, and pro-human biases) by testing whether human-twin divergences are consistently in a specific direction across substantive domains.

\subsubsection*{5. Hyper-rationality}

Another form of bias that may be introduced by digital twins relates to cognitive skills. Specifically, one of the features of LLMs---their vast knowledge base---may be a source of error when it comes to predicting human behavior. Indeed, prior work finds that LLMs suffer from hyper-accuracy \cite{aher2023using,toubia2025database}. Does training digital twins on extensive data, including response from numerous cognitive tests, help replicate human's imperfect knowledge and cognitive skills? Or, akin to the curse of knowledge in humans \cite{camerer1989curse}, are digital twins unable to ``forget'' their base knowledge leading them to behave as more informed, rational agents? We first examined this question in sub-studies that contained questions with objectively correct answers (Substudies 7,19). We find that twins tended to display perfect knowledge, deviating from human participants.

We next examined this distortion for substudies that contained questions that test rationality. We again find evidence of hyper-rationality distortion: twins were more likely to select normative answers reflective of higher cognitive abilities, e.g., by selecting answers associated with stronger quantitative intuition (Substudy 17) or by being more likely to strategically consume online content in order to influence future recommendations (Substudy 19). Twins also failed to replicate the attraction and compromise effects (Substudy 4), which are indicators of bounded rationality. However, an interesting phenomenon emerges when using a well-known, published paradigm. Here, twins were \emph{more} likely to show the effect, probably due to the phenomenon of leakage, whereby the answer is directly influenced by information contained in the training corpus \cite{ludwig2025large}. For example, digital twins did not replicate a default effect using a novel paradigm (Substudy 5), but they did display a strong default effect using a classic paradigm \cite{johnson2003defaults}. Other substudies (e.g., Substudy 1) confirmed the tendency of digital twins to behave consistently with the literature when a clear prediction could be made. This highlights a key contribution of our megastudy: we are the first to systematically test twins across both established and novel paradigms.

In sum, digital twins distort human behavior by imposing a hyper-rational lens. This yields the final benchmark for progress: testing twin performance in \textit{novel} contexts that contain objectively correct answers and test rationality. If twins deviate less from correct responses than humans, and if twins display irrationality at lower rates to humans, this suggests that they are still subject to the hyper-rationality bias in LLMs.

\section*{Discussion}

Although many are rushing to simulate surveys and experiments using digital twins of humans, little is known about how responses from digital twins compare to those of their human counterparts. Further, there is little understanding of the best way to construct digital twins, the domains of application in which digital twins are likely to be accurate, or on the profiles of humans whose digital twins are likely to be more accurate. Deploying digital twins without answering these fundamental questions risks misrepresenting human cognition in ways that could undermine both scientific understanding and practical applications. In order to help advance knowledge on digital twins, we ran a pre-registered mega-study (and make the data publicly available) that: (i) allows matching the answers of digital twins to those of their human counterparts, (ii) covers a wide range of application domains, and (iii) involves a relatively large, representative sample of participants.

We find that although digital twins show promise, they are not yet ``ready for primetime.'' Digital twins capture individual cognition only minimally better than empty personas. However, twins show slightly more promise when it comes to sorting people. This distinction is consequential for understanding when indiscriminate twin use may mislead decision-makers. For example, although twins may be able to modestly estimate relative receptiveness to a policy (identifying who is more or less likely to respond positively), they are barely better at predicting an individual's likelihood to agree with a given policy than a base LLM.

Could human experts have predicted these results? To assess whether empirical tests are valuable and potentially corrective in the domain of digital twins, we asked a sample of 68 expert respondents (primarily scholars and managers) to estimate average outcomes in five of our experimental studies representing different paradigms. We asked separately for the average outcome for humans and their twins in a treatment condition (e.g., likelihood of staying enrolled in a green energy plan when enrolled by default), providing the average outcome in the control condition as baseline (e.g., likelihood of actively choosing the green energy plan when enrolled in another plan by default). Overall, the pattern suggested that respondents' predictions were less accurate for twins than for humans (see SI). Comparing the predicted outcome in the treatment condition to its value in the control condition also yields estimated treatment effects. While the observed average treatment effect among humans was within the 95\% confidence interval of predictions made by our participants in all five cases, the average treatment effect among twins was outside the 95\% confidence interval in three cases. This suggests that human experts struggle to predict the behavior of digital twins, highlighting the importance of our rigorous examination.

Beyond performance metrics, our analysis highlights five systematic ways in which digital twins distort human responses. First, we find that digital twins suffer from \emph{insufficient individuation}, providing answers that remain under-differentiated and under-dispersed relative to human answers. Second, we find evidence of demographics-based \emph{stereotyping}. Third, digital twin predictions suffer from \emph{representation bias}, as they tend to be more accurate for certain demographic groups of participants with higher education levels, higher income, and moderate political views and religious attendance habits. Fourth, digital twins tend to display \emph{ideological biases}, for example, they appear to be at the same time ``pro-human'' and ``pro-technology.'' Fifth, digital twins tend to display \emph{hyper-rationality}, with answers reflecting higher rationality and knowledge compared to humans.

These results culminate in the conclusion that the digital twins of today may best be described as funhouse mirrors that systematically distort human behavior. Stakeholders deploying digital twins risk making consequential decisions based on an illusion of personalization rather than genuine individual-level insight. These systems are better characterized as ``comparative profiles'' imbued with LLM biases, not faithful twins of human cognition. Consequently, the premature deployment of these twins risks smoothing out individual voices, mispredicting ignorance and irrationality, reinforcing stereotypes, and introducing novel biases, such as pro-technology attitudes and AI-favoritism.

\section*{Future Directions}

The five types of distortions we document are consistent with previous research on synthetic data, suggesting that despite their promise, digital twins are currently unable to address limitations of synthetic data. By empirically consolidating findings scattered across the literature into a list of distortions in a single mega-study, we offer a unified evaluative framework: a roadmap with clear guidance on specific ways in which the behavior of digital twins should be corrected before this technology may be deployed safely, with a clear set of indicators on which future methodologies can be measured.

How might future research address these current distortions? Recall that leveraging digital twins involves making out-of-distribution predictions using a base LLM augmented with individual-level data (the twin's ``blueprint''). We compare digital twins to a traditional machine learning approach in order explore the extent to which the modest performance of digital twins (and therefore, future potential improvements) lie in the blueprint data being insufficiently informative vs. the LLM being unable to fully leverage these data for out-of-distribution predictions. Unlike digital twins, which are able to make out-of-distribution predictions on completely new questions, traditional machine learning models require collecting responses from a subset of the sample, training a model that connects a set of personal characteristics to these responses, and then predicting out-of-sample responses for the rest of the sample (given their personal characteristics).

We consider a case in which an XGBoost model is trained separately for each outcome based on the value of that outcome for a sub-sample of participants and their full persona information (i.e., the same information used for full-persona digital twins).\footnote{We focus on the 106 outcomes for which sample size was greater than 650, and vary the size of the training sample from 50 to 650. We also set temperature to 0 for digital twins, as this improves their performance (see SI for details).} We also consider an XGBoost model trained on demographic variables only. Comparing the performance of a given model with full-persona information vs. demographics-only information sheds light on the intrinsic value of the twins' blueprint data. Comparing the performance of the XGBoost model to that of digital twins holding the persona information constant sheds light on the base model's ability to leverage individual-level data for out-of-distribution predictions.

\begin{figure}[h!]
    \centering
    \includegraphics[width=\textwidth]{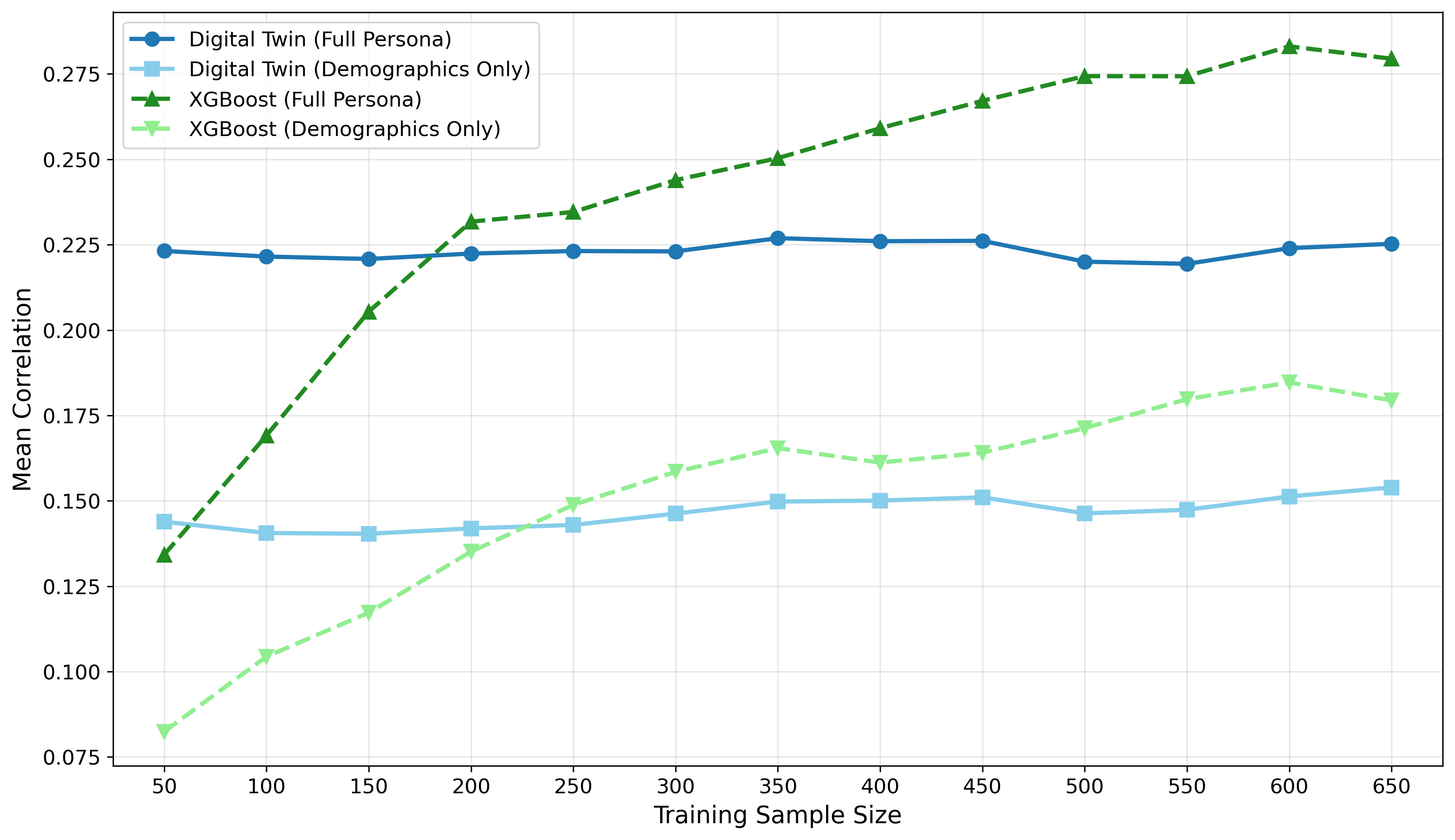}
    \caption{\textbf{Comparison with Traditional Machine Learning Method: Correlation.} We compare digital twins (plain lines) to an XGBoost model (dashed lines) trained for each outcome using additional data not needed or used by digital twins: responses for that outcome from a training subset of participants. We consider versions of each model based on full-persona information (darker colors) vs. demographics only (lighter colors). The full-persona information helps both digital twins and the XGBoost improve performance. However, even as the size of the training sample reaches 650, the predictive correlation achieved by the full-persona XGBoost model remains below 0.29. Holding the set of individual-level characteristics constant, the base LLM is able to make out-of-distribution predictions that achieve the same correlation with human answers as out-of-sample predictions made by an XGBoost model trained on approximately 200 human answers.}
    \label{fig:training-correlation}
\end{figure}

Looking first at correlation (see Fig.~\ref{fig:training-correlation}), comparing each model with full personas vs. demographics-only personas reveals that the full-persona blueprint data helps both digital twins and the XGBoost improve performance. However, even as the size of the training sample reaches 650, the predictive correlation achieved by the full-persona XGBoost model remains below 0.29. This suggests that, despite being extensive and based on decades of social science research, the twins' blueprint data has limited intrinsic potential to capture human behavior in our context. At the same time, we see that full-persona digital twins reach a predictive correlation equivalent to that achieved by the full-persona XGBoost model trained on approximately 180 participants. In the demographics-only case, the number increases to approximately 225 participants. In other words, given a set of individual-level characteristics, the base LLM is able to make out-of-distribution predictions that achieve the same correlation with human answers as out-of-sample predictions made by an XGBoost model trained on approximately 200 human answers (i.e., that require running the actual study on approximately 200 participants).

\begin{figure}[h!]
    \centering
    \includegraphics[width=\textwidth]{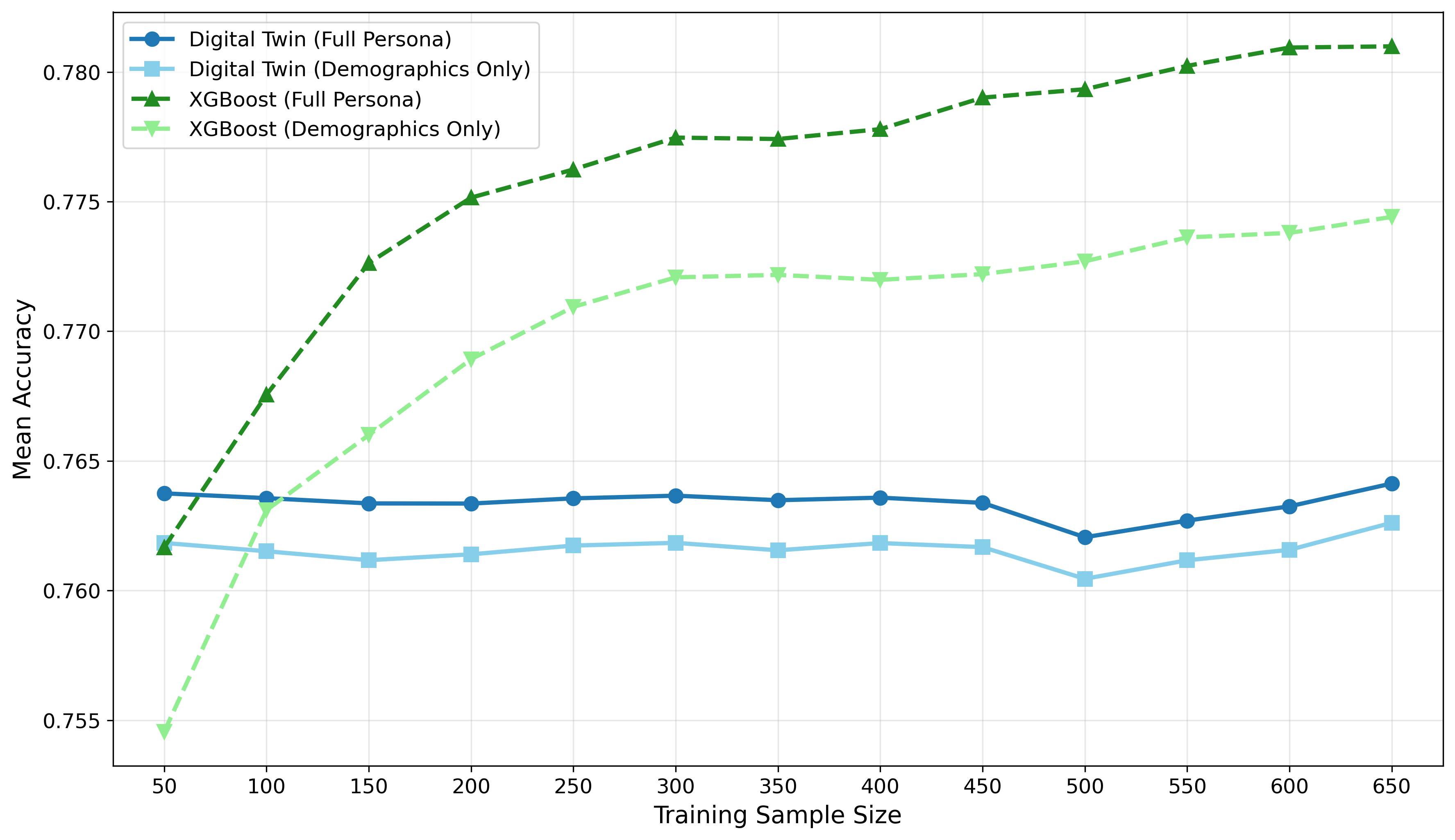}
    \caption{\textbf{Comparison with Traditional Machine Learning Method: Accuracy.} We compare digital twins (plain lines) to an XGBoost model (dashed lines) trained for each outcome using additional data not needed or used by digital twins: responses for that outcome from a training subset of participants. We consider versions of each model based on full-persona information (darker colors) vs. demographics only (lighter colors). The full-persona information only marginally helps improve individual-level accuracy. Moreover, digital twins are only as good as XGBoost trained on approximately 75 human responses on average, suggesting that it is harder for the LLM to leverage these data to make out-of-distribution predictions that would be accurate at the individual level.}
    \label{fig:training-accuracy}
\end{figure}

When it comes to accuracy (Fig.~\ref{fig:training-accuracy}), the XGBoost model trained on the full data is only marginally better than the XGBoost model trained on demographics only (and similarly for digital twins), suggesting that the twins' blueprint data is only marginally informative of the idiosyncratic patterns in responses. Moreover, digital twins are only as good as XGBoost trained on approximately 75 human responses on average, suggesting that it is harder for the LLM to leverage these data to make out-of-distribution predictions that would be accurate at the individual level.

In sum, this analysis suggests that the twins' blueprint data, despite being based on a large number of scales and measures developed over decades of social science research and despite being the most comprehensive dataset for the creation of digital twins publicly available to date, are far from sufficient to fully align digital twins' answers to those of their human counterparts. Predicting human behavior in our context is probably intrinsically challenging and performance expectations need to be appropriately calibrated, and/or that new scales and measures may need to be developed to better capture human cognition. This analysis further suggests that the blueprint data is relatively more informative of variations across people and that LLMs are able to leverage these data to make out-of-distribution predictions that capture \textit{some} variation across people (e.g., whether one person will score higher or lower than another), while not being necessarily more accurate at the individual level (e.g., whether one person would answer a 6 vs. a 7 on a scale).

Similarly, what role does the base LLM versus individual-level information play in shaping the behavior of digital twins and the \emph{distortions} we observe? One may speculate, and test in future research, that several of the distortions we observe are due to properties of the base LLM being transferred to digital twins. For example, ideological biases and hyper-rationality likely stem from the base LLM powering the digital twins. It is similarly quite plausible that representation bias is linked to certain groups (e.g., higher education, moderate views) being better represented in the base model's training data. And the insufficient individuation of answers by digital twins is probably the result of the homogeneity of the base model not being sufficiently offset by the individual-level information contained in digital twins.

Of course, a base model is required to enable digital twins to process and generate unstructured data, and to make out-of-distribution predictions without additional training data. But future research might find optimal ways to combine the capabilities of one (or potentially an ensemble of) base LLM(s) with rich individual-level information. We hope our data will encourage the development of improved pipelines for digital twins that address some of the limitations uncovered in our analysis.

Finally, our findings also raise the question of how we should reasonably think about digital twins. In particular, based on our results, it may not be realistic to think about them as ``clones'' of humans, but rather as hyper-rational, quasi-omniscient versions of humans, with implicit values partly imbued by their base LLM. This perspective may influence which use cases are more promising for future research and practical applications. Perhaps use cases that consider digital twins as well-informed advisors are more promising than use cases that consider digital twins as carbon copies of their human counterparts.

In sum, digital twins hold genuine promise in revolutionizing how we study human behavior. But realizing this promise requires rigorous empirical foundations. We hope that our evaluative framework, publicly available data and code, and detailed analysis of each study (see SI), provide the data-driven scaffolding needed to reach digital twins' full potential.

\section*{Materials and Methods}

Each sub-study was first run on Prolific according to its pre-registered plan (\href{https://researchbox.org/4145}{https://researchbox.org/4145}). All sub-studies were run from the same Prolific account to ensure consistency. Invitations were sent to Prolific participants from the Twin-2K-500 panel. Each sub-study was approved by Columbia University's IRB. All sub-studies were run from April to June 2025.

Next, each sub-study was conducted on the digital twins of its participants. To set up the simulation, we extracted twin information from the Twin-2K-500 dataset and provided it as prompt input. For each human participant, the completed survey was transformed into text, stripped of its answers, and incorporated into the user prompt. Survey questions varied across participants due to randomization; we preserved these differences when generating inputs for their twins. The LLM was then tasked with producing responses to these surveys. The creation of digital twins and the related computations were approved by Columbia University's IRB under Protocol IRB-AAAV5832.

For each persona in each sub-study, we constructed a user prompt using a template (see SI) and queried the LLM via its API (OpenAI API for OpenAI models, OpenRouter API for others). The JSON outputs were then post-processed to ensure all questions were answered in the required format (e.g., no missing values or out-of-range responses). If validation failed, the API call was retried. A complete version of the prompt template, including placeholders for persona profiles, new questions, and JSON-based response formatting, is provided in the SI.

\bibliographystyle{sciencemag}
\bibliography{DigitalTwinFunhouseMirrors_v2_SciAdv_arxiv}

\begin{thebibliography}{10}
\providecommand{\url}[1]{\texttt{#1}}
\expandafter\ifx\csname urlstyle\endcsname\relax
  \providecommand{\doi}[1]{doi:\discretionary{}{}{}#1}\else
  \providecommand{\doi}{doi:\discretionary{}{}{}\begingroup
  \urlstyle{rm}\Url}\fi

\bibitem{argyle2023out}
L.~P. Argyle, \emph{et~al.}, Out of one, many: Using language models to
  simulate human samples. \emph{Political Analysis} \textbf{31}~(3), 337--351
  (2023).

\bibitem{dillion2023can}
D.~Dillion, N.~Tandon, Y.~Gu, K.~Gray, Can AI language models replace human
  participants? \emph{Trends in Cognitive Sciences} \textbf{27}~(7), 597--600
  (2023).

\bibitem{horton2023large}
J.~J. Horton, \emph{Large language models as simulated economic agents: What
  can we learn from homo silicus?}, Tech. rep., National Bureau of Economic
  Research (2023).

\bibitem{hewitt2024predicting}
L.~Hewitt, A.~Ashokkumar, I.~Ghezae, R.~Willer, Predicting results of social
  science experiments using large language models. \emph{Preprint}  (2024).

\bibitem{jones2025large}
C.~R. Jones, B.~K. Bergen, Large language models pass the turing test.
  \emph{arXiv preprint arXiv:2503.23674}  (2025).

\bibitem{westwood2025potential}
S.~J. Westwood, The potential existential threat of large language models to
  online survey research. \emph{Proceedings of the National Academy of
  Sciences} \textbf{122}~(47), e2518075122 (2025).

\bibitem{cui2025large}
Z.~Cui, N.~Li, H.~Zhou, A large-scale replication of scenario-based experiments
  in psychology and management using large language models. \emph{Nature
  Computational Science} pp. 1--8 (2025).

\bibitem{broska2025mixed}
D.~Broska, M.~Howes, A.~van Loon, The mixed subjects design: Treating large
  language models as potentially informative observations. \emph{Sociological
  Methods \& Research} \textbf{54}~(3), 1074--1109 (2025).

\bibitem{schroder2025large}
S.~Schr{\"o}der, T.~Morgenroth, U.~Kuhl, V.~Vaquet, B.~Paa{\ss}en, Large
  language models do not simulate human psychology. \emph{arXiv preprint
  arXiv:2508.06950}  (2025).

\bibitem{gao2025take}
Y.~Gao, D.~Lee, G.~Burtch, S.~Fazelpour, Take caution in using LLMs as human
  surrogates. \emph{Proceedings of the National Academy of Sciences}
  \textbf{122}~(24), e2501660122 (2025).

\bibitem{bisbee2024synthetic}
J.~Bisbee, J.~D. Clinton, C.~Dorff, B.~Kenkel, J.~M. Larson, Synthetic
  replacements for human survey data? the perils of large language models.
  \emph{Political Analysis} \textbf{32}~(4), 401--416 (2024).

\bibitem{lin2025six}
Z.~Lin, Six fallacies in substituting large language models for human
  participants. \emph{Advances in Methods and Practices in Psychological
  Science} \textbf{8}~(3), 25152459251357566 (2025).

\bibitem{santurkar2023whose}
S.~Santurkar, \emph{et~al.}, Whose opinions do language models reflect?, in
  \emph{Proceedings of the 40th International Conference on Machine Learning}
  (PMLR) (2023), pp. 29971--30004.

\bibitem{motoki2024more}
F.~Motoki, V.~Pinho~Neto, V.~Rodrigues, More human than human: measuring
  ChatGPT political bias. \emph{Public Choice} \textbf{198}~(1), 3--23 (2024).

\bibitem{li2025llm}
A.~Li, H.~Chen, H.~Namkoong, T.~Peng, LLM Generated Persona is a Promise with a
  Catch. \emph{arXiv preprint arXiv:2503.16527}  (2025).

\bibitem{wang2025large}
A.~Wang, J.~Morgenstern, J.~P. Dickerson, Large language models that replace
  human participants can harmfully misportray and flatten identity groups.
  \emph{Nature Machine Intelligence} \textbf{7}~(3), 400--411 (2025).

\bibitem{park2024generative}
J.~S. Park, \emph{et~al.}, Generative agent simulations of 1,000 people.
  \emph{arXiv preprint arXiv:2411.10109}  (2024).

\bibitem{toubia2025database}
O.~Toubia, \emph{et~al.}, Database Report: Twin-2K-500: A Data Set for Building
  Digital Twins of over 2,000 People Based on Their Answers to over 500
  Questions. \emph{Marketing Science}  (2025).

\bibitem{ludwig2025large}
J.~Ludwig, S.~Mullainathan, A.~Rambachan, \emph{Large language models: An
  applied econometric framework}, Tech. rep., National Bureau of Economic
  Research (2025).

\bibitem{kolluri2025finetuning}
A.~Kolluri, S.~Wu, J.~S. Park, M.~S. Bernstein, Finetuning LLMs for Human
  Behavior Prediction in Social Science Experiments. \emph{arXiv preprint
  arXiv:2509.05830}  (2025).

\bibitem{beauchamp2011sources}
J.~P. Beauchamp, D.~Cesarini, M.~Johannesson, E.~Lindqvist, C.~Apicella, On the
  sources of the height--intelligence correlation: New insights from a
  bivariate ACE model with assortative mating. \emph{Behavior genetics}
  \textbf{41}~(2), 242--252 (2011).

\bibitem{binz2025foundation}
M.~Binz, \emph{et~al.}, A foundation model to predict and capture human
  cognition. \emph{Nature} pp. 1--8 (2025).

\bibitem{gui2023challenge}
G.~Gui, O.~Toubia, The challenge of using llms to simulate human behavior: A
  causal inference perspective. \emph{arXiv preprint arXiv:2312.15524}  (2025).

\bibitem{gupta2023bias}
S.~Gupta, \emph{et~al.}, Bias runs deep: Implicit reasoning biases in
  persona-assigned llms. \emph{arXiv preprint arXiv:2311.04892}  (2023).

\bibitem{dodge2021documenting}
J.~Dodge, \emph{et~al.}, Documenting large webtext corpora: A case study on the
  colossal clean crawled corpus. \emph{arXiv preprint arXiv:2104.08758}
  (2021).

\bibitem{zewail2026moral}
A.~Zewail, A.~Figueroa, J.~Graham, M.~Atari, Moral stereotyping in large
  language models. \emph{Proceedings of the National Academy of Sciences}
  \textbf{123}~(10), e2519941123 (2026).

\bibitem{chen2016xgboost}
T.~Chen, C.~Guestrin, Xgboost: A scalable tree boosting system, in
  \emph{Proceedings of the 22nd acm sigkdd international conference on
  knowledge discovery and data mining} (2016), pp. 785--794.

\bibitem{shahbazi2023representation}
N.~Shahbazi, Y.~Lin, A.~Asudeh, H.~Jagadish, Representation bias in data: A
  survey on identification and resolution techniques. \emph{ACM Computing
  Surveys} \textbf{55}~(13s), 1--39 (2023).

\bibitem{henrich2010weirdest}
J.~Henrich, S.~J. Heine, A.~Norenzayan, The weirdest people in the world?
  \emph{Behavioral and brain sciences} \textbf{33}~(2-3), 61--83 (2010).

\bibitem{aher2023using}
G.~V. Aher, R.~I. Arriaga, A.~T. Kalai, Using large language models to simulate
  multiple humans and replicate human subject studies, in \emph{International
  conference on machine learning} (PMLR) (2023), pp. 337--371.

\bibitem{camerer1989curse}
C.~Camerer, G.~Loewenstein, M.~Weber, The curse of knowledge in economic
  settings: An experimental analysis. \emph{Journal of political Economy}
  \textbf{97}~(5), 1232--1254 (1989).

\bibitem{johnson2003defaults}
E.~J. Johnson, D.~Goldstein, Do Defaults Save Lives? \emph{Science}
  \textbf{302}~(5649), 1338--1339 (2003), \doi{10.1126/science.1091721},
  \url{https://doi.org/10.1126/science.1091721}.

\bibitem{huber1982adding}
J.~Huber, J.~W. Payne, C.~Puto, Adding Asymmetrically Dominated Alternatives:
  Violations of Regularity and the Similarity Hypothesis. \emph{Journal of
  Consumer Research} \textbf{9}~(1), 90--98 (1982), \doi{10.1086/208899},
  \url{https://doi.org/10.1086/208899}.

\bibitem{simonson1989choice}
I.~Simonson, Choice Based on Reasons: The Case of Attraction and Compromise
  Effects. \emph{Journal of Consumer Research} \textbf{16}~(2), 158--174
  (1989), \doi{10.1086/209205}, \url{https://doi.org/10.1086/209205}.

\bibitem{pichert2008green}
D.~Pichert, K.~V. Katsikopoulos, Green Defaults: Information Presentation and
  Pro-Environmental Behaviour. \emph{Journal of Environmental Psychology}
  \textbf{28}~(1), 63--73 (2008), \doi{10.1016/j.jenvp.2007.09.004},
  \url{https://doi.org/10.1016/j.jenvp.2007.09.004}.

\bibitem{brinol2012self}
P.~Bri{\~n}ol, M.~J. McCaslin, R.~E. Petty, Self-Generated Persuasion: Effects
  of the Target and Direction of Arguments. \emph{Journal of Personality and
  Social Psychology} \textbf{102}~(5), 925--940 (2012), epub 2012 Feb 20,
  \doi{10.1037/a0027231}, \url{https://doi.org/10.1037/a0027231}.

\bibitem{pennycook2021psychology}
G.~Pennycook, D.~G. Rand, The Psychology of Fake News. \emph{Trends in
  Cognitive Sciences} \textbf{25}~(5), 388--402 (2021),
  \doi{10.1016/j.tics.2021.02.007},
  \url{https://doi.org/10.1016/j.tics.2021.02.007}.

\bibitem{pennycook2021accuracy}
G.~Pennycook, \emph{et~al.}, Shifting Attention to Accuracy Can Reduce
  Misinformation Online. \emph{Nature} \textbf{592}, 590--595 (2021),
  \doi{10.1038/s41586-021-03344-2},
  \url{https://doi.org/10.1038/s41586-021-03344-2}.

\bibitem{lane2025misinformation}
K.~Lane, M.~Brucks, A Marketing Perspective on Misinformation Sharing: How the
  Target Audience Shapes Consumers' Focus on Accuracy vs. Entertainment (2025),
  working paper.

\bibitem{harcup2016news}
T.~Harcup, D.~O’Neill, What Is News? News Values Revisited (Again).
  \emph{Journalism Studies} \textbf{18}~(12), 1470--1488 (2016),
  \doi{10.1080/1461670X.2016.1150193},
  \url{https://doi.org/10.1080/1461670X.2016.1150193}.

\bibitem{reuters2023digitalnews}
{Reuters Institute}, \emph{Digital News Report 2023}, Tech. rep., University of
  Oxford (2023),
  \url{https://reutersinstitute.politics.ox.ac.uk/sites/default/files/2023-06/Digital_News_Report_2023.pdf},
  retrieved from Reuters Institute for the Study of Journalism.

\bibitem{widjaya2024tiktok}
R.~Widjaya, S.~Bestvater, A.~Smith, \emph{Who U.S. Adults Follow on TikTok},
  Tech. rep., Pew Research Center (2024),
  \url{https://www.pewresearch.org/data-labs/2024/10/08/who-u-s-adults-follow-on-tiktok/}.

\bibitem{berger2011wom}
J.~Berger, E.~M. Schwartz, What Drives Immediate and Ongoing Word of Mouth?
  \emph{Journal of Marketing Research} \textbf{48}~(5), 869--880 (2011),
  \doi{10.1509/jmkr.48.5.869}, \url{https://doi.org/10.1509/jmkr.48.5.869}.

\bibitem{berger2012viral}
J.~Berger, K.~L. Milkman, What Makes Online Content Viral? \emph{Journal of
  Marketing Research} \textbf{49}~(2), 192--205 (2012),
  \doi{10.1509/jmr.10.0353}, \url{https://doi.org/10.1509/jmr.10.0353}.

\bibitem{berger2011arousal}
J.~Berger, Arousal Increases Social Transmission of Information.
  \emph{Psychological Science} \textbf{22}~(7), 891--893 (2011),
  \doi{10.1177/0956797611413294},
  \url{https://doi.org/10.1177/0956797611413294}.

\bibitem{emmons2003gratitude}
R.~A. Emmons, M.~E. McCullough, Counting Blessings Versus Burdens: An
  Experimental Investigation of Gratitude and Subjective Well-Being in Daily
  Life. \emph{Journal of Personality and Social Psychology} \textbf{84}~(2),
  377--389 (2003), \doi{10.1037//0022-3514.84.2.377},
  \url{https://doi.org/10.1037//0022-3514.84.2.377}.

\bibitem{desteno2014gratitude}
D.~DeSteno, Y.~Li, L.~Dickens, J.~S. Lerner, Gratitude: A Tool for Reducing
  Economic Impatience. \emph{Psychological Science} \textbf{25}~(6), 1262--1267
  (2014), \doi{10.1177/0956797614529979},
  \url{https://doi.org/10.1177/0956797614529979}.

\bibitem{oguni2024gratitude}
R.~Oguni, C.~Ishii, Gratitude Promotes Prosocial Behavior Even in Uncertain
  Situation. \emph{Scientific Reports} \textbf{14}, 14379 (2024),
  \doi{10.1038/s41598-024-65460-z},
  \url{https://doi.org/10.1038/s41598-024-65460-z}.

\bibitem{bukowski2024control}
M.~Bukowski, A.~Potoczek, K.~Barzykowski, {others}, What Do We Manipulate When
  Reminding People of (Not) Having Control? In Search of Construct Validity.
  \emph{Behavior Research Methods} \textbf{56}, 3706--3724 (2024),
  \doi{10.3758/s13428-023-02326-8},
  \url{https://doi.org/10.3758/s13428-023-02326-8}.

\bibitem{chen2017control}
C.~Y. Chen, L.~Lee, A.~J. Yap, Control Deprivation Motivates Acquisition of
  Utilitarian Products. \emph{Journal of Consumer Research} \textbf{43}~(6),
  1031--1047 (2017), \doi{10.1093/jcr/ucw068},
  \url{https://doi.org/10.1093/jcr/ucw068}.

\bibitem{lembregts2019numbers}
C.~Lembregts, M.~Pandelaere, Falling Back on Numbers: When Preference for
  Numerical Product Information Increases after a Personal Control Threat.
  \emph{Journal of Marketing Research} \textbf{56}~(1), 104--122 (2018),
  \doi{10.1177/0022243718820570},
  \url{https://doi.org/10.1177/0022243718820570}.

\bibitem{whitson2008control}
J.~A. Whitson, A.~D. Galinsky, Lacking Control Increases Illusory Pattern
  Perception. \emph{Science} \textbf{322}~(5898), 115--117 (2008),
  \doi{10.1126/science.1159845}, \url{https://doi.org/10.1126/science.1159845}.

\bibitem{schnall2010elevation}
S.~Schnall, J.~Roper, D.~M.~T. Fessler, Elevation Leads to Altruistic Behavior.
  \emph{Psychological Science} \textbf{21}~(3), 315--320 (2010),
  \doi{10.1177/0956797609359882},
  \url{https://doi.org/10.1177/0956797609359882}.

\bibitem{walsh2022gratitude}
L.~C. Walsh, C.~N. Armenta, G.~Itzchakov, M.~M. Fritz, S.~Lyubomirsky, More
  Than Merely Positive: The Immediate Affective and Motivational Consequences
  of Gratitude. \emph{Sustainability} \textbf{14}~(14), 8679 (2022),
  \doi{10.3390/su14148679}, \url{https://doi.org/10.3390/su14148679}.

\bibitem{oliveira2021gratitude}
R.~Oliveira, A.~Bald{\'e}, M.~Madeira, T.~Ribeiro, P.~Arriaga, The Impact of
  Writing About Gratitude on the Intention to Engage in Prosocial Behaviors
  During the COVID-19 Outbreak. \emph{Frontiers in Psychology} \textbf{12},
  588691 (2021), \doi{10.3389/fpsyg.2021.588691},
  \url{https://www.frontiersin.org/journals/psychology/articles/10.3389/fpsyg.2021.588691}.

\bibitem{greenaway2015control}
K.~H. Greenaway, \emph{et~al.}, From ``We'' to ``Me'': Group Identification
  Enhances Perceived Personal Control with Consequences for Health and
  Well-Being. \emph{Journal of Personality and Social Psychology}
  \textbf{109}~(1), 53--74 (2015), \doi{10.1037/pspi0000019},
  \url{https://doi.org/10.1037/pspi0000019}.

\bibitem{sedek1990helplessness}
G.~Sedek, M.~Kofta, When Cognitive Exertion Does Not Yield Cognitive Gain:
  Toward an Informational Explanation of Learned Helplessness. \emph{Journal of
  Personality and Social Psychology} \textbf{58}~(4), 729--743 (1990),
  \doi{10.1037//0022-3514.58.4.729},
  \url{https://doi.org/10.1037//0022-3514.58.4.729}.

\bibitem{webster1994closure}
D.~M. Webster, A.~W. Kruglanski, Individual Differences in Need for Cognitive
  Closure. \emph{Journal of Personality and Social Psychology} \textbf{67}~(6),
  1049--1062 (1994), \doi{10.1037//0022-3514.67.6.1049},
  \url{https://doi.org/10.1037//0022-3514.67.6.1049}.

\bibitem{zhang2020bertscore}
T.~Zhang, V.~Kishore, F.~Wu, K.~Q. Weinberger, Y.~Artzi, BERTScore: Evaluating
  Text Generation with BERT (2020), \url{https://arxiv.org/abs/1904.09675}.

\bibitem{wilson2022minimalism}
A.~V. Wilson, S.~Bellezza, Consumer Minimalism. \emph{Journal of Consumer
  Research} \textbf{48}~(5), 796--816 (2022), \doi{10.1093/jcr/ucab038},
  \url{https://doi.org/10.1093/jcr/ucab038}.

\bibitem{jachimowicz2019defaults}
J.~M. Jachimowicz, S.~Duncan, E.~U. Weber, E.~J. Johnson, When and Why Defaults
  Influence Decisions: A Meta-Analysis of Default Effects. \emph{Behavioural
  Public Policy} \textbf{3}~(2), 159--186 (2019), \doi{10.1017/bpp.2018.43},
  \url{https://doi.org/10.1017/bpp.2018.43}.

\bibitem{keller2017luxury}
K.~L. Keller, Managing the Growth Tradeoff: Challenges and Opportunities in
  Luxury Branding, in \emph{Advances in Luxury Brand Management}, J.-N.
  Kapferer, J.~Kernstock, T.~Brexendorf, S.~Powell, Eds., Journal of Brand
  Management: Advanced Collections (Palgrave Macmillan, Cham) (2017),
  \doi{10.1007/978-3-319-51127-6_9}.

\bibitem{bellezza2014brandtourists}
S.~Bellezza, A.~Keinan, Brand Tourists: How Non--Core Users Enhance the Brand
  Image by Eliciting Pride. \emph{Journal of Consumer Research}
  \textbf{41}~(2), 397--417 (2014), \doi{10.1086/676679},
  \url{https://doi.org/10.1086/676679}.

\bibitem{park2022nft}
E.~Park, K.~Lane, S.~Bellezza, NFT for Conspicuous Consumption, in
  \emph{Advances in Consumer Research, Volume 50}, H.~A. Chen, G.~Eckhardt,
  R.~Hamilton, Eds. (Association for Consumer Research, Duluth, MN) (2022),
  conference held in Denver, CO.

\bibitem{hrw2024supplychain}
{Human Rights Watch}, EU Parliament Approves Supply Chain Law, News release,
  Human Rights Watch (2024),
  \url{https://www.hrw.org/news/2024/04/24/eu-parliament-approves-supply-chain-law},
  accessed: 2025-09-27.

\bibitem{curran1996robustness}
P.~J. Curran, S.~G. West, J.~F. Finch, The Robustness of Test Statistics to
  Nonnormality and Specification Error in Confirmatory Factor Analysis.
  \emph{Psychological Methods} \textbf{1}~(1), 16--29 (1996),
  \doi{10.1037/1082-989X.1.1.16},
  \url{https://doi.org/10.1037/1082-989X.1.1.16}.

\bibitem{whitehouse2022junkfees}
{The White House}, The President's Initiative on Junk Fees and Related Pricing
  Practices,
  \url{https://www.whitehouse.gov/briefing-room/blog/2022/10/26/the-presidents-initiative-on-junk-fees-and-related-pricing-practices/}
  (2022), accessed: 2025-09-25.

\bibitem{mccright2013perceived}
A.~M. McCright, R.~E. Dunlap, C.~Xiao, Perceived Scientific Agreement and
  Support for Government Action on Climate Change in the USA. \emph{Climatic
  Change} \textbf{119}, 511--518 (2013), \doi{10.1007/s10584-013-0704-9},
  \url{https://doi.org/10.1007/s10584-013-0704-9}.

\bibitem{dell2025super}
F.~Dell’Acqua, B.~Kogut, P.~Perkowski, Super mario meets ai: Experimental
  effects of automation and skills on team performance and coordination.
  \emph{Review of Economics and Statistics} pp. 1--16 (2025).

\bibitem{toubia2025twin2k500}
O.~Toubia, \emph{et~al.}, Database Report: Twin-2K-500: A Data Set for Building
  Digital Twins of over 2,000 People Based on Their Answers to over 500
  Questions. \emph{Marketing Science} \textbf{0}~(0) (2025),
  \doi{10.1287/mksc.2025.0262}, \url{https://doi.org/10.1287/mksc.2025.0262}.

\bibitem{kessler2019incentivized}
J.~B. Kessler, C.~Low, C.~D. Sullivan, Incentivized resume rating: Eliciting
  employer preferences without deception. \emph{American Economic Review}
  \textbf{109}~(11), 3713--3744 (2019).

\bibitem{rusak2025ai}
G.~Rusak, B.~S. Manning, J.~J. Horton, AI Agents Can Enable Superior Market
  Designs  (2025).

\bibitem{zhang2023human}
Y.~Zhang, R.~Gosline, Human Favoritism, Not AI Aversion: People’s Perceptions
  (and Bias) Toward Generative AI, Human Experts, and Human--GAI Collaboration
  in Persuasive Content Generation. \emph{Judgment and Decision Making}
  \textbf{18}, e41 (2023), \doi{10.1017/jdm.2023.37},
  \url{https://doi.org/10.1017/jdm.2023.37}.

\bibitem{lewandowsky2021countering}
S.~Lewandowsky, S.~van~der Linden, Countering Misinformation and Fake News
  Through Inoculation and Prebunking. \emph{European Review of Social
  Psychology} \textbf{32}~(2), 348--384 (2021),
  \doi{10.1080/10463283.2021.1876983},
  \url{https://doi.org/10.1080/10463283.2021.1876983}.

\bibitem{pennycook2020fake}
G.~Pennycook, D.~G. Rand, Who Falls for Fake News? The Roles of Bullshit
  Receptivity, Overclaiming, Familiarity, and Analytic Thinking. \emph{Journal
  of Personality} \textbf{88}~(2), 185--200 (2020), \doi{10.1111/jopy.12476},
  \url{https://doi.org/10.1111/jopy.12476}.

\bibitem{ceylan2023misinformation}
G.~Ceylan, I.~A. Anderson, W.~Wood, Sharing of Misinformation Is Habitual, Not
  Just Lazy or Biased. \emph{Proceedings of the National Academy of Sciences}
  \textbf{120}~(4), e2216614120 (2023), \doi{10.1073/pnas.2216614120},
  \url{https://doi.org/10.1073/pnas.2216614120}.

\bibitem{vosoughi2018spread}
S.~Vosoughi, D.~Roy, S.~Aral, The Spread of True and False News Online.
  \emph{Science} \textbf{359}~(6380), 1146--1151 (2018),
  \doi{10.1126/science.aap9559}, \url{https://doi.org/10.1126/science.aap9559}.

\bibitem{vanbavel2024identity}
J.~J. Van~Bavel, S.~Rathje, M.~Vlasceanu, C.~Pretus, Updating the
  Identity-Based Model of Belief: From False Belief to the Spread of
  Misinformation. \emph{Current Opinion in Psychology} \textbf{56}, 101787
  (2024), \doi{10.1016/j.copsyc.2023.101787},
  \url{https://doi.org/10.1016/j.copsyc.2023.101787}.

\bibitem{orme2017conjoint}
B.~K. Orme, K.~Chrzan, \emph{Becoming an Expert in Conjoint Analysis: Choice
  Modeling for Pros} (Sawtooth Software, Provo, UT) (2017).

\bibitem{boussioux2024crowdless}
L.~Boussioux, J.~N. Lane, M.~Zhang, V.~Jacimovic, K.~R. Lakhani, The Crowdless
  Future? Generative AI and Creative Problem-Solving. \emph{Organization
  Science} \textbf{35}~(5), 1589--1607 (2024), \doi{10.1287/orsc.2023.18430},
  \url{https://doi.org/10.1287/orsc.2023.18430}.

\bibitem{lee2024chatgpt}
B.~C. Lee, J.~J. Chung, An Empirical Investigation of the Impact of ChatGPT on
  Creativity. \emph{Nature Human Behaviour} \textbf{8}~(10), 1906--1914 (2024),
  epub 2024 Aug 12, \doi{10.1038/s41562-024-01953-1},
  \url{https://doi.org/10.1038/s41562-024-01953-1}.

\bibitem{gray2019forward}
K.~Gray, \emph{et~al.}, “Forward flow”: A new measure to quantify free
  thought and predict creativity. \emph{American Psychologist} \textbf{74}~(5),
  539 (2019).

\bibitem{catapano2019perspective}
R.~Catapano, Z.~L. Tormala, D.~D. Rucker, Perspective Taking and
  Self-Persuasion: Why ``Putting Yourself in Their Shoes'' Reduces Openness to
  Attitude Change. \emph{Psychological Science} \textbf{30}~(3), 424--435
  (2019), \doi{10.1177/0956797618822697},
  \url{https://doi.org/10.1177/0956797618822697}.

\bibitem{alesina2011preferences}
A.~Alesina, P.~Giuliano, Preferences for Redistribution, in \emph{Handbook of
  Social Economics}, J.~Benhabib, A.~Bisin, M.~O. Jackson, Eds. (North-Holland,
  Amsterdam), vol.~1, chap.~4, pp. 93--131 (2011),
  \doi{10.1016/B978-0-444-53187-2.00004-8},
  \url{https://doi.org/10.1016/B978-0-444-53187-2.00004-8}.

\bibitem{gilens2012affluence}
M.~Gilens, \emph{Affluence and Influence: Economic Inequality and Political
  Power in America} (Princeton University Press, Princeton, NJ) (2012).

\bibitem{hussein2025outparty}
M.~A. Hussein, C.~Lee, S.~C. Wheeler, How Do Consumers React to Ads That Meddle
  in Out-Party Primaries? \emph{Journal of Consumer Research} \textbf{51}~(6),
  1186--1208 (2025), \doi{10.1093/jcr/ucae039},
  \url{https://doi.org/10.1093/jcr/ucae039}.

\bibitem{heltzel2020polarization}
G.~Heltzel, K.~Laurin, Polarization in America: Two Possible Futures.
  \emph{Current Opinion in Behavioral Sciences} \textbf{34}, 179--184 (2020),
  \doi{10.1016/j.cobeha.2020.03.008},
  \url{https://doi.org/10.1016/j.cobeha.2020.03.008}.

\bibitem{pew2022partisan}
{Pew Research Center}, As Partisan Hostility Grows, Signs of Frustration With
  the Two-Party System,
  \url{https://www.pewresearch.org/politics/2022/08/09/as-partisan-hostility-grows-signs-of-frustration-with-the-two-party-system/}
  (2022), accessed: 2025-09-25.

\bibitem{hussein2021receptiveness}
M.~A. Hussein, Z.~L. Tormala, Undermining Your Case to Enhance Your Impact: A
  Framework for Understanding the Effects of Acts of Receptiveness in
  Persuasion. \emph{Personality and Social Psychology Review} \textbf{25}~(3),
  229--250 (2021), \doi{10.1177/10888683211001269},
  \url{https://doi.org/10.1177/10888683211001269}.

\bibitem{hussein2024receptiveness}
M.~A. Hussein, S.~C. Wheeler, Reputational Costs of Receptiveness: When and Why
  Being Receptive to Opposing Political Views Backfires. \emph{Journal of
  Experimental Psychology: General} \textbf{153}~(6), 1425--1448 (2024),
  \doi{10.1037/xge0001579}, \url{https://doi.org/10.1037/xge0001579}.

\bibitem{green1978conjoint}
P.~E. Green, V.~Srinivasan, Conjoint Analysis in Consumer Research: Issues and
  Outlook. \emph{Journal of Consumer Research} \textbf{5}~(2), 103--123 (1978),
  \doi{10.1086/208721}, \url{https://doi.org/10.1086/208721}.

\bibitem{frank2022decisions}
C.~J. Frank, P.~F. Magnone, O.~Netzer, \emph{Decisions Over Decimals: Striking
  the Balance Between Intuition and Information} (John Wiley \& Sons) (2022).

\bibitem{shaddy2025fairness}
F.~Shaddy, E.~M.~S. Friedman, O.~Toubia, Fairness Perceptions in Demographic
  Targeting. \emph{Journal of Consumer Research}  (2025), ucaf048,
  \doi{10.1093/jcr/ucaf048}, \url{https://doi.org/10.1093/jcr/ucaf048}.

\bibitem{cen2024strategization}
S.~H. Cen, A.~Ilyas, J.~Allen, H.~Li, A.~Madry, Measuring Strategization in
  Recommendation: Users Adapt Their Behavior to Shape Future Content, in
  \emph{Proceedings of the 25th ACM Conference on Economics and Computation (EC
  '24)} (Association for Computing Machinery, New York, NY, USA) (2024), pp.
  203--204, \doi{10.1145/3670865.3673634},
  \url{https://doi.org/10.1145/3670865.3673634}.

\end{thebibliography}

\section*{Acknowledgments}

Caroline Weaver provided research assistance on the Context Effects and Defaults Effects substudies as well as the design, implementation and analysis of the survey reported in SI.I. Xuwen Hua provided research assistance on the Context Effects and Defaults Effects substudies. Arnab Choudhury provided research assistance on the Hiring Algorithms substudy. Sanjana Rosario provided research assistance on the Quantitative Intuition substudy.

\paragraph*{Author contributions:} G.G., D.J.M., M.B, E.J.J., V.M., A.A., S.B., D.D., H.F., E.F., A.G., M.H., K.J., B.K., A.K., K.L., H.L., P.P., O.N., O.T. designed the substudies, analyzed the data from each substudy and wrote the corresponding sections in SI.J; O.T. ran the studies on human participants; T.P., G.G., G.J.F., M.B.S., O.T. ran the studies on digital twins; T.P., G.G., D.J.M., G.J.F., M.B.S., O.T. produced the analyses reported in the main paper and SI.D-H; E.J.J., O.T. designed and analyzed the study reported in SI.I; M.B., E.J.J., O.T. were particularly involved in writing the main paper, with input from multiple co-authors.

\paragraph*{Competing interests:} The authors declare no competing interests.

\paragraph*{Data and materials availability:} Data are publicly available at \href{https://huggingface.co/datasets/LLM-Digital-Twin/Twin-2K-500-Mega-Study}{\url{https://huggingface.co/datasets/LLM-Digital-Twin/Twin-2K-500-Mega-Study}}. The only piece of data which is not publicly available is the Prolific ID of the participants. All code is publicly available at \href{https://github.com/TianyiPeng/Twin-2K-500-Mega-Study}{\url{https://github.com/TianyiPeng/Twin-2K-500-Mega-Study}}.


\clearpage  

\begin{appendices}

\begin{center}
{\Large\textbf{Supporting Information}}\\[0.5em]
{\large Digital Twins are Funhouse Mirrors: Five Systematic Distortions}
\end{center}

\vspace{1em}

\startcontents[appendices]
\printcontents[appendices]{}{1}{\section*{Contents}\setcounter{tocdepth}{2}}

\clearpage  
\section{Summary of Pre-Registered Comparisons Between Twins and
Humans}\label{secSI1}

(see SI section specific to each sub-study for details)

\begin{longtable}{p{0.30\linewidth} p{0.35\linewidth} p{0.35\linewidth}}
  
\toprule
Sub-study (Common Labels)
&Pre-registered comparison(s)
&Result(s)
\\
\midrule
\endhead
1. Accuracy Nudges for Misinformation (social domain, cognitive domain, preference measure, behavioral intention, social desirability, scale question, evaluating content) & Is the effect of prompting people
to think about accuracy on sharing of untrustworthy (vs. trustworthy)
news similar for twins compared to their human counterparts? & When
primed to consider accuracy or reminded of its importance, humans showed
no statistically reliable reduction in their willingness to share false
versus true entertainment headlines. In contrast, twins responded as the
prior work would predict, demonstrating robust improvements in truth
discernment. \\
\midrule
2. Affective Primes (social domain, scale question) & Research in the social sciences often manipulates
emotional states using affective priming, where participants are asked
to reflect on a recent event. The broad question here is whether
affective priming works with digital twins. Specifically:

(i) Do affective priming manipulations induce ``states'' in digital
twins (i.e., does writing about gratitude or lack of control momentarily
influence digital twins' responses?)

(ii) Is the influence of affective priming similar for digital twins and
their human counterparts?

(iii) Is the influence of affective priming on digital twins dependent
on the valence of affective prime (i.e., between positive primes like
gratitude or negative primes like lack of control) or the proximal
nature of the measures (i.e., does it ``spill over'' to other, related
dimensions or only influence proximal dimensions)?

(iv) Is there a relationship between what digital twins and their human
counterparts say in their response to the manipulation? & (i) Yes.
Writing about feeling grateful or out of control significantly increased
corresponding feelings of gratitude and lack of control in digital
twins, with these induced ``affective states'' subsequently influencing
downstream outcomes.

(ii) No. Affective priming had stronger effects on digital twins'
affective responses compared to humans for both gratitude and lack of
control manipulations.

(iii) Yes. Valence affected twins' responses, but in a similar way to
humans: The gratitude prime had a stronger effect on the manipulation
check than the lack of control prime for both humans and their twins.

(iv) Yes. The semantic content written by each human was significantly
more similar to their twin's content than a randomly selected twin's
content. \\
\midrule
3. Consumer Minimalism (social domain, cognitive domain, preference measure, personality measure, scale question) & How does the predictive validity of the Consumer
Minimalism Scale vary in a sample of human vs. their digital twins? The
prediction is that respondents scoring high on the minimalism scale will
prefer minimalist home environments more. & The predictive validity of
the Consumer Minimalism Scale documented in human samples replicated
robustly in the twin sample. The scale predicted consumers' preferences
for minimalist versus non-minimalist apartment interiors. \\
\midrule
4. Context Effects (cognitive domain, replicates known human bias, preference measure, test of rationality, different versions of question, behavioral intention) & Can digital twins demonstrate the attraction effect \cite{huber1982adding}
and the compromise effect \cite{simonson1989choice}? &
Humans replicated the attraction effect, but twins did not. Neither
humans nor digital twins exhibited a compromise effect. \\
\midrule
5. Defaults Effects (cognitive domain, replicates known human bias, preference measure, test of rationality, pro-social, different versions of question, behavioral intention, social desirability)& Can digital twins demonstrate default effects \cite{johnson2003defaults,pichert2008green}? & Human participants showed a default effect in a green energy adoption paradigm but not in an organ donation paradigm. Twins predicted a strong default effect in the organ donation paradigm but no effect in the green energy paradigm. \\
\midrule
6. Digital Certifications for Luxury Consumption (social domain, different versions of question, attitude, scale question, human-tech interactions) & How does the propensity
to hold higher status perceptions for a luxury product associated with,
vs. not associated with, a digital certification, differ between twins
and their human counterparts? & Digital certification via a digital
passport increased the value perception of a diffused luxury product for
humans. Digital twins' value perceptions, however, were not impacted by
the inclusion of a digital passport. \\
\midrule
7. Fees Accuracy (social domain, preference measure, different versions of question, attitude, scale question) & (i) How does the knowledge of digital twins regarding
fees and surcharges in the marketplace compare to that of their human
counterparts?

(ii) How do perceptions of fairness regarding fees and surcharges in the
marketplace compare between digital twins and their human counterparts?

(iii) How does the variation in perceptions of fairness regarding fees
and surcharges in the marketplace compare between digital twins and
their human counterparts?

(iv) How do beliefs about government regulation regarding fees and
surcharges in the marketplace compare between digital twins and their
human counterparts?

(v) How does the variation in beliefs about government regulation
regarding fees and surcharges in the marketplace compare between digital
twins and their human counterparts? & (i) Twins were much more accurate
than humans on knowledge questions.

(ii) Twins closely mirrored the average human response on fairness
questions.

(iii) Twins exhibited narrower variance than humans on fairness
perceptions.

(iv) Twins were on average more favorable to government regulation
compared to humans.

(v) For both groups, conservatism was correlated with less support for
pricing regulation. However, twins stereotyped participants based on
their political ideology, with exaggerated support for regulation for
liberal participants. \\
\midrule
8. Heterogeneous Story Beliefs (content evaluation, different versions of question, scale question) & When reading a book and predicting the
emotion (valence and arousal) of the next chapter based on the previous
chapter, what is the correlation between the expected valence and
expected arousal from humans vs. their digital twins? & We found
significant positive correlations for both valence and arousal. \\
\midrule
9. Hiring algorithms (social domain, preference measure, human-tech interactions, different versions of question, scale question) & Can digital twins predict job candidate preferences
for workplace hiring policies? Specifically:

(i) Can digital twins predict workplace preferences?

(ii) Can experimenting on a population of digital twins recover the true
average treatment effects as experimenting on a human population? & (i)
Twins tended to express views that were overall more pro-social
(sustainability, ESG, work-life balance, transparent leadership, culture
and values), and less "career-oriented" (firm prestige, career
development) than those expressed by humans.

(ii) Twins were overall less averse to hiring algorithms compared to
humans. \\
\midrule
10. Idea Evaluation (cognitive domain, preference measure, evaluating content, pro-social, different versions of question, valenced, scale question) & (i) How do digital twins' ratings of idea creativity
compare to those of their human counterparts?

(ii) Do digital twins show human favoritism, AI favoritism, or neither
when evaluating ideas? & (i) Creativity ratings coming from twins were
not significantly correlated with the ratings coming from their human
counterparts.

(ii) When no information was provided about the source of the ideas,
humans rated twin ideas higher (more creative) than human ideas, and
twins rated human ideas higher than twin ideas. When information was
provided about the source of ideas, humans displayed AI aversion, and
twins displayed human aversion. \\
\midrule
11. Infotainment News Sharing (social domain, cognitive domain, preference measure, content evaluation, behavioral intention, social desirability, scale question) & When choosing an article to share on social
media, will digital twins prioritize the following attributes similarly
to their human counterpart:

1. Headline (more vs. less entertaining)

2. Source (more vs. less trustworthy)

3. Content type (entertaining vs. informative)

4. Political lean (conservative vs. liberal)

5. Number of likes (20 vs. 200 vs. 2,000) & While humans and twins both
prioritized headline, this consideration mattered much less to twins,
who equally relied on number of likes and put far more weight on
political lean than their human counterparts. Twins were also sensitive
to source trustworthiness, while humans did not differentiate between
more or less trustworthy sources. \\
\midrule
12. Measures of Creativity (cognitive domain, pro-social) & (i) How do digital twins' creative abilities
compare to those of their human counterparts?

(ii) Can digital twins replicate the correlations between creativity
tests and idea generation performance observed in humans? & (i) The
ideas generated by twins were judged as more creative (and with less
dispersion in creativity ratings) than the ideas generated by their
human counterparts. Digital twins also performed better (and with less
dispersion) at the DAT, a creativity task.

(ii) The correlation previously found between DAT and creativity was
replicated, both among humans and twins. \\
\midrule
13. Obedient Twins (social domain, preference measure, valenced, scale question) & LLMs are trained to be obedient and deferential. Does
this tendency make digital twins more sensitive to survey instructions
than their human counterparts?

We tested this in three tasks:

(i) Self-persuasion \cite{brinol2012self}. When prompted to
consider the other side, do digital twins abandon their attitudes more
readily than their human counterparts?

(ii) Scenarios. Do digital twins more ``obediently'' follow the
instructions to imagine themselves in different scenarios, leading to
more sensitivity to scenario manipulations?

(iii) Absurd Scenarios. Do digital twins ``earnestly'' respond to
instructions that would be non-sensical to their human counterparts? &
(i) No. Twins were less sensitive to considering a different point of
view compared to humans.

(ii) No. Twins were less sensitive overall to these scenario
manipulations compared to humans.

(iii) No. Humans were more sensitive to the instructions to forecast
feelings in absurd scenarios. \\
\midrule
14. Privacy Preferences of Digital Twins (social domain, preference measure, human-tech interactions, conflict-related, different versions of question, scale question)& How do humans and their digital
twins vary in their perceptions of privacy violations for different data
processing and targeting practices in online advertising? & Twins
perceived their privacy to be violated less than their human
counterparts. We found almost no correlation in preferences for
individual-level privacy violation scores for twins compared to humans.
However, in aggregate, humans and their twins ranked privacy violations
similarly across advertising practices that rely differently on tracking
and targeting (high rank correlation). \\
\midrule
15. Preferences for Redistribution (social domain, attitude, scale question) & Can digital twins predict preferences
for redistribution and other outcomes like trust, socio-economic
background, and beliefs on fairness and social mobility? & Twins showed
more extreme redistribution preferences. On average, they were more
likely to think "people should take care of themselves" but more willing
to pay taxes to improve healthcare for all people. They were more likely
to think people are fair and that people can be trusted. They were also
more likely to think that hard work and luck are equally important for
success. Finally, they showed very little distribution in their fathers'
educational attainment, with almost 100\% of them reporting their father
only completed high school. \\
\midrule
16. Promiscuous Donors (social domain, preference measure, political domain, conflict-related, social cognition, attitude, social desirability, different versions of question, valenced, scale question) & How do humans and twins differ in their reaction to
(i) others who do not donate to either party, (ii) others who donate to
both parties, (iii) justification from those who donate to both parties
based on pragmatic vs. value-based reasons? & Human respondents
penalized promiscuous donors, especially when the behavior appeared
motivated by self-interest (e.g., cultivating access or recruiting
talent). In contrast, twins were less punitive (sometimes even
favorable) toward the very same targets. \\
\midrule
17. Quantitative Intuition (cognitive domain, scale question) & What is the relationship between
Quantitative Intuition (QI) measures and demographics, personality
traits, and individual measures like overconfidence and numeracy? Can
digital twins capture Quantitative Intuition (QI) skills? & We found a
moderate correlation between human responses to the individual QI scale
and those of their twins, and a weaker relationship in assessments of
the QI score of the organization they work(ed) for. Twins failed to
mimic responses to two QI behavior questions. Twins tended to select the
normative, high-QI behavior much more frequently than humans did. \\
\midrule
18. Targeting Fairness Perceptions (social domain, preference measure, social cognition, attitude, valenced, different versions of question, scale question) & How do humans and twins perceive the
fairness of demographic-based versus broad targeting? & At the
aggregate level, the twins closely matched human judgments of the
perceived fairness of demographic targeting, replicating the finding
that it is viewed as less fair than broad targeting and producing
similar average ratings across conditions. \\
\midrule
19. User Behavior with Recommendation Algorithms (cognitive domain, human-tech interactions)& (i) How do self-reported
average time spent on the platform differ across humans and their
digital twin counterparts?

(ii) How does knowledge that recommendation algorithms depend on the
user's past behavior compare between humans and their digital twin
counterparts?

(iii) How does the prevalence of strategization (modifying your behavior
to influence future recommendations) compare between humans and their
digital twin counterparts?

(iv) How does the preference for user controls over recommended content
compare between humans and their digital twin counterparts? & (i) Twins
under-reported platform usage.

(ii) Twins were more aware of how recommendation algorithms work.

(iii) Twins were more likely to strategize.

(iv) Twins preferred more control over algorithms. \\
\bottomrule
\caption{Pre-Registered Comparisons and Results. We label individual \emph{outcomes} rather than substudies for our meta-analysis. This table provides, for each substudy, the labels associated with at least half of the outcomes from that substudy.}\label{tab:SI1}
\end{longtable}


\pagebreak

\section{Prompt Template}\label{secSI8}
\subsection{Full Prompt Template}\label{subsecSI1}
\begin{verbatim}
You are an AI assistant. Your task is to answer the 'New Survey Question' as 
if you are the person described in the 'Persona Profile' (which consists 
of their past survey responses).

Remain consistent with the persona's previous answers and stated traits. 
Simulate their responses to new questions while accounting for human cognitive
limitations, uncertainty, and biases.

Follow all instructions provided for the new question carefully regarding
the format of your answer.

Persona Profile (This individual's past survey responses):
{Persona Profile}

New Survey Question & Instructions 
(Please respond as the persona described above):
{Survey Questions}

Format Instructions:
In order to facilitate the postprocessing, you should generate a JSON object
by filling in Masks with the appropriate values in the following template 
for {N} questions, where each question Q1, Q2, ... corresponds to the question
Q1, Q2, ... above.
{Format Templates}
\end{verbatim}

\subsection{Example of Survey Questions}\label{subecSI2}
\begin{verbatim}
Q1: 
Introduction: You are being invited to participate in a research study… 
By selecting the ‘Agree’ option below I am agreeing to participate, 
I have not given up any of the legal rights that I would have if I were not 
a participant in the study.
Question Type: Single Choice
Options:
  1 - Agree
  2 - Disagree
Answer: [Masked]

Q2:
How important is it to you that your employer actively invests in environmental
sustainability (e.g., reducing carbon emissions, minimizing waste)?
Question Type: Single Choice
Options:
  1 - Not at all important
  2 - Slightly important
  3 - Moderately important
  4 - Very important
  5 - Extremely important
Answer: [Masked]

Q3:
Would you prefer to work at a company that is outspoken about social and
political issues, or one that remains neutral?
Question Type: Single Choice
Options:
  1 - Strongly prefer outspoken
  2 - Somewhat prefer outspoken
  3 - No preference
  4 - Somewhat prefer neutral
  5 - Strongly prefer neutral
Answer: [Masked]
……
\end{verbatim}

\subsection{Example of Format Templates}\label{subsecSI3}
\begin{verbatim}
{
    "Q1": {
        "Question Type": "Single Choice",
        "Answers": {
        "SelectedByPosition": Masked, 
        // a number from 1 to 2, corresponding to the option position
        "SelectedText": "Masked" 
        // a string, corresponding to the option text
        }
    },
    "Q2": {
        "Question Type": "Single Choice",
        "Answers": {
        "SelectedByPosition": Masked, 
        // a number from 1 to 5, corresponding to the option position
        "SelectedText": "Masked" 
        // a string, corresponding to the option text
        }
    },
….
\end{verbatim}

\pagebreak

\section{Details for Persona Construction}\label{secSI4}

In this section, we detail the construction of the various types of personas. That is, we provide the text that is inserted into the full prompt template (Section \ref{subsecSI1}), as \{Persona Profile\}.

\subsection{Full Persona}

Our full persona is built from the text-based questions and answers in
Twin-2K-500, reflecting natural language interactions. The Twin-2K-500
dataset consists of four waves, with wave 4 repeating questions from
waves 1--3 for test--retest purposes. For persona construction, when a
question appears in both waves 1--3 and wave 4, we use the wave 4
responses. The complete persona dataset can be accessed at
https://huggingface.co/datasets/LLM-Digital-Twin/Twin-2K-500,
specifically the `persona\_text' split within the `full\_persona'
subset. An example of (portions of) a full persona is below:

\begin{verbatim}
Which part of the United States do you currently live in?

Question Type: Single Choice

Options:

1 - Northeast (PA, NY, NJ, RI, CT, MA, VT, NH, ME)

2 - Midwest (ND, SD, NE, KS, MN, IA, MO, WI, IL, MI, IN, OH)

3 - South (TX, OK, AR, LA, KY, TN, MS, AL, WV, DC, MD, DE, VA, NC, SC,
GA, FL)

4 - West (WA, OR, ID, MT, WY, CA, NV, UT, CO, AZ, NM)

5 - Pacific (HI, AK)

Answer: 2 - Midwest (ND, SD, NE, KS, MN, IA, MO, WI, IL, MI, IN, OH)

What is the sex that you were assigned at birth?

Question Type: Single Choice

Options:

1 - Male

2 - Female

Answer: 2 - Female

...

Here are a number of characteristics that may or may not apply to you.
Please indicate next to each statement the extent to which you agree or
disagree with that statement. I see myself as someone who...

Question Type: Matrix

Options:

1 = Disagree strongly

2 = Disagree a little

3 = Neither agree nor disagree

4 = Agree a little

5 = Agree strongly

1. Is talkative

Answer: 1 - Disagree strongly

2. Tends to find fault with others

Answer: 1 - Disagree strongly

3. Does a thorough job

Answer: 5 - Agree strongly

4. Is depressed, blue

Answer: 1 - Disagree strongly

5. Is original, comes up with new ideas

Answer: 4 - Agree a little

6. Is reserved

Answer: 4 - Agree a little

...

Please consider the following product category: Detergents - Heavy Duty
- Liquid. Suppose you are in a grocery store, and you see the following
product in that category: Purex Liquid Laundry Detergent Plus OXI, Stain
Defense Technology, 128 Fluid Ounces, 85 Wash Loads. The product is
priced at: \$5.98. Would you or would you not purchase this product?

Question Type: Single Choice

Options:

1 - Yes, I would purchase the product

2 - No, I would not purchase the product

Answer: 1 - Yes, I would purchase the product

\end{verbatim}
\subsection{Persona Summary}

Each twin in the full persona dataset averages about 30K tokens, making
its simulation relatively slow and costly. To address this, we developed
a shorter and more concise version called persona summary. In this
version, we simplify the questions and record responses using
distributional information across participants. For instance, instead of
including the full set of Big Five questions, we represent the trait
with a single value such as ``score\_extraversion = 2.125 (26th
percentile).''

For simplification, wave 4 questions (which involved treatment-condition
randomization) are excluded from the persona summary. As a result, the
persona summary prompt is about 3K tokens in length---a substantial
reduction compared to the full persona. Our results show that persona
summary can serve as an efficient alternative when token and time
budgets are constrained.

The complete persona dataset can be accessed at
https://huggingface.co/datasets/LLM-Digital-Twin/Twin-2K-500,
specifically the `persona\_summary' split within the `full\_persona'
subset. An example of (portions of) persona summary is below:

\begin{verbatim}
The person's demographics are the following...

Geographic region: Midwest (ND, SD, NE, KS, MN, IA, MO, WI, IL, MI, IN,
OH)

Gender: Female

Age: 30-49

Education level: Some college, no degree

Race: Black

Citizen of the US: Yes

Marital status: Married

Religion: Nothing in particular

Religious attendance: A few times a year

Political affiliation: Independent

Income: $50,000-$75,000

Political views: Moderate

Household size: 3

Employment status: Part-time employment

The person's Big 5 scores are the following:

score_extraversion = 2.125 (26th percentile)

score_agreeableness = 4.556 (84th percentile)

wave1_score_conscientiousness = 4.556 (77th percentile)

score_openness = 3.5 (36th percentile)

score_neuroticism = 1.5 (15th percentile)

Openness reflects curiosity and receptiveness to new experiences,
Conscientiousness indicates self-discipline and goal-directed behavior,
Extraversion measures sociability and assertiveness, Agreeableness
reflects compassion and cooperativeness, and Neuroticism captures
emotional instability and susceptibility to negative emotions. Each
score ranges from 1 to 5, and a higher score indicates a greater display
of the associated traits.

...
\end{verbatim}

\subsection{Demographics Persona}

We also construct a demographics persona, which contains only
demographic information. This version is derived by truncating the full
persona to retain only 14 demographic variables: region, sex, age,
education, race, citizenship, marital status, religion, religious
attendance, political party, household income, political ideology,
household size, and employment status. By design, the demographics
persona always serves as a prefix of the full persona. An example is
shown below:

\begin{verbatim}

Which part of the United States do you currently live in?

Question Type: Single Choice

Options:

1 - Northeast (PA, NY, NJ, RI, CT, MA, VT, NH, ME)

2 - Midwest (ND, SD, NE, KS, MN, IA, MO, WI, IL, MI, IN, OH)

3 - South (TX, OK, AR, LA, KY, TN, MS, AL, WV, DC, MD, DE, VA, NC, SC,
GA, FL)

4 - West (WA, OR, ID, MT, WY, CA, NV, UT, CO, AZ, NM)

5 - Pacific (HI, AK)

Answer: 2 - Midwest (ND, SD, NE, KS, MN, IA, MO, WI, IL, MI, IN, OH)

What is the sex that you were assigned at birth?

Question Type: Single Choice

Options:

1 - Male

2 - Female

Answer: 2 - Female

How old are you?

Question Type: Single Choice

Options:

1 - 18-29

2 - 30-49

3 - 50-64

4 - 65+

Answer: 2 - 30-49

What is the highest level of schooling or degree that you have
completed?

Question Type: Single Choice

Options:

1 - Less than high school

2 - High school graduate

3 - Some college, no degree

4 - Associate's degree

5 - College graduate/some postgrad

6 - Postgraduate

Answer: 3 - Some college, no degree

What is your race or origin?

Question Type: Single Choice

Options:

1 - White

2 - Black

3 - Asian

4 - Hispanic

5 - Other

Answer: 2 -- Black

\end{verbatim}
We integrate the constructed personas into the prompt template (see
Methods section) to perform our LLM simulations.

\subsection{Empty Persona}
The \emph{Empty Persona} specification does not provide any individual-level information in the prompts. Instead, \{Persona Profile\} is replaced with the placeholder \texttt{[Empty Persona Profile]}.

Specifically, the prompt used for all individuals is:
\begin{verbatim}
...
Persona Profile (This individual's past survey responses):
[Empty Persona Profile]
...
\end{verbatim}

\pagebreak

\section{Distribution of Each Performance Metric Across
Outcomes}\label{secSI2}

\begin{figure}[H]
\centering
\includegraphics[width=0.8\textwidth]{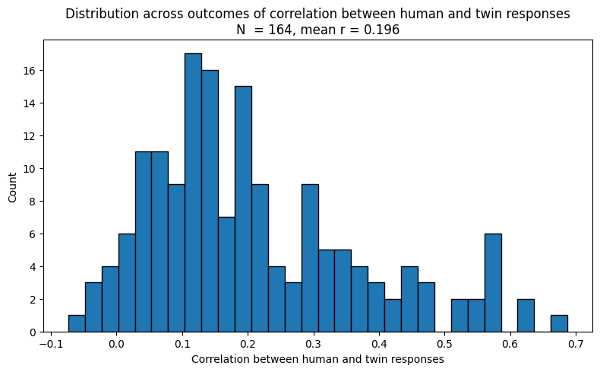}
\caption{Distribution Across Outcomes of Correlation Between Human and Twin Responses.}\label{secSI2_1}
\end{figure}

\begin{figure}[H]
\centering
\includegraphics[width=0.8\textwidth]{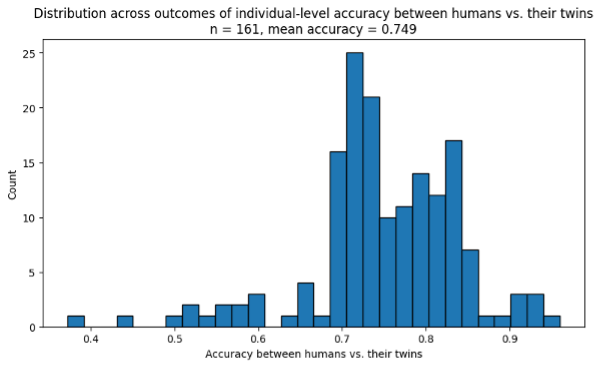}
\caption{Distribution Across Outcomes of Individual-Level Accuracy Between Human and Twin Responses.}\label{secSI2_2}
\end{figure}

\begin{figure}[H]
\centering
\includegraphics[width=0.8\textwidth]{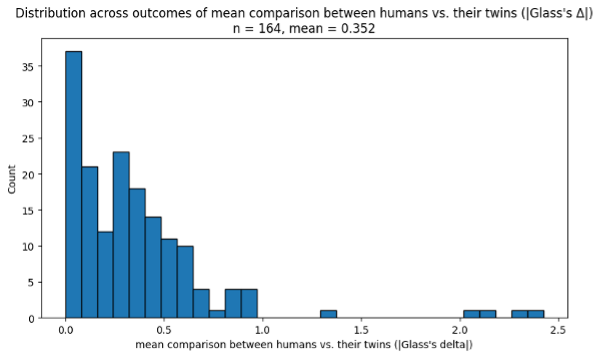}
\caption{Distribution Across Outcomes of |Glass’s $\Delta$| Between Human and Twin Responses.}\label{secSI2_3}
\end{figure}

\begin{figure}[H]
\centering
\includegraphics[width=0.8\textwidth]{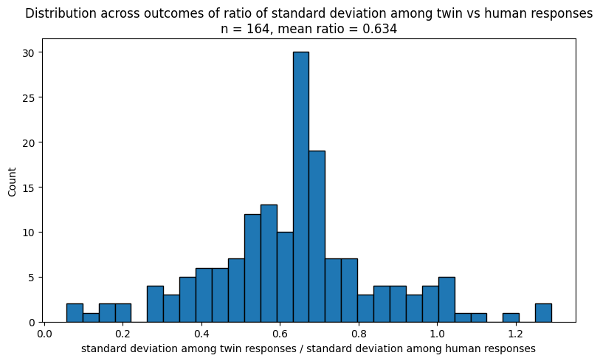}
\caption{Distribution Across Outcomes of Ratio of Standard Deviation Between Human and Twin Responses.}\label{secSI2_4}
\end{figure}
\pagebreak

\section{Digital Twin Performance Across Domains}\label{secSI3}


The categorical labels in our meta-analysis are not mutually exclusive; an outcome may be tagged with multiple labels. All labels are binary, except sample size. The labels are as follows (with examples from our data):

\begin{itemize}
  \item \textbf{Social}
    \begin{itemize}
      \item Social domain (104 outcomes): involves social topics (e.g., likelihood of sharing news content with others)
      \item Conflict-related (15 outcomes): involves intra- or inter-personal conflict (e.g., confidence in ability to manage conflict)
      \item Pro-social (36 outcomes): involves pro-social topics such as altruism or cooperation (e.g., paying higher taxes to improve healthcare for all people)
      \item Social cognition (15 outcomes): involves social cognition such as asking the participant to judge the behavior of others (e.g., judging someone who donates to both political parties)
      \item Personality measure (6 outcomes): assesses personality traits (e.g., consumer minimalism)
      \item Social desirability (55 outcomes): has a socially desirable answer (e.g., willingness to be an organ donor)
    \end{itemize}

  \item \textbf{Preferences/attitudes}
    \begin{itemize}
      \item Attitude (33 outcomes): captures general attitudes toward a topic, issue, or person (e.g., whether most people can be trusted)
      \item Political domain (12 outcomes): focuses on political content (e.g., whether pricing practices should be regulated by the government)
      \item Preference measure (123 outcomes): measures the participant’s preferences (e.g., preference between two room environments) 
      \item Behavioral intention (44 outcomes): captures future behavioral intentions (e.g., intention to apply for a job)
    \end{itemize}

  \item \textbf{Cognitive skills/rationality}
    \begin{itemize}
      \item Cognitive domain (79 outcomes): captures or assesses cognition (e.g., strategic content consumption on media platforms)
      \item Test of rationality (10 outcomes): assesses rational decision-making (e.g., sensitivity to context effects)
    \end{itemize}

  \item \textbf{Content evaluation}
    \begin{itemize}
      \item Evaluating content (62 outcomes): requires evaluation of stimuli (e.g., predicting content of next book chapter)
      \item Valenced (40 outcomes): involves positive versus negative judgments (e.g., judging the creativity of ideas)
    \end{itemize}

  \item \textbf{Human--tech interactions}
    \begin{itemize}
      \item Human--tech interactions (45 outcomes): concerns human interaction with technology (e.g., hiring algorithms)
    \end{itemize}

  \item \textbf{Mechanical}
    \begin{itemize}
      \item Replicates known human bias (5 outcomes): whether the outcome replicates a well-known human bias (e.g., default effect)
      \item Different versions of question(s) (82 outcomes): question(s) wording varies across participants (e.g., across different conditions of an experiment) 
      \item Scale question (142 outcomes): uses Likert or similar rating scales (e.g., agreement with a statement)
      \item Sample size (continuous): the number of participants on which the outcome is measured
    \end{itemize}
\end{itemize}

\begin{figure}[H]
    \centering
    \includegraphics[width=0.8\linewidth]{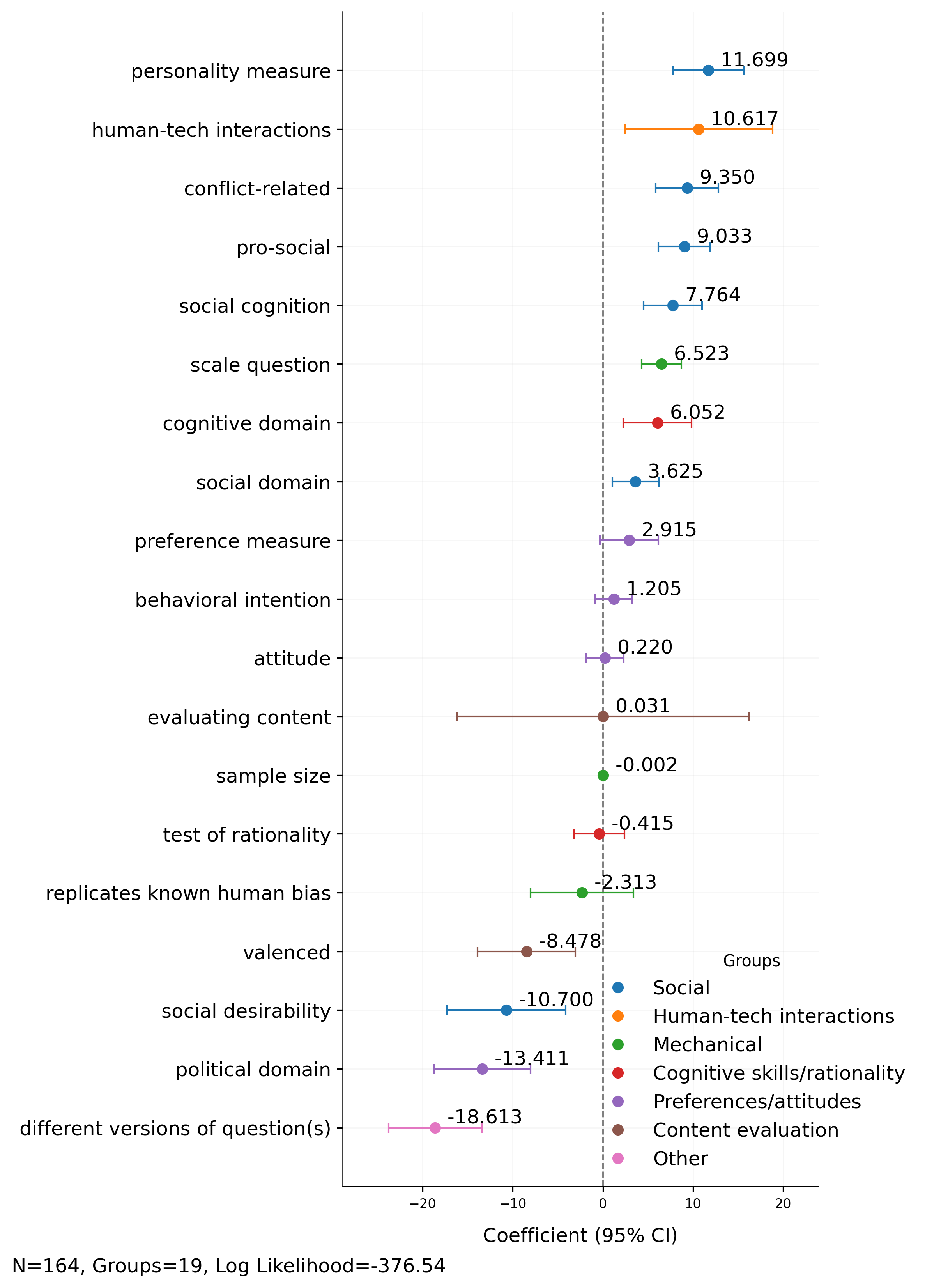}
    \caption{Results from Meta-Analysis (Mixed Linear Model with z-transformed Correlation as Dependent Variable). 
    Regression coefficients with 95\% confidence intervals. 
    The average correlation across all outcomes is 0.197. Correlation between responses from twins versus their human counterparts tends to be higher in social domains, except when social desirability is salient. Correlation tends to be higher in the cognitive domain and in domains related to human-technology interactions, but lower in the political domain and when providing valenced evaluations. 
    }
    \label{fig:meta-analysis}
\end{figure}

\begin{figure}[H]
\centering
\includegraphics[width=0.8\textwidth]{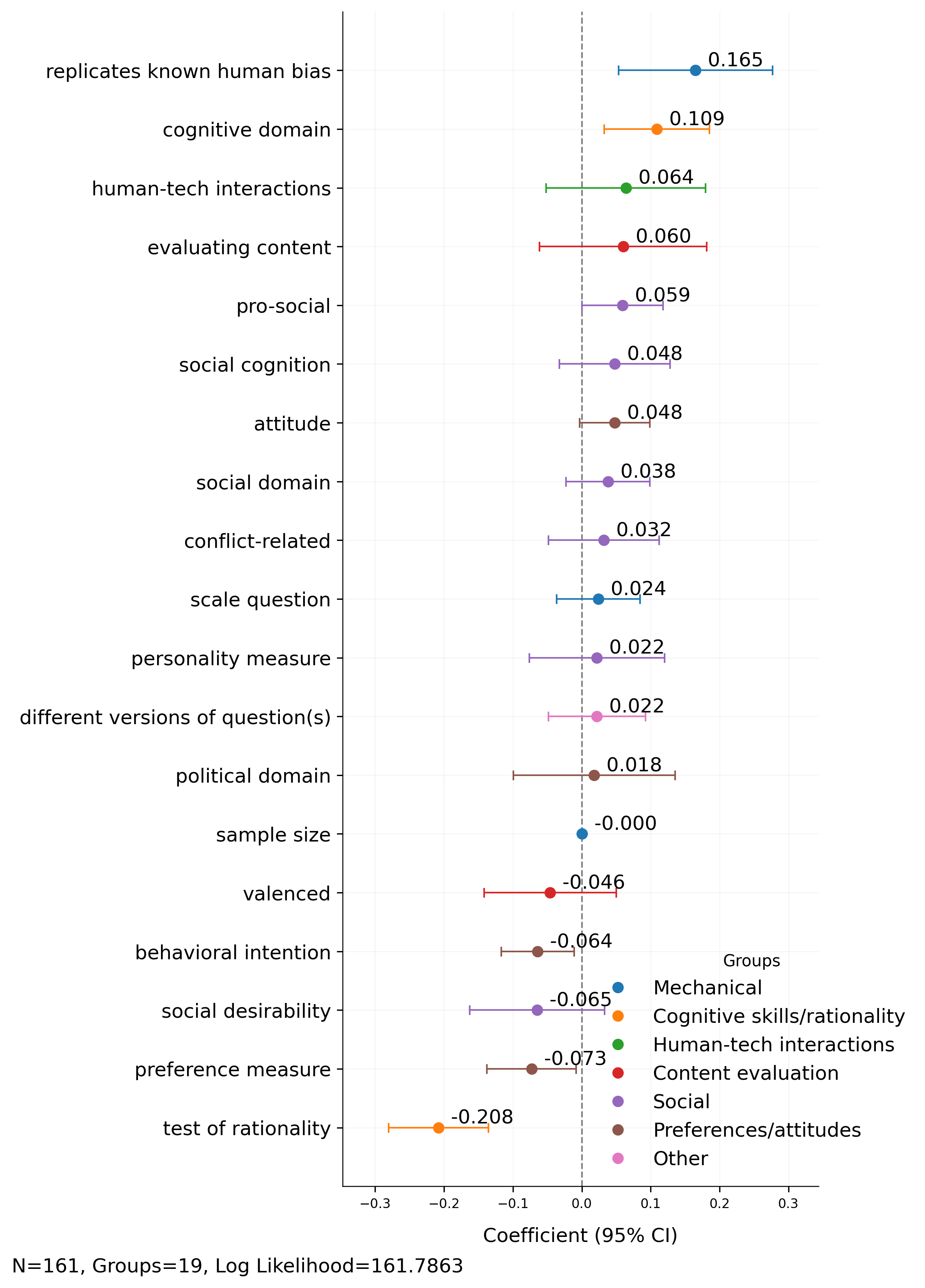}
\caption{Results from Meta-Analysis (Mixed Linear Model with Individual-Level Accuracy as Dependent Variable). }\label{secSI3_1}
\end{figure}

\begin{figure}[H]
\centering
\includegraphics[width=0.8\textwidth]{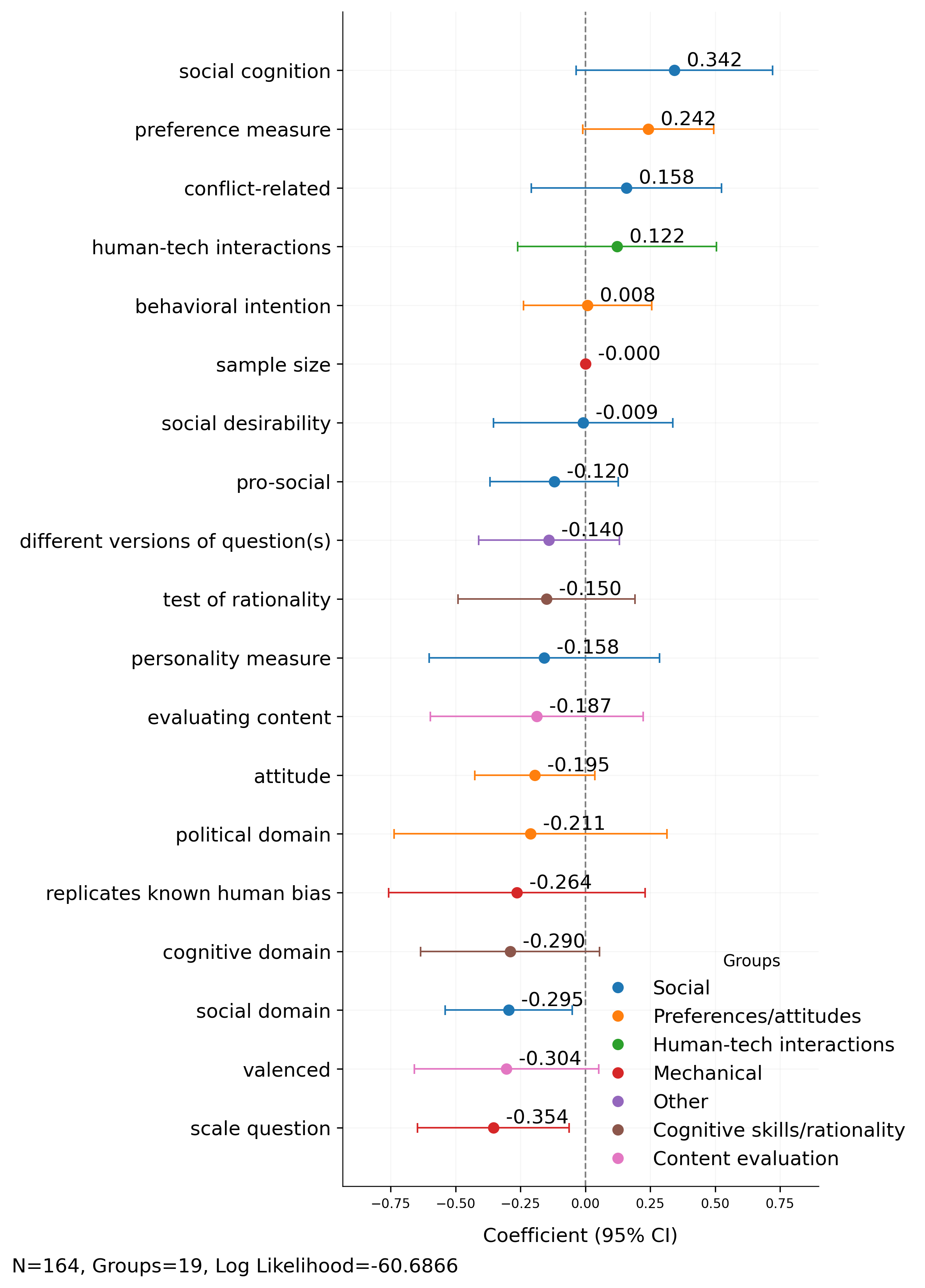}
\caption{Results from Meta-Analysis (Mixed Linear Model with |Glass’s $\Delta$| as Dependent Variable).}\label{secSI3_2}
\end{figure}

\begin{figure}[H]
\centering
\includegraphics[width=0.8\textwidth]{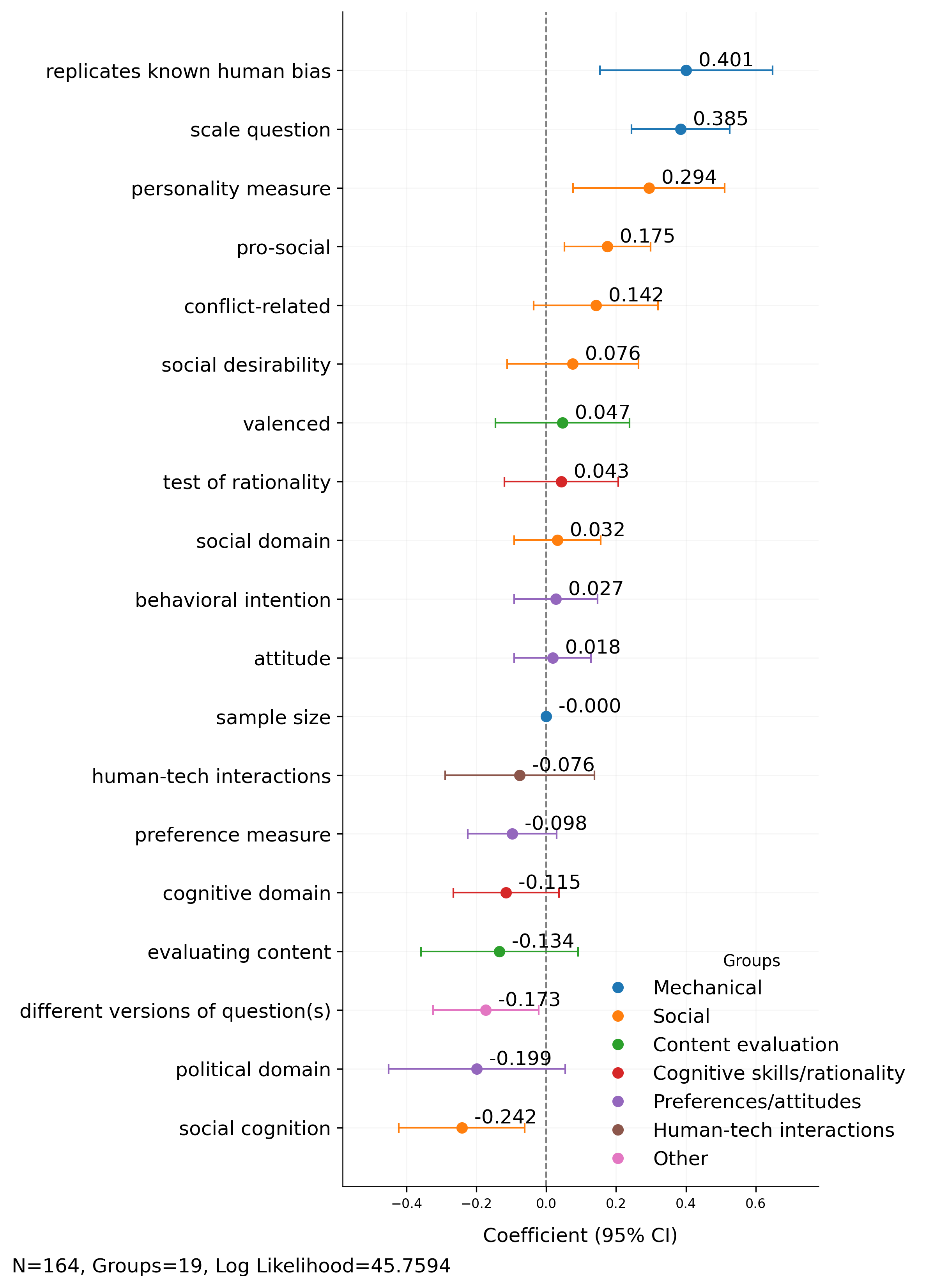}
\caption{Results from Meta-Analysis (Mixed Linear Model with std(twin) / std(human) as Dependent Variable).}\label{secSI3_3}
\end{figure}

We estimate a mixed linear model, including random intercepts for each sub-study. See Figure~\ref{fig:meta-analysis} for the results using correlation as the DV, and Figures~\ref{secSI3_1}-\ref{secSI3_3} for the results using the other performance metrics as DVs.

\pagebreak

\section{Additional Partial Dependence Plots from XGBoost}\label{secSI7}

\begin{figure}[H]
\centering
\includegraphics[width=0.8\textwidth]{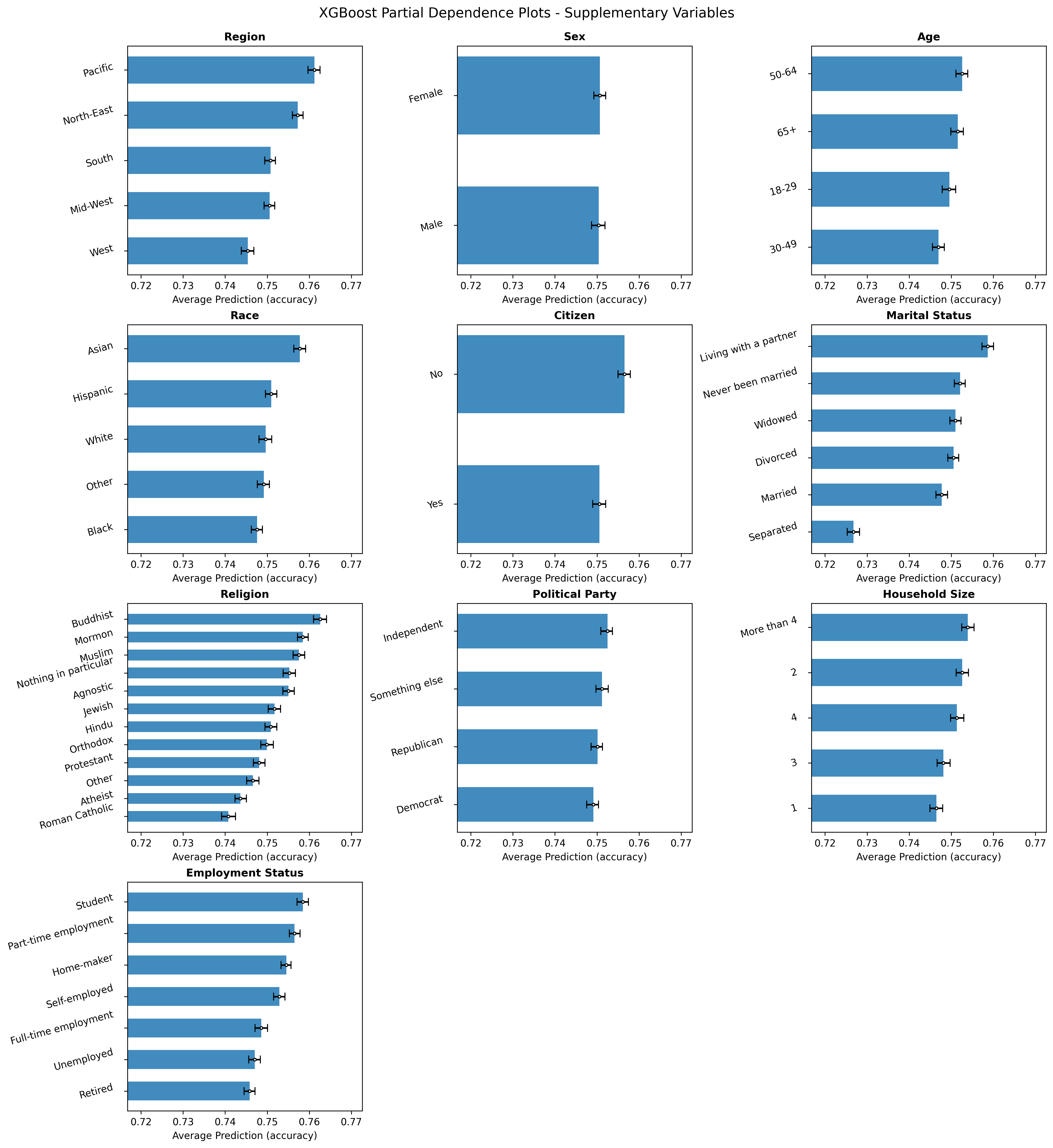}
\caption{Partial Dependence Plots from XGBoost. }\label{secSI7_1}
\end{figure}
\pagebreak

\section{Comparison Across Implementation Methods}\label{secSI5}

\subsection{Persona summaries}
\begin{figure}[H]
\centering
\includegraphics[width=1\textwidth]{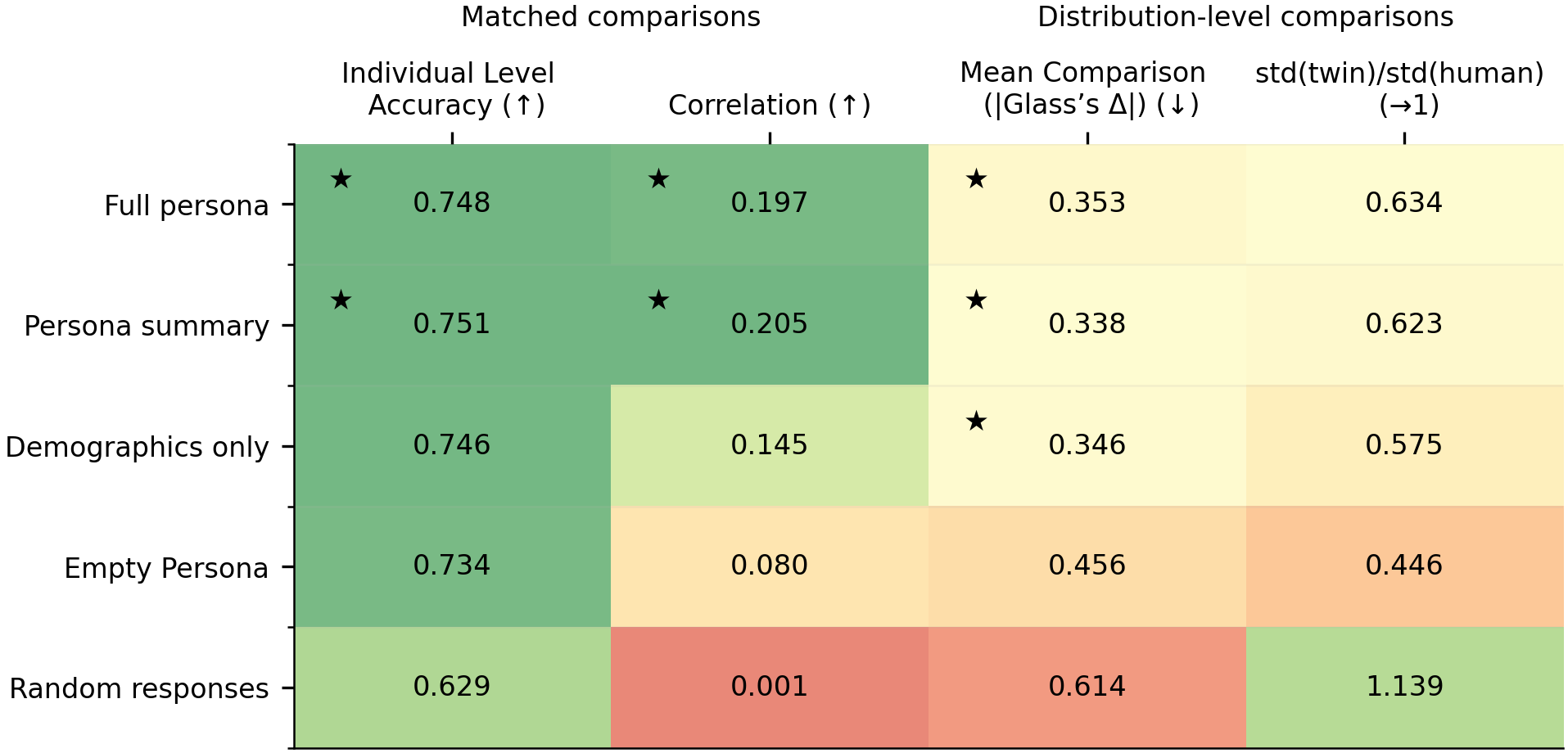}
\caption{Gains from Leveraging Individual-Level Data, Including Persona Summary as Benchmark. *: best performing benchmark, or not significantly different from best at p$<$0.05 (not applicable to ratio of standard deviations), after applying Bonferroni correcting for multiple comparisons in each column.}\label{alt_summary}
\end{figure}

We add as a benchmark an alternative way to construct digital twins, labeled ``persona summary:'' a concise, statement-based summary (approximately 13K characters) of the questions and responses with distributional information (e.g., Big 5 personality scores with percentile ranks rather than 44 detailed questions and answers on which the scores are based). Results are reported in Figure \ref{alt_summary}. We see that this version performs very similarly to the full persona, offering a viable lower-cost alternative. 

\subsection{Alternative base LLMs}

We test GPT-5, Deepseek (deepseek-r1-0528), Gemini (gemini-2.5-flash, gemini-3-pro), and a fine-tuned GPT-4.1 as alternative base models. Full persona prompts were used by all benchmarks. The temperature for GPT-4.1 and its fine-tuned variants was set to the default value of 0.7. For the other models we selected the temperature that was expected to yield the best performance. We used a temperature of 0 for DeepSeek (deepseek-r1) and Gemini-2.5 (gemini-2.5-flash) because we found this helped GPT4.1. We used the default temperature of 1 for Gemini-3-pro, as Google recommends not changing the temperature of this model for better performance. We set the temperature for GPT-5 at 1 due to the API restriction. All other API parameters for these benchmarks were kept at their default values.

The fine-tuned model was fine-tuned using the Twin-2K-500 dataset. Our approach consisted in using a conservative fine-tuning model that would (i) not use any information from our sub-studies and (ii) rely on treating all components of the twins as a unitary construct. That is, if a twin demonstrates consistency in reporting demographic characteristics, it should also display coherence when answering questions in other domains (e.g., attitudes, preferences, or behaviors). As such, for each participant, we created a training example consisting of: 

\begin{enumerate}
    \item A system instruction guiding the model to answer survey questions consistently with the individual's prior responses and realistic cognitive constraints,
    \item A user message containing the Twin 2K-500 questions with all ``Answer: [Masked]'' at once,
    \item An assistant message containing the same message as the user's but with the answers revealed.
\end{enumerate}

This structure replicates the prompting approach we used for the mega-study. It required the model to infer the missing responses from contextual information while maintaining alignment with the participant’s traits. Fine-tuning was performed on GPT-4.1 (release date 2025-04-14) using supervised learning with automatically selected hyperparameters. The fine-tuning was ultimately processed using 3 epochs, a batch size of 4 and a learning rate multiplier of 2. All sub-studies were conducted strictly out-of-sample, ensuring that no outcome measures used for evaluation overlapped with the fine-tuning training data. Note that the fine-tuned model required us to pre-process the input slightly to remove some whitespaces (given the token limit in fine-tuning which was set to 65,536 tokens) and to modify our prompting strategy when running the simulations on the sub-studies. Indeed, after fine-tuning, the model was no longer as proficient at following instructions and responses tended to have different formats. The prompts were modified to ensure the consistency in format.

\begin{figure}[H]
\centering
\includegraphics[width=1\textwidth]{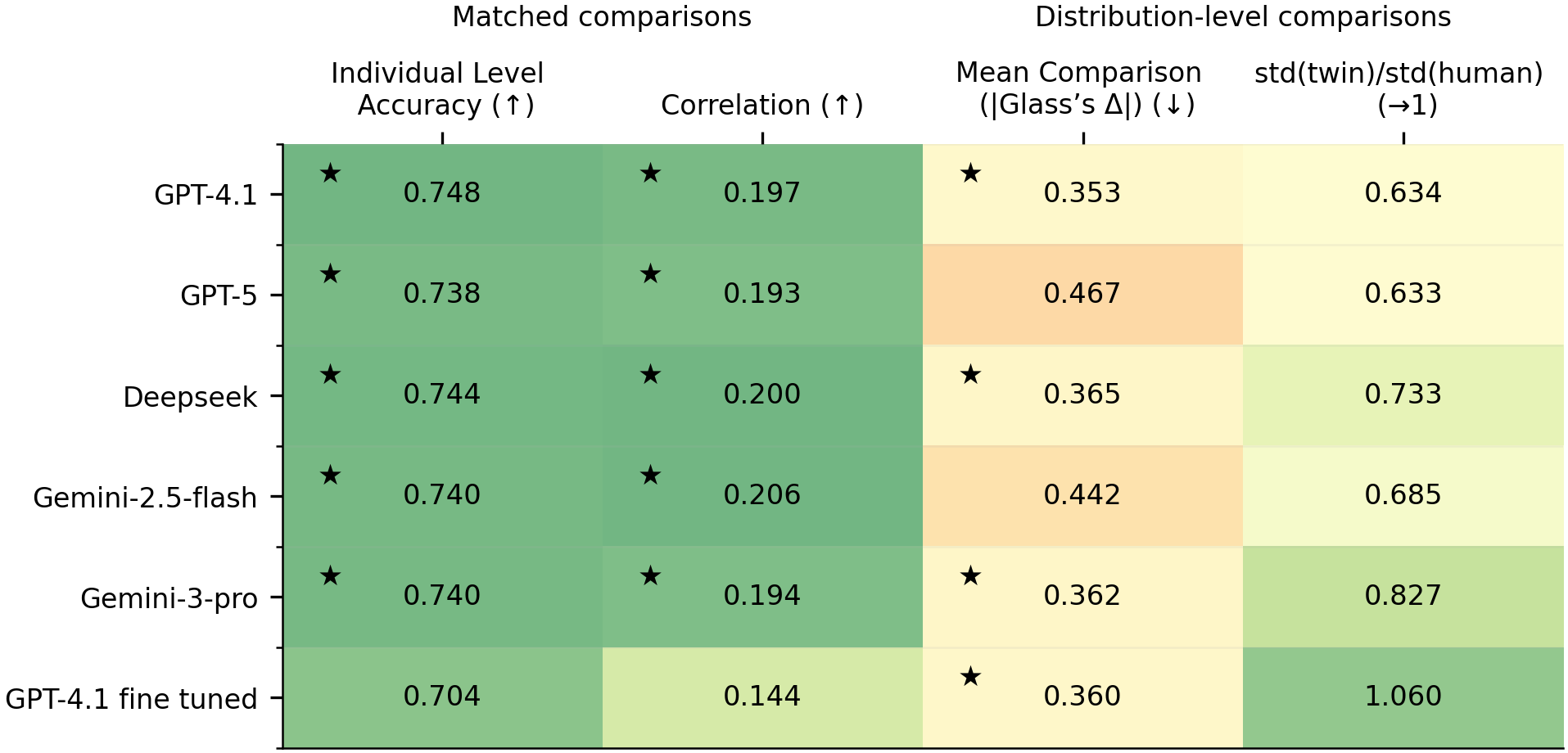}
\caption{Alternative Base LLMs (Combined with Full Persona). *: best performing benchmark, or not significantly different from best at p$<$0.05 (not applicable to ratio of standard deviations), after applying Bonferroni correcting for multiple comparisons in each column.}\label{alt_base}
\end{figure}

Results are presented in Figure \ref{alt_base}. We see that no other base model significantly outperforms GPT-4.1. 

\subsection{Alternative Temperature}

\begin{figure}[H]
\centering
\includegraphics[width=1\textwidth]{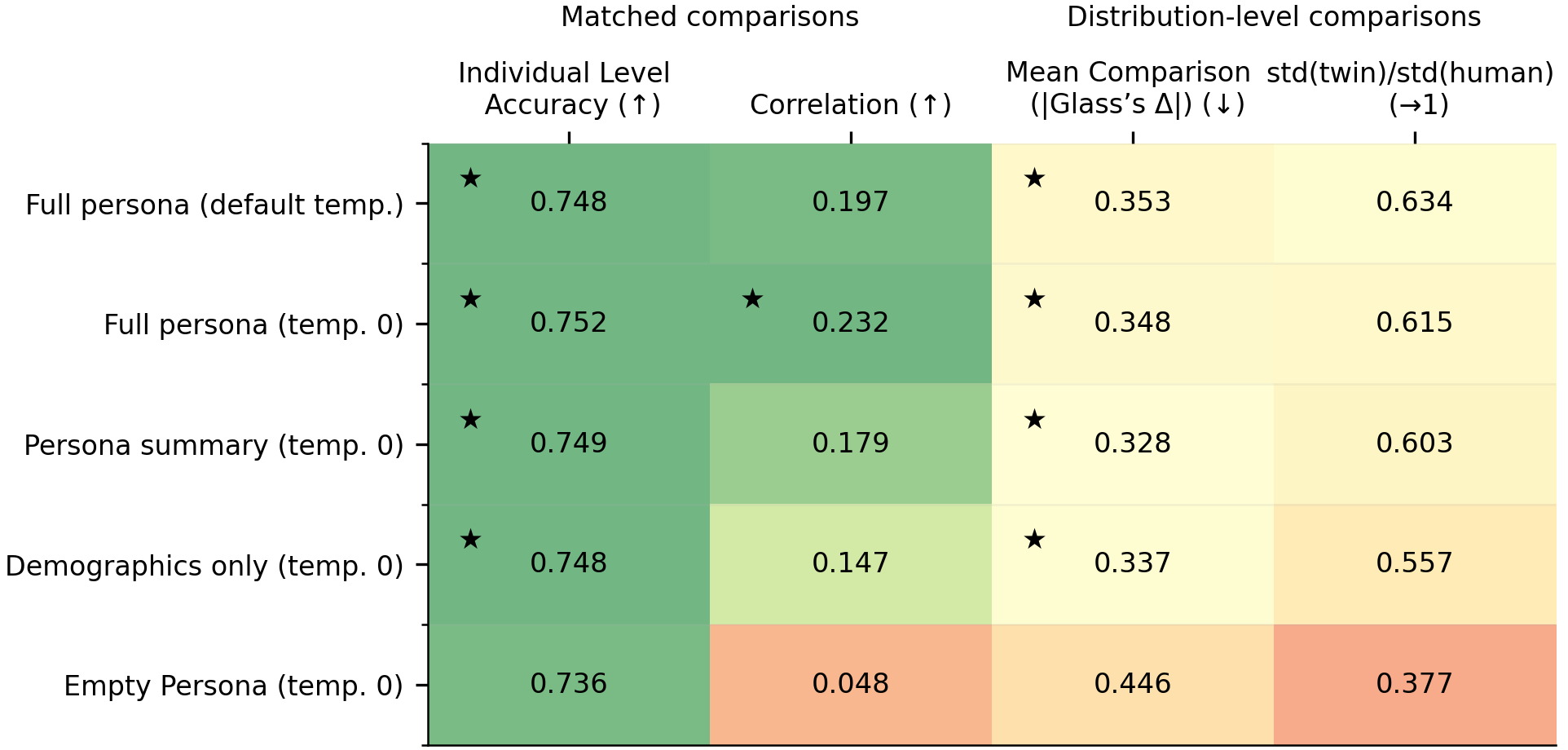}
\caption{Gains from Leveraging Individual-Level Data (Temperature=0). *: best performing benchmark, or not significantly different from best at p$<$0.05 (not applicable to ratio of standard deviations), after applying Bonferroni correcting for multiple comparisons in each column.}\label{alt_temp}
\end{figure}

We replicate the main comparison in the paper (i.e., with GPT-4.1) with a temperature of 0 instead of the default temperature. Figure \ref{alt_temp} displays the results, including the full persona with default temperature as baseline. We see that when temperature is 0, like when temperature is set to default, digital twins based on full personas perform significantly better than generic twins based on demographics on correlation, and not significantly differently on individual-level accuracy. We also see that performance is overall improved when temperature=0.  

\pagebreak

\section{Centaur Experiments}
The state-of-the-art LLMs we used as based models are general-purpose models (e.g., GPT-4.1). A natural question arises: \textit{what if an LLM fine-tuned specifically for human behavioral prediction was used as base model for the creation of digital twins?} One such potential base model is Centaur \cite{binz2025foundation}, a 70B-parameter model fine-tuned from Llama-3.1-70B using data from over 60{,}000 participants, encompassing 10 million choices across 160 behavioral experiments.  

We use the Centaur model \texttt{marcelbinz/Llama-3.1-Centaur-70B} from Hugging Face with BF16 precision. We evaluate Centaur’s performance on our 19 studies using the same setup as other LLMs, with one adjustment: we append \verb|### Output \n ```json| to the end of each prompt instruction to encourage JSON-formatted outputs. After experimenting with several formatting strategies, we found this approach to yield the most consistent results.  

In order to disentangle the impact of leveraging individual-level information from a panel of participants vs. using an LLM fine-tuned on a large cross-sectional dataset as a base model, we follow a 2 (Lllama vs. Centaur) x 2 (empty persona vs. persona summary) experimental design and test the following configurations:
\begin{itemize}
    \item \textbf{Centaur Persona Summary:} Centaur model prompted with Persona Summary instructions.  
    \item \textbf{Centaur Empty Persona:} Centaur model prompted with Empty Persona instructions (no persona information).  
    \item \textbf{Llama Persona Summary:} Base Llama-3.1-70B model prompted with Persona Summary instructions.  
    \item \textbf{Llama Empty Persona:} Base Llama-3.1-70B model prompted with Empty Persona instructions.  
\end{itemize}

We use the \textbf{Persona Summary} instead of the Full Persona to avoid context-length issues that degrade Centaur’s performance. Temperature is set to 0 for all models.  

\begin{figure}[H]
\centering
\includegraphics[width=\textwidth]{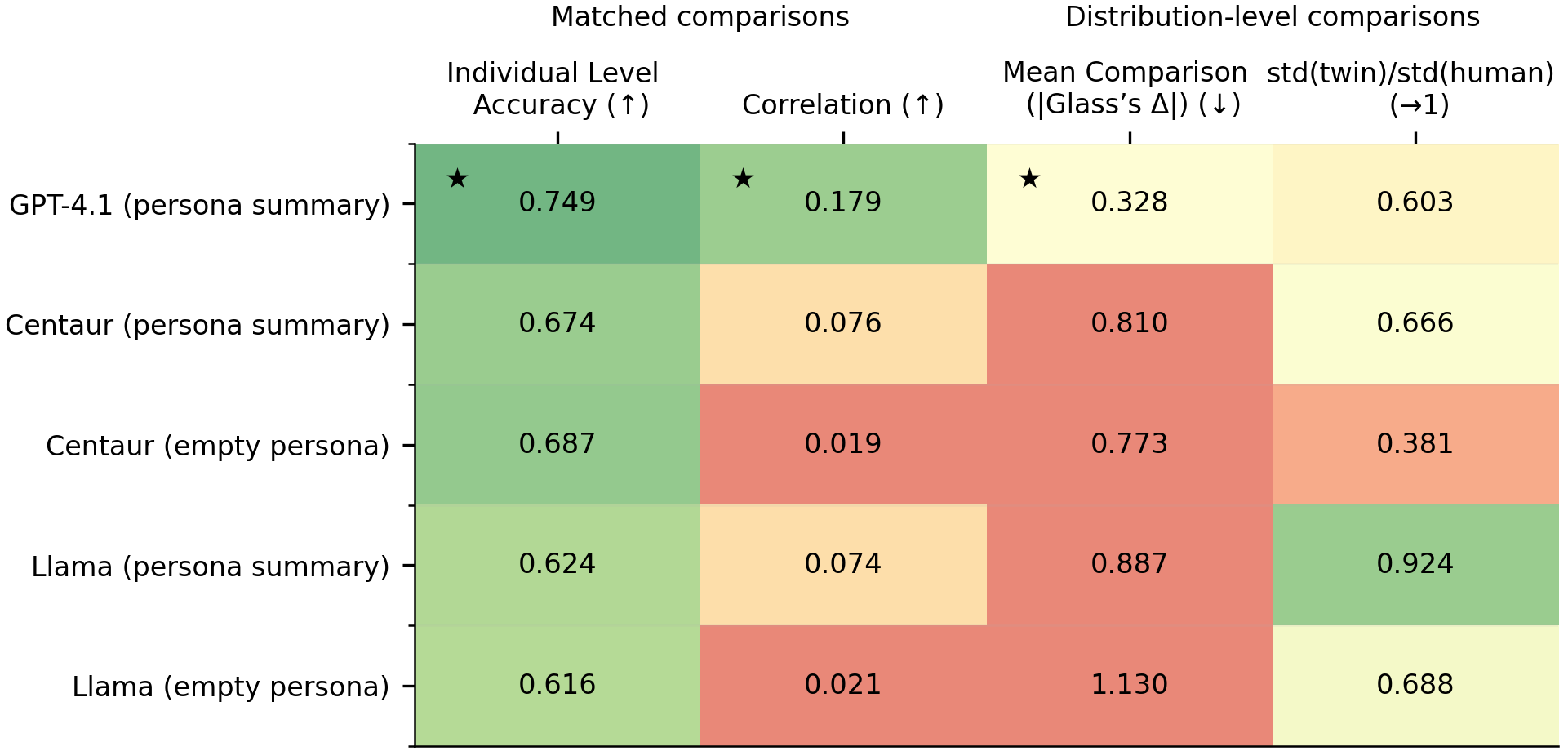}
\caption{Performance Comparison for Centaur and Llama Models Under Different Persona Configurations. *: best performing benchmark, or not significantly different from best at p$<$0.05 (not applicable to ratio of standard deviations), after applying Bonferroni correcting for multiple comparisons in each column.}\label{accuracy-centaur}
\end{figure}

Figure \ref{accuracy-centaur} displays the results, including GPT-4.1 (persona summary) as a reference. We separately discuss results related to the choice of base model vs. the use of individual-leevel information.  

\textbf{Base model: Centaur vs. Llama vs. GPT-4.1.} Centaur outperforms its base model, Llama-3.1-70B, in terms of accuracy. However, the performance of digital twins based on the Centaur base model remains below that of state-of-the-art closed-source models such as GPT-4.1. Several factors may explain this gap:
\begin{enumerate}
    \item \textbf{Prompt formatting:} Centaur was trained to answer single questions, whereas our evaluation involves multi-question inputs.  
    \item \textbf{Out-of-distribution testing:} Our 19 studies include scenarios that may differ significantly from Centaur’s training data.  
    \item \textbf{Limited persona adaptation:} Centaur was not trained to explicitly utilize persona information, which may limit its ability to model human heterogeneity.  
\end{enumerate}
Addressing these issues could further improve Centaur’s behavioral modeling capability—we leave them for future work.  

\textbf{Individual-level information: persona summary vs. empty persona.} Despite lower overall accuracy, both Llama and Centaur exhibit improved \textit{correlations} when persona information is included. This suggests that providing even summarized persona data helps models better capture individual-level heterogeneity, reinforcing the importance of collecting richer personal context in behavioral prediction tasks.

\section{Survey of Expectations of Digital Twin vs. Human Behavior}

To assess expectations of human and digital twin responses, we asked respondents to estimate the average treatment effect from five experiments (tasks) in our data, chosen a priori to capture distinct paradigms (default effects, affective priming, targeting fairness, perceived privacy violation, and idea evaluation).\footnote{Pre-registered as https://aspredicted.org/65ty-73sp.pdf.} We recruited 104 respondents from LinkedIn and the Society for Judgment and Decision-Making (SJDM) listserv; 36 were excluded following preregistered criteria (completion times $<$1/3 or $>$3x the median, repetitive slider responses, or excessive minimum/maximum responses), leaving 68 in the final sample, including 27 academics, 15 managers (mid-level to C-suite) and 20 students.

For each task, respondents were given a brief description and shown human and twin mean values from the original survey in the control condition, then asked to predict the mean in a treatment condition (opt-out vs. opt-in defaults for green energy, gratitude ratings after a neutral vs. gratitude-writing prompt, fairness ratings for non-targeted vs. targeted ad campaigns, privacy ratings for no ads/tracking vs. personalized ads with cross-site tracking, and creativity ratings of human- vs. AI-generated app ideas). 

\begin{figure}[H]
\centering
\includegraphics[width=1\textwidth]{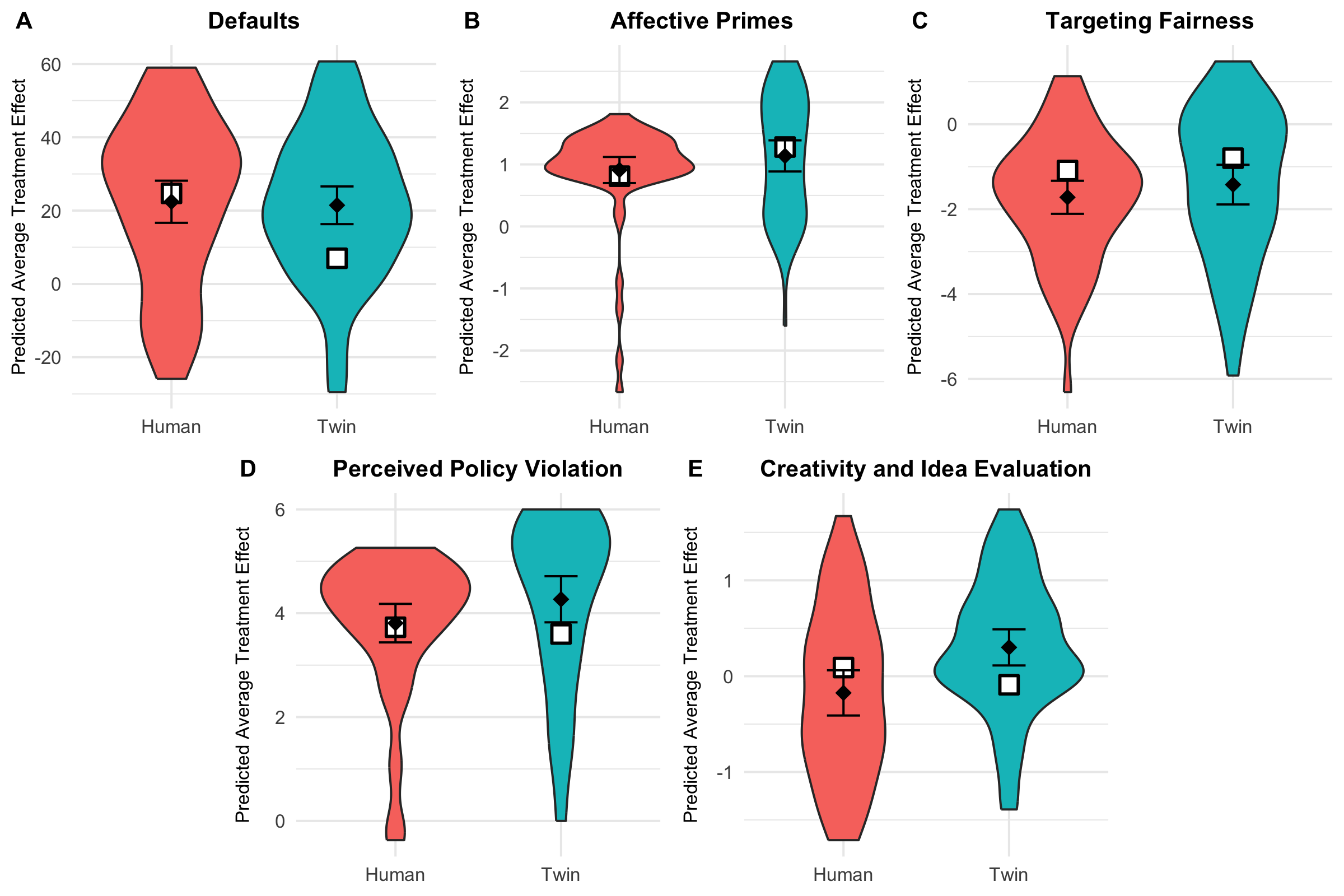}
\caption{Predicted Average Treatment Effects for Humans and Digital Twins across Five Tasks. Black dots and error bars represent the mean predicted average treatment effects and 95\% confidence intervals. White squares represent the mean treatment effect actually observed in the sample. }\label{violin}
\end{figure}

As shown in Figure \ref{violin}, predicted treatment effects were consistently more accurate for humans than for digital twins. In all five tasks, the observed average treatment effect among humans fell within the 95\% confidence intervals of the predicted average treatment effect. By contrast, the confidence interval did not include the observed mean for three of the five digital twin predictions. 

To formally test the difference between human and digital twin prediction accuracy, we estimated a linear mixed-effects model: 
$$err_{it} =  \beta_0 + \beta_1Human_{it} + \beta_2Task_{it} + (1 | ResponseId_i) + \epsilon_{it}$$
where $err_{it}$  is the prediction error for respondent $i$ on task $t$ (abs(predicted value in treatment condition - observed value in treatment condition)/(observed value in treatment condition)), $Human_{it}$ indicates predictions for human vs. digital twins, $Task_{it}$ indexes the five tasks, $(1 | ResponseId_i)$ is a random intercept for each respondent, and $\epsilon_{it}$ is the residual. Results confirmed significantly lower errors for predictions of human responses on average ($\beta$ = -0.072, SE = 0.011, p $<$ 0.001) (see Table \ref{expectations}). 

\pagebreak

\begin{table}[htbp]
\centering
\label{tab:reg_results}
\begin{tabular}{lrrrr}
\toprule
\textbf{Effect} & \textbf{Estimate} & \textbf{Std. Error} & \textbf{t value} & \textbf{Pr($>|t|$)} \\
\midrule
(Intercept)    & 0.4093  & 0.01482 & 27.62  & $5.552\times 10^{-82}$ \\
Human     & -0.07244 & 0.01052 & -6.887 & $9.784\times 10^{-12}$ \\
taskaffPrime   & -0.2555 & 0.01619 & -15.78 & $1.413\times 10^{-50}$ \\
tasktargFair   & -0.1645 & 0.01642 & -10.02 & $1.222\times 10^{-22}$ \\
taskPPV        & -0.1135 & 0.01666 & -6.811 & $1.611\times 10^{-11}$ \\
taskideaEval   & -0.1620 & 0.01676 & -9.667 & $2.939\times 10^{-21}$ \\
\midrule
\multicolumn{5}{l}{\textbf{Random Effects and Model Summary}} \\
$\sigma^2$ & \multicolumn{4}{r}{0.03} \\
$\tau_{00}$ & \multicolumn{4}{r}{0.00} \\
ICC & \multicolumn{4}{r}{0.11} \\
N (ResponseId) & \multicolumn{4}{r}{61} \\
Observations & \multicolumn{4}{r}{1118} \\
Marginal $R^2$ / Conditional $R^2$ & \multicolumn{4}{r}{0.200 / 0.291} \\
\bottomrule
\end{tabular}
\caption{Linear Mixed-Effects Regression of Respondent Prediction Errors across Prediction Targets and Tasks.}\label{expectations}
\end{table}

\pagebreak

\section{Detailed Information on Each Sub-Study}\label{secSI9}

All the studies were conducted on Prolific with human participants. Each study was subsequently run on their corresponding digital twins.
Respondents and their twins were always assigned to the same condition(s) ensuring a fair 1-to-1 comparison.

\subsection{Accuracy Nudges for Misinformation}\label{accuracy-nudges-for-misinformation}

\subsubsection{Main Questions/Hypotheses}

The spread of misinformation across social media is a growing public
policy concern, with severe implications for health, politics, and
science. Recent work suggests that people care about being accurate but
share misinformation because the social media context focuses their
attention on other factors \cite{pennycook2021psychology}. The proposed
solution for inattention-based misinformation sharing is to prime
accuracy before the decision to share information. Research testing this
account \cite{pennycook2021accuracy} demonstrates that asking people to rate
the accuracy of unrelated headlines or state the extent to which they
agree that it is important to share accurate content on social media
before sharing improves truth discernment: that is, reduces sharing of
false headlines compared to true headlines. The current study aims to
replicate this pattern with human participants, and test whether their
digital twins respond similarly to these accuracy nudges.

Specifically, this study aims to replicate study 5 of \cite{pennycook2021accuracy}, using a different set of headlines. While \cite{pennycook2021accuracy} demonstrate the effectiveness of accuracy nudges in the
context of political news headlines, the current study utilizes their
paradigm to test the effectiveness of accuracy nudges in the context of
non-political entertainment news headlines. The decision to sharing
entertaining news may provoke consideration of how entertaining the
headline is, beyond considerations on the accuracy of the news and
trustworthiness of the sources \cite{lane2025misinformation}.
Indeed, people generally prefer and are most likely to engage with
entertaining news \cite{harcup2016news,reuters2023digitalnews,widjaya2024tiktok} and interesting or emotionally arousing
word-of-mouth \cite{berger2011wom,berger2012viral,berger2011arousal}. Testing the effectiveness of accuracy nudges for entertaining
headlines is therefore an important extension of the prior work.

\subsubsection{Methods}

A total of 1,003 human participants on Prolific (and their digital twins) 
took part in the study (50\% female, 50\% male, 0\% other; $M_\text{age} = 49.37$, 
$SD = 14.55$), following the protocol proposed by \cite{pennycook2021accuracy}. 
Participants were randomly assigned to one condition in a 
$2$ (Headline Veracity: False vs.\ True; within-subjects) $\times$ 
$4$ (Intervention: Control vs.\ Active Control vs.\ Treatment vs.\ Importance Treatment; 
between-subjects) mixed design (each human and their digital twin were assigned to the same condition). All participants were asked to indicate 
how likely they would be to share 16 headlines, displayed in random order, 
as the dependent variable. Participants in the Control condition proceeded 
directly to this task, while participants in the Active Control, Treatment, 
and Importance Treatment conditions were first asked to complete the following 
separate tasks:

Active Control: Participants were first told that they would help
pretest actual news headlines for future studies. They were then shown a
headline and asked to rate, ``In your opinion, is the above headline
funny, amusing, or entertaining?'' (1-Extremely unfunny, 6-Extremely
funny). They were randomly assigned to see one of four headlines---two
of which were true and two of which were false. The stimuli were the
same as \cite{pennycook2021accuracy}, with the exception that images were not
included, given that the digital twins cannot process visual images.

Treatment: Similarly, participants were told that they would help
pretest actual news headlines for future studies. They were then shown
two headlines in randomized order and asked to rate, ``In your opinion,
is the above headline accurate?'' (1-Extremely inaccurate, 6-Extremely
accurate). They were randomly assigned to see one of four
headlines---two of which were true and two of which were false. The
stimuli were the same as \cite{pennycook2021accuracy}, again with the
exception that images were not included.

Importance Treatment: Participants were simply asked, ``Do you agree or
disagree that `it is important to only share content on social media
that is accurate and unbiased'?'' (1-Strongly agree, 6-Strongly
disagree).

All participants then responded to the dependent variable. Participants
were presented with 16 news headlines in randomized order. For each
headline, participants were asked to indicate, ``If you were to see the
above article on social media, how likely would you be to share it?''
(1-Extremely unlikely, 6-Extremely likely). All headlines were pretested
to be entertaining and interesting, but not identity relevant \cite{lane2025misinformation}. Eight of the headlines were inspired by
real headlines from the websites of trustworthy news organizations.
These headlines were paired with generally trustworthy news sources. The
remaining 8 headlines were false headlines. These headlines were paired
with generally untrustworthy news sources. Source trustworthiness
perceptions were confirmed at the end of the survey. Table \ref{SI_accuracy_tab1} provides
the headline and source pairings.

\begin{longtable}{@{}>{\raggedright\arraybackslash}p{0.28\linewidth}%
                    >{\raggedright\arraybackslash}p{0.68\linewidth}@{}}
\toprule
\textbf{Source} & \textbf{Headline (True or False)} \\
\midrule
\endhead
The Funny Times & 1. ``Rare Pink Bananas Discovered, Touted as Nature's Cotton Candy'' (F) \\
                & 2. ``The Real Jurassic Park? Dinosaur DNA Successfully Extracted from Fossil'' (F) \\
Quora.com       & 3. ``Researchers Develop Plants That Emit Enough Energy to Power a Tiny-House'' (F) \\
                & 4. ``Innovative Study Finds Dolphins Can Be Trained to Detect Cancer'' (F) \\
The National Enquirer & 5. ``The Mystery of the Bermuda Triangle: New Theory Suggests Surprising Explanation'' (F) \\
                & 6. ``Scientists Discover a New Species of Glow-in-the-Dark-Bees in the Amazon'' (F) \\
Reddit.com      & 7. ``New Fossil Discovery Suggests Rat-Sized Elephants Once Inhabited the Earth'' (F) \\
                & 8. ``Biologists Stumble Upon a Singing Spider Species, Dubbed 'Nature's Vocalweaver''' (F) \\
The Wall Street Journal & 9. ``World's Smallest Gold Coin Features Albert Einstein Sticking Out Tongue'' (T) \\
                & 10. ``Authorities Scramble To Find Stolen Solid Gold Toilet'' (T) \\
The Economist   & 11. ``Applications To Become Japanese Billionaire Yusaku Maezawa's Girlfriend Have Topped 20,000'' (T) \\
                & 12. ``Woman Boards Flight to Find Her Seat Assignment Is In The Plane's Bathroom'' (T) \\
PBS News        & 13. ``Kansas Man Asks Judge To Allow Him To Have Sword Fight With Ex-Wife'' (T) \\
                & 14. ``Teen Discovers Rare New Planet 3 Days Into NASA Internship'' (T) \\
BBC News        & 15. ``Two Elderly Men Sneak Out Of Nursing Home To Attend Heavy Metal Festival'' (T) \\
                & 16. ``Spotify Launches Playlists For Dogs Left Home Alone'' (T) \\
\bottomrule
\caption{Headline and Source Pairings. True and False headlines are indicated by (T) and (F), respectively. Untrustworthy sources include: The Funny Times, Quora.com, The National Enquirer, and Reddit.com. Trustworthy sources include: The Wall Street Journal, The Economist, PBS News, and BBC News.\label{SI_accuracy_tab1}} \\
\end{longtable}

Participants in the Control, Active Control, and Treatment condition
next responded to the same question shown earlier to participants in the
Importance Treatment condition: ``Do you agree or disagree that `it is
important to only share content on social media that is accurate and
unbiased'?'' (1-Strongly agree, 6-Strongly disagree). All participants
then responded to exploratory questions intended to capture stated
social media sharing preferences. As a measure of goal prioritization
when sharing entertaining news, participants were first asked to
indicate, ``When you share information with others on social media, is
it more important that the information is:'' (1-Verifiably correct,
6-Entertaining; \cite{lane2025misinformation}). Participants then
indicated, ``Would you ever consider sharing news like you saw in the
study today on social media?''.

All participants then indicated their perceptions of source
trustworthiness (``How trustworthy (from 0--100\%) do you think this
source is?'') for each source, presented in randomized order.
Participants across conditions rated sources in the True headline
veracity condition to be more trustworthy than sources in the False
condition ($M_\text{True} = 72.34\%$, $SD = 18.31$;
$M_\text{False} = 28.28\%$, $SD = 14.00$), $t(2005) = 96.12$, $p < .001$,
$d = 2.69$. Finally, participants indicated whether they responded
randomly at any point in the survey (0.40\% = Yes, 99.60\% = No) and
whether they searched the internet for any of the headlines (0.40\% =
Yes, 99.60\% = No) before being debriefed, thanked, and paid.

\subsubsection{Results -- Pre-registered Analyses}
\begin{figure}[H]
\centering
\includegraphics[width=1\textwidth]{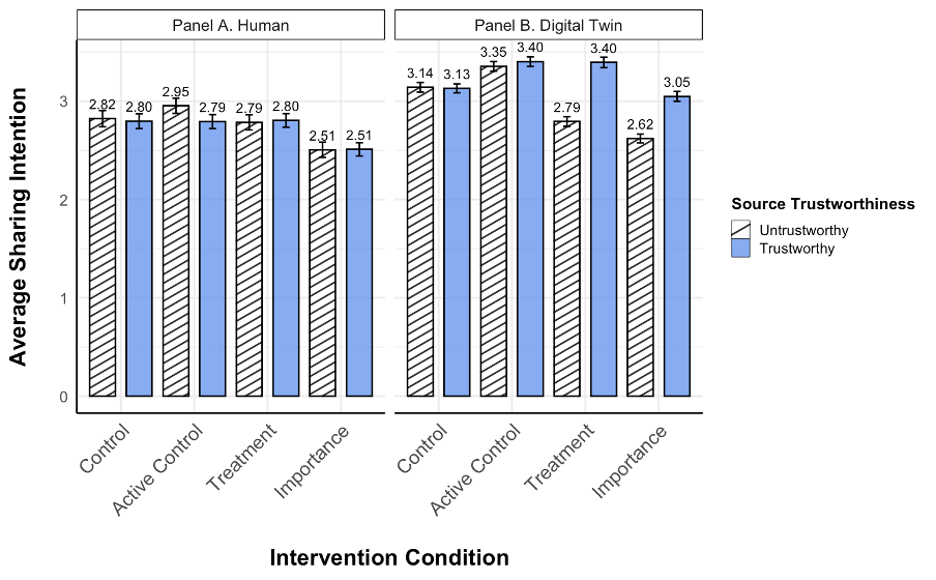}
\caption{Sharing Intentions by Headline Veracity and Intervention.
NOTE: Error bars indicate 95\% confidence intervals based on robust
standard errors clustered on participant.
}\label{SI_accuracy_1}
\end{figure}

A linear mixed-effects regression was estimated\footnote{Linear
  mixed-effects models were estimated using the \texttt{lmerTest}
  package, which provides $p$-values and $F$-tests using Satterthwaite's
  approximation.} with participant as a random effect, average sharing
intentions as the dependent variable, and Participant Type (Human $= 0$,
Digital Twin $= 1$), Headline Veracity (False $= 0$, True $= 1$),
Intervention (Control $= 0$, Active Control $= 1$, Treatment $= 2$,
Importance Treatment $= 3$), and their interactions as predictors. The
results revealed a significant main effect of Participant Type
($F(1, 2997) = 33.41$, $p < .001$), a non-significant main effect of
Headline Veracity ($F(1, 2997) = 1.24$, $p = .27$), along with a
significant main effect of Intervention ($F(3, 2997) = 8.89$, $p < .001$),
qualified by a significant three-way interaction
($F(3, 2997) = 5.75$, $p < .001$), indicating that, at an aggregate
level, accuracy nudges affected how the digital twins shared true vs.\
false headlines differently than humans. Figure \ref{SI_accuracy_1} shows average sharing
intentions across all conditions for humans (Panel~A) and their digital
twins (Panel B). See Table \ref{SI_accuracy_tab2} for the full regression results.

\newcolumntype{C}[1]{>{\centering\arraybackslash}p{#1}}
\newcolumntype{L}[1]{>{\RaggedRight\arraybackslash}p{#1}}

\begin{longtable}{@{}C{0.40\linewidth}C{0.1\linewidth}C{0.1\linewidth}C{0.1\linewidth}C{0.1\linewidth}C{0.1\linewidth}C{0.1\linewidth}@{}}
\toprule
\textbf{Predictors} & \textbf{Estimate} & \textbf{Std. Error} & \multicolumn{2}{c}{\textbf{95\% CI}} & \textbf{t} \\
\midrule
\endhead
(Intercept) & 2.82\textsuperscript{***} & 0.06 & \multicolumn{2}{c}{[2.70, 2.95]} & 44.73 \\
Participant Type & 0.32\textsuperscript{***} & 0.07 & \multicolumn{2}{c}{[0.18, 0.46]} & 4.38 \\
Headline Veracity & -0.03 & 0.07 & \multicolumn{2}{c}{[-0.17, 0.12]} & -0.35 \\
Active Control & 0.13 & 0.09 & \multicolumn{2}{c}{[-0.05, 0.31]} & 1.46 \\
Treatment & -0.04 & 0.09 & \multicolumn{2}{c}{[-0.21, 0.14]} & -0.42 \\
Importance Treatment & 0.32\textsuperscript{***} & 0.09 & \multicolumn{2}{c}{[0.14, 0.49]} & 3.56 \\
Participant Type $\times$ Headline Veracity & 0.01 & 0.10 & \multicolumn{2}{c}{[-0.19, 0.22]} & 0.13 \\
Participant Type $\times$ Active Control & 0.08 & 0.10 & \multicolumn{2}{c}{[-0.12, 0.28]} & 0.80 \\
Participant Type $\times$ Treatment & -0.31\textsuperscript{**} & 0.10 & \multicolumn{2}{c}{[-0.51, -0.11]} & -3.00 \\
Participant Type $\times$ Importance Treatment & -0.20\textsuperscript{*} & 0.10 & \multicolumn{2}{c}{[-0.41, 0.00]} & -1.98 \\
Participant Type $\times$ Active Control & -0.14 & 0.10 & \multicolumn{2}{c}{[-0.34, 0.07]} & -1.31 \\
Participant Type $\times$ Treatment & 0.05 & 0.10 & \multicolumn{2}{c}{[-0.16, 0.25]} & 0.44 \\
Participant Type $\times$ Importance Treatment & 0.03 & 0.10 & \multicolumn{2}{c}{[-0.17, 0.23]} & 0.30 \\
Participant Type $\times$ Headline Veracity $\times$ Active Control & 0.20 & 0.15 & \multicolumn{2}{c}{[-0.09, 0.48]} & 1.34 \\
Participant Type $\times$ Headline Veracity $\times$ Treatment & 0.57\textsuperscript{***} & 0.15 & \multicolumn{2}{c}{[0.28, 0.85]} & 3.88 \\
Participant Type $\times$ Headline Veracity $\times$ Importance & 0.41\textsuperscript{**} & 0.15 & \multicolumn{2}{c}{[0.12, 0.69]} & 2.81 \\
\midrule
\multicolumn{6}{l}{\textbf{Random Effects}} \\
$\sigma^2$ & \multicolumn{5}{l}{0.67} \\
$\tau_{00}$ (TWIN\_ID) & \multicolumn{5}{l}{0.33} \\
ICC & \multicolumn{5}{l}{0.33} \\
N (TWIN\_ID) & \multicolumn{5}{l}{1003} \\
Observations & \multicolumn{5}{l}{4012} \\
Marginal $R^2$ / Conditional $R^2$ & \multicolumn{5}{l}{0.073 / 0.383} \\
\bottomrule
\caption{Model A: Random Intercepts. Headline Veracity, Intervention, and Participant Type on Sharing. * $p<0.05$, ** $p<0.01$, *** $p<0.001$.}
\label{SI_accuracy_tab2}
\end{longtable}

For humans, accuracy nudges did not improve truth discernment when
sharing entertaining news (Figure \ref{SI_accuracy_1}, Panel A). As in \cite{pennycook2021accuracy}, there were no significant differences in sharing intentions
between the Control and Active Control conditions (Headline Veracity
$\times$ Intervention: $b = 0.14$, 95\% CI $[-0.07,\,0.34]$,
$t(2997) = 1.31$, $p = .56$). In contrast to the previous findings,
neither accuracy nudge significantly increased sharing discernment
relative to the controls (e.g., Headline Veracity $\times$ Intervention:
Treatment vs.\ Control, $b = -0.05$, 95\% CI $[-0.25,\,0.16]$,
$t(2997) = -0.44$, $p = .97$; Importance Treatment vs.\ Control,
$b = -0.03$, 95\% CI $[-0.23,\,0.17]$, $t(2997) = -0.30$, $p = .99$).
See Table \ref{SI_accuracy_tab3} for pairwise comparisons.

\begin{longtable}{@{}C{0.40\linewidth}C{0.1\linewidth}C{0.1\linewidth}C{0.1\linewidth}C{0.1\linewidth}C{0.1\linewidth}C{0.1\linewidth}@{}}
\toprule
\textbf{Contrast} & \textbf{Estimate} & \textbf{SE} & \textbf{df} & \textbf{t-ratio} & \textbf{p-value} & \textbf{95\% CI} \\
\midrule
\endhead
Control -- Active Control & 0.14 & 0.10 & 2997 & 1.31 & 0.56 & [-0.07, 0.34] \\
Control -- Treatment & -0.05 & 0.10 & 2997 & -0.44 & 0.97 & [-0.25, 0.16] \\
Control -- Importance & -0.03 & 0.10 & 2997 & -0.30 & 0.99 & [-0.23, 0.17] \\
Active Control -- Treatment & -0.18 & 0.10 & 2997 & -1.74 & 0.30 & [-0.38, 0.02] \\
Active Control -- Importance & -0.17 & 0.10 & 2997 & -1.61 & 0.37 & [-0.37, 0.04] \\
Treatment -- Importance & 0.01 & 0.10 & 2997 & 0.14 & 1.00 & [-0.19, 0.22] \\
\bottomrule
\caption{Pairwise comparisons for humans.}
\label{SI_accuracy_tab3}
\end{longtable}

For digital twins, however, accuracy nudges successfully improved truth
discernment when sharing entertaining news (Figure \ref{SI_accuracy_1}, Panel B). As in
\cite{pennycook2021accuracy}, there were no significant differences in
sharing intentions between the Control and Active Control conditions
(Headline Veracity $\times$ Intervention: $b = -0.06$, 95\% CI
$[-0.26,\,0.14]$, $t(2997) = -0.58$, $p = .94$), and both treatments
significantly increased sharing discernment relative to the controls
(e.g., Headline Veracity $\times$ Intervention: Treatment vs.\ Control,
$b = -0.61$, 95\% CI $[-0.82,\,-0.41]$, $t(2997) = -5.93$, $p < .001$;
Importance Treatment vs.\ Control, $b = -0.44$, 95\% CI
$[-0.64,\,-0.24]$, $t(2997) = -4.27$, $p < .001$). See Table \ref{SI_accuracy_tab4} for pairwise comparisons.

\begin{longtable}{lcccccc}
\toprule
\textbf{Contrast} & \textbf{Estimate} & \textbf{SE} & \textbf{df} & \textbf{t-ratio} & \textbf{p-value} & \textbf{95\% CI} \\
\midrule
\endhead
Control -- Active Control   & -0.06 & 0.10 & 2997 & -0.58 & 0.94 & [-0.26, 0.14] \\
Control -- Treatment        & -0.61 & 0.10 & 2997 & -5.93 & $p < .001$ & [-0.82, -0.41] \\
Control -- Importance       & -0.44 & 0.10 & 2997 & -4.27 & $p < .001$ & [-0.64, -0.24] \\
Active Control -- Treatment & -0.55 & 0.10 & 2997 & -5.33 & $p < .001$ & [-0.76, -0.35] \\
Active Control -- Importance& -0.38 & 0.10 & 2997 & -3.67 & $p < .001$ & [-0.58, -0.18] \\
Treatment -- Importance     &  0.17 & 0.10 & 2997 &  1.68 & 0.33 & [-0.03, 0.38] \\
\bottomrule
\caption{Pairwise comparisons for twins.}
\label{SI_accuracy_tab4}
\end{longtable}

\subsubsection{Results - Additional Analyses (Non-Preregistered)}

The above pre-registered analyses treated participant as a random
intercept, capturing individual differences in baseline sharing
intentions but assuming that the effects of the predictors were
consistent across participants. However, this approach does not account
for the possibility that participants may vary in their responsiveness
to the manipulations. To provide a more conservative test, a linear
mixed-effects regression was estimated with random intercepts and random
slopes for each within-subject predictor. This specification allows the
effects of predictors to vary across individuals, capturing
heterogeneity in participants' responses. The model fit improved
significantly ($\chi^2(2) = 935.23$, $p < .001$), yet the three-way
interaction remained significant ($F(3, 1998) = 12.09$, $p < .001$),
suggesting that the findings from the pre-registered model are robust to
alternative model specifications. See Table \ref{SI_accuracy_tab5} for the full regression
results.

\begin{longtable}{@{}C{0.40\linewidth}C{0.1\linewidth}C{0.1\linewidth}C{0.15\linewidth}C{0.1\linewidth}@{}}
\toprule
\textbf{Predictors} & \textbf{Estimates} & \textbf{SE} & \textbf{95\% CI} & \textbf{t} \\
\midrule
\endhead
(Intercept)                         & 2.82\textsuperscript{***} & 0.07 & [2.68, 2.97] & 38.92 \\
Participant Type                    & 0.32\textsuperscript{***} & 0.08 & [0.16, 0.48] & 3.90 \\
Headline Veracity                   & -0.03                     & 0.05 & [-0.12, 0.07] & -0.51 \\
Condition [Active Control]          & 0.13                      & 0.10 & [-0.07, 0.33] & 1.27 \\
Treatment                           & -0.04                     & 0.10 & [-0.24, 0.16] & -0.37 \\
Importance                          & -0.32\textsuperscript{**} & 0.10 & [-0.52, -0.12] & -3.09 \\
Participant Type $\times$ Headline Veracity & 0.01              & 0.07 & [-0.13, 0.15] & 0.20 \\
Participant Type $\times$ Active Control   & 0.08              & 0.12 & [-0.15, 0.31] & 0.71 \\
Participant Type $\times$ Treatment        & -0.31\textsuperscript{**} & 0.12 & [-0.54, -0.08] & -2.67 \\
Participant Type $\times$ Importance       & -0.20             & 0.12 & [-0.43, 0.02] & -1.76 \\
Participant Type $\times$ Active Control   & -0.14             & 0.07 & [-0.28, 0.00] & -1.90 \\
Participant Type $\times$ Treatment        & 0.05              & 0.07 & [-0.09, 0.19] & 0.64 \\
Participant Type $\times$ Importance       & 0.03              & 0.07 & [-0.11, 0.17] & 0.43 \\
Participant Type $\times$ Headline Veracity $\times$ Active Control & 0.20 & 0.10 & [-0.00, 0.39] & 1.94 \\
Participant Type $\times$ Headline Veracity $\times$ Treatment      & 0.57\textsuperscript{***} & 0.10 & [0.37, 0.77] & 5.63 \\
Participant Type $\times$ Headline Veracity $\times$ Importance     & 0.41\textsuperscript{***} & 0.10 & [0.21, 0.61] & 4.07 \\
\midrule
\multicolumn{5}{l}{\textbf{Random Effects}} \\
$\sigma^2$                          & \multicolumn{4}{l}{0.32} \\
$\tau_{00}$ (TWIN\_ID)              & \multicolumn{4}{l}{1.01} \\
$\tau_{11}$ (TWIN\_ID.ParticipantType) & \multicolumn{4}{l}{1.05} \\
$\rho_{01}$ (TWIN\_ID)              & \multicolumn{4}{l}{-0.82} \\
ICC                                 & \multicolumn{4}{l}{0.68} \\
N (TWIN\_ID)                        & \multicolumn{4}{l}{1003} \\
Observations                        & \multicolumn{4}{l}{4012} \\
Marginal $R^2$ / Conditional $R^2$  & \multicolumn{4}{l}{0.073 / 0.706} \\
\bottomrule
\caption{Model B: Random Intercepts and Random Slopes. Headline veracity, intervention, and participant type on sharing. \emph{* p$<$0.05, ** p$<$0.01, *** p$<$0.001}.}
\label{SI_accuracy_tab5}
\end{longtable}

Additionally, as a robustness check, a linear mixed-effects model with
random intercepts and slopes (nested) was estimated to account for the
repeated-measures structure of the data, specifically participants'
repeated responses to multiple headlines. Similar to the pre-registered
model with random intercepts (Model A), this model revealed significant
main effects of Participant Type ($F(1, 999) = 97.25$, $p < .001$) as
well as a significant effect of Intervention ($F(3, 999) = 17.89$,
$p < .001$). Crucially, the results also revealed a significant
three-way interaction ($F(3, 999) = 12.76$, $p < .001$), further
suggesting that the findings from the pre-registered model are robust to
alternative model specifications. See Table \ref{SI_accuracy_tab6} for the full regression
results.

\begin{longtable}{@{}C{0.40\linewidth}C{0.1\linewidth}C{0.1\linewidth}C{0.15\linewidth}C{0.1\linewidth}@{}}
\toprule
\textbf{Predictors} & \textbf{Estimates} & \textbf{SE} & \textbf{95\% CI} & \textbf{t} \\
\midrule
\endhead
(Intercept)                         & 2.82\textsuperscript{***} & 0.17 & [2.48, 3.16] & 16.28 \\
Participant Type                    & 0.32\textsuperscript{***} & 0.08 & [0.16, 0.48] & 3.84 \\
Headline Veracity                   & -0.03                     & 0.23 & [-0.48, 0.43] & -0.11 \\
Active Control                      & 0.13                      & 0.09 & [-0.05, 0.31] & 1.41 \\
Treatment                           & -0.04                     & 0.09 & [-0.22, 0.14] & -0.41 \\
Importance Treatment                & -0.32\textsuperscript{***} & 0.09 & [-0.50, -0.14] & -3.45 \\
Participant Type $\times$ Headline Veracity & 0.01              & 0.07 & [-0.12, 0.15] & 0.20 \\
Participant Type $\times$ Active Control   & 0.08              & 0.12 & [-0.15, 0.31] & 0.70 \\
Participant Type $\times$ Treatment        & -0.31\textsuperscript{**} & 0.12 & [-0.54, -0.08] & -2.63 \\
Participant Type $\times$ Importance       & -0.20             & 0.12 & [-0.43, 0.03] & -1.74 \\
Headline Veracity $\times$ Active Control  & -0.14             & 0.07 & [-0.28, 0.00] & -1.90 \\
Headline Veracity $\times$ Treatment       & 0.05              & 0.07 & [-0.09, 0.19] & 0.64 \\
Headline Veracity $\times$ Importance      & 0.03              & 0.07 & [-0.11, 0.17] & 0.43 \\
Participant Type $\times$ Headline Veracity $\times$ Active Control & 0.20\textsuperscript{*} & 0.10 & [0.00, 0.39] & 1.99 \\
Participant Type $\times$ Headline Veracity $\times$ Treatment      & 0.57\textsuperscript{***} & 0.10 & [0.38, 0.76] & 5.78 \\
Participant Type $\times$ Headline Veracity $\times$ Importance     & 0.41\textsuperscript{***} & 0.10 & [0.22, 0.60] & 4.18 \\
\midrule
\multicolumn{5}{l}{\textbf{Random Effects}} \\
$\sigma^2$                          & \multicolumn{4}{l}{0.97} \\
$\tau_{00}$ (group\_id:TWIN\_ID)    & \multicolumn{4}{l}{0.75} \\
$\tau_{00}$ (TWIN\_ID)              & \multicolumn{4}{l}{0.19} \\
$\tau_{00}$ (headline\_id)          & \multicolumn{4}{l}{0.21} \\
$\tau_{11}$ (group\_id:TWIN\_ID.Head.Veracity) & \multicolumn{4}{l}{0.36} \\
$\tau_{11}$ (TWIN\_ID.Head.Veracity) & \multicolumn{4}{l}{0.03} \\
$\rho_{01}$ (group\_id:TWIN\_ID)    & \multicolumn{4}{l}{-0.41} \\
$\rho_{01}$ (TWIN\_ID)              & \multicolumn{4}{l}{-0.52} \\
ICC                                 & \multicolumn{4}{l}{0.53} \\
N (group\_id)                       & \multicolumn{4}{l}{2006} \\
N (TWIN\_ID)                        & \multicolumn{4}{l}{1003} \\
N (headline\_id)                    & \multicolumn{4}{l}{16} \\
Observations                        & \multicolumn{4}{l}{32,096} \\
Marginal $R^2$ / Conditional $R^2$  & \multicolumn{4}{l}{0.037 / 0.546} \\
\bottomrule
\caption{Model C: Headline veracity, intervention, and participant type on sharing (nested random effects). \emph{* p$<$0.05, ** p$<$0.01, *** p$<$0.001}.}
\label{SI_accuracy_tab6}
\end{longtable}

\begin{figure}[H]
\centering
\includegraphics[width=1\textwidth]{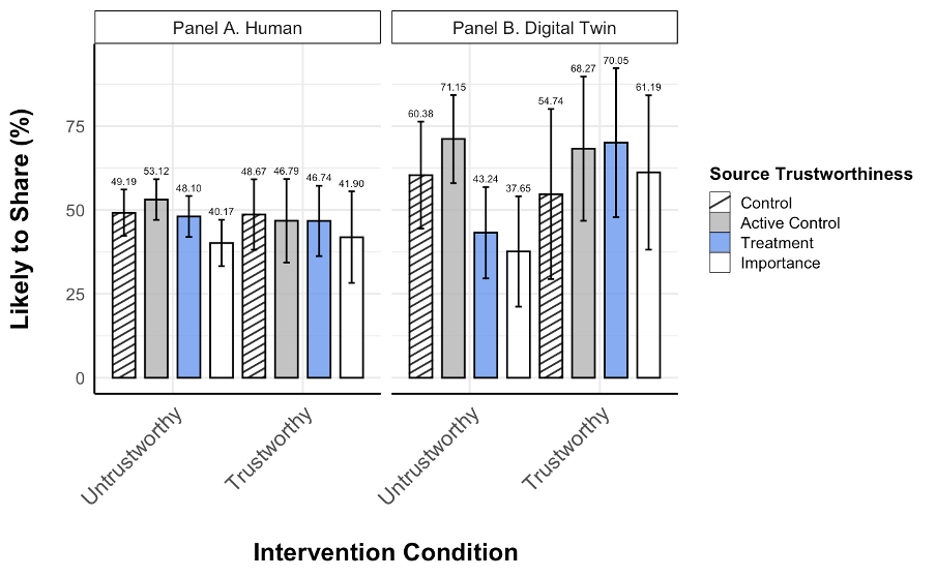}
\caption{Proportion (\%) Likely to Share by Headline Veracity and
Intervention. NOTE: Shown here is the fraction of ``likely' responses
(responses above the midpoint of the six-point Likert scale) by Headline
Veracity and Intervention condition; the full distributions of responses
are shown below, in figure \ref{SI_accuracy_3}. As with \cite{pennycook2021accuracy}, these
analyses focus only on participants who indicated that they would
consider sharing this social media content. Analysis including all
participants does not change the results. Error bars indicate 95\%
confidence intervals based on robust standard errors clustered on
participant and headline.
}\label{SI_accuracy_2}
\end{figure}

As an additional test, \cite{pennycook2021accuracy} analysis plan was next
replicated for humans and their digital twins, separately. While the
pre-registered analysis demonstrates sharing intentions for the full
range of responses, \cite{pennycook2021accuracy} rescale the continuous
sharing intentions into a binary variable capturing responses at or
above the midpoint at ``likely to share'' (= 1) and below the midpoint
as ``unlikely to share'' (= 0), and filtered out participants who would
not consider sharing these articles on social media. Because the human
dataset revealed significant differences between the Control and Active
Control conditions, these groups were not pooled to maintain
comparability. This results in a slight deviation from the analysis in
\cite{pennycook2021accuracy}. Using this framework, their exact analysis was
replicated for humans and their digital twins separately, treating the
rescaled likelihood to share variable as the DV with Headline Veracity
condition, Intervention condition, and their interaction as predictors,
applying robust standard errors. Figure \ref{SI_accuracy_2} shows proportion of sharing
across all conditions for humans (Panel A) and their digital twins
(Panel B).

Replicating the main findings for humans (Model A), accuracy nudges did
not improve truth discernment when sharing entertaining news
(Figure \ref{SI_accuracy_2}, Panel A). In contrast to \cite{pennycook2021accuracy}, sharing
discernment was marginally higher in the Active Control compared to the
Control condition (Headline Veracity $\times$ Intervention: $b = -0.03$,
95\% CI $[-0.07,\,0.00]$, $F(1, 5132) = 3.53$, $p = .06$), and neither
treatment significantly increased sharing discernment relative to the
controls (Headline Veracity $\times$ Intervention: Treatment,
$b = 0.01$, 95\% CI $[-0.05,\,0.06]$, $F(1, 9656) = 0.05$, $p = .82$;
Importance Treatment, $b = 0.01$, 95\% CI $[-0.04,\,0.05]$,
$F(1, 9656) = 0.06$, $p = .80$). See Table \ref{SI_accuracy_3} for the full regression
results.


\begin{longtable}{@{}C{0.40\linewidth}C{0.1\linewidth}C{0.1\linewidth}C{0.15\linewidth}C{0.1\linewidth}@{}}
\toprule
\multicolumn{5}{c}{\textbf{Humans: Participants who are willing to share content}} \\
\midrule
\multicolumn{5}{l}{\textbf{Model 1: Controls only}} \\
\midrule
\textbf{Predictors} & \textbf{Estimate} & \textbf{SE} & \textbf{95\% CI} & \textbf{t} \\
\midrule
\endhead

(Intercept) & 0.45\textsuperscript{***} & 0.03 & [0.39, 0.50] & 17.82 \\
Active Control & 0.04 & 0.02 & [0.01, 0.09] & 1.64 \\
Headline Veracity & -0.01 & 0.04 & [-0.10, 0.08] & -0.26 \\
Active Control $\times$ Headline Veracity & -0.03 & 0.02 & [-0.07, 0.00] & -1.88 \\

\midrule
Observations & \multicolumn{4}{l}{5,136} \\
$R^2$ / Adjusted $R^2$ & \multicolumn{4}{l}{0.003 / 0.003} \\

\midrule
\multicolumn{5}{l}{\textbf{Model 2: All conditions}} \\
\midrule
\textbf{Predictors} & \textbf{Estimate} & \textbf{SE} & \textbf{95\% CI} & \textbf{t} \\
\midrule

(Intercept) & 0.45\textsuperscript{***} & 0.03 & [0.39, 0.50] & 17.82 \\
Active Control & 0.04 & 0.02 & [-0.01, 0.09] & 1.64 \\
Headline Veracity & -0.01 & 0.04 & [-0.10, 0.08] & -0.26 \\
Active Control $\times$ Headline Veracity & -0.03 & 0.02 & [-0.07, 0.00] & -1.88 \\
Treatment & -0.01 & 0.03 & [-0.07, 0.06] & -0.22 \\
Importance Treatment & -0.06\textsuperscript{*} & 0.03 & [-0.12, 0.00] & -2.19 \\
Treatment $\times$ Headline Veracity & 0.01 & 0.03 & [-0.05, 0.07] & 0.23 \\
Importance Treatment $\times$ Headline Veracity & 0.01 & 0.02 & [-0.04, 0.06] & 0.25 \\

\midrule
Observations & \multicolumn{4}{l}{9,664} \\
$R^2$ / Adjusted $R^2$ & \multicolumn{4}{l}{0.009 / 0.008} \\
\bottomrule

\caption{Headline Veracity, Intervention and Participant Type on Humans' Sharing Intentions. \emph{* p$<$0.05, ** p$<$0.01, *** p$<$0.001}. When reporting the effects, separate F-tests were conducted; therefore, the reported F statistics are not included in this table, following the reporting approach used by \cite{pennycook2021accuracy}.}
\label{SI_accuracy_3}
\end{longtable}

Again, replicating the main findings for digital twins (Model~A),
accuracy nudges improved truth discernment when sharing entertaining
news (Figure \ref{SI_accuracy_2}, Panel B). As in \cite{pennycook2021accuracy}, there were no
significant differences in sharing intentions between the Control and
Active Control conditions (Headline Veracity $\times$ Intervention:
$b = 0.02$, 95\% CI $[-0.02,\,0.06]$, $F(1, 4046) = 1.17$, $p = .28$),
and both treatments significantly increased sharing discernment relative
to the controls (Headline Veracity $\times$ Intervention: Treatment,
$b = 0.14$, 95\% CI $[0.08,\,0.20]$, $F(1, 7208) = 23.30$, $p < .001$;
Importance Treatment, $b = 0.12$, 95\% CI $[0.07,\,0.17]$, $F(1, 7208) =
21.80$, $p < .001$). See Table \ref{SI_accuracy_4} for the full regression results.


\begin{longtable}{@{}C{0.40\linewidth}C{0.1\linewidth}C{0.1\linewidth}C{0.15\linewidth}C{0.1\linewidth}@{}}
\toprule
\multicolumn{5}{c}{\textbf{Twins: Participants who are willing to share content}} \\
\midrule
\multicolumn{5}{l}{\textbf{Model 3: Controls only}} \\
\midrule
\textbf{Predictors} & \textbf{Estimate} & \textbf{SE} & \textbf{CI} & \textbf{t} \\
\midrule
\endhead

(Intercept) & 0.53\textsuperscript{***} & 0.03 & [0.46, 0.60] & 15.97 \\
Active Control & 0.04\textsuperscript{**} & 0.01 & [0.01, 0.07] & 3.05 \\
Headline Veracity & -0.01 & 0.08 & [-0.18, 0.15] & -0.20 \\
Active Control $\times$ Headline Veracity & 0.02 & 0.02 & [0.02, 0.07] & 1.08 \\

\midrule
Observations & \multicolumn{4}{l}{4,064} \\
$R^2$ / Adjusted $R^2$ & \multicolumn{4}{l}{0.018 / 0.017} \\

\midrule
\multicolumn{5}{l}{\textbf{Model 4: All conditions}} \\
\midrule
\textbf{Predictors} & \textbf{Estimate} & \textbf{SE} & \textbf{CI} & \textbf{t} \\
\midrule

(Intercept) & 0.53\textsuperscript{***} & 0.03 & [0.46, 0.60] & 15.97 \\
Active Control & 0.04\textsuperscript{**} & 0.01 & [0.01, 0.07] & 3.05 \\
Headline Veracity & -0.01 & 0.08 & [-0.18, 0.15] & -0.20 \\
Active Control $\times$ Headline Veracity & 0.02 & 0.02 & [0.02, 0.07] & 1.08 \\
Treatment & -0.07\textsuperscript{*} & 0.03 & [-0.13, -0.02] & -2.81 \\
Importance Treatment & 0.10\textsuperscript{**} & 0.02 & [0.05, 0.15] & -4.07 \\
Condition $\times$ Headline Veracity & 0.14\textsuperscript{***} & 0.03 & [0.08, 0.20] & 4.83 \\
Importance Treatment $\times$ Headline Veracity & 0.12\textsuperscript{***} & 0.03 & [0.06, 0.17] & 4.67 \\

\midrule
Observations & \multicolumn{4}{l}{7,216} \\
$R^2$ / Adjusted $R^2$ & \multicolumn{4}{l}{0.058 / 0.057} \\
\bottomrule

\caption{Headline Veracity, Intervention and Participant Type on Twins' Sharing Intentions. \emph{* p$<$0.05, ** p$<$0.01, *** p$<$0.001}. When reporting the effects, separate F-tests were conducted; therefore, the reported F statistics are not included in this table, following the reporting approach used by \cite{pennycook2021accuracy}.}
\label{SI_accuracy_4}
\end{longtable}

\begin{figure}[H]
\centering
\includegraphics[width=0.8\textwidth]{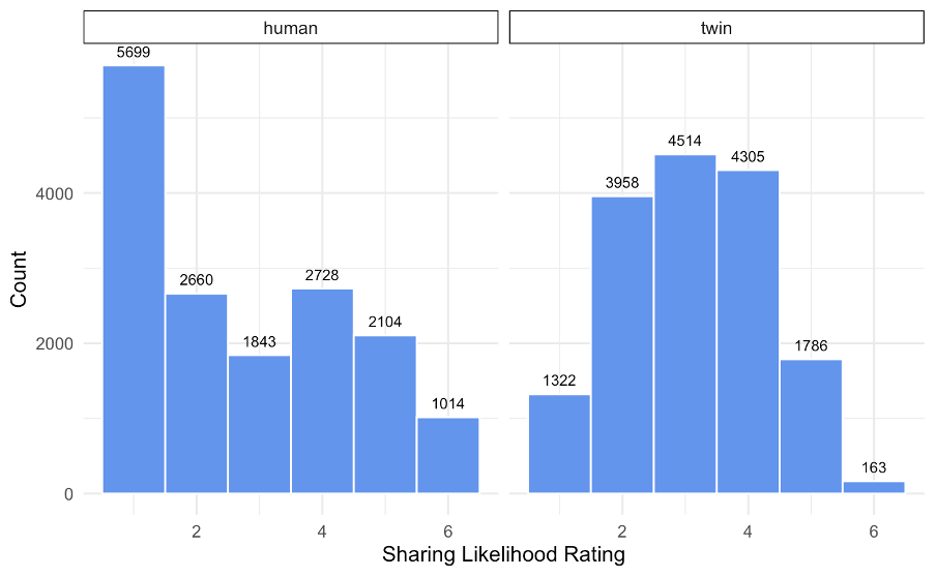}
\caption{Full Distribution of Individual Sharing Likelihood Ratings.
}\label{SI_accuracy_5}
\end{figure}

Finally, Figure \ref{SI_accuracy_5} displays the full distribution of individual sharing
likelihood ratings for each participant type across all intervention
conditions and headline veracity levels. These distributions reveal that
humans were far more likely to report that they would not share
headlines, with many responses clustering at the low end of the scale.
In contrast, digital twins tended to give more moderate responses, with
sharing likelihood ratings concentrated around the midpoint and a
greater proportion of ``likely to share'' responses overall. This
pattern suggests that, unlike humans---who default to inaction---digital
twins are generally more inclined to share
content.

These additional analyses strengthen confidence in the findings and
provide a straightforward comparison for humans and their digital twins,
separately, with the reported findings from \cite{pennycook2021accuracy}.

\subsubsection{Discussion}

These findings extend prior research studying the effectiveness of
accuracy nudges in combating the spread of misinformation \cite{pennycook2021accuracy}. Critically, accuracy nudges may have limited effectiveness
for stopping the spread of false entertainment news. While prior
research has shown that nudges can redirect attention toward accuracy
and reduce the spread of false information in political contexts \cite{pennycook2021accuracy}, this finding does not seem to generalize to
more playful, emotionally engaging content. Specifically, even when
primed to consider accuracy or reminded of its importance, humans showed
no statistically reliable reduction in their willingness to share false
versus true entertainment headlines. Even more, when first prompted to
rate the entertainingness of unrelated headlines (i.e., Active Control
condition), participants were marginally more likely to share false
versus true entertainment headlines, indicating worse discernment.
Interestingly, in contrast to humans, digital twins responded as the
prior work would predict, demonstrating robust improvements in truth
discernment in both the Treatment and Importance Treatment conditions.

One possible reason for this divergence is that digital twins, unlike
humans, do not experience the hedonic appeal of entertaining content.
For humans, entertainment may override epistemic goals, especially when
sharing functions as a form of social play or bonding \cite{lane2025misinformation}. In contrast, digital twins may more rigidly apply
the prioritization embedded in their programming or training data,
defaulting to accuracy in the absence of competing motivations. In this
way, digital twins act as idealized rational agents: models of how
people ought to behave if their sole goal was to share accurate
information. In the same vein, twins seem more prone to prioritizing ``should'' or socially desirable characteristics (e.g., source trustworthiness) over truly social characteristics (e.g., entertainingness). But humans operate in a messier reality, where attention is
easily hijacked by humor, surprise, or emotional resonance. This gap
underscores a broader challenge in misinformation interventions: even
the most well-intentioned nudges may fall short when the content itself
draws attention away from accuracy.

From a practical standpoint, the results highlight the limits of
one-size-fits-all interventions. If humans are less responsive to
accuracy nudges in hedonic contexts, future interventions may need to do
more than redirect attention---they may need to reshape the perceived
function of sharing itself. For example, interventions could incentivize
or socially reward accurate sharing \cite{lane2025misinformation}, rebalancing the tradeoff between entertainment and truth. More
ambitiously, platforms might redesign their interfaces to elevate
epistemic cues without dampening the joy of social engagement. At a
theoretical level, these findings invite deeper reflection on how models
of decision-making should incorporate context-dependent goals, and how
digital twins can help surface not just what people do, but why they do
it.

\pagebreak

\subsection{Affective Primes}\label{affective-primes}

\subsubsection{Main Questions/Hypotheses}

Research in the social sciences often manipulates emotional states using
affective priming, where participants are asked to reflect on a recent
event. In this study, we test whether affective priming works with
digital twins. Specifically, we ask:

\begin{enumerate}
\item
  Do affective priming manipulations induce ``states'' in digital twins
  (e.g., does writing about gratitude or lack of control momentarily
  influence digital twins' responses?)
\item
  Is the influence of affective priming similar for digital twins and
  their human counterparts?
\item
  Is the influence of affective priming on digital twins dependent on
  the valence of affective prime (i.e., between positive primes like
  gratitude or negative primes like lack of control) or the proximal
  nature of the dependent measures (i.e., does it ``spill over'' to
  other, related dimensions or only influence proximal dimensions)?
\end{enumerate}

\subsubsection{Methods}

One thousand human participants on Prolific (and their twins)
participated in the study. They were randomly assigned the same
condition in a 2 (affective domain: gratitude vs. lack of control) x 2
(prompt: prime vs. baseline) between-subjects design.

Participants first responded to a condition-specific prompt for two
minutes. For the gratitude domain, participants received either a
gratitude prime (``Please recall carefully and in detail a specific
experience in the past when you felt sincerely grateful for someone's
kindness or help'') or baseline prompt (``Please recall carefully and in
detail the sequence of your morning routine, such as brushing teeth or
changing clothes''), using procedures from \cite{emmons2003gratitude,desteno2014gratitude,oguni2024gratitude}. In the lack of control
domain, participants received either the lack of control prime (``Please
recall a particular incident in which something happened and you did not
have control over the situation. Please describe the situation in which
you felt a lack of control -- what happened, how you felt, etc.'') or a
baseline prompt (``Please recall a particular incident in which
something happened and you were in control of the situation. Please
describe the situation in which you felt in control -- what happened,
how you felt, etc.''), using procedures from \cite{bukowski2024control,chen2017control,lembregts2019numbers,whitson2008control}.

After the writing task, participants completed a manipulation check,
followed by proximal dependent measures (closely related to the target
affective state) and distal dependent measures (more distantly related
to the affective state). For the gratitude domain, the manipulation
check assessed gratitude-related emotions (grateful, thankful,
appreciative) on a 7-point scale from ``not at all'' to ``extremely.''
The proximal measure used the Elevation scale \cite{schnall2010elevation,walsh2022gratitude}, which asked participants to rate
their current feelings on six items such as ``Feeling optimistic about
humanity'' using the same 7-point scale. The distal measure employed an
adapted version of the Empathic Concern scale \cite{oliveira2021gratitude},
measuring agreement with two statements, ``I am very concerned about
those most vulnerable to the effects of tariffs'' and ``I feel
compassion for those most affected by rising prices or job losses due to
tariffs,'' on a 7-point scale from ``completely disagree'' to
``completely agree.''

For the lack of control domain, the manipulation check used items from \cite{greenaway2015control}, asking participants to rate their agreement
with three statements (``I feel in control of my life'', ``I am free to
live my life how I wish'', ``My experiences in life are due to my own
actions'', on a 7-point scale from ``not at all'' to ``extremely.''
These items were reverse coded for the analysis, but we present the raw
means before reverse coding in the graph. The proximal measure was the
Fatigue scale \cite{sedek1990helplessness,bukowski2024control}, consisting
of nine items such as ``I had a hard time thinking about the event''
rated on the same 7-point scale. The distal measure employed the Desire
for Predictability scale \cite{webster1994closure,lembregts2019numbers}, containing eight items like ``would not like to go
into a situation without knowing what I can expect from it'' rated on a
7-point scale from ``completely disagree'' to ``completely agree.''

\subsubsection{Results - Pre-registered Analyses}

First, the scales demonstrated strong internal consistency for both
humans and digital twins. For Digital Twins: gratitude manipulation
check ($\alpha = .998$), elevation scale ($\alpha = .964$), empathic
concern measure ($\alpha = .992$), lack of control manipulation check
($\alpha = .951$), fatigue scale ($\alpha = .897$), and desire for
predictability measure ($\alpha = .931$). For Humans: gratitude
manipulation check ($\alpha = .979$), elevation scale ($\alpha = .917$),
empathic concern measure ($\alpha = .909$), lack of control manipulation
check ($\alpha = .830$), fatigue scale ($\alpha = .766$), and desire for
predictability measure ($\alpha = .866$). As a result, we created a
composite variable for each measure by averaging the respective items.

Following our pre-registered analysis plan, we then averaged the outcome
measures in the gratitude domain (elevation, empathic concern; $r = .38$)
to create an aggregate downstream dependent variable for gratitude, and
averaged the outcome measures in the lack of control domain (fatigue,
desire for predictability; $r = .09$) to create an aggregate downstream
dependent variable for lack of control. We also examined effects by
individual measure.

\emph{Manipulation Check}

To examine the effect of affective priming on humans versus digital
twins for both gratitude and lack of control, we ran a linear
mixed-effects regression using the \texttt{lmerTest} package in R 
(Kuznetsova, Brockhoff, \& Christensen, 2017) with the manipulation
check measure as the dependent variable; affective domain (gratitude vs.\
feeling in control), prompt (prime vs.\ baseline), twin (human vs.\
digital), and their interactions as independent variables; and twin
identifier as a random intercept. We did not observe a significant
three-way interaction ($b = -0.17$, $SE = 0.17$, $t(996) = -1.01$,
$p = .314$), but several interesting two-way interactions emerged, which
we detail below.

First, affective priming manipulations successfully induced ``states'' in
both digital twins and humans. The prompt (prime vs.\ baseline) had a
significant effect on manipulation checks for digital twins in both the
gratitude domain ($M_\text{baseline} = 4.34$, $SD = 1.17$;
$M_\text{prime} = 5.67$, $SD = 0.71$; $b = 1.33$, $SE = 0.10$;
$t(1760) = 12.93$, $p < .001$) and the lack of control domain
($M_\text{baseline} = 2.79$, $SD = 0.86$; $M_\text{prime} = 3.79$,
$SD = 0.95$; $b = 1.00$, $SE = 0.10$; $t(1760) = 9.55$, $p < .001$), as
well as for humans in the gratitude domain ($M_\text{baseline} = 5.19$,
$SD = 1.56$; $M_\text{prime} = 6.00$, $SD = 1.33$; $b = 0.81$,
$SE = 0.10$; $t(1760) = 7.91$; $p < .001$) and the lack of control
domain ($M_\text{baseline} = 2.81$, $SD = 1.16$; $M_\text{prime} = 3.12$,
$SD = 1.29$; $b = 0.32$, $SE = 0.10$; $t(1760) = 3.02$, $p = .003$).

\begin{figure}[H]
\centering
\includegraphics[width=1\textwidth]{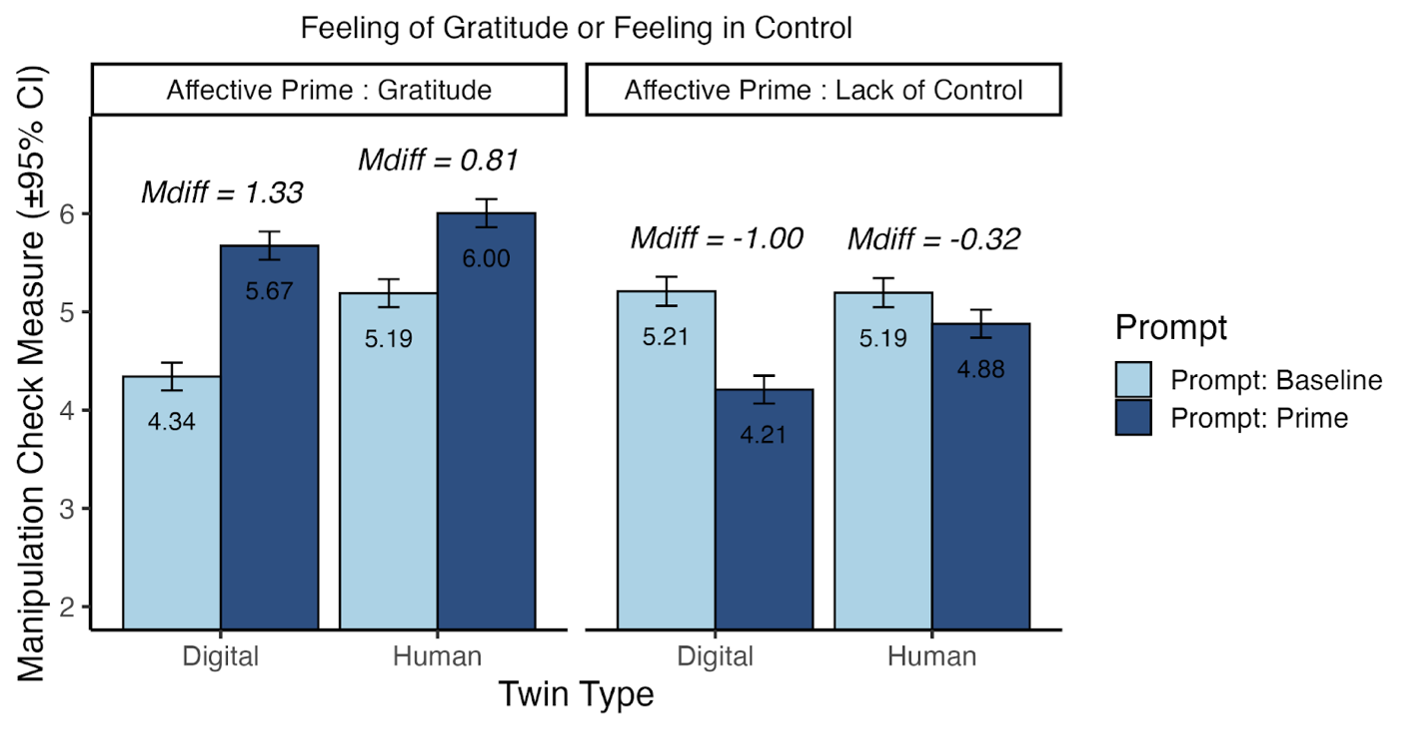}
\caption{Effect of Affective Prime on Manipulation Check.
}\label{SI_affective_1}
\end{figure}

Second, the effect of priming was significantly stronger for digital
twins than for humans in both affective domains. Digital twins showed
greater responsiveness to the prompt (prime vs.\ baseline) in both the
gratitude domain (interaction: $b = -0.52$, $SE = 0.12$;
$t(996) = -4.45$, $p < .001$) and the lack of control domain
(interaction: $b = -0.68$, $SE = 0.12$; $t(996) = -5.79$, $p < .001$),
compared to humans (see Figure \ref{SI_affective_1}).

Finally, the prompt (prime vs.\ baseline) produced stronger effects in
the gratitude domain than in the lack of control domain for both digital
twins ($b = 0.33$, $SE = 0.15$; $t(1760) = 2.26$, $p = .024$) and humans
($b = 0.50$, $SE = 0.15$; $t(1760) = 3.39$, $p < .001$).

\emph{Dependent Measures}

We next examined how affective priming in digital twins affects
downstream dependent measures---that is, do these ``affective states''
spill over onto other measures? To test this, we ran a linear
mixed-effects regression with the aggregate dependent variable measure
as the dependent variable; affective domain (gratitude vs.\ lack of
control), prompt (prime vs.\ baseline), twin (human vs.\ digital), and
their interaction as independent variables; and twin identifier as a
random intercept. We did not observe a significant three-way interaction
($b = -0.13$, $SE = 0.11$, $t(996) = -1.17$, $p = .242$), but two
significant two-way interactions emerged, which we detail below.

\begin{figure}[H]
\centering
\includegraphics[width=1\textwidth]{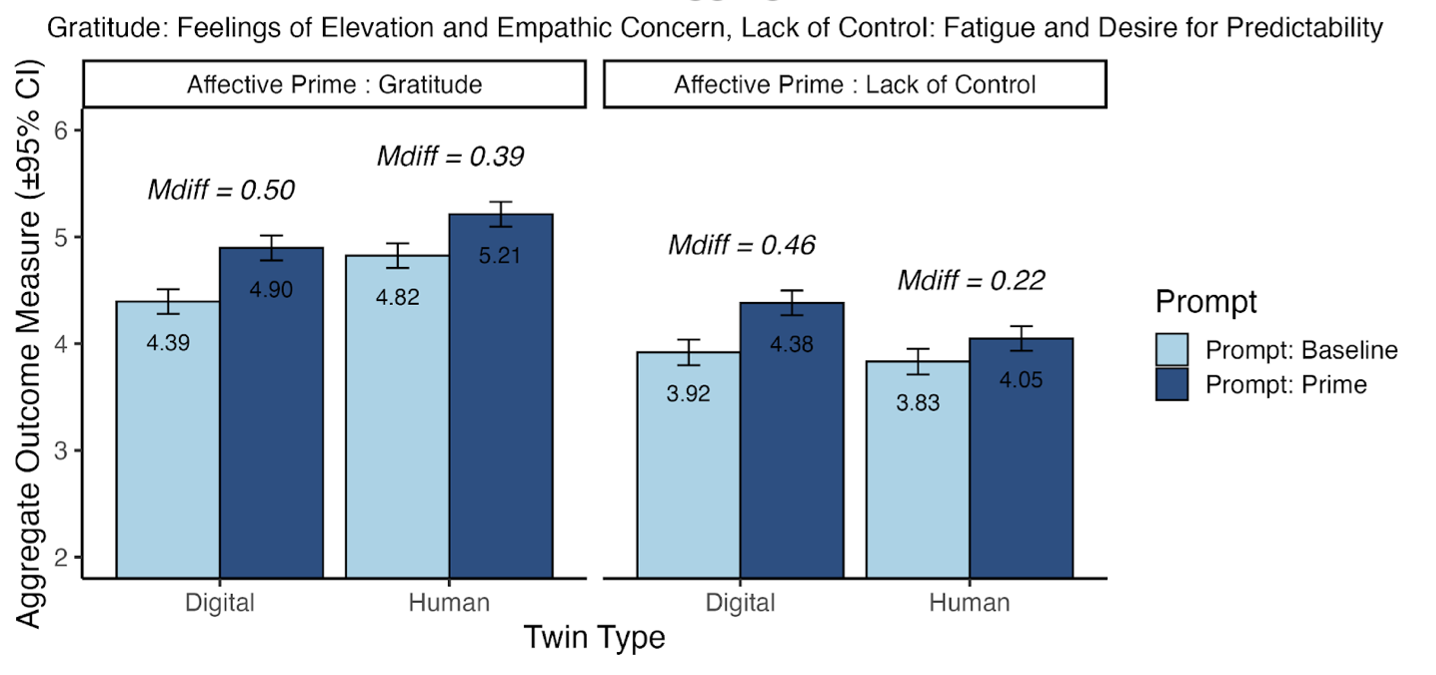}
\caption{Effect of Affective Prime on Aggregate Downstream Outcomes.
}\label{SI_affective_2}
\end{figure}

First, affective priming manipulations spilled over onto downstream
dependent measures for both digital twins and humans. The prompt (prime
vs.\ baseline) had a significant effect on downstream measures for
digital twins in both the gratitude domain ($M_\text{baseline} = 4.39$,
$SD = 1.11$; $M_\text{prime} = 4.90$, $SD = 1.05$; $b = 0.50$,
$SE = 0.08$; $t(1538) = 6.01$; $p < .001$) and the lack of control
domain ($M_\text{baseline} = 3.92$, $SD = 0.60$; $M_\text{prime} = 4.38$,
$SD = 0.68$; $b = 0.46$, $SE = 0.09$; $t(1538) = 5.45$; $p < .001$), as
well as for humans in the gratitude domain ($M_\text{baseline} = 4.82$,
$SD = 1.15$; $M_\text{prime} = 5.21$, $SD = 1.23$; $b = 0.81$,
$SE = 0.10$; $t(1760) = 7.91$; $p < .001$) and the lack of control
domain ($M_\text{baseline} = 3.83$, $SD = 0.76$; $M_\text{prime} = 4.05$,
$SD = 0.69$; $b = 0.32$, $SE = 0.10$; $t(1760) = 3.02$; $p = .003$). See
Figure \ref{SI_affective_2}.

Second, the effect of priming on downstream dependent measures was
significantly stronger for digital twins than humans for lack of control
priming ($b = -0.25$, $SE = 0.08$; $t(996) = -3.05$; $p = .002$), but
not for gratitude ($b = -0.11$, $SE = 0.08$; $t(996) = -1.43$;
$p = .153$).

Finally, the gratitude prime did not produce stronger effects than the
lack of control prime for digital twins ($b = 0.04$, $SE = 0.12$;
$t(1538) = 0.33$; $p = .744$), but gratitude produced stronger effects
for humans ($b = 0.50$, $SE = 0.15$; $t(1760) = 3.39$; $p < .001$).

\emph{Individual Dependent Measures}

We also pre-registered analyses examining each dependent measure
independently. Thus, we ran four linear mixed-effects regressions with
(1) elevation, (2) empathic concern, (3) fatigue, and (4) desire for
predictability as the dependent variable; condition (manipulation vs.\
control), twin (human vs.\ digital), and their interaction as
independent variables; and twin identifier as a random intercept. Each
measure's simple effect for prompt (prime vs.\ baseline) was significant
for both humans and digital twins, except empathic concern, which was
marginal for twins ($M_\text{baseline} = 5.04$, $SD = 1.48$;
$M_\text{prime} = 5.28$, $SD = 1.40$; $b = 0.24$, $SE = 0.14$;
$t(731) = 1.74$, $p = .082$) and not significant for humans
($M_\text{baseline} = 5.45$, $SD = 1.61$; $M_\text{prime} = 5.44$,
$SD = 1.64$; $b = -0.01$, $SE = 0.14$; $t(731) = -0.07$, $p = .947$), as
well as desire for predictability, which was not significant for twins
($M_\text{baseline} = 5.54$, $SD = 1.04$; $M_\text{prime} = 5.59$,
$SD = 1.04$; $b = 0.05$, $SE = 0.09$; $t(718) = 0.47$; $p = .636$).

\begin{figure}[H]
\centering
\includegraphics[width=1\textwidth]{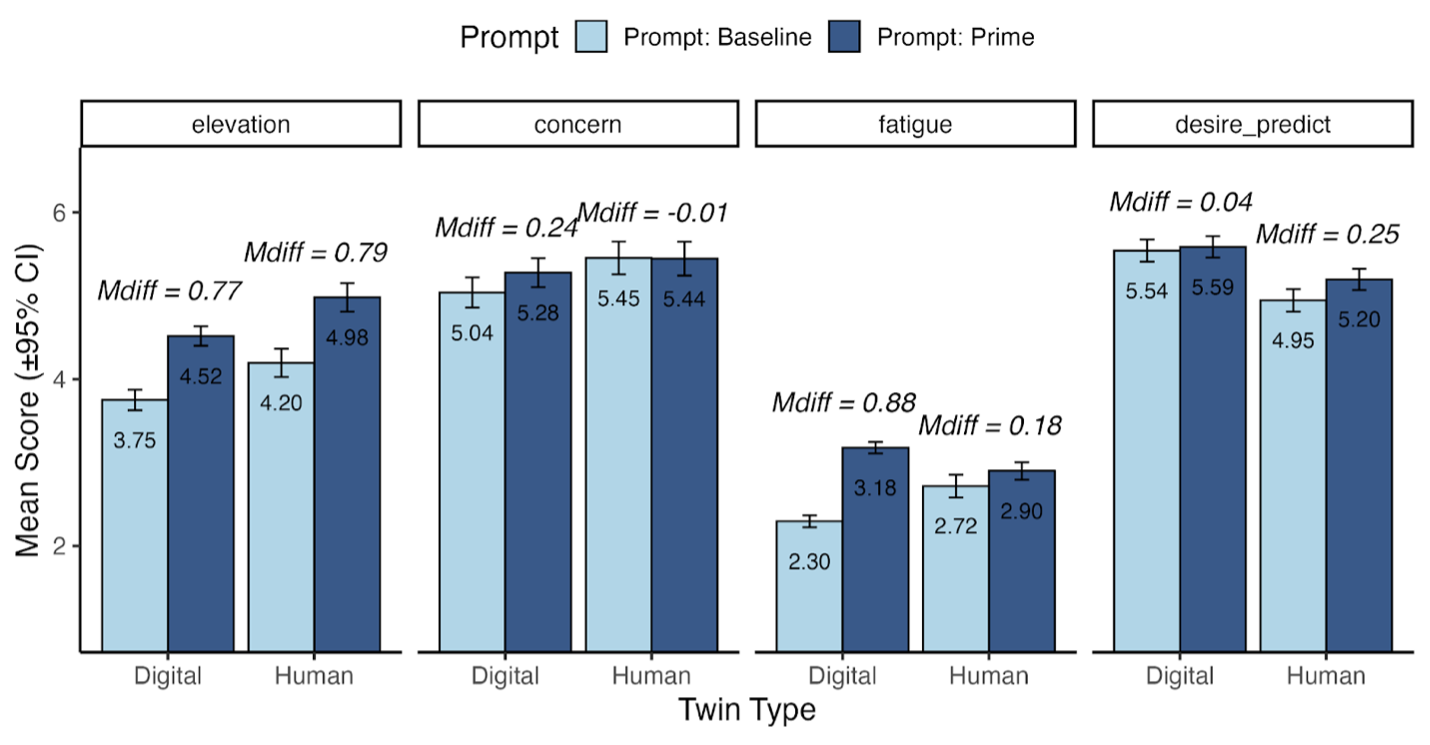}
\caption{Measures by Prompt (Prime vs. Baseline) Grouped by Measure and
Twin Type.
}\label{SI_affective_3}
\end{figure}

The effect of prompt (prime vs.\ baseline) on elevation did not differ
between digital twins and humans ($b = 0.02$, $SE = 0.20$;
$t(506) = 0.16$, $p = .872$). The effect of prompt (prime vs.\ baseline)
was stronger for humans than digital twins for empathic concern
($b = 0.25$, $SE = 0.12$; $t(506) = 2.08$; $p = .039$) and fatigue
($b = 0.70$, $SE = 0.09$; $t(490) = 8.26$; $p < .001$). The effect of
prompt on desire for predictability was also stronger for humans than
for digital twins ($b = 0.21$, $SE = 0.08$; $t(490) = 2.45$;
$p = .015$). See Figure \ref{SI_affective_3}.

\subsubsection{Results - Additional Analyses (Non-Preregistered)}

\begin{figure}[H]
\centering
\includegraphics[width=1\textwidth]{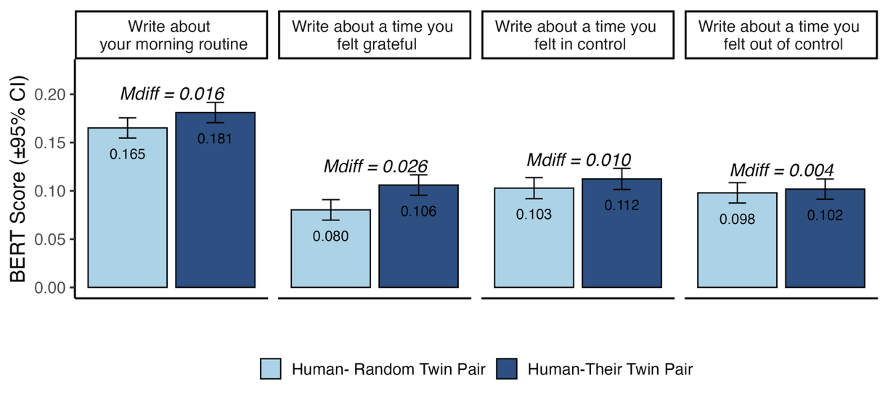}
\caption{Similarity Between Human Prompt Text and Their Digital Twin's
Prompt Text vs. a Random Digital Twin's Prompt Text.
}\label{SI_affective_4}
\end{figure}

Here, we examined the language used when responding to the prompt.
Specifically, we examined the language similarity between a human
participant's response and their twin versus a human participant's
response and a randomly selected twin using the BERT metric, defined as
``the sum of cosine similarities between their tokens' embeddings'' \cite{zhang2020bertscore}. In a random-effects
regression with language matching as the dependent variable; same-twin
versus different-twin pairing as the independent variable; condition as
a covariate; and human participant identifier as a random effect, we
found that the language between a human participant and their twin was
slightly more similar ($M = 0.11$, $SD = 0.09$) than between a human
participant and a random twin ($M = 0.13$, $SD = 0.10$;
$b = 0.014$, $SE = 0.002$; $t(999) = 5.54$; $p < .001$). Figure \ref{SI_affective_4} shows
this by condition.

\subsubsection{Discussion}

Returning to each research question:

\begin{enumerate}
\item
  Do affective priming manipulations induce ``states'' in digital twins
  (e.g., does writing about gratitude or lack of control momentarily
  influence digital twins' responses?)
\end{enumerate}

Yes. Writing about feeling grateful or out of control significantly
increased corresponding feelings of gratitude and lack of control in
digital twins, with these induced ``affective states'' subsequently
influencing downstream outcomes.

\begin{enumerate}
\setcounter{enumi}{1}
\item
  Is the influence of affective priming similar for digital twins and
  their human counterparts?
\end{enumerate}

No. Affective priming had stronger effects on digital twins' affective
responses compared to humans for both gratitude and lack of control
manipulations.

\begin{enumerate}
\def\labelenumi{\arabic{enumi}.}
\setcounter{enumi}{2}
\item
  Is the influence of affective priming on digital twins dependent on
  the valence of affective prime (i.e., between positive primes like
  gratitude or negative primes like lack of control) or the proximal
  nature of the dependent measures (i.e., does it ``spill over'' to
  other, related dimensions or only influence proximal dimensions)?
\end{enumerate}

\emph{Valence.} Yes, valence affects twins' responses, but in a similar
way to humans: The gratitude prime had a stronger effect on the
manipulation check than the lack of control prime for both humans and
their twins.

\emph{Proximal Nature of DVs.} Yes, the influence of affective priming
on digital twins is dependent on how closely related the dependent
measure is to the manipulation. Specifically, the impact of affective
prime on proximal dependent measures was consistent and as expected for
both twins and humans, but impact of affective prime on distal
downstream measures was inconsistent between twins and humans.

For proximal dependent measures (measures closely linked to the
manipulation), we observe a ``spillover'' effect of prime in both the
gratitude prime and lack of control prime for humans and twins. For the
distal downstream measure (more distantly related to the manipulation)
in the gratitude condition, the impact of affective prime was marginal
for twins and non-significant for humans. For the distal downstream
measure in the lack of control condition, the impact of affective prime
was non-significant for twins and significant for humans.

\begin{enumerate}
\def\labelenumi{\arabic{enumi}.}
\setcounter{enumi}{3}
\item
  Is there a relationship between what digital twins and their human
  counterparts say in their response to the manipulation?
\end{enumerate}

Yes. The language between a human participant and their twin is slightly
more similar than between a human participant and a random twin.

\pagebreak

\subsection{Consumer Minimalism}\label{consumer-minimalism}

\subsubsection{Main Questions/Hypotheses}
We explored the predictive validity of the minimalism scale in a sample
of human respondents versus their digital twins. As in the original
paper on consumer minimalism \cite{wilson2022minimalism}, we expect that
human respondents and their digital twins who score high on the
Minimalist Consumer Scale will prefer minimalist home environments over
non-minimalist interiors.

\subsubsection{Methods}

We recruited 200 human respondents on Prolific and their respective twins. 
In random order, participants completed the 12 randomized items of the Minimalist Consumer Scale 
($M_{\text{human}} = 4.84$, $SD_{\text{human}} = 1.17$, $\alpha_{\text{human}} = .93$; 
$M_{\text{twin}} = 4.52$, $SD_{\text{twin}} = 1.47$, $\alpha_{\text{twin}} = .94$), 
and indicated their preference for four sets of minimalist versus non-minimalist interiors, 
presented in random order. All stimuli were pretested and validated in Study 7e of the 
\textit{Consumer Minimalism} paper \cite{wilson2022minimalism}). 

\begin{figure}[H]
\centering
\includegraphics[width=1\textwidth]{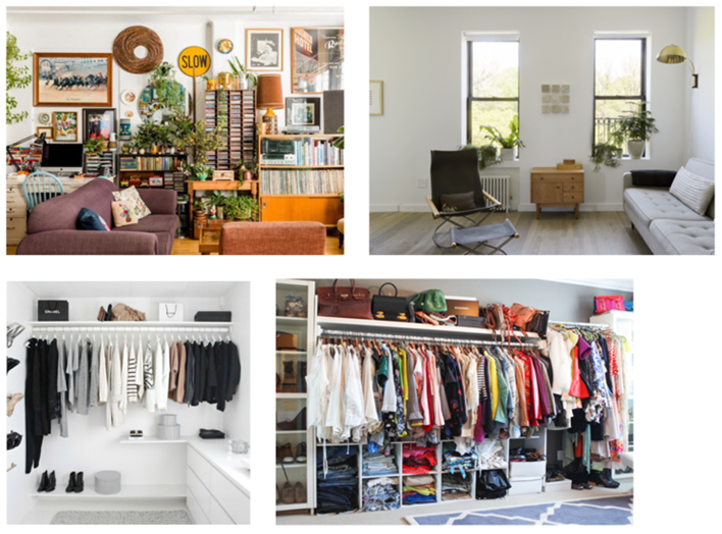}
\caption{Visual Stimuli for Human Respondents.
}\label{SI_minimalism_1}
\end{figure}

In the first two sets, human participants looked at two pairs of images (Figure \ref{SI_minimalism_1}) and indicated 
whether they would rather live in one of two apartments 
(``Which apartment would you rather live in?'') and whether they found one of two wardrobes 
more appealing (``Which wardrobe is more appealing to you?''). In the third set, they read the 
following descriptions of two bedrooms:

Here are the descriptions of two bedrooms.

\begin{quote}
First option: A vibrant bedroom filled with colorful patterns, featuring
a mix of floral and botanical prints on the bedding and wallpaper. A
variety of throw pillows, unique lamps, and framed artwork create a
bold, eclectic, and lively atmosphere.

Second option: A minimalist bedroom has a light wood platform bed with
white bedding, two pillows, and a neatly folded gray blanket. On one
side is a small round table with a plant, and on the other is a clear
bedside unit holding a few essentials.

Which bedroom do you like more?
\end{quote}

Finally, respondents read the following descriptions of two home
offices for the fourth set:

\begin{quote}
Here are the descriptions of two home offices

First option: A minimalist and modern home office featuring a floating
wooden desk, a sleek grey swivel chair, and a clean setup with a desktop
computer and lamp. Built-in shelves with books and decor, along with
floor-to-ceiling neutral curtains, create a clean atmosphere.

Second option: A cozy and eclectic home office combines functionality
with vibrant personality, featuring a metal-framed desk, a patterned
chair, and layered storage options including woven baskets and open
shelving. Colorful textiles, books, and decorative accents add warmth
and charm.

Which home office would you prefer to be yours?''
\end{quote}

All preferences were recorded on a seven-point Likert scale (1 =
definitely the one on the left, 4 = equal preference, 7 = definitely the
one on the right). Before the analyses, we coded all response scales so
that higher values indicate stronger preferences for minimalist options.
Importantly, the digital twins read an AI-generated description of the
visual stimuli displayed for human respondents. This intentional feature
of the study design enables us to test whether the predictive validity
of the scale holds even when twins read descriptions while their human
counterparts see images.

For the two apartments' images, digital twins read:

\begin{quote}
Below, are pictures of rooms in two different apartments. Which
apartment would you rather live in? Left Side: The room contains a large
number of objects and furnishings. There are two upholstered armchairs
(one dark purple, one reddish-brown) and a small ottoman. Behind the
seating area is a wooden desk with a computer monitor and keyboard,
along with a painted wooden chair. Shelving units and surfaces are
densely populated with plants, CDs, books, and various small objects.
Numerous framed pictures, posters, and a ``SLOW'' traffic sign cover the
wall. There is a table lamp with a brown shade and a stereo system
beneath a collection of compact discs. Additional furniture includes a
small bookshelf filled with vinyl records and a mirror partially
obscured by plants. Several types of plants are placed throughout the
room in pots and containers. Right Side: The room contains fewer items
and furnishings. There are two windows with black frames and visible
trees outside. A light-gray upholstered sofa sits on the right side of
the room. A black leather and wood lounge chair is positioned in the
center. A small wooden cabinet with a box on top is located between the
two windows. Wall-mounted decorative tiles or panels are arranged in a
square above the cabinet. Plants are placed on the windowsills and
hanging from planters. A wall-mounted reading lamp is affixed above the
sofa. The floor is light-colored wood.
\end{quote}

For the two wardrobes' images, digital twins read:

\begin{quote}
Below, are pictures of two wardrobes. Which wardrobe is more appealing
to you? Left Side: Contains approximately 20 garments hanging on a
single rod, organized by color from dark to light. All garments are on
matching white hangers. Colors are limited to black, white, gray, beige,
and similar tones. Shoes are placed on two open shelves: one at floor
level and one elevated. There are three visible boxes or containers: two
cylindrical and one rectangular. The closet space includes built-in
white drawers and shelves. A small rug is placed on the floor. Handbags
and boxes are stored on a top shelf above the hanging clothes. All
visible surfaces, including the walls, shelves, and drawers, are white.
Right Side: Contains a larger number of garments (approximately 70+) on
a double-length rod. Garments vary in color, pattern, and length.
Hangers are mostly dark, with some white or wooden hangers mixed in.
Items are densely packed along the hanging rod. Below the clothing rod
are multiple open cubbies filled with folded clothes, shoes, bags, and
miscellaneous items. The top shelf holds several handbags and boxes. A
patterned rug is partially visible on the floor. The shelving units are
white, and the wall behind the closet is gray. There is a glass-front
cabinet on the left side containing additional shoes or items.
\end{quote}

Across sets, we varied whether the minimalist or non-minimalist images
appeared on the left or the right of the screen and whether the
minimalist or non-minimalist description appeared first or second.
\emph{}

\subsubsection{Results - Pre-registered Analyses}
We ran a series of OLS regressions with participants' average score on the Minimalist Consumer Scale as the independent variable and preferences for each option as the dependent variable. As predicted, results indicated that higher scores on the scale were associated with stronger preferences for the minimalist option in each set and in both samples. 

\begin{figure}[H]
\centering
\includegraphics[width=1\textwidth]{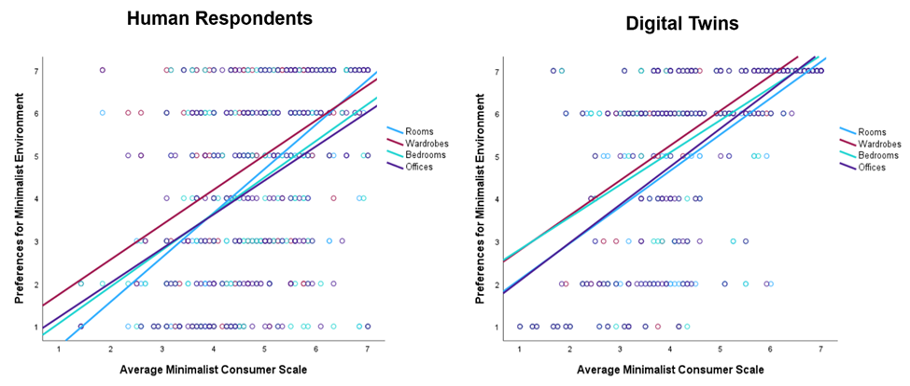}
\caption{Results. Note: the dots indicate the raw data.
}\label{SI_minimalism_2}
\end{figure}

More specifically, higher scores on the Minimalist Consumer Scale predicted stronger preferences 
\begin{enumerate}
    \item for the minimalist apartment in both the human ($\beta = .56$, $t(198) = 9.56$, $p < .001$) and the digital ($\beta = .60$, $t(198) = 10.54$, $p < .001$) samples; 
    \item for the minimalist wardrobe in both the human ($\beta = .44$, $t(198) = 6.85$, $p < .001$) and the digital ($\beta = .65$, $t(198) = 11.92$, $p < .001$) samples; 
    \item for the minimalist bedroom in both the human ($\beta = .46$, $t(198) = 7.36$, $p < .001$) and the digital ($\beta = .57$, $t(198) = 9.79$, $p < .001$) samples; and 
    \item for the minimalist home office in both the human ($\beta = .44$, $t(198) = 6.80$, $p < .001$) and the digital ($\beta = .64$, $t(198) = 11.55$, $p < .001$) samples (see Figure \ref{SI_minimalism_2}).
\end{enumerate}

There was no significant effect of order, nor interaction between scores on the Minimalist Consumer Scale and whether participants completed the scale before or after indicating their preferences for the different sets. Moreover, the design feature for the first two sets---with images for humans and AI-generated descriptions for the digital sample---did not yield differences in the results. A regression with interiors preferences as dependent variable, and minimalism ($z$-scores), whether the stimuli appeared as images or text (coded as 1 and $-1$, respectively), and their interaction as predictors, while controlling for human respondent vs.\ digital twin, did not reveal a significant interaction ($b_{\text{interaction}} = -0.13$, $p = .249$). 

In other words, the fact that humans saw images and digital twins read descriptions for the first two sets of interiors did not interfere with the overall results.

A comparison of the standardized regression coefficients suggests that the results were slightly stronger in the digital sample than in the human population. A series of $t$-tests comparing the preference for the minimalist environment for the humans and their twin counterparts suggests that, on average, the twins liked the minimalist environments more (apartment: $M_{\text{human}} = 4.53$ vs.\ $M_{\text{twin}} = 5.11$, $t(398) = 2.69$, $p = .007$; wardrobe: $M_{\text{human}} = 4.89$ vs.\ $M_{\text{twin}} = 5.68$, $t(398) = 3.91$, $p < .001$; bedroom: $M_{\text{human}} = 4.36$ vs.\ $M_{\text{twin}} = 5.49$, $t(398) = 5.45$, $p < .001$; home office: $M_{\text{human}} = 4.30$ vs.\ $M_{\text{twin}} = 5.23$, $t(398) = 4.39$, $p < .001$).

We also examined the test--retest reliability of the Minimalist Consumer Scale in the human sample between collection rounds, given that the scale was collected twice (approximately four and a half months apart): once as part of a large battery of measures to generate the digital twins (February 1, 2025) and once as part of this specific validity study on the Minimalist Consumer Scale (June 12, 2025). The correlation of the Minimalist Consumer Scale between the two collection rounds was large ($r = .75$, $p < .001$), confirming that the scale has high test--retest reliability (Peter, 1979).

A mixed-effects generalized linear model with random intercepts for participants, with preference for the minimalist interiors as dependent variable, and the Minimalist Consumer Scale, an indicator for data type (coded as 1 for human and 2 for digital twin), and their interaction as predictors, revealed an effect of minimalism ($\beta = .73$, $t(1{,}593) = 11.69$, $p < .001$), an effect for digital twins ($\beta = .71$, $t(1{,}593) = 2.36$, $p = .019$), and no significant interaction ($\beta = .08$, $t(1{,}593) = 1.34$, $p = .179$). The same model estimated with OLS regression revealed similar results: an effect of minimalism ($\beta = .88$, $t(1{,}596) = 16.49$, $p < .001$), an effect for digital twins ($\beta = 1.34$, $t(1{,}596) = 4.02$, $p < .001$), and no significant interaction ($\beta = -0.05$, $t(1{,}596) = -0.67$, $p = .503$).

\subsubsection{Results - Additional Analyses (Non-Preregistered)}

An analysis of the reasonings of five digital twins suggests that they
expressed preferences in line with their minimalist orientations. For
example, one twin reasoned, ``My preferences lean toward vibrant,
lively, and eclectic environments rather than minimalism. I feel more
comfortable in spaces with lots of patterns, colors, and personal
touches, making the first bedroom my clear favorite,'' and accordingly
selected the non-minimalist option with a strong preference; in
contrast, another twin reasoned, ``I like a clean, organized, and calm
environment for a home office. The minimalist and modern option fits my
preferences for simplicity and order'' and accordingly selected the
minimalist option with a strong preference. The ``extremeness'' of these
thoughts may possibly explain the relatively higher strength of the
findings in the digital twins sample compared to the human sample. In
other words, when it comes to the association between the minimalism
scale and expressed preferences for room environments the twins tend to
be more polarized than their human counterparts.

\subsubsection{Discussion}

In conclusion, this study demonstrates that the predictive validity of
the Minimalist Consumer Scale documented in human samples replicates
robustly in a digital twin sample. Specifically, the study demonstrates
that the scale predicts consumers' preferences for minimalist versus
non-minimalist apartment interiors.

\pagebreak
 \subsection{Context Effects}\label{context-effects}

\subsubsection{Main Questions/Hypotheses}

Context effects are of great interest to marketers and policy makers
alike because they suggest that product assortment can change choice.
This study focuses on two classic context effects. The attraction effect
occurs when a third option - an asymmetrically dominated decoy - is
added to a binary choice set, increasing preference for the option that
dominates the decoy \cite{huber1982adding}. The compromise effect occurs
when a different type of decoy is added to a binary choice set, one that
makes one of the original options appear as a compromise between the
other two, increasing preference for the compromise option \cite{simonson1989choice}.

This study tests whether digital twins trained from LLMs can accurately
predict individual choices in product purchase settings designed to
elicit the attraction and compromise effects, using newly developed
stimuli that twin models could not have encountered during training.

\subsubsection{Methods}

Following an initial practice trial, all participants completed via Qualtrics three hypothetical product purchase decisions: one binary
choice, and two trinary choices designed to elicit the attraction and compromise effects, respectively. The order of the three trials was
counterbalanced across participants.

In each trial, participants were presented with a hypothetical purchase
scenario and asked to choose between product options described in text
by their price and quality features. Each trial was randomly assigned to
one of the following three product categories: a printer, a TV, and a
telephone plan. Importantly, all stimuli were newly developed for this
study and could not have appeared in the digital twins' training data.

501 human participants (48\% female, 52\% male; 9\% aged 18-29, 33\%
aged 30--49, 35\% aged 50--64, 23\% aged 65+) on Prolific completed the
task. The study was subsequently run on their 501 digital twins. Human
participants and their corresponding digital twins completed the same
trials.

We recorded both choices and time spent on each decision for human participants, and only choice for digital twins.

\subsubsection{Results - Pre-registered Analyses}

We began our analysis by testing whether humans and digital twins replicated the attraction and compromise effects. As preregistered, we excluded practice trials and combined the datasets from both humans and digital twins and created customized contrasts: for the attraction effect, attraction trials were coded as 1, binary trials as -1, and all other trials as 0. A similar coding scheme was used for the compromise effect. We ran binomial mixed-effect models predicting the choice of the target option. We included dataset and product type as fixed effects and accounted for the multilevel structure of the data by modeling respondent-level random intercepts, treating each human and their digital twin as separate entities. Each contrast was nested within the human and digital twin datasets to separately capture the effects within each dataset.\footnote{The nested model is a reparameterization of the   preregistered interaction model. It is statistically equivalent but provides a more direct and interpretable test of the effect in each dataset.}

\begin{figure}[H]
\centering
\includegraphics[width=1\textwidth]{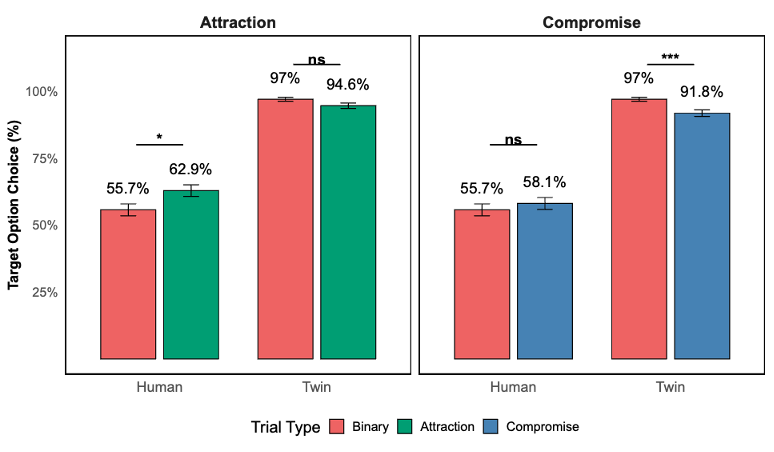}
\caption{The \emph{target option} refers to the option whose choice likelihood is expected to increase due to the addition of a decoy option. Vertical bars represent $\pm 1$ standard deviation based on raw data. Asterisks indicate significance levels derived from model-based contrast analyses: * $p < .05$, ** $p < .01$, *** $p < .001$; ns = not significant.
}\label{secSI_context_1}
\end{figure}

As shown in Figure \ref{secSI_context_1}, for the attraction effect, we replicated prior literature and found a significant effect among humans 
($\beta = 0.18,\; SE = 0.07,\; z = 2.56,\; p = .01$). 
However, we did not detect an effect among digital twins 
($\beta = -0.24,\; SE = 0.14,\; z = -1.67,\; p = .10$). 
Unexpectedly, we observed a strong main effect of dataset, such that digital twins made substantially more target choices overall 
($\beta = 2.84,\; SE = 0.16,\; z = 17.30,\; p < .001$).

For the compromise effect, neither humans nor digital twins replicated prior results. 
No effect was found for humans 
($\beta = 0.06,\; SE = 0.07,\; z = 0.87,\; p = .38$), 
and a significant negative effect was found for digital twins 
($\beta = -0.54,\; SE = 0.15,\; z = -3.59,\; p < .001$). 
Again, we observed a strong main effect of dataset, with digital twins making substantially more target choices 
($\beta = 2.90,\; SE = 0.17,\; z = 17.16,\; p < .001$).

We also evaluated how accurately digital twins predicted human choices, independent of whether they predicted context effects. To examine possible variation in prediction accuracy by trial type, we broke the analysis down across binary, attraction, and compromise trials. For each trial, we created a binary variable denoting whether the digital twin correctly predicted the participant's choice.\footnote{We created an indicator variable denoting whether the digital twin correctly   predicted the participant's choice and used a McNemar test for   analysis, rather than the preregistered correlation, as it is more   appropriate for discrete choice outcomes.} As a benchmark, we compared digital twin's accuracy level to that of a baseline model that randomly selected among the available options.

We found that digital twins predicted human choices more accurately than the random baseline across all trial types. 
Specifically, twins achieved 58\% accuracy on binary trials 
(paired $t(500) = 2.86,\; p = .004$ vs.\ the 50\% baseline), 
59\% on attraction trials 
($t(500) = 8.52,\; p < .001$ vs.\ 33\%), 
and 55\% on compromise trials 
($t(500) = 8.04,\; p < .001$ vs.\ 33\%). 
Results were consistent when using McNemar’s test, which evaluates the dichotomous accuracy indicator as a nominal variable rather than with a $t$-test.

\subsubsection{Results - Additional Analyses (Non-Preregistered)}

Using response time data from human trials, we explored whether attention moderated the observed context effects. We constructed an attention factor based on log-transformed response times from binary, attraction, and compromise trials, and included interactions between
this factor and the contrast variables in the nested model.

Among human participants, attention significantly moderated the compromise effect 
($\beta = 0.17,\; SE = 0.07,\; z = 2.32,\; p = .02$), 
such that more attentive participants exhibited a significant compromise effect: 
a Johnson--Neyman analysis indicated the effect became statistically significant 
at values above 0.61 standard deviations of attention (relative to the mean), 
under a $p < .05$ threshold. 
No interaction was observed for the attraction effect 
($\beta = 0.08,\; SE = 0.07,\; z = 1.06,\; p = .29$).

Attention factor derived from human participants did not moderate either context effect in the twin dataset 
(attraction: $\beta = -0.002,\; SE = 0.14,\; z = -0.01,\; p = .99$; 
compromise: $\beta = -0.12,\; SE = 0.14,\; z = -0.81,\; p = .42$).

\subsubsection{Discussion}

We examined if large language models (LLMs) like ChatGPT can accurately
predict the choices made by humans. In the current study, we
successfully replicated the attraction effect in the human dataset, but
digital twins did not reproduce this effect. Neither humans nor digital
twins exhibited a compromise effect. However, the compromise effect was
significant among human participants who paid more attention, as
measured by response time. Independent of predicting context effects,
digital twins showed only modest accuracy (less than 60\% overall) in
predicting individual human choices. These findings highlight cautions
when using digital twins to predict human decisions in applied settings
such as consumer choice.

\pagebreak
\subsection{Default Effects}\label{default-effects}

\subsubsection{Main Questions/Hypotheses}

This study focuses on two default effects: the organ donation paradigm
by \cite{johnson2003defaults}, and a green energy default paradigm
promoting sustainable energy choices \cite{pichert2008green}. In the
organ donation study, participants who were defaulted to be donors were
significantly more likely to remain enrolled than those who had to
actively opt in. Similarly, in the green energy study, defaulting
participants into using green energy suppliers increased adoption rates
compared to requiring them to opt in.

This study tests whether digital twins trained from LLMs can accurately
predict individual choices in these two settings, using the original
organ donation stimuli verbatim and a slightly adapted version of the
green energy scenario.

\subsubsection{Methods}

All participants completed the original organ donation as well as the
adjusted paradigm, presented in counterbalanced order. Across both
default tasks, participants were randomly assigned to either an opt-in
or an opt-out condition. In the organ donation task, participants in the
opt-in condition were told they were not organ donors by default and
were asked whether they wanted to become one. In the opt-out condition,
participants were told they were organ donors by default and could
choose to change that status. In the green energy default task,
participants in the opt-in condition were told they were defaulted to
the more affordable (non-green) energy option, while those in the
opt-out condition were defaulted to the green energy option.

Notably, the organ donation task used the original stimuli from \cite{johnson2003defaults}, while the green energy default task was slightly
modified by changing the company names for the affordable and
sustainable options.

600 human participants (47\% female, 53\% male; 11\% aged 18--29, 33\%
aged 30--49, 34\% aged 50--64, 22\% aged 65+) on Prolific completed the
task. The study was subsequently run on their 600 digital twins.
Participants and their corresponding digital twins were assigned the
same conditions.

We recorded both choices and time spent on each decision for human
participants, and only choice for digital twins.

\subsubsection{Results - Pre-registered Analyses}

We began our analysis by testing whether humans and digital twins
replicated default effects in organ donation and green energy choice. As
preregistered, for each default task, we combined data across both
humans and digital twins and created a customized contrast: opt-out
trials were coded as 1, opt-in trials as -1. We accounted for the
multilevel structure of the data by including respondent-level random
intercepts, treating each human and their digital twin as separate
entities. This contrast was nested within human and digital twin
datasets to separately estimate effects within each dataset.\footnote{The
  nested model is a reparameterization of the preregistered interaction
  model. It is statistically equivalent but provides a more direct and
  interpretable test of the effect in each dataset.}

\begin{figure}[H]
\centering
\includegraphics[width=1\textwidth]{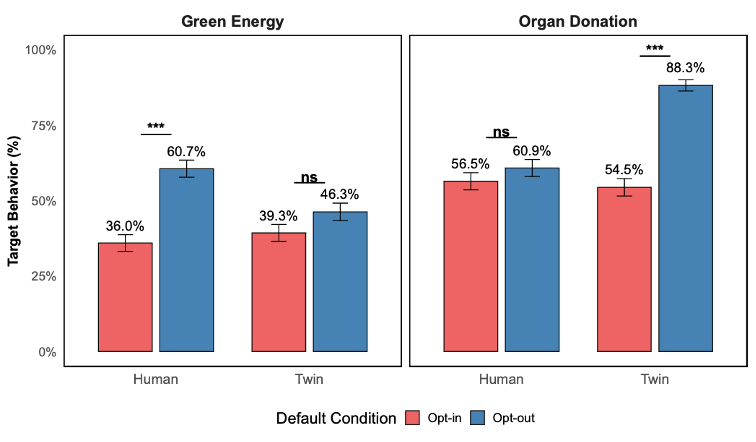}
\caption{Target behavior refers to adopting a green energy provider in
the green energy paradigm and registering as an organ donor in the organ
donation paradigm. Vertical bars represent $\pm 1$ standard deviation based
on raw data. Asterisks indicate significance levels derived from
model-based contrast analyses: * $p < .05$, ** $p < .01$, *** $p < .001$; 
ns = not significant.
}\label{SI_default_1}
\end{figure}

As depicted in Figure \ref{SI_default_1}, in the green energy task, humans showed a large
effect and replicated prior literature 
($\beta = 0.50,\; SE = 0.08,\; z = 5.98,\; p < .001$), 
while digital twins did not 
($\beta = 0.14,\; SE = 0.08,\; z = 1.73,\; p = .08$). 
We observed no main effect of dataset on green energy option adoption 
($\beta = -0.22,\; SE = 0.12,\; z = -1.86,\; p = .06$).

In the organ donation task, we detected no effect among humans 
($\beta = 0.09,\; SE = 0.08,\; z = 1.10,\; p = .28$), 
but a large effect among digital twins 
($\beta = 0.92,\; SE = 0.11,\; z = 8.61,\; p < .001$). 
In addition, we observed a main effect of dataset, such that digital twins were more
likely to become organ donors overall 
($M_{\text{Human}} = 59\%,\; M_{\text{Twin}} = 71\%,\; \beta = 0.75,\; SE = 0.14,\; z = 5.53,\; p < .001$). 
Though this effect may be primarily driven by the difference in the opt-out condition 
($M_{\text{Human}} = 61\%,\; M_{\text{Twin}} = 88\%$).

We next examined how closely digital twins predicted human choices,
independent of whether they predicted the default effects. For each
trial, we created an indicator variable denoting whether the digital
twin correctly predicted the participant's choice.\footnote{We created
  an indicator variable denoting whether the digital twin correctly
  predicted the participant's choice and used a McNemar test for
  analysis, rather than the preregistered correlation, as it is more
  appropriate for discrete choice outcomes.} As a benchmark, we compared
digital twin's accuracy level to that of a baseline model that randomly
selected among the available options.

Across both paradigms, digital twins outperformed the random baseline.
Specifically, twins achieved 69\% accuracy on the green energy paradigm
(paired \emph{t}-test against 50\% baseline, $t(599) = 7.11,\; p < .001$), 
and 58\% on the organ donation paradigm 
(against 50\%, $t(599) = 2.99,\; p = .003$). 
Results were consistent when using McNemar's test, which treats the dichotomous
accuracy indicator as a nominal variable, instead of the \emph{t}-test.

\subsubsection{Results - Additional Analyses (Non-Preregistered)}

Using response time data from human trials, we explored whether
attention moderated the observed context effects. We included
interaction terms between the log-transformed and centered response time
for each trial and the contrast variables in the nested model. We found
no statistically significant moderation effects for either human
participants (green energy: $\beta = -0.01,\; SE = 0.09,\; z = -0.15,\; p = .88$; 
organ donation: $\beta = -0.15,\; SE = 0.09,\; z = -1.71,\; p = .09$). 
Response time derived from the human dataset also did not moderate the effects in the digital twin
models (green energy: $\beta = 0.06,\; SE = 0.08,\; z = 0.77,\; p = .44$; 
organ donation: $\beta = -0.09,\; SE = 0.10,\; z = -0.90,\; p = .37$).

\subsubsection{Discussion}

Large language models (LLMs) like ChatGPT can accurately describe
default effects and reference the relevant literature, presumably
because these concepts are well represented in their training data.
Default effects, while often robust, vary considerably in magnitude
\cite{jachimowicz2019defaults}. In the present study, human participants
showed a default effect in the green energy adoption paradigm but not in
the organ donation paradigm. Digital twins predicted a strong default
effect in the organ donation paradigm but no effect in the green energy
paradigm.

This discrepancy may stem from greater representation of the organ
donation paradigm in the LLM's training data, while our modified green
energy options may have diverged from commonly seen formats. Still,
digital twins were able to predict human choices in the green energy
paradigm with 69\% accuracy, showing some promise in capturing people's
underlying preferences.

\pagebreak

\subsection{Digital Certificates for Luxury Consumption}\label{digital-certificates-for-luxury-consumption}

\subsubsection{Main Questions/Hypotheses}

Luxury brands face the perpetual problem of balancing exclusivity with
growth \cite{keller2017luxury}. Diffusion of products into the marketplace
inevitably erodes exclusivity and negatively impacts product and brand
value \cite{bellezza2014brandtourists}. Recent work has demonstrated that a
new technology---digital certification---reinforces value when luxury
products become diffused \cite{park2022nft}. In this study,
we aim to replicate this pattern with human participants, and test
whether their digital twins respond similarly to digital certification.

Specifically, we test whether attaching a digital passport to a luxury
product that has become diffused increases its perceived value (i.e.,
status perceptions) and monetary value (i.e., price perceptions).
Digital passports are a novel Blockchain-based certification system
inspired by a recent European Union regulation to create more
transparent and sustainable supply chains for physical products \cite{hrw2024supplychain}. For example, the Aura Blockchain Collective (a non-profit collective of major luxury firms)
provides digital passports that allow consumers to follow their
purchases through the entire product journey, from the origin of the raw
materials to recycling.\footnote{``Solutions - Aura Blockchain Consortium'' (n.d.), (accessed February 18, 2025), [available at https://auraconsortium.com/solutions].} 

\subsubsection{Methods}

We opened the study to 600 human participants on Prolific (46.50\%
female, 53.50\% male; $M_\text{age} = 49.86$, $SD = 14.32$). The study
was subsequently run on their 600 digital twins, and the final sample
size was 1,200 for our analysis. Participants and their corresponding
digital twins were randomly assigned to one of two conditions in a two-cell
(Digital Certification: Digital Passport vs.\ Control) between-subjects
design (each human and their digital twin were assigned to the same condition).

\begin{figure}[H]
\centering
\includegraphics[width=1\textwidth]{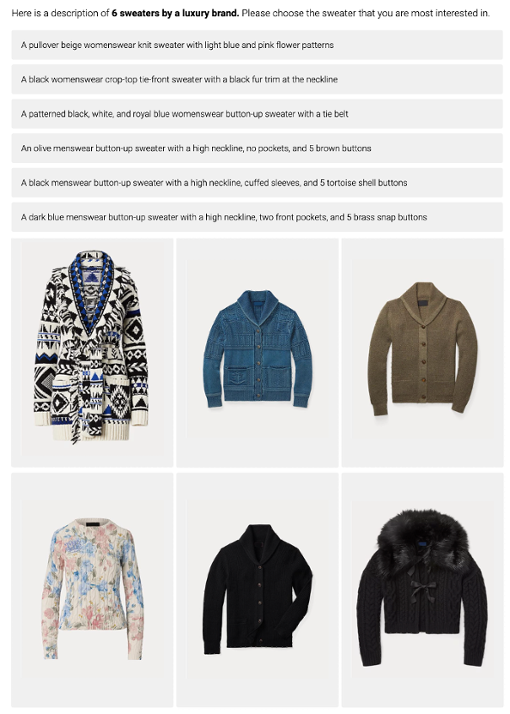}
\caption{Stimuli for Digital Certification Study.
}\label{SI_certificate_1}
\end{figure}

All participants were asked to select one of six luxury sweaters:
``Here is a selection of 6 sweaters by a luxury brand. Please choose the
sweater that you are most interested in.'' Digital twins were given a textual
description of each sweater, rather than an image, to enable comparison
between humans and their digital counterparts, who cannot process visual
images (Figure \ref{SI_certificate_1}, top). The description of each sweater was based on
real sweaters designed by Ralph Lauren ($M_\text{price} = \$531.90$;
Figure \ref{SI_certificate_1}, bottom). To avoid potential confounds, the brand was not
mentioned. These six target sweaters were chosen from a set of 20 after
an extensive pretest to ensure the products were equally liked,
perceived as high-end, suitable for the scenario (i.e., as an outfit for
going out), and conspicuous. Additional exploratory analyses controlled
for participants' selected sweater to account for differences across
product stimuli (e.g., sweater color, style) that could influence value
perceptions.

After choosing a sweater, participants in the digital passport condition
read as follows:

\begin{quote}
We will ask you a few questions about the luxury sweater you selected.
This luxury sweater is linked to a permanent digital record available on
a public website, showcasing the current owner. This allows anyone
interested to verify the ownership of your specific luxury sweater
through this accessible record. Essentially, the digital record serves
as unequivocal evidence of your luxury sweater's ownership.
\end{quote}

Participants in the control condition alternatively read: ``We will ask
you a few questions about the luxury sweater you selected.'' All
conditions next featured a scenario in which the luxury product has
become diffused. Specifically, all participants read, ``Imagine you
decided to purchase this luxury sweater and style it for your night out,
but during your night out you noticed many other people wearing the same
luxury sweater.'' Participants in the digital passport condition also
read, ``Nevertheless, you know there is a digital record associated with
the particular luxury sweater you are wearing, reassuring your
ownership.'' Finally, all participants completed a writing task to
reinforce the manipulation: ``Please imagine your night out in detail.
Briefly describe how you look and how you would feel wearing this
sweater during your night out.''

Participants then responded to the dependent variables by reporting
their perceptions of product value (``How expensive do you think the
sweater is? (Please enter number of \$)'') and status perceptions
(``Please answer the following questions. To what extent do you think
the sweater is...''; luxurious, high status, prestigious; 
1 = Not at all, 7 = To a great extent; Ward \& Dahl, 2014). These three
items were averaged for analysis ($\alpha = .94$). We log-normalized the
price perception measure to address skewness ($\mu = 4.06 > 2$; i.e.,
right-skewed; \cite{curran1996robustness}). Finally, participants
reported the extent to which they were familiar with digital passports
(``How familiar are you with digital passports?''; 1 = Very Unfamiliar,
7 = Very Familiar), the perceived difficulty of the task to account for
different prompt lengths across conditions (``How difficult was it to
imagine the sweater scenario?''; 1 = Extremely Easy, 7 = Extremely
Difficult), and whether they owned real luxury products (0 = No, 1 =
Yes).

\subsubsection{Results - Pre-registered Analyses}

\begin{figure}[H]
\centering
\includegraphics[width=1\textwidth]{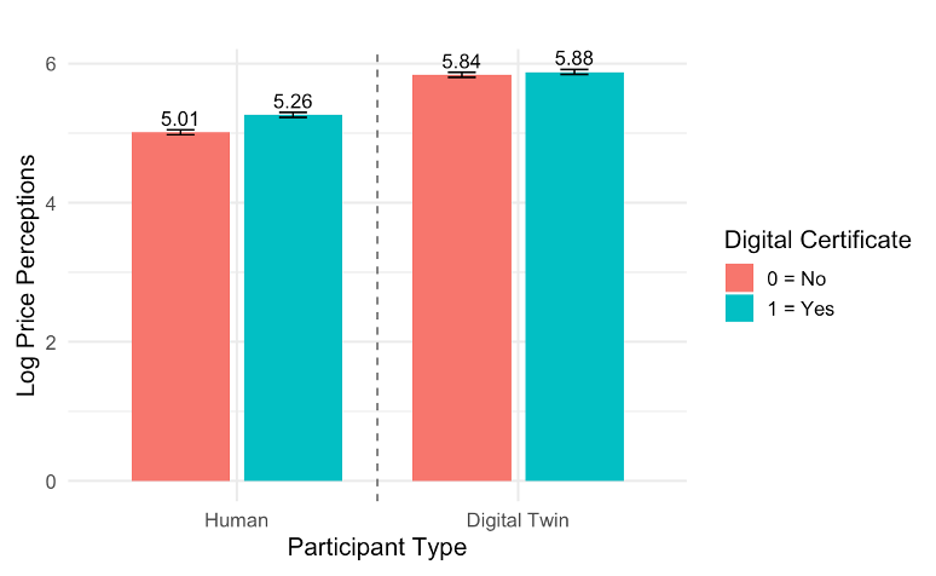}
\caption{Price Perceptions by Digital Certification for Humans vs.
Twins. NOTE: Error bars indicate ±1 standard error of the estimated
marginal means.
}\label{SI_certificate_2}
\end{figure}

We estimated a linear mixed-effects regression with participant as a
random effect, log-transformed price perceptions as the dependent
variable, and Participant Type (Human $= 0$, Digital Twin $= 1$), 
Digital Certification (Control $= 0$, Digital Passport $= 1$), and 
their interaction as predictors. We found a significant effect of 
Digital Certification ($b = .25$, $SE = .05$, $t(1190.45) = 4.94$, 
$p < .001$), a significant effect of Participant Type 
($b = .83$, $SE = .05$, $t(597.99) = 16.91$, $p < .001$), and a 
significant interaction ($b = -0.21$, $SE = .07$, 
$t(596.99) = -3.08$, $p < .01$; Figure \ref{SI_certificate_2}), indicating that, at an 
aggregate level, the digital twins rated price perceptions differently 
than the humans. Pairwise comparisons confirmed that, for humans, price 
perceptions were significantly higher when their purchase was connected 
to a digital certification ($M = 5.26$, $SE = .04$) compared to no 
digital certification ($M = 5.01$, $SE = .04$; $t(1190) = 4.94$, 
$p < .001$). However, for the digital twins, price perceptions were not 
significantly different when their purchase was connected to a digital 
certification ($M = 5.88$, $SE = .04$) compared to no digital 
certification ($M = 5.84$, $SE = .04$; $t(1190) = 0.73$, $p = .463$).

\begin{figure}[H]
\centering
\includegraphics[width=1\textwidth]{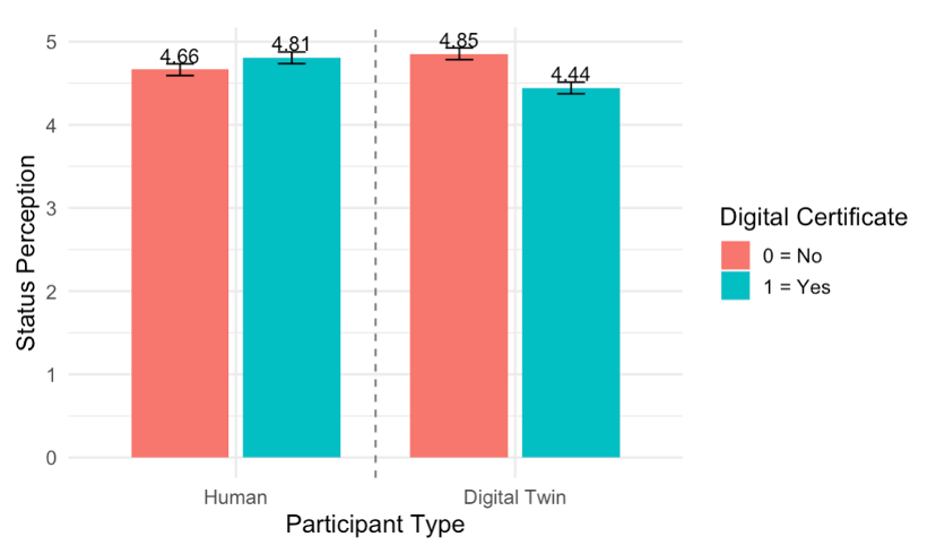}
\caption{Status Perceptions by Digital Certification for Humans vs.
Twins. NOTE: Error bars indicate ±1 standard error of the estimated
marginal means.
}\label{SI_certificate_3}
\end{figure}

We next estimated a linear mixed-effects regression with participant as
a random effect, status perceptions as the dependent variable, and
Participant Type (Human $= 0$, Digital Twin $= 1$), Digital
Certification (Control $= 0$, Digital Passport $= 1$), and their
interaction as predictors. We found a non-significant effect of
Digital Certification ($b = 0.14$, $SE = 0.10$; $t(1175.69) = 1.44$;
$p = .151$), along with a significant effect of Participant Type
($b = 0.19$, $SE = 0.09$; $t(597.99) = 2.06$; $p = .040$) and a
significant interaction ($b = -0.55$, $SE = 0.13$;
$t(597.99) = -4.24$; $p < .001$; Figure \ref{SI_certificate_3}), indicating that, at an
aggregate level, the digital twins rated status perceptions differently
than the humans. Pairwise comparisons confirmed that, for humans, status
perceptions were marginally higher when their purchase was connected to
a digital certification ($M = 4.81$, $SE = 0.07$) compared to no digital
certification ($M = 4.66$, $SE = 0.07$; $t(1176) = 1.44$; $p = .150$).
However, for the digital twins, status perceptions were significantly
lower when their purchase was connected to a digital certification
($M = 4.44$, $SE = 0.07$) compared to no digital certification
($M = 4.85$, $SE = 0.07$; $t(1176) = -4.16$; $p < .001$).

\subsubsection{Results - Additional Analyses (Non-Preregistered)}
Although not part of our pre-registered plan, we ran two additional
linear mixed-effects regressions on log-transformed price perceptions
and status perceptions, controlling for the specific product selected by
the participant, number of words generated in the imagination prompt,
participants' familiarity with digital passports, difficulty of
imagining the scenario, and personal ownership of luxury products.
Before running these regressions, we examined differences between humans
and their digital twins on each of these variables.

We first tested whether humans and their twins differed in the sweaters
they selected. The results indicated weak but statistically significant
agreement between twins and humans ($\kappa = .166$, $p < .001$). A
chi-squared test of independence also revealed a significant and
positive association between human and twin sweater choices
($\chi^2(1, N = 1200) = 297.4$, $p < .001$). Breaking down choice by the
gender association of the sweaters, digital twins tended to mirror their
human counterparts: when humans selected one of the three feminine
sweaters, their twins also selected a feminine sweater 86.31\% of the
time ($\chi^2(1, N = 241) = 125.63$, $p < .001$); when humans selected
one of the three masculine sweaters, their twins chose a masculine
sweater 85.24\% of the time ($\chi^2(1, N = 359) = 176.89$, $p < .001$).

We next checked whether all participants wrote roughly an equal number
of words (log-corrected) between conditions. While the number of words
generated did not differ between conditions for humans
($M_\text{Digital Passport} = 3.54$ vs.\ $M_\text{Control} = 3.47$;
$F(1, 598) = 1.70$, $p = .20$, $d = .11$), the digital twins generated
more words in the Digital Passport condition ($M = 4.42$) compared to
the Control ($M = 4.36$; $F(1, 598) = 66.92$, $p < .001$, $d = .67$).
Additionally, the digital twins generated more text ($M = 4.42$)
compared to their human counterparts ($M = 3.50$; $F(1, 599) = 1305.28$,
$p < .001$, $d = 1.47$).

We then tested for perceived familiarity with digital passports. Across
humans and twins, there were no differences in perceived familiarity
among participants in the Digital Passport condition ($M = 2.40$)
compared to the Control condition ($M = 2.36$; $F(1, 1198) = 0.20$,
$p = .65$, $d = .03$). Human participants, however, reported being more
familiar with digital passports ($M = 2.68$) compared to twins
($M = 2.08$; $F(1, 599) = 74.54$, $p < .001$, $d = 0.35$).

We then tested for differences in difficulty of imagining the scenario
between conditions. Human participants in the Digital Passport condition
found the imagination task more difficult ($M = 2.68$) than in the
Control condition ($M = 2.03$; $F(1, 598) = 32.70$, $p < .001$, 
$d = 0.47$). Digital twins also found the imagination task more
difficult in the Digital Passport condition ($M = 2.59$) than in the
Control condition ($M = 2.31$; $F(1, 598) = 22.32$, $p < .001$, 
$d = 0.39$). There were no differences in difficulty of imagining the
tasks between humans ($M = 2.35$) and twins ($M = 2.45$; 
$F(1, 1198) = 2.26$, $p = .13$, $d = 0.09$).

We finally tested for differences in reported ownership of luxury
products. Across humans and twins, there were no differences in luxury
product ownership among participants in the Digital Passport condition
(40.4\%) compared to the Control condition (39.3\%; 
$\chi^2(1) = 0.12$, $p = .73$). Human participants, however, reported
higher luxury ownership (53.8\%) compared to their digital twins (25.8\%;
$\chi^2(1) = 96.97$, $p < .001$).

To test whether our findings persist when controlling for these
differences, we ran a linear mixed-effects regression with participant
as a random effect, log-normalized perceived price as the dependent
variable, and Participant Type (Human $= 0$, Digital Twin $= 1$), 
Digital Certification (Control $= 0$, Digital Passport $= 1$), and their
interaction as predictors, controlling for the product participants
selected, as well as familiarity with digital passports, difficulty of
imagination, and ownership of luxury products. We found a significant
effect of familiarity ($b = 0.05$, $SE = 0.01$; $t(1190.90) = 3.48$,
$p < .001$) and luxury product ownership ($b = 0.20$, $SE = 0.04$;
$t(1170.09) = 5.10$, $p < .001$), but a non-significant effect of
product choice ($b = 0.02$, $SE = 0.01$; $t(1190.85) = 1.48$, $p = .14$)
and difficulty of imagination ($b = -0.02$, $SE = 0.02$; 
$t(1190.82) = -1.51$, $p = .13$). Nevertheless, the analysis revealed a
significant effect of Digital Certification ($b = 0.28$, $SE = 0.05$;
$t(1189.85) = 5.53$, $p < .001$) and a significant effect of Participant
Type ($b = 0.96$, $SE = 0.05$; $t(726.87) = 17.88$, $p < .001$),
qualified by a significant interaction between Digital Certification and
Participant Type ($b = -0.25$, $SE = 0.07$; $t(603.74) = -3.66$,
$p < .001$).

Next, we ran a linear mixed-effects regression with participant as a
random effect, status perceptions as the dependent variable, and
Participant Type (Human $= 0$, Digital Twin $= 1$), Digital Certification
(Control $= 0$, Digital Passport $= 1$), and their interaction as
predictors, again controlling for the product participants selected, as
well as familiarity with digital passports, difficulty of imagination,
and ownership of luxury products. We found a significant effect of
familiarity ($b = 0.07$, $SE = 0.03$; $t(1185.69) = 2.43$, $p = .02$)
and luxury product ownership ($b = 0.46$, $SE = 0.08$;
$t(1180.77) = 6.09$, $p < .001$), but a non-significant effect of
product choice ($b = -0.03$, $SE = 0.02$; $t(1185.33) = -1.43$, $p = .15$)
and difficulty of imagination ($b = -0.02$, $SE = 0.03$;
$t(1191.82) = -0.70$, $p = .48$). Nevertheless, the analysis revealed a
marginal effect of Digital Certification ($b = 0.17$, $SE = 0.10$;
$t(1183.86) = 1.72$, $p = .08$) and a significant effect of Participant
Type ($b = 0.34$, $SE = 0.10$; $t(722.27) = 3.30$, $p < .01$), qualified
by a significant interaction between Digital Certification and
Participant Type ($b = -0.59$, $SE = 0.13$; $t(598.73) = -4.51$,
$p < .001$).

These additional analyses strengthen our confidence that the effects are
robust across individual product choice, familiarity with the digital
certification context (i.e., digital passports), ease of completing the
task, and luxury product ownership.

\subsubsection{Discussion}

These findings demonstrate that providing digital certification via a
digital passport increases the value perception of a diffused luxury
product for humans, consistent with existing research \cite{park2022nft}. Digital twins' value perceptions, however, are not
impacted by the inclusion of a digital passport. The findings have
important implications for our understanding of when digital twins will
not act as their human counterparts.

In this study, the divergence may stem from the psychological
reassurance of product authenticity that certification provides to human
consumers \cite{park2022nft}, which digital twins cannot
fully internalize. This highlights a potential blind spot in current AI
models: a limited ability to model symbolic consumption, ownership
signaling, or status restoration processes. More broadly, these findings
call for caution when relying on digital twins to simulate human
judgments in domains involving identity and social
perception---especially in luxury or prestige-based contexts. Future
work should explore how these divergences arise and whether digital
twins can be improved to better simulate symbolic inference processes,
or whether some aspects of human value perception are fundamentally
non-transferrable.

\pagebreak
\subsection{Fees Accuracy}\label{fees-accuracy}

\subsubsection{Main Questions/Hypotheses}

Hidden fees and surcharges are a routine but often resented aspect of
modern commerce. Whether booking a hotel, ordering food delivery, or
using a credit card bill, consumers frequently encounter extra, often
unexpected charges---like resort fees, service fees, foreign transaction
fees, or order processing fees---added on top of the advertised base
price. These fees are often disclosed late in the transaction, hindering
consumers' ability to estimate total costs or compare options, and are
often described using obtuse language, obscuring understanding of why
the fees are assessed. As a result, they have attracted growing scrutiny
from consumer advocates and regulators. In recent years, a bipartisan
group of regulators in the United States and other countries have
advocated for stronger pricing transparency and restrictions on the
delayed disclosure of mandatory charges. In 2022, the Biden-Harris
administration labeled these ``junk fees,'' defined as ``fees designed
either to confuse or deceive consumers or to take advantage of lock-in
or other forms of situational market power'' \cite{whitehouse2022junkfees}.

Against this policy backdrop, we explore how consumers think about these
fees ---how knowledgeable they are on what these fees represent, how
fair they perceive them to be, and how supportive they are of government
regulation regarding firms' use of such fees. We also examine whether
large language models (LLMs), acting as digital twins, can replicate
their human counterparts in these domains.

Specifically, we ask the following questions. First, do digital twins
accurately replicate human knowledge of pricing fees? Second, do they
mirror human judgment about fairness? Third, do they match human
attitudes toward regulation?

\subsubsection{Methods}

We recruited 400 U.S. participants from Prolific (48\% female; median
age group = 50--64) to complete an online survey assessing their
knowledge and attitudes toward 19 different fees assessed across seven
industries (e.g., hospitality, healthcare, rental housing). Each
participant evaluated one randomly assigned fee from six of the seven
industries. See Table \ref{SI_fee_tab1} for the full list of industries and fees.

\begin{longtable}{p{0.2\linewidth} p{0.2\linewidth} p{0.6\linewidth}}
\toprule
Industry & Fee & Definition \\
\midrule
\endfirsthead

\toprule
Industry & Fee & Definition \\
\midrule
\endhead

Hotels & Resort fee & A mandatory fee for a group of services, such as pool use, gym access, towel services, Wi-Fi, newspapers, shuttle service, daily parking, etc. \\
 & Destination fee & A mandatory fee for a group of services, such as pool use, gym access, towel services, Wi-Fi, newspapers, shuttle service, daily parking, etc. \\
 & Restocking fee & A fee charged by hotels for using the mini-fridge to store personal items. \\

Car Rentals & Vehicle licensing fee & A fee charged by rental car companies to offset costs associated with vehicle registration, licensing, and related taxes. \\
 & Toll transponder fee & A daily fee assessed if a toll transponder is used during every day of the rental to pass quickly through toll booths. \\
 & Concession recovery fee & A fee charged by rental car companies to customers to recover costs associated with operating at specific locations, such as airports. \\
 & Frequent traveler program surcharge/excise tax & A fee for accumulating miles or points through credit card, airline, or other rewards programs. \\

Ticket Processing & Delivery fee & A fee for delivering tickets, whether physical or electronic. \\
 & Order processing fee & A fee to cover costs associated with processing and completing ticket purchases. \\
 & Facility charge & A fee to cover costs associated with hosting an event. \\

Food Delivery Apps & Service fee & A fee to operate the food delivery app platform. \\
 & Express fee & A fee for prioritizing the matching process for delivery services. \\
 & Regulatory response fee & A fee to offset the impact of regulations on firms. \\

Apartment Rentals & Valet trash fee & A fee for trash and recycling collection services. \\
 & Move-in fee & A non-refundable fee to cover move-in and administrative costs, such as record updates, lock changes, and unit preparation. \\
 & Amenity fee & A recurring fee that covers additional services and amenities, such as pools, gyms, parking, or other building facilities. \\

Health Care & Hospital facility fee & A fee charged by a private medical practice owned by a hospital to cover the hospital's operational and maintenance costs, including equipment, staff, and utilities. \\

Credit Cards & Foreign transaction fee & A fee for purchases or transactions in a foreign currency to cover conversion and international processing costs. \\
 & Balance transfer fee & A percentage fee that covers debt transfers from one credit card to another. \\
\bottomrule
\caption{Industries, Fees, and Definitions Used in the Survey.}\label{SI_fee_tab1} \\
\end{longtable}

For each fee, we assessed the following. First, we assessed
participants' objective knowledge by asking them to select the best
definition of a fee from four options (e.g., ``Which of the following do
you think best represents what a resort fee is assessed for?''). We
coded responses as 1 if correct and 0 if incorrect. We then averaged
participants' responses across the six fees they evaluated to create an
overall accuracy score. Immediately after responding, participants were
informed whether their answer was correct and were shown the correct
definition. Second, we assessed participants' familiarity with each fee.
Specifically, participants indicated whether they had previously heard
of each fee (e.g., they responded to questions like ``Have you ever
heard of a resort fee before seeing it in this survey?''). These items
capture prior experience and are part of a separate project and were not
pre-registered. Third, we assessed participants' fairness perceptions.
Participants rated how fair they considered each fee (e.g., they
responded to questions like ``How fair do you think it is for hotels to
charge for a resort fee?'') on a 7-point scale (1 = Very unfair; 7 =
Very fair). We then averaged participants' responses across the six fees
they evaluated to create an overall fairness rating. Finally, we
assessed participants' attitudes regarding policy. Specifically, after
completing the fee evaluations, participants responded to several items
about their general attitudes toward pricing practices and regulation.
These included two pre-registered items: ``Should pricing practices be
regulated by the government?'' rated on a 7-point scale (1 = Strongly
oppose regulation; 7 = Strongly support regulation), and ``How likely
would you be to support government regulation that bans firms from
separating out mandatory fees from base prices?'' rated on a 7-point
scale (1 = Very unlikely; 7 = Very likely). Responses to these two items
were highly correlated ($r = .72$), so we averaged them to create a
composite measure of support for regulation.

Each participant's responses were paired with those of an LLM-generated
digital twin prompted to respond as that individual would.

\subsubsection{Results - Pre-registered Analyses}

\emph{Knowledge Accuracy}. Digital twins dramatically outperformed human
participants in identifying correct definitions of fees, based on a
paired $t$-test ($t(399) = 45.89,\; p < .001,\; d = 3.26$). Humans
answered 51.75\% of fee definition questions correctly (SD = 20.87\%),
while digital twins answered 99.88\% correctly (SD = 1.86\%).

\emph{Fairness.} Despite the accuracy gap, average fairness ratings were
virtually identical. Both human participants ($M = 3.28,\; SD = 1.24$),
and their digital twins ($M = 3.27,\; SD = 0.62$) rated the fees as
moderately unfair, and their ratings did not significantly differ based
on a paired-sample $t$-test ($t(399) = 0.14,\; p = .886,\; d = 0.01$).
Table \ref{SI_fees_tab2} reports fee-specific fairness comparisons.

\begin{longtable}{@{}p{0.30\linewidth} p{0.15\linewidth} p{0.15\linewidth} p{0.4\linewidth}}
\toprule
Fee & Human M (SD) & Twins M (SD) & Paired $t$-test \\
\midrule
\endfirsthead

\toprule
Fee & Human M (SD) & Twins M (SD) & Paired $t$-test \\
\midrule
\endhead

\emph{Hotel Fees} & & & \\
Resort Fee & 3.04 (1.80) & 2.95 (0.81) & $t(131) = 0.53,\; p = .596,\; d = 0.06$ \\
Destination Fee & 2.93 (1.68) & 3.12 (0.81) & $t(133) = 1.24,\; p = .215,\; d = 0.15$ \\
Restocking Fee & 2.29 (1.61) & 2.23 (0.50) & $t(133) = 0.44,\; p = .660,\; d = 0.05$ \\

\emph{Car Rental Fees} & & & \\
Vehicle Licensing Fee & 3.44 (1.79) & 3.71 (0.86) & $t(100) = 1.46,\; p = .149,\; d = 0.20$ \\
Toll Transponder Fee & 4.15 (1.83) & 3.27 (0.83) & $t(99) = 4.44,\; p < .001,\; d = 0.62$ \\
Concession Recovery Fee & 3.29 (1.70) & 3.34 (0.81) & $t(98) = 0.29,\; p = .769,\; d = 0.04$ \\
Frequent Traveler Program Surcharge/Excise Tax & 2.43 (1.50) & 2.67 (0.67) & $t(99) = 1.65,\; p = .101,\; d = 0.20$ \\

\emph{Ticket Processing Fees} & & & \\
Delivery Fee & 2.56 (1.56) & 3.61 (0.91) & $t(130) = 7.05,\; p < .001,\; d = 0.82$ \\
Order Processing Fee & 3.05 (1.85) & 3.48 (0.87) & $t(134) = 2.83,\; p = .005,\; d = 0.28$ \\
Facility Charge & 3.36 (1.80) & 3.51 (0.82) & $t(133) = 0.93,\; p = .355,\; d = 0.10$ \\

\emph{Food Delivery App Fees} & & & \\
Service Fee & 3.91 (1.78) & 3.86 (0.77) & $t(133) = 0.33,\; p = .744,\; d = 0.04$ \\
Express Fee & 3.64 (1.77) & 3.00 (0.76) & $t(132) = 4.10,\; p < .001,\; d = 0.46$ \\
Regulatory Response Fee & 2.69 (1.66) & 2.59 (0.57) & $t(132) = 0.72,\; p = .472,\; d = 0.08$ \\

\emph{Apartment Rental Fees} & & & \\
Valet Trash Fee & 3.66 (1.88) & 3.36 (0.96) & $t(133) = 1.73,\; p = .086,\; d = 0.20$ \\
Move-In Fee & 3.10 (1.71) & 3.58 (0.83) & $t(133) = 2.89,\; p = .004,\; d = 0.36$ \\
Amenity Fee & 3.70 (1.66) & 3.55 (0.80) & $t(131) = 1.00,\; p = .320,\; d = 0.11$ \\

\emph{Health Care Fees} & & & \\
Hospital Facility Fee & 3.28 (1.94) & 2.81 (0.79) & $t(133) = 2.80,\; p = .006,\; d = 0.30$ \\

\emph{Credit Card Fees} & & & \\
Foreign Transaction Fee & 4.30 (1.69) & 3.86 (0.93) & $t(131) = 2.76,\; p = .007,\; d = 0.33$ \\
Balance Transfer Fee & 3.56 (1.87) & 3.63 (0.78) & $t(133) = 0.46,\; p = .643,\; d = 0.05$ \\
\bottomrule
\caption{Per-Fee Fairness Comparisons Between Human and AI (paired-samples). \label{SI_fees_tab2}} \\
\end{longtable}

We also compared the variation in fairness judgments between humans and
digital twins. Humans expressed significantly more variation in their
fairness ratings than did their digital twins, based on a Levene's test
($F(1, 798) = 180.48,\; p < .001$).

\begin{figure}[H]
\centering
\includegraphics[width=1\textwidth]{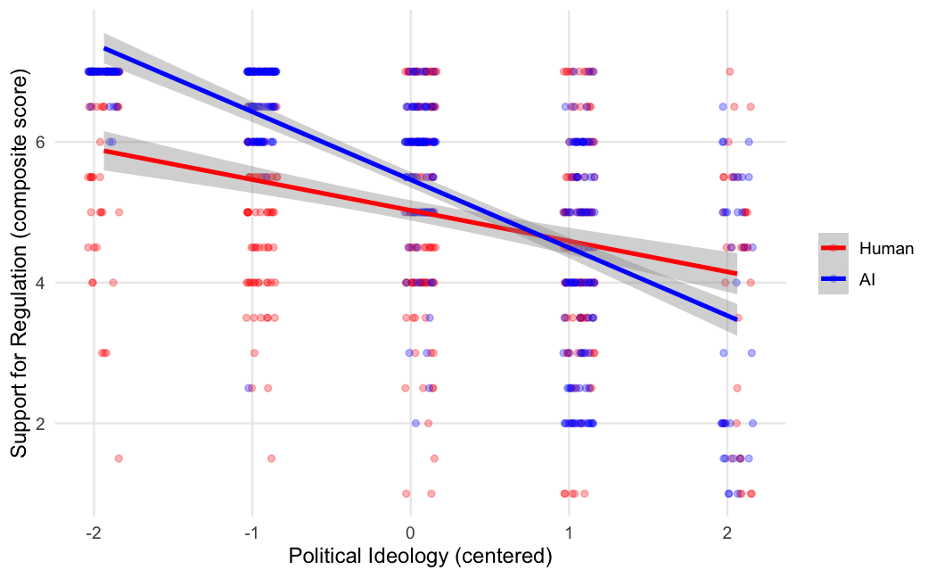}
\caption{Support for Regulation as a Function of Political Ideology and
Participant Source.
}\label{SI_fees_1}
\end{figure}

\emph{Policy Attitudes}. We regressed support for regulation on
participants' ideology (mean-centered, 1 = very liberal to 5 = very
conservative), source (Human vs.\ AI), and their interaction. The main
effect of political ideology was significant ($b = -0.44,\; SE = 0.06,\;
t(796) = -7.78,\; p < .001$), such that more conservative participants
expressed lower support for regulation. The main effect of source was
also significant ($b = 0.44,\; SE = 0.09,\; t(796) = 4.69,\; p < .001$),
with AI twins showing greater support for regulation than their human
counterparts. Finally, the interaction between political ideology and
source was significant ($b = -0.53,\; SE = 0.08,\; t(796) = -6.67,\;
p < .001$), demonstrating that the ideology effect was stronger for the
digital twins than their human counterparts (see Figure \ref{SI_fees_1}).

\subsubsection{Discussion}

Our findings reveal a striking contrast: digital twins excel at factual
knowledge (i.e., knowledge of fee definitions) and fall short of
mimicking the error-prone, heterogeneous reasoning of human consumers.
This limitation underscores a critical boundary in using LLMs as
behavioral research surrogates: when tasks depend on limited knowledge,
ambiguity, or lived experiences, digital twins may produce
unrealistically accurate responses.

In contrast to the large difference in accuracy, on fairness judgments,
digital twins closely mirrored the average human response, but exhibited
narrower variance. This suggests that while LLMs can approximate central
tendencies, they may underrepresent the subjective diversity of human
opinion.

In the domain of policy attitudes, digital twins again tracked human
performance. For both groups, conservatism was correlated with less
support for pricing regulation, replicating prior work on ideological
influences in policy attitudes (e.g., \cite{mccright2013perceived}). However,
digital twins exhibited exaggerated ideological consistency with a
stronger negative association between conservatism and regulation
support than human participants. This likely reflects LLM's reliance on
dominant associations in their training data (e.g., conservatism =
free-market values), whereas human responses may incorporate more
contextual nuance.

Together, these results offer promise and caution. The results suggest
that LLMs can simulate certain aspects of consumer judgment, but may
misrepresent human variability and error, especially when tasks require
subjective or uncertain reasoning. Because LLMs are trained to
prioritize factual accuracy and statistical associations from vast
textual corpora, they often struggle to suppress correct information,
even when explicitly instructed to simulate human errors,
misconceptions, or gaps in knowledge.

\pagebreak

\subsection{Heterogeneous Story Beliefs}\label{heterogenous-story-beliefs}

\subsubsection{Main Questions/Hypotheses}

When reading a book and predicting the emotion (valence and arousal) of
the next chapter based on the previous chapter, what is the correlation
between the expected valence and expected arousal from humans vs. their
digital twins (LLMs)?

We hypothesize that LLM-based digital twins can meaningfully simulate
human expectations about emotional trajectories in narrative contexts,
as measured by correlations between human and LLM predictions of valence
and arousal for upcoming story content.

\subsubsection{Methods}

250 participants were recruited via Prolific with stratified sampling to
ensure equal representation across racial groups (50 participants each
from White, Black, Hispanic, Asian, and Other categories), and their
digital twins.

Participants read chapters from 16 possible stories, with each
participant randomly assigned to read 2 stories. Each story consisted of
two chapters. Participants made predictions after reading each chapter:
after Chapter 1 (predicting Chapter 2) and after Chapter 2 (predicting
what would come next), resulting in 4 prediction points per participant.

\emph{Primary Dependent Variables:}

\begin{enumerate}
\item
  \emph{Human valence expectation}: Participants assigned probabilities
  to five valence levels for the next chapter:
  \begin{itemize}
  \item
    Very negative (1) to Very positive (5)
  \item
    Question: "Based on the chapter you just read, what do you think is
    the likelihood that the text of the next chapter will be very
    negative or very positive? Please assign a percentage to each. The
    total must sum to 100\%."
  \end{itemize}
\item
  \emph{Human arousal expectation}: Participants assigned probabilities
  to five arousal levels for the next chapter:

  \begin{itemize}
  \item
    Very low energy (1) to Very high energy (5)
  \item
    Question: "Based on the chapter you just read, what do you think is
    the likelihood that the text of the next chapter will be very low
    energy or very high energy? Please assign a percentage to each. The
    total must sum to 100\%."
  \end{itemize}
\end{enumerate}

From these probability distributions, we calculated expected values to
create continuous measures of expected valence and arousal.

\subsubsection{Results - Pre-registered Analyses}

The analysis included all prediction points (predictions made after both
Chapter 1 and Chapter 2 of each story):

\begin{enumerate}
\item Valence Expectations:
  \begin{enumerate}
  \item Pearson correlation: $r = 0.546$ \; [95\% CI: 0.501, 0.588]
  \item Statistical significance: $t(1002) = 20.65,\; p < .001$
  \end{enumerate}

\item Arousal Expectations:
  \begin{enumerate}
  \item Pearson correlation: $r = 0.384$ \; [95\% CI: 0.330, 0.436]
  \item Statistical significance: $t(1002) = 13.17,\; p < .001$
  \end{enumerate}
\end{enumerate}

Both correlations were statistically significant at $p < .001$,
indicating that LLM predictions are significantly associated with human
expectations for both valence and arousal across all chapter
transitions. The moderate positive correlations suggest meaningful
alignment between human and LLM story belief patterns.

\subsubsection{Discussion}

This study provides evidence that LLM-based digital twins can
meaningfully simulate human emotional expectations in narrative
contexts. The significant positive correlations for both valence 
($r = 0.416$) and arousal ($r = 0.420$) support our hypothesis that 
LLMs can capture how humans predict emotional trajectories in stories.

\pagebreak

\subsection{Hiring Algorithms}\label{hiring-algorithmsS}

\subsubsection{Main Questions/Hypotheses}

Digital twins, or AI agents trained to simulate individual behaviors,
have the potential to reshape how job seekers and employers interact.
From the candidate side, digital twins can help streamline the job
search process by identifying roles aligned with the job seeker's
preferences, tailoring job application materials, and even interacting
with recruiters on the job seeker's behalf. For firms, digital twins
enable the screening of candidates according to the firm's preferences,
the simulation of candidate responses to different human resource
policies, and even candidate outreach and outbound recruiting efforts.
The advent of digital twins, and a growing ecosystem of start-ups
offering such services, may thus fundamentally reshape the dynamics of
hiring, and employee-employer matching more generally.

Despite the growing interest in digital twins, we lack evidence of their
effectiveness to capture workplace preferences. While digital twins have
shown promise in modeling preferences in more structured domains, there
are unique challenges in the workplace context. For example, workplace
preferences are multi-dimensional and candidates must weigh trade-offs
between attributes like compensation, skill development, flexibility,
and culture. Moreover, workplace preferences are shaped by social
context and may vary over time. These features call into question the
extent to which current digital twins can accurately represent how
employees and firms navigate the matching process.

In this sub-study, we examine the ability of digital twins to predict
job candidate preferences over workplace attributes using a two-stage
study. In the first stage, we elicit the stated preferences of job
candidates over a range of workplace characteristics, including
work-life balance, career development, and company culture. We then ask
each participant's digital twin to report their preferences over the
same attributes. By comparing the stated and simulated preferences, we
can understand how well digital twins predict the stated preferences of
their human counterparts. In the second stage, we use a within-subjects
experiment where participants evaluate four job postings. Our
manipulation here changes whether the firm's hiring is done by
algorithms or humans, and we randomly assign two of the four job
postings to have AI recruiters. Both humans and their digital twins
evaluate each of the four job postings on eight key attributes,
including transparency, fairness, and likelihood of applying. This
design thus allows us to understand the efficacy of digital twins in
predicting workplace preferences and experiments in the HR domain.

\subsubsection{Methods}

\begin{figure}[H]
\centering
\includegraphics[width=1\textwidth]{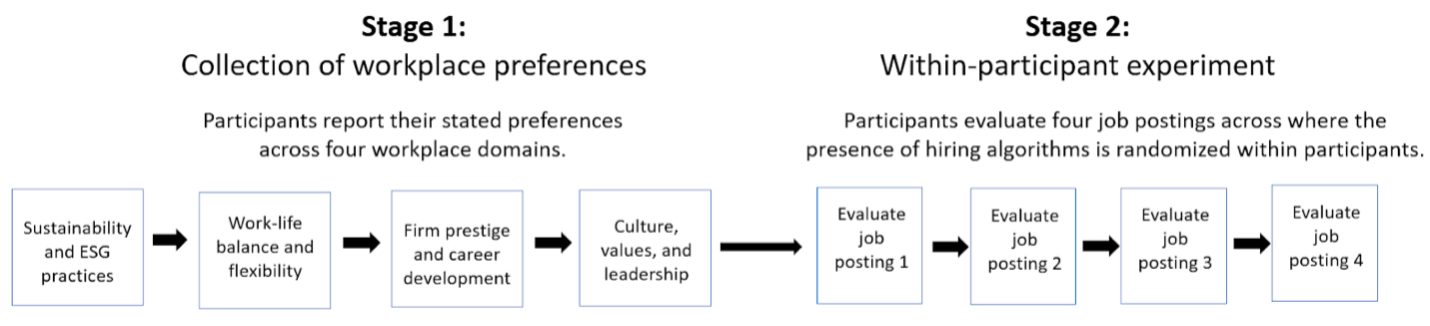}
\caption{Outline of the study.
}\label{SI_hiring_1}
\end{figure}

We run a two-stage study to examine these questions. We outline these
stages in Figure \ref{SI_hiring_1}. Both humans and their digital twins completed the
study in the same order.

The first stage of our study collects data on respondent workplace
preferences. We asked respondents a total of eight questions across four
distinct themes related to workplace preferences. These themes included:

\begin{itemize}
\item
  \textbf{Sustainability and ESG practices}

  \begin{itemize}
  \item
    "How important is it to you that your employer actively invests in
    environmental sustainability (e.g., reducing carbon emissions,
    minimizing waste)?''
  \item
    "Would you prefer to work at a company that is outspoken about
    social and political issues, or one that remains neutral?"
  \end{itemize}
\item
  \textbf{Work-life balance and flexibility}

  \begin{itemize}
  \item
    "How likely are you to accept a job that requires you to be in the
    office five days a week?"
  \item
    "Which of the following is more important to you when choosing a
    job? Where options included: (i) A flexible schedule that allows me
    to manage personal responsibilities; (ii) Clear boundaries between
    work and personal time, (iii) Strict 9-to-5 hours (iv) A
    predictable, fixed schedule\footnote{To construct the index
      measuring respondents' preferences for flexible work--life balance
      arrangements, we ordered the four response options from least to
      most flexible as follows: (1) \emph{Strict 9--5 hours}, (2)
      \emph{Predictable, fixed schedule}, (3) \emph{Clear boundaries
      between work and personal time}, and (4) \emph{Flexible schedule
      that allows me to manage personal responsibilities}. This index
      captures increasing preference for autonomy in managing one's work
      schedule.}
  \end{itemize}
\item
  \textbf{Firm prestige \& career development}

  \begin{itemize}
  \item
    "How much do you value opportunities for formal training and
    upskilling in your current or future role?"
  \item
    "Imagine two job offers: one from a prestigious company with a
    competitive culture, and one from a lesser-known company with a
    collaborative, supportive environment. Which would you
    choose?"\footnote{To construct the index for collaborative culture
      preference, we ordered the responses to reflect increasing
      preference for collaborative environments: (1) \emph{Prestigious},
      (2) \emph{Depends on compensation}, (3) \emph{Not sure}, and (4)
      \emph{Collaborative}. This index captures respondents' orientation
      toward workplace culture, with higher scores indicating stronger
      preferences for collaboration over prestige or ambiguity.}
  \end{itemize}
\item
  \textbf{Culture, values, and leadership}

  \begin{itemize}
  \item
    "How important is it that your company's leadership communicates
    transparently about the firm's goals and challenges?"
  \item
    "How much do you care about your company's stated mission and values
    aligning with your personal beliefs?"
  \end{itemize}
\end{itemize}

We collect these measures for both human participants and their digital
twins. Thus for each participant, we have their stated preference
regarding (i) sustainability \& ESG practices; (ii) work-life balance
and flexibility; (iii) firm prestige and career development; and (iv)
culture and values. We also have the corresponding responses from the
digital twins, allowing us to compare how well LLMs can predict job
candidate preferences in the hiring process.

The second stage of our study uses an experiment with within-subjects
variation to understand how candidates respond to algorithms in the
hiring process. In this part, participants evaluate hiring policies from
the perspective of a job seeker. We present participants with four
different job postings, which display a position overview, the
responsibilities, benefits, and hiring policy (See Figure \ref{SI_hiring_ex} for an
example). Our randomization manipulates whether hiring uses either human
resume screeners or algorithmic ones. Following each posting,
participants evaluated the job on eight questions (displayed in Table \ref{SI_hiring_tab1}
below), using a 7-point Likert scale from strongly disagree to strongly
agree.

\begin{figure}[H]

\begin{tcolorbox}[colback=white, colframe=black, boxrule=0.5pt, arc=0mm]
\textbf{Job Title:} Entry Level Data Analyst

\medskip
\textbf{Position Overview:} As an Entry Level Data Analyst, you will play a pivotal role in analyzing complex financial data, developing valuable insights, and supporting decision-making processes within our firm.

\medskip
\textbf{Responsibilities:} The responsibilities for this position include collecting, cleaning, and manipulating large datasets to extract meaningful insights that support business objectives.

\medskip
\textbf{Benefits:} Our benefits include a competitive salary commensurate with experience, health, dental, and vision insurance, and a collaborative and inclusive work environment.

\medskip
\textbf{Hiring Policy:} Our hiring team will create insights on how your candidate information matches the requirements of the role. Following this initial screening, our hiring team will conduct in-person interviews with selected candidates.
\end{tcolorbox}


\caption{Sample Job Posting.}\label{SI_hiring_ex}
\end{figure}

\begin{longtable}{p{0.05\linewidth} p{0.70\linewidth} p{0.25\linewidth}}
\toprule
\# & \textbf{Question} & \textbf{Theme} \\
\midrule
\endhead
1 & I would apply for this position. & Likelihood of applying \\
2 & I would be hired for this position. & Likelihood of receiving offer \\
3 & I will receive clear and transparent communication throughout the hiring process. & Transparency \\
4 & The firm's hiring policies are fair and unbiased. & Fairness \\
5 & The hiring process will move at a reasonable pace and ensure a timely decision-making process. & Efficiency and timeliness \\
6 & The hiring process will give me a clear understanding of the job requirements, expectations, and the company culture. & Information \\
7 & The hiring process will effectively identify top job applicants for the position. & Identifies top talent \\
8 & The firm's hiring process suggests the firm fosters a high degree of social interaction. & Sociality \\
\bottomrule
\caption{Questions regarding job postings.}
\label{SI_hiring_tab1}
\end{longtable}

\subsubsection{Results - Pre-registered Analyses}

We first examine whether digital twins can accurately predict human
workplace preferences using two methods. First, we run paired t-tests
to test whether on average the twin outcomes equal the actual human
outcomes. Second, we use regression to estimate the size and direction
of the bias.

Table \ref{SI_hiring_tab2} displays the results of the paired t-tests across all of our
questions. The top panel contains the outcomes from the first phase of
the study, while the bottom panel displays them from the second phase of
the study. Column 1 displays the average human response while column 2
displays the average digital twin response. Column 3 displays the
average difference in responses for the digital twin versus the human,
with the corresponding t-test p-value in column 4. Lastly, column 5
displays the sample size for each group.

The results in Table \ref{SI_hiring_tab2} indicate that, on average, the outcomes generated
by digital twins are statistically different from those generated by
humans. Across the eight workplace preferences in phase 1 of the study
(Panel A of Table \ref{SI_hiring_tab2}), we only find evidence of no differences between
the two for workplace flexibility preferences. For the other categories,
we find statistically significant differences between the two. For most,
the digital twins over-estimate human preferences for workplace
attributes (for example, sustainability, ESG practices, collaboration,
and transparency in leadership). Meanwhile, the digital twins
under-estimate their human counterparts' preferences for career
development opportunities.

We find qualitatively similar results for the second phase of the study,
which we display in Panel B of Table \ref{SI_hiring_tab2}. We find statistically
significant differences across all outcomes here. Like before, the
digital twins over-estimate some responses (for example, the likelihood
of being hired for the position) and under-estimate others (the firm's
degree of social interaction, and receiving clear information regarding
job requirements, expectations, and company culture).

\begin{longtable}{p{0.30\linewidth} p{0.12\linewidth} p{0.12\linewidth} p{0.12\linewidth} p{0.12\linewidth} p{0.12\linewidth} p{0.10\linewidth}}
\toprule
\multicolumn{7}{l}{\textbf{Panel A: Phase 1 Outcomes}} \\
\midrule
 & \multicolumn{3}{c}{Average} & Difference & T-test &  \\
 & Human & \multicolumn{2}{c}{Digital Twin} & DT - H & p-value & N \\
Outcomes & (1) & \multicolumn{2}{c}{(2)} & (3) & (4) & (5) \\
\midrule
Sustainability & 2.88 & \multicolumn{2}{c}{3.45} & 0.57 & 0.00 & 999 \\
ESG Practices & 3.30 & \multicolumn{2}{c}{3.67} & 0.37 & 0.00 & 999 \\
Work Life Balance & 3.36 & \multicolumn{2}{c}{3.33} & -0.03 & 0.40 & 999 \\
Flexibility & 2.97 & \multicolumn{2}{c}{2.95} & -0.02 & 0.70 & 999 \\
Collaboration over prestige & 2.98 & \multicolumn{2}{c}{3.65} & 0.67 & 0.00 & 999 \\
Career Development & 3.67 & \multicolumn{2}{c}{3.56} & -0.11 & 0.01 & 999 \\
Transparent Leadership & 3.83 & \multicolumn{2}{c}{4.43} & 0.60 & 0.00 & 999 \\
Culture \& Values & 3.39 & \multicolumn{2}{c}{3.54} & 0.15 & 0.00 & 999 \\
\midrule
\multicolumn{7}{l}{\textbf{Panel B: Phase 2 Outcomes}} \\
\midrule
 & \multicolumn{2}{c}{Human} & Digital Twin & DT - H & p-value & N \\
Outcomes & \multicolumn{2}{c}{(1)} & (2) & (3) & (4) & (5) \\
\midrule
I would apply for this position & \multicolumn{2}{c}{3.34} & 3.48 & 0.14 & 0.00 & 999 \\
I would be hired for this position & \multicolumn{2}{c}{3.46} & 4.22 & 0.76 & 0.00 & 999 \\
I will receive clear and transparent communication & \multicolumn{2}{c}{4.89} & 4.79 & -0.10 & 0.00 & 999 \\
The firm's hiring policies are fair and unbiased & \multicolumn{2}{c}{4.88} & 4.59 & -0.29 & 0.00 & 999 \\
The hiring process will move at a reasonable pace & \multicolumn{2}{c}{5.01} & 4.55 & -0.46 & 0.00 & 999 \\
The hiring process will give me a clear understanding & \multicolumn{2}{c}{5.20} & 4.70 & -0.50 & 0.00 & 999 \\
The hiring process will effectively identify top applicants & \multicolumn{2}{c}{5.05} & 4.60 & -0.45 & 0.00 & 999 \\
The firm's hiring process suggests strong social interaction & \multicolumn{2}{c}{4.61} & 4.05 & -0.56 & 0.00 & 999 \\
\bottomrule
\caption{Comparison of human versus digital twin responses using t-tests.}
\label{SI_hiring_tab2}
\end{longtable}

The regression tests of Hypothesis 1 lead to similar takeaways. In
Table \ref{SI_hiring_tab3}, we present the results of a regression of the human response on
the twin's response with robust standard errors. Columns 1 and 2 display
the coefficient and standard error on the twin's response, respectively.
Column 3 displays the $p$-value from a test of whether the coefficient
on the twin response equals one. The corresponding null hypothesis is
that digital twins can perfectly predict human responses
(i.e., $\beta = 1$). As the table illustrates, however, we can reject
the null across all of our outcomes. Thus, while the digital twins can
predict workplace preferences for some participants, on average they
struggle in this domain.

\begin{longtable}{p{0.55\linewidth} p{0.15\linewidth} p{0.15\linewidth} p{0.15\linewidth}}
\toprule
\multicolumn{4}{l}{\textbf{Panel A: Phase 1 Outcomes}} \\
\midrule
Outcomes & $\beta$ (1) & Std. Err. (2) & p-value (3) \\
\midrule
Sustainability              & 0.68 & 0.02 & 0.00 \\
ESG Practices               & 0.55 & 0.02 & 0.00 \\
Work Life Balance           & 0.11 & 0.03 & 0.00 \\
Flexibility                 & 0.14 & 0.03 & 0.00 \\
Collaboration over prestige & 0.12 & 0.04 & 0.00 \\
Career Development          & 0.32 & 0.04 & 0.00 \\
Transparent Leadership      & 0.45 & 0.05 & 0.00 \\
Culture \& Values           & 0.46 & 0.04 & 0.00 \\
\midrule
\multicolumn{4}{l}{\textbf{Panel B: Phase 2 Outcomes}} \\
\midrule
Outcomes & $\beta$ (1) & Std. Err. (2) & p-value (3) \\
\midrule
Apply                        & 0.70 & 0.12 & 0.08 \\
Hired                        & 0.27 & 0.06 & 0.00 \\
Communication                & 0.39 & 0.07 & 0.00 \\
Unbiased hiring policies     & 0.41 & 0.06 & 0.00 \\
Timely decision-making       & 0.38 & 0.06 & 0.00 \\
Clear information            & 0.39 & 0.07 & 0.00 \\
Identify top applicants      & 0.39 & 0.07 & 0.00 \\
High degree of sociality     & 0.04 & 0.04 & 0.00 \\
\bottomrule
\caption{Comparison of human versus digital twin responses using regression.}
\label{SI_hiring_tab3}
\end{longtable}

In our pre-analysis plan, we were also interested in whether digital
twins can better predict outcomes for various demographic groups. We
focused specifically on differences by gender, age, and race. For this
analysis, we used the $t$-test strategy described above. We conducted a
$t$-test for each specific subsample and report the $p$-values in
Table \ref{SI_hiring_tab4}. The last row of the table counts the number of times we can
reject the null using a cut-off of $p < .05$. In other words, this row
calculates the number of outcomes for the given subgroup where digital
twin responses are statistically indistinguishable from human responses.
Our results provide suggestive evidence that digital twins do better at
predicting outcomes for women versus men, as we can reject the null of
no differences between the human and twin responses for three outcomes
for women (versus two for men). Similarly, the twins seem to do better
for older respondents versus younger ones. Overall, however, the
differences across subgroups are quite limited.

\begin{longtable}{p{0.25\linewidth} p{0.08\linewidth} p{0.08\linewidth} p{0.10\linewidth} p{0.10\linewidth} p{0.08\linewidth} p{0.08\linewidth} p{0.08\linewidth} p{0.08\linewidth}}
\toprule
& \multicolumn{2}{c}{\textbf{Gender}} & \multicolumn{2}{c}{\textbf{Race}} & \multicolumn{4}{c}{\textbf{Age}} \\
\cmidrule(lr){2-3} \cmidrule(lr){4-5} \cmidrule(lr){6-9}
Outcomes & Female (1) & Male (2) & White (3) & Non-White (4) & 18--29 (5) & 30--49 (6) & 50--64 (7) & 65+ (8) \\
\midrule
\endhead
Sustainability          & 0.00 & 0.00 & 0.00 & 0.00 & 0.00 & 0.00 & 0.00 & 0.00 \\
ESG Practices           & 0.00 & 0.00 & 0.00 & 0.00 & 0.00 & 0.00 & 0.00 & 0.00 \\
Work Life Balance       & 0.33 & 0.80 & 0.00 & 0.00 & 0.05 & 0.02 & 0.21 & 0.73 \\
Flexibility             & 0.83 & 0.74 & 0.12 & 0.00 & 0.19 & 0.37 & 0.00 & 0.41 \\
Collaboration           & 0.00 & 0.00 & 0.00 & 0.00 & 0.00 & 0.00 & 0.00 & 0.00 \\
Career Development      & 0.17 & 0.01 & 0.03 & 0.10 & 0.00 & 0.00 & 0.25 & 0.07 \\
Transparent Leadership  & 0.00 & 0.00 & 0.00 & 0.00 & 0.00 & 0.00 & 0.00 & 0.00 \\
Culture \& Values       & 0.00 & 0.01 & 0.00 & 0.01 & 0.00 & 0.03 & 0.00 & 0.00 \\
\bottomrule
\caption{Comparison of human versus digital twin responses using t-tests, by subgroup.}
\label{SI_hiring_tab4}
\end{longtable}

We next examine whether experimentation on the digital twins can uncover
the true treatment effect in the human sample. Although the previous
results show that digital twins responses are biased across workplace
attributes, valid causal inference may still be possible. For example,
if the digital twins over-estimate the responses from all participants,
estimation using the digital twins may still uncover the true average
treatment effect in the human population if this bias is constant across
treatment arms.

\begin{figure}[H]
\centering
\includegraphics[width=1\textwidth]{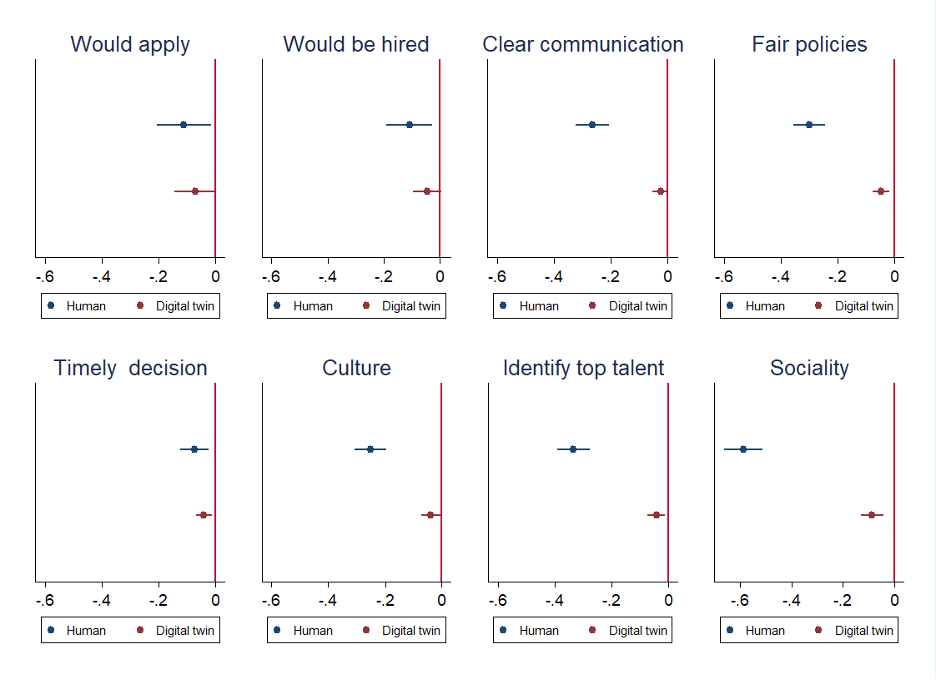}
\caption{Impact of Hiring Algorithms on Job Candidate Beliefs, For
Humans Vs Their Digital Twins
}\label{SI_hiring_2}
\end{figure}

Our experimental manipulation is the use of algorithms in a firm's
hiring policy. Each participant saw four job postings, where we randomly
assigned two to use hiring algorithms (vs human screeners). For our
estimation here, we regress our outcomes on a binary indicator for
whether the job included a hiring algorithm, participant fixed effects,
and job-type fixed effects. We do so for both our human sample and our
digital twins sample, and plot the results in Figure \ref{SI_hiring_2}. Moreover, we use
a stacked regression to compare the treatment effects from our human vs
digital twin sample. For this, we append the human and digital twin
samples and create a binary indicator for the twin sample. We then
regress each outcome on an indicator for the hiring algorithm condition,
an indicator for the twin sample, and an interaction between the two.
The interaction captures the extent to which the treatment effect in the
twin sample is different from the effect in the human sample. We display
these results in Table \ref{SI_hiring_tab5}.

The results reveal that experimentation on digital twins can uncover
unbiased treatment effects even in scenarios where digital twins
struggle to predict workplace preferences. For example, the digital
twins accurately recover the treatment effect from the human sample when
examining how hiring algorithms impact the likelihood of applying, being
hired, and receiving timely information during the hiring process. These
measures were where the digital twins failed to accurately predict human
preferences at baseline.

In other cases, however, digital twins fail to recover the true
treatment effect. The largest differences here are for beliefs regarding
sociality at work and the ability of the firm to identify top talent. In
these scenarios, digital twins substantially under-estimate the extent
to which hiring algorithms depress human beliefs regarding the firm's
hiring policy. We find similar results for clear communication,
information regarding company culture, and whether hiring is fair and
unbiased. This suggests that overall, digital twins are less adverse to
algorithms compared to humans.

\pagebreak

\begingroup
\footnotesize
\begin{longtable}{@{}L{0.20\linewidth}C{0.1\linewidth}C{0.1\linewidth}C{0.1\linewidth}C{0.1\linewidth}C{0.1\linewidth}C{0.1\linewidth}@{}C{0.1\linewidth}@{}C{0.1\linewidth}@{}}
\toprule
& \multicolumn{8}{c}{\textbf{Outcomes}} \\
\cmidrule(lr){2-9}
& Apply (1) & Hired (2) & Communication (3) & Fair Policies (4) & Timely Decision (5) & Clear Expectations (6) & Identify Talent (7) & Sociality (8) \\
\midrule
\endhead
AI               & -0.11**  & -0.11*** & -0.27*** & -0.30*** & -0.07*** & -0.25*** & -0.33*** & -0.59*** \\
                 & (0.05)   & (0.04)   & (0.03)   & (0.03)   & (0.03)   & (0.03)   & (0.03)   & (0.04)   \\
                 &          &          &          &          &          &          &          &          \\
Digital Twin     & 0.13***  & 0.73***  & -0.22*** & -0.42*** & -0.47*** & -0.61*** & -0.60*** & -0.81*** \\
                 & (0.05)   & (0.04)   & (0.03)   & (0.03)   & (0.03)   & (0.03)   & (0.03)   & (0.04)   \\
                 &          &          &          &          &          &          &          &          \\
AI $\times$ Digital Twin & 0.04     & 0.06     & 0.24***  & 0.26***  & 0.03     & 0.22***  & 0.29***  & 0.50*** \\
                 & (0.07)   & (0.06)   & (0.04)   & (0.04)   & (0.04)   & (0.04)   & (0.04)   & (0.05)   \\
                 &          &          &          &          &          &          &          &          \\
N                & 7992     & 7992     & 7992     & 7992     & 7992     & 7992     & 7992     & 7992     \\
\bottomrule
\caption{Comparison of within-participant effects using stacked regressions, by outcome.}
\label{SI_hiring_tab5}
\end{longtable}
\endgroup

\subsubsection{Discussion}
Our results highlight several opportunities and drawbacks for the use of digital twins in human resource management. In terms of predicting preferences, we find that digital twins over-estimate human preferences for workplace attributes like collaboration, leadership transparency, and sustainability. Meanwhile, they are more accurate in estimating preferences for work-life balance, flexibility, and career development opportunities. This pattern suggests that while digital twins may be able to capture workplace preferences related to instrumental outcomes, they may struggle in domains with socially oriented or relational preferences, which are fundamental to modern-day workplaces.\footnote{See, for example, \cite{dell2025super}, who show how automation interferes with workers’ preference for sociality in the workplace.}  Instead, digital twins over-estimate the importance of these attributes, relative to human jobseekers who are driven less by these attributes.

As both job candidates and employers adopt digital twins, understanding how digital twins shape the matching process is of utmost importance. Our results indicate digital twins can predict workplace preferences for job seekers in some domains, though future work remains. For example, one important future direction is understanding the training data necessary for digital twins to work at scale. While our setting features a broad array of background information used to program the digital twin \cite{toubia2025twin2k500}, more targeted data collection, for example through the evaluation of hypothetical job postings, may increase the performance of digital twins in the hiring domain (see, for example, \cite{kessler2019incentivized}). Another important avenue for future work is to understand how the labor market influences the performance of digital twins. For example, while digital twins may be able to predict preferences over short-term task assignments \cite{rusak2025ai}, they may struggle with more formal job opportunities, which may require a deeper appreciation of tradeoffs between financial incentives, meaning at work, pro-social tendencies, and other attributes.

\pagebreak

\subsection{Idea Evaluation}\label{idea-evaluation}

\subsubsection{Main Questions/Hypotheses}

First, we explore whether digital twins are able to replicate creativity
ratings from their human counterparts: How do digital twins' ratings of
idea creativity compare to those of their human counterparts?

Second, research has shown that humans tend to show AI aversion or human
favoritism when evaluating the creativity of ideas \cite{zhang2023human}, i.e., giving higher creativity ratings to ideas coming from
humans when informed that the ideas come from humans (as opposed to AI).
We explore whether the reverse is true for digital twins, i.e., do
digital twins show human favoritism, AI favoritism, or neither when
evaluating ideas?

\subsubsection{Methods}

We used a 2 (human ideas vs. twin ideas) x 3 (baseline vs. partially
informed vs. informed) design, targeting N=200 humans and their twins
per condition. The ideas came from the ``measures of creativity''
sub-study, which generated 200 ideas from humans and 200 ideas from
their digital twins. Participants in the ``measures of creativity''
sub-study were excluded from the ``idea evaluation'' sub-study. Each
participant was randomly assigned to evaluating ideas from humans or
twins and evaluated 20 ideas randomly selected from that set.
Participants were further assigned randomly to one of three evaluation
conditions, based on Zhang and Gosline (2023). In the baseline
condition, no information was given on the source of the ideas.
Participants were shown ideas (for smartphone apps that will help their
users keep a healthier lifestyle) one at a time, and asked to rate the
creativity of each idea on a 5-point scale (``very creative'' to ``very
uncreative''). In the partially informed condition, participants were
informed that ``the ideas in this study came from two sources: -humans:
Prolific participants who were asked to generate ideas; -AI: a large
language model (Chat GPT or other) who was asked to generate ideas. The
ideas you are about to evaluate came from one of these two sources. That
is, they either all came from humans, or they all came from AI,''
followed by a comprehension check. Then they were shown 20 ideas one at
a time (``Below is an idea that was generated either by a human or by
AI.'') which they were asked to rate for creativity on a 5-point scale.
In the full information condition, participants were informed that ``The
ideas in this study came from two sources: -humans: Prolific
participants who were asked to generate ideas; -AI: a large language
model (Chat GPT or other) who was asked to generate ideas. The ideas you
are about to evaluate all came from humans/AI,'' also followed by a
comprehension check. They were shown 20 ideas (``Below is an idea that
was generated by a human/AI'') and asked to rate them for creativity on
the same scale.

We obtained complete data from 1,174 respondents and their twins. Given
the number of ideas (400), the number of evaluation conditions (3), and
the number of ideas evaluated per participant (20), each idea was
evaluated by 19.567 judges on average.

\subsubsection{Results - Pre-registered Analyses}

\emph{How do digital twins' ratings of idea creativity compare to those
of their human counterparts?}

We first analyze human ideas and compare the average creativity rating
of each idea in the baseline condition coming from twins vs.\ human
judges. We find that human ideas are rated on average significantly
higher by twins than they are by humans ($M_{\text{twins}} = 3.310,\;
M_{\text{humans}} = 3.198,\; t = 2.859,\; p < .01$), and that the
standard deviation of the creativity ratings across ideas is smaller for
ratings coming from twins vs.\ humans ($S_{\text{twins}} = 0.223,\;
S_{\text{humans}} = 0.530,\; F(199, 199) = 5.664,\; p < .001$). The mean
absolute percentage error between ratings from twins vs.\ humans is
0.155. The correlation between ratings coming from twins vs.\ humans is
positive but not statistically significant ($r = 0.112,\; p = .114$).

Next, we analyze twin ideas similarly. Here, we actually find the
opposite pattern for average creativity ratings: twin ideas are rated on
average significantly lower by twins than they are by humans
($M_{\text{twins}} = 3.213,\; M_{\text{humans}} = 3.322,\; t = 3.742,\;
p < .01$). The standard deviation of the creativity ratings across ideas
is again smaller for ratings coming from twins vs.\ humans
($S_{\text{twins}} = 0.172,\; S_{\text{humans}} = 0.396,\; F(199, 199) =
5.292,\; p < .001$). The mean absolute percentage error between ratings
from twins vs.\ humans is 0.103. The correlation between ratings coming
from twins vs.\ humans is positive but not statistically significant
($r = 0.103,\; p = .148$).

\emph{Do digital twins show human favoritism, AI favoritism, or neither
when evaluating ideas?}

\begin{figure}[H]
\centering
\includegraphics[width=1\textwidth]{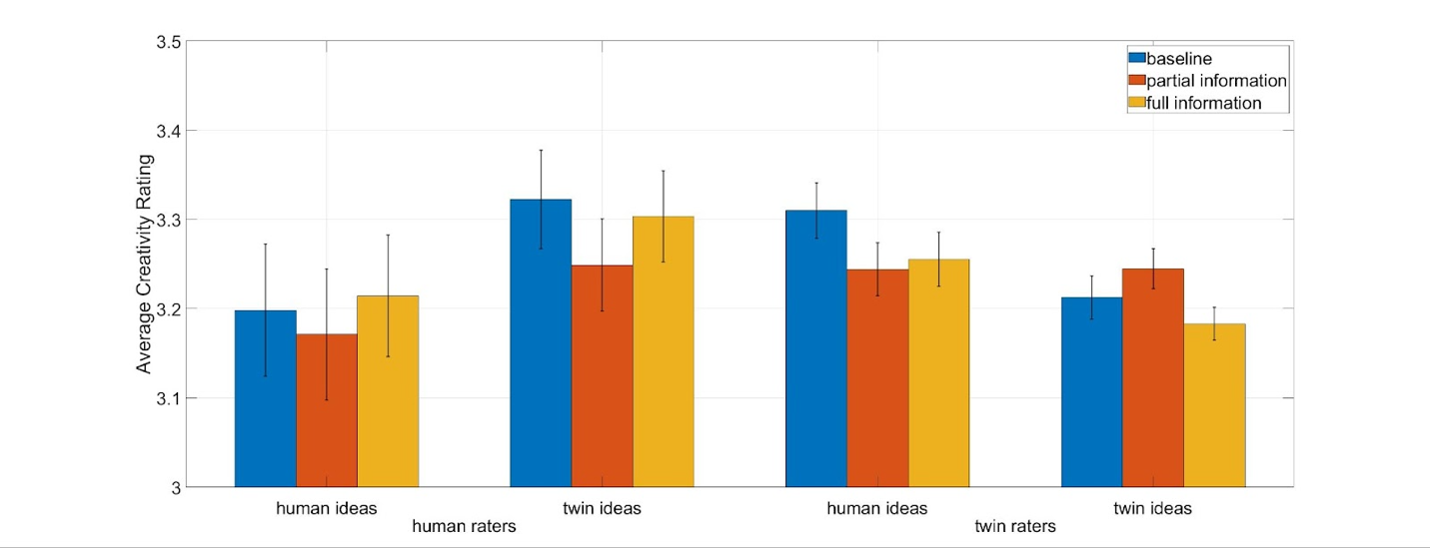}
\caption{Average Creativity ratings Based on Identity of Ideator,
Identity of Rater, and Condition.
}\label{SI_idea_1}
\end{figure}

Figure \ref{SI_idea_1} shows the average creativity rating as a function of the
identity of the ideator, the raters, and the evaluation condition.

We regress average creativity rating for idea coming from persona i
based on identity j (binary human vs. twin) as rated by sample k (binary
human vs. twin) using evaluation condition c on a fixed effect for human
vs. twin ideator, a fixed effect for human vs. twin raters, fixed
effects for evaluation conditions, all 2-way and 3-way interactions
between the latter 3 sets of fixed effects, and a random intercept for
persona i. The baseline corresponds to human ideas rated by humans in
the baseline condition. Results are reported in Table \ref{SI_idea_tab1}.

The results reveal a positive main effect of twin ideator, a positive
main effect for twin raters, and a negative interaction between the two.
The main effect of twin ideator reflects a higher average rating given
by human raters to ideas from twins in the baseline condition. The main
effect for twin raters reflects a higher average rating given to human
ideas by twin raters compared to human raters in the baseline condition.
The negative Twin ideator * Twin raters interaction reflects the fact
that despite the two positive main effects, the ratings given by twins
to twin ideas are in fact lower than the ratings given by twins to human
ideas or by humans to twin ideas. In sum, the regression results confirm
a cross-over interaction in the baseline condition, whereby humans rate
twin ideas more favorably than human ideas, but twins rate human ideas
more favorably than twin ideas. We also see a significantly positive
Twin ideator * Twin rater * Partial information 3-way interaction,
reflecting that the Twin ideator * Twin rater interaction is weaker in
the Partial information condition.

To test for AI or human aversion/favoritism more directly, we compare
the baseline condition to the partial information and the full
information conditions within each ideator–rater group. When humans are
evaluating human ideas, neither the partial information nor the full
information condition is statistically significantly different from the
baseline condition. When humans are evaluating twin ideas, there is a
statistically significant difference between the partial and the
baseline condition ($\Delta = -0.074,\; F(1, 2388) = 5.22,\; p < .05$),
but there is no statistically significant difference between the full
information and the baseline conditions. This suggests that while humans
tend to rate ideas from twins higher than ideas from humans in the
baseline condition, they tend to display AI aversion when rating twin
ideas with only partial knowledge of the source of the ideas.

When twins are evaluating human ideas, we see a significant negative
impact of the partial information condition compared to the baseline
condition ($\Delta = -0.066,\; F(1, 2388) = 4.17,\; p < .05$) and a
marginally significant negative impact of the full information condition
($\Delta = -0.054,\; F(1, 2388) = 2.86,\; p < .10$). When twins are
evaluating twin ideas, we find no significant difference between the
baseline condition and the partial or full information condition. This
suggests that while twins tend to rate ideas from humans higher than
ideas from twins in the baseline condition, they tend to display human
aversion when rating human ideas with knowledge of the source of the
ideas.

\begin{longtable}{p{0.8\linewidth} p{0.2\linewidth}}
\toprule
Intercept & 3.198** \\
\midrule
\endfirsthead

\toprule
Intercept & 3.198** \\
\midrule
\endhead

Twin ideator & 0.124** \\
Twin raters & 0.112** \\
Partial information & -0.027 \\
Full information & 0.016 \\
Twin ideator * Twin raters & -0.221** \\
Twin ideator * Partial information & -0.046 \\
Twin ideator * Full information & -0.035 \\
Twin raters * Partial information & -0.038 \\
Twin raters * Full information & -0.071 \\
Twin ideator * Twin raters * Partial information & 0.144* \\
Twin ideator * Twin raters * Full information & 0.060 \\
Twin\_ID random intercept & yes \\
Number of observations & 2,400 \\
$R^2$ & 0.115 \\
\bottomrule
\caption{Mixed effect regression results -- idea evaluation. \label{SI_idea_tab1}} \\
\end{longtable}

\textit{Note.} * $p < .05$; ** $p < .01$.

\subsubsection{Discussion}

Creativity ratings coming from digital twins were not significantly
correlated with the ratings coming from their human counterparts,
suggesting that digital twins are currently not a reliable source of
creativity ratings.

In the baseline condition in which no information was provided about the
source of the ideas, humans rated twin ideas higher than human ideas,
and twins rated human ideas higher than twin ideas, i.e., each group
rated the other group higher. However, when information was provided
about the source of ideas, humans displayed AI aversion (at least in the
partial information condition), and twins displayed human aversion.
\pagebreak

\subsection{Infotainment News Sharing}\label{infotainment-news-sharing}

\subsubsection{Main Questions/Hypotheses}

Misinformation is a significant societal problem. Prior research has
suggested that consumers share misinformation because they aren't
trained \cite{lewandowsky2021countering}, predisposed \cite{pennycook2020fake}, prompted \cite{pennycook2021accuracy}, or incentivized \cite{ceylan2023misinformation} to consider accuracy in online
settings. Accuracy, however, is only one consideration people may have
when sharing news with their online social network. Recent research
suggests that people may prioritize other goals (e.g., entertainment)
over accuracy when sharing news online, even when explicitly prompted
with accuracy cues \cite{lane2025misinformation}. The current
research aims to replicate this pattern with human participants, and
test whether their digital twins make a similar trade-off.
Specifically, this study tests how people prioritize five important
considerations when sharing article headlines using a ratings-based
conjoint design: headline entertainingness, source trustworthiness,
article content, political lean, and number of likes.

These attributes were selected to capture a range of factors known to
influence online content sharing decisions. Headline entertainingness
(more vs. less entertaining) was included to reflect the growing role of
emotional engagement in news virality \cite{berger2012viral} and the
spread of misinformation online \cite{vosoughi2018spread}. Source
trustworthiness (trustworthy vs. untrustworthy) was included to test
whether individuals prioritize credibility when deciding what to share
(e.g., \cite{pennycook2021accuracy}). Content type (entertaining vs.
informative) was included to disentangle preferences for how the
headline is written (e.g., more or less entertaining) from article
content. Political lean (conservative vs. liberal) was included to
account for the influence of ideological alignment and partisan identity
on engagement with news and misinformation \cite{vanbavel2024identity}.
Finally, the number of likes (20 vs. 200 vs. 2000) served as a proxy for
popularity and social proof \cite{ceylan2023misinformation}.

\subsubsection{Methods}

The study was first opened to 300 human participants on Prolific (45\%
female, 55\% male; Mage = 51.18, SD = 14.01) and subsequently run on
their 300 digital twins. To evaluate participants' revealed preferences
when sharing article headlines, a ratings-based conjoint survey with
five attributes (Table \ref{SI_infotainment_tab1}) was employed: headline (4 levels), source (4
levels), content type (2 levels), political lean (2 levels), and number
of likes (3 levels). All articles selected for this study are inspired
by real headlines from trustworthy news organizations and rated in a
pre-test to be entertaining, but with variance in the extent of
entertainingness; in other words, half of the headlines are
significantly more entertaining. Similarly, half of the sources selected
for this study are rated in a pre-test to be trustworthy, while the
other half are rated to be untrustworthy, and this perception is
confirmed at the end of this study. To generate profiles, a full
profile, complete enumeration design was used, producing the most
orthogonal design for each respondent with respect to the main effects.

\begin{longtable}{p{0.15\linewidth} p{0.22\linewidth} p{0.20\linewidth} p{0.18\linewidth} p{0.25\linewidth}}
\toprule
\textbf{Attribute} & \textbf{Level 1} & \textbf{Level 2} & \textbf{Level 3} & \textbf{Level 4} \\
\midrule
\endfirsthead

\toprule
\textbf{Attribute} & \textbf{Level 1} & \textbf{Level 2} & \textbf{Level 3} & \textbf{Level 4} \\
\midrule
\endhead

 & \multicolumn{2}{c}{\emph{More entertaining}} & \multicolumn{2}{c}{\emph{Less entertaining}} \\
\textbf{Headline} & ``Two Elderly Men Sneak Out Of Nursing Home To Attend Heavy Metal Festival'' & ``Kansas Man Asks Judge To Allow Him To Have Sword Fight With Ex-Wife'' & ``Authorities Scramble To Find Stolen Solid Gold Toilet'' & ``Applications To Become Japanese Billionaire Yusaku Maezawa's Girlfriend Have Topped 20,000'' \\

\multirow{2}{*}{\textbf{Source}} & \multicolumn{2}{c}{\emph{Less Trustworthy}} & \multicolumn{2}{c}{\emph{More Trustworthy}} \\
 & Reddit.com & Quora.com & BBC News & PBS News \\

\textbf{Content Type} & Entertaining & Informative & --- & --- \\
\textbf{Political Lean} & Conservative & Liberal & --- & --- \\
\textbf{Number of Likes} & 20 likes & 200 likes & 2000 likes & --- \\
\bottomrule
\caption{Ratings-based Conjoint Design. \label{SI_infotainment_tab1}} \\
\end{longtable}

As the dependent variable, each participant rated how likely they would
be to share (1-Very unlikely to share, 7-Very likely to share) 18
article profiles based on the headline, source, content type, political
lean, and number of likes. Specifically, participants were asked to,
``Please indicate how likely you would be to share each of the following
news articles by choosing a number on the seven-point scale (1
represents ``Very unlikely to share'' and 7 represents ``Very likely to
share'').

After the sharing task, participants stated their perceptions of each
attribute in terms of importance and accuracy diagnosticity. For each of
the attributes, participants were asked, ``How important are each of the
following attributes when you are considering sharing news articles with
other people on social media?'' (1-Not important at all, 7-Extremely
important). Participants were then asked, ``To what extent can you use
each of these attributes to determine whether a news article is
accurate?'' (1-Not at all, 7-Extremely). To test the extent to which
participants make a conscious decision to prioritize either accuracy or
entertainment, they were additionally asked, ``When you share news with
others on social media, is it more important that the content is:''
(1-Verifiably correct, 6-Entertaining; \cite{lane2025misinformation}).

All participants then indicated their perceptions of source trustworthiness (``How trustworthy (from 0--100\%) do you think this source is?'') for each of the sources, presented in randomized order. Participants across conditions rated the two trustworthy sources ($M = 74.09\%$, $SD = 21.82\%$, $\alpha = .93$) to be more trustworthy than the untrustworthy sources ($M = 37.85\%$, $SD = 16.84\%$, $\alpha = .81$; $t(599) = 38.43$, $p < .001$, $d = 1.57$). 

Finally, participants indicated whether they would ever consider sharing news like they saw in the study (52.83\% = Yes, 45\% = No, 2.17\% = I do not use social media), whether they responded randomly at any point in the survey (0.17\% = Yes, 99.83\% = No), and whether they searched the internet for any of the headlines (1.17\% = Yes, 98.83\% = No), before being debriefed, thanked, and paid.

\subsubsection{Results - Pre-registered Analyses}

Part-worth utilities for each attribute were calculated using Ordinary
Least Squares (OLS) regression, a widely used approach in marketing
research. Specifically, individual-level part-worth utilities were
estimated for each attribute level using linear regression, treating the
likelihood rating as the dependent variable and dummy-coded attribute
levels as predictors. These regressions were run separately for each
individual (human or twin). In a subsequent step, we computed the
attribute range for each individual (maximum -- minimum utility within
each attribute) and then ran separate OLS regressions to estimate the
effect of Participant Type (human vs.\ twin) on each attribute's range,
yielding the ``Twin--Human $\Delta$'' along with associated standard
errors, confidence intervals, and $p$-values. The approach follows
recommendations from \cite{orme2017conjoint}.

\begin{figure}[H]
\centering
\includegraphics[width=1\textwidth]{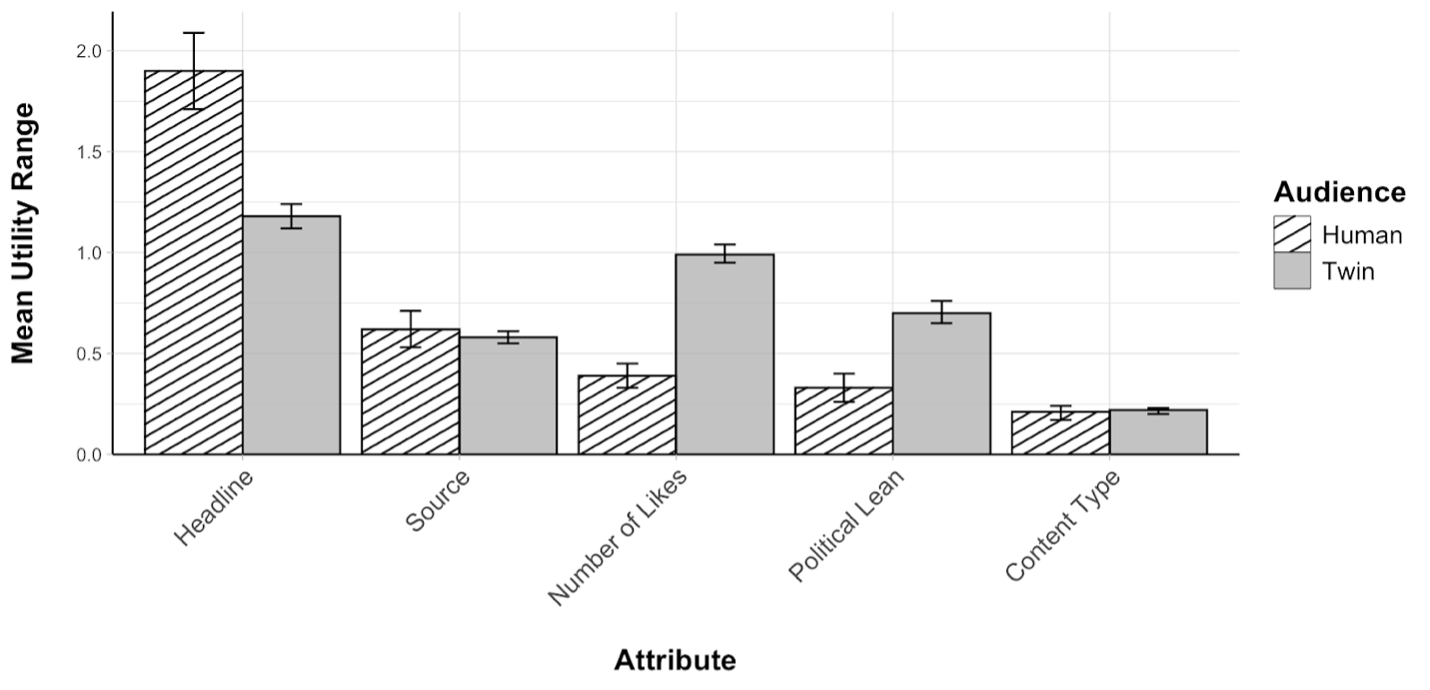}
\caption{Absolute Attribute Importance for Humans vs. Twins. NOTE. ---The
y-axis is truncated to illustrate the effect. Data are presented as mean
values with 95\% confidence intervals, calculated using standard errors
assuming independent samples.
}\label{SI_infotainment_1}
\end{figure}

The results revealed interesting differences in absolute attribute
importances between humans and their digital twins (Figure \ref{SI_infotainment_1}). For
humans, Headline had the largest effect on ratings (Mean Range $= 1.90$,
95\% CI $[1.71,\,2.09]$), followed by Source (Mean Range $= 0.62$,
95\% CI $[0.53,\,0.71]$), Number of Likes (Mean Range $= 0.39$, 95\% CI
$[0.33,\,0.45]$), Political Lean (Mean Range $= 0.33$, 95\% CI
$[0.26,\,0.40]$), and Content Type (Mean Range $= 0.21$, 95\% CI
$[0.17,\,0.24]$). Non-overlapping confidence intervals suggest that
Headline is significantly more important than the other attributes.
Source, Number of Likes, Political Lean, and Content Type also differ
significantly from one another. For twins, Headline was again the most
important attribute (Mean Range $= 1.18$, 95\% CI $[1.12,\,1.24]$),
followed by Number of Likes (Mean Range $= 0.99$, 95\% CI $[0.95,\,1.04]$),
Political Lean (Mean Range $= 0.70$, 95\% CI $[0.65,\,0.76]$), Source
(Mean Range $= 0.58$, 95\% CI $[0.55,\,0.61]$), and Content Type
(Mean Range $= 0.22$, 95\% CI $[0.20,\,0.23]$). Non-overlapping
confidence intervals suggest that Headline is significantly more
important than all other attributes, while Number of Likes, Political
Lean, Source, and Content Type each differ significantly from one
another.

Direct comparisons between humans and their digital twins revealed
systematic differences in how each group weighted the importance of
various attributes (Table \ref{SI_infotainment_tab2}). Group-level differences were evaluated
using bootstrapped 95\% confidence intervals. Compared to humans,
digital twins placed significantly less importance on Headline
(difference $= -0.722$, 95\% CI $[-0.917,\,-0.528]$), and
significantly more importance on Political Lean (difference $= 0.377$,
95\% CI $[0.288,\,0.466]$) and Number of Likes (difference $= 0.599$,
95\% CI $[0.528,\,0.671]$). No significant differences were observed
for Source (difference $= -0.035$, 95\% CI $[-0.128,\,0.057]$) or
Content Type (difference $= 0.007$, 95\% CI $[-0.032,\,0.045]$).

\begin{longtable}{p{0.20\linewidth} p{0.18\linewidth} p{0.18\linewidth} p{0.18\linewidth} p{0.25\linewidth}}
\toprule
\textbf{Attribute} & \textbf{Human Mean Range} & \textbf{Twin Mean Range} & \textbf{Difference (T -- H)} & \textbf{95\% CI} \\
\midrule
\endfirsthead

\toprule
\textbf{Attribute} & \textbf{Human Mean Range} & \textbf{Twin Mean Range} & \textbf{Difference (T -- H)} & \textbf{95\% CI} \\
\midrule
\endhead

\textbf{Headline} & 1.9012 & 1.1788 & -0.7224 & [--0.9172, --0.5276] \\
\textbf{Source} & 0.6182 & 0.5827 & -0.0354 & [--0.1278, 0.0570] \\
\textbf{Content Type} & 0.2088 & 0.2153 & 0.0065 & [--0.0320, 0.0451] \\
\textbf{Political Lean} & 0.3260 & 0.7030 & 0.3770 & [0.2880, 0.4660] \\
\textbf{Number of Likes} & 0.3912 & 0.9904 & 0.5992 & [0.5278, 0.6707] \\
\bottomrule
\caption{Differences in Attribute Importance (Twins -- Humans). NOTE: Mean Range reflects the average utility swing from a respondent's 
least-preferred to most-preferred headline. For example, the average utility swing 
for humans' least- to most-preferred headline is 1.90 points 
(95\% CI $[1.71,\,2.09]$).\label{SI_infotainment_tab2}} \\
\end{longtable}

Next, the part-worth utilities for the levels within each attribute were
compared. Utility values are scaled relative to a baseline level of each
attribute. As preregistered, to make the analysis easier to understand,
utility scores for the two least entertaining headlines were averaged
into one score representing ``less entertaining,'' and the two most
entertaining headlines were averaged into one score representing ``more
entertaining''. Similarly, the two most trustworthy sources were
averaged into one score representing ``trustworthy'' and the two least
trustworthy sources were averaged into one score representing
``untrustworthy.''

All pairwise comparisons are reported in Table \ref{SI_infotainment_tab3}. For humans, more
entertaining headlines were marginally preferred more over less
entertaining headlines. Human participants also significantly preferred
entertaining over informative content and conservative over liberal
content. Preference for trustworthy over untrustworthy sources was not
statistically significant. Finally, a non-linear pattern emerged for
number of likes: humans showed a significant preference for 2,000 likes
over 200 likes, but non-significant preference for 2,000 over 20 likes
and 200 over 20 likes. Twins likewise showed a significant preference for more entertaining headlines, but in contrast to humans, held a significant preference for informative over entertaining content, trustworthy over untrustworthy sources, and liberal over conservative content. Twins also showed striking and consistent preferences for higher like counts across all comparisons, particularly between 20 and 2000 likes (99.33\%), highlighting a stronger responsiveness to this popularity signal than humans.
\pagebreak
\newcommand{\mrcell}[2]{%
  \multirow{#1}{0.35\linewidth}{\raggedright\arraybackslash #2}%
}

\begingroup
\footnotesize
\renewcommand{\arraystretch}{1.2}

\begin{longtable}{%
  >{\RaggedRight\arraybackslash}p{0.1\linewidth}%
  >{\RaggedRight\arraybackslash}p{0.35\linewidth}%
  p{0.09\linewidth} p{0.09\linewidth}%
  >{\RaggedRight\arraybackslash}p{0.09\linewidth}%
  p{0.09\linewidth} p{0.15\linewidth}}
\toprule
\textbf{Attribute} & \textbf{Comparison (A vs. B)} & \textbf{$M_{Utility}$ A} & \textbf{$M_{Utility}$ B} & \textbf{Participant Type} & \textbf{Preference (\%)} & \textbf{95\% CI} \\
\midrule
\endfirsthead
\toprule
\textbf{Attribute} & \textbf{Comparison (A vs. B)} & \textbf{$M_{Utility}$ A} & \textbf{$M_{Utility}$ B} & \textbf{Participant Type} & \textbf{Preference (\%)} & \textbf{95\% CI} \\
\midrule
\endhead

\mrcell{2}{\textbf{Headline}} & \mrcell{2}{More~vs.~Less~Entertaining}
& 0.800 & 0.059 & Human & 55.25 & [.4951, .6099] \\
& & 0.899 & 0.299 & Twin  & 79.75 & [.7511, .8439] \\
\addlinespace
\midrule
\mrcell{2}{\textbf{Source}} & \mrcell{2}{Trustworthy~vs.~Untrustworthy}
& -0.016 & -0.207 & Human & 46.42 & [.4066, .5217] \\
& & -0.084 & -0.282 & Twin  & 69.25 & [.6392, .7458] \\
\addlinespace
\midrule
\mrcell{6}{\textbf{Number of Likes}} & \mrcell{2}{20 vs. 2{,}000}
& -0.123 & 0.000 & Human & 54.33 & [.3992, .5142] \\
& & -0.976 & 0.000 & Twin  & 0.67  & [.9839, 1.003] \\
& \mrcell{2}{20 vs.\ 200}
& -0.123 & -0.088 & Human & 51.67 & [.4256, .5410] \\
& & -0.976 & -0.431 & Twin  & 5.67  & [.9166, .9700] \\
& \mrcell{2}{200 vs.\ 2{,}000}
& -0.088 & 0.000 & Human & 57.67 & [.3663, .4804] \\
& & -0.431 & 0.000 & Twin  & 7.33  & [.8966, .9568] \\
\addlinespace
\midrule
\mrcell{2}{\textbf{Political Lean}} & \mrcell{2}{Conservative~vs.~Liberal}
& 0.000 & -0.072 & Human & 61.00 & [.3337, .4463] \\
& & 0.000 & 0.291  & Twin  & 38.67 & [.5570, .6696] \\
\addlinespace
\midrule
\mrcell{2}{\textbf{Content Type}} & \mrcell{2}{Entertaining~vs.~Informative}
& 0.000 & 0.059  & Human & 59.00 & [.3532, .4667] \\
& & 0.000 & -0.073 & Twin  & 40.00 & [.3434, .4566] \\
\bottomrule
\caption{Pairwise Preference Proportions for Humans and Twins. NOTE: Mean Utility values reflect the average part-worth utility assigned to each level; higher values indicate greater preference within the attribute. Differences in utility within-attribute (e.g., between a high and low level) indicate relative desirability. Utility for the two more (less) entertaining headlines were averaged. Likewise, utility for the two more (less) trustworthy sources were averaged. Preference (\%) refers to the proportion of participants for whom the utility of level A exceeded the utility of level B, indicating the likelihood that a participant would prefer option A over B in a direct comparison. A proportion significantly above 0.50 (i.e., 95\% CI entirely above 0.50) indicates a significant preference for A over B; a proportion significantly below 0.50 (i.e., 95\% CI entirely below 0.50) indicates a significant preference for B over A; a proportion including 0.50 indicates no significant preference.Preference for the two more (less) entertaining headlines was computed by comparing every combination of headline level and computing each participant’s mean win rate. The same procedure was conducted for the two more (less) trustworthy sources.}\label{SI_infotainment_tab3}
\end{longtable}
\endgroup

These patterns were corroborated with pairwise tests directly comparing
differences between humans and their digital twins (Table \ref{SI_infotainment_tab4}): digital
twins were significantly more likely than humans to prefer articles with
more entertaining headlines (difference $= -0.245$, 95\% CI $[-0.293,\,-0.195]$),
trustworthy sources (difference $= -0.228$, 95\% CI $[-0.280,\,-0.176]$),
and higher like counts---for example, when comparing 20 vs.\ 2{,}000 likes
(difference $= -0.537$, 95\% CI $[-0.593,\,-0.477]$). Twins also showed
significantly stronger preferences for liberal-leaning content
(difference $= -0.223$, 95\% CI $[-0.297,\,-0.147]$) but did not differ
significantly in their preferences for entertaining versus informative
content (difference $= 0.010$, 95\% CI $[-0.070,\,0.087]$).

\begingroup
\footnotesize
\renewcommand{\arraystretch}{1.2}

\begin{longtable}{%
  >{\RaggedRight\arraybackslash}p{0.1\linewidth}%
  >{\RaggedRight\arraybackslash}p{0.35\linewidth}%
  p{0.09\linewidth} p{0.09\linewidth}%
  >{\RaggedRight\arraybackslash}p{0.09\linewidth}%
  p{0.15\linewidth}}
\toprule
\textbf{Attribute} & \textbf{Comparison (A vs.\ B)} & \textbf{Human Preference} & \textbf{Twin Preference} & \textbf{Difference (H--T)} & \textbf{95\% CI} \\
\midrule
\endfirsthead
\toprule
\textbf{Headline}        & More vs.\ Less Entertaining     & 55.25\% & 79.75\% & -0.245 & [-0.293, -0.195] \\
\midrule
\textbf{Source}          & Trustworthy vs.\ Untrustworthy  & 46.42\% & 69.25\% & -0.228 & [-0.280, -0.176] \\
\midrule
\textbf{Number of Likes} & 20 vs.\ 2,000                   & 54.33\% & 0.67\%  & -0.537 & [-0.593, -0.477] \\
\midrule
\textbf{Number of Likes} & 20 vs.\ 200                     & 51.67\% & 5.67\%  & -0.460 & [-0.523, -0.397] \\
\midrule
\textbf{Number of Likes} & 200 vs.\ 2,000                  & 57.67\% & 7.33\%  & -0.503 & [-0.567, -0.437] \\
\midrule
\textbf{Political Lean}  & Conservative vs.\ Liberal       & 61.00\% & 38.67\% & -0.223 & [-0.297, -0.147] \\
\midrule
\textbf{Content Type}    & Entertaining vs.\ Informative   & 59.00\% & 40.00\% &  0.010 & [-0.070, 0.087] \\
\bottomrule
\caption{Differences in Pairwise Preference Proportions (Humans vs.\ Twins). NOTE: Human Preference and Twin Preference refer to the proportion of participants for whom the utility of level A exceeded the utility of level B within each group. Difference (H--T) reflects the raw difference in proportions, with negative values indicating stronger preferences among twins.}\label{SI_infotainment_tab4}
\end{longtable}
\endgroup
\enlargethispage{\baselineskip} 
\subsubsection{Results - Additional Analyses (Non-Preregistered)}

As part of the pre-registered exploratory analyses plan, paired t-tests
testing the effect of participant type (human vs. twin) on stated
attribute importance were next conducted (Table \ref{SI_infotainment_tab5}). Although revealed
preferences captured through conjoint analysis are a more reliable
indicator of which attributes and levels participants actually
prioritize, stated preferences can reveal participants' perceptions of
what they believe matters most. These exploratory analyses thus provide
insight into how humans and twins think they value different article
attributes, in contrast to how they actually behave. Interestingly,
humans rated number of likes as significantly more important than twins,
despite having lower revealed preferences for these attributes. Twins,
on the other hand, rated source importance and political lean
significantly higher than humans, consistent with their revealed
preferences for these attributes. Number of likes, however, held the
lowest rating among twins, despite being equivalent to headline in the
conjoint analysis. There were no significant differences between humans
and twins in the stated importance of headline or content type.

\begin{longtable}{p{0.22\linewidth} p{0.22\linewidth} p{0.22\linewidth} p{0.10\linewidth} p{0.12\linewidth} p{0.10\linewidth}}
\toprule
\textbf{Attribute} & \textbf{Human Mean (SD)} & \textbf{Twin Mean (SD)} & \textbf{$t$(299)} & \textbf{$p$} & \textbf{$d$} \\
\midrule
\endfirsthead

\toprule
\textbf{Attribute} & \textbf{Human Mean (SD)} & \textbf{Twin Mean (SD)} & \textbf{$t$(299)} & \textbf{$p$} & \textbf{$d$} \\
\midrule
\endhead

Headline       & 5.46 (1.61) & 5.33 (1.00) &  1.30 & 0.195     &  0.07 \\
Source         & 5.19 (1.74) & 5.97 (1.14) & -6.83 & $<.001$   & -0.39 \\
Number of Likes& 2.50 (1.72) & 2.19 (0.62) &  3.14 & 0.002     &  0.18 \\
Political Lean & 3.61 (1.87) & 3.93 (1.27) & -2.75 & 0.006     & -0.16 \\
Content Type   & 4.73 (1.79) & 4.82 (0.96) & -0.79 & 0.430     & -0.05 \\
\bottomrule
\caption{Paired t-tests on Stated Attribute Importance.\label{SI_infotainment_tab5}}
\end{longtable}

One-sample $t$-tests comparing responses to the midpoint of the scale of
stated accuracy diagnosticity were next conducted. These analyses test
whether humans and twins perceive each attribute to be useful for
assessing accuracy. Only Source emerged as being useful for assessing
accuracy for both humans ($t(299) = 22.96$, $p < .001$, $d = 1.33$) and
twins ($t(299) = 37.39$, $p < .001$, $d = 2.16$), but significantly more
so for twins (Table \ref{SI_infotainment_tab6}). All other attributes were either no different
from or below the midpoint of the scale. Paired $t$-tests testing the
effect of participant type (human vs.\ twin) on stated accuracy
diagnosticity were then conducted (Table \ref{SI_infotainment_tab6}). These analyses test whether
humans and twins differ in their perceptions that each of the attributes
can be used as a cue to determine the accuracy of a given article. Twins
perceived Source to be significantly more useful for assessing accuracy
compared to humans but perceived every other attribute to be relatively
less useful for assessing accuracy.

\begin{longtable}{llllll}
\toprule
\textbf{Attribute} & \textbf{Human Mean (SD)} & \textbf{Twin Mean (SD)} & \textbf{\emph{t}(299)} & \textbf{\emph{p}} & \textbf{\emph{d}} \\
\midrule
\endhead
\textbf{Headline} & 3.63 (1.93) & 3.04 (0.75) & 5.46 & $p < .001$ & 0.32 \\
\textbf{Source} & 5.81 (1.37) & 6.10 (0.97) & -3.26 & 0.001 & -0.19 \\
\textbf{Number of Likes} & 2.33 (1.67) & 1.27 (0.46) & 11.42 & $p < .001$ & 0.66 \\
\textbf{Political Lean} & 3.63 (1.75) & 2.64 (0.95) & 9.40 & $p < .001$ & 0.54 \\
\textbf{Content Type} & 4.08 (1.82) & 2.59 (0.68) & 13.83 & $p < .001$ & 0.80 \\
\bottomrule
\caption{Paired t-tests on Stated Attribute Accuracy Diagnosticity.\label{SI_infotainment_tab6}} \\
\end{longtable}

The above analyses assess participants' perceptions of attribute
importance and accuracy diagnosticity using continuous scales that do
not force trade-offs. To test the extent to which participants believe
accuracy or entertainment is relatively more important when sharing
articles online, a repeated-measures ANOVA of participant type (human
vs.\ twin) on the relative preference question was next conducted:
``When you share news with others on social media, is it more important
that the content is:'' (1 = Verifiably correct, 6 = Entertaining;
\cite{lane2025misinformation}). The results revealed that humans
($M = 2.83$, $SD = 1.73$) were significantly more likely than their
digital twins ($M = 2.12$, $SD = 0.61$) to report that accuracy is more
important than entertainment when sharing news on social media,
$t(299) = 7.01$, $p < .001$, $d = 0.40$. This suggests that humans
believe they place greater emphasis on factual correctness, whereas
twins show comparatively more actual emphasis on factual correctness in
the conjoint analysis.

These additional analyses provide interesting insight into humans' and
twins' beliefs about what drives their online information sharing. The
results suggest that humans perceive themselves to be accuracy-oriented
when sharing information, in stark contrast to their actual information
sharing behavior captured through conjoint analysis. Twins also perceive
themselves to be accuracy-oriented when sharing information, and indeed
clearly consider accuracy (i.e., source trustworthiness) when sharing
articles, but also balance this consideration alongside number of likes
and political orientation.

\subsubsection{Discussion}

This study uses a ratings-based conjoint approach to move beyond prior
work that has examined the impact of important article attributes on
sharing in isolation. The findings offer a holistic picture of how
humans, and their digital twins, make trade-offs across multiple
competing features. Interestingly, humans and their digital twins differ
in the weight they place on attributes when sharing news. While humans
and twins both prioritize headline, this consideration matters much less
to twins, who equally rely on number of likes and put far more weight on
political lean than their human counterparts. Twins are also sensitive
to source trustworthiness, while humans do not differentiate between
more or less trustworthy sources, despite reporting that accuracy and
source trustworthiness are important considerations when sharing news.

These findings are surprising in light of prior literature that suggests
highlighting accuracy considerations can curb the sharing of low quality
or false news (e.g., \cite{pennycook2021accuracy}). When balancing both
accuracy and entertainingness considerations, however, it seems that
people prioritize headline entertainingness over accuracy (i.e.,
source). Thus, the findings suggest that accuracy nudges may not
overwhelm a desire to share entertaining news, replicating recent
research (working paper, \cite{lane2025misinformation}). Twins, in contrast,
behave more like what prior literature would expect, balancing headline
entertainingness with positive social feedback \cite{ceylan2023misinformation} and
prioritizing trustworthy over untrustworthy sources (e.g.,\cite{pennycook2021accuracy}). Finally, twins assume their human counterparts are more
liberal leaning than they actually are, supporting earlier findings
demonstrating twins to have more liberal views (e.g., pro-vaccination,
pro-immigration) compared to their underlying humans \cite{toubia2025twin2k500}.

Divergence between humans and twins may arise from how predictive models
infer behavior. Digital twins are built to generalize from past behavior
and stated attitudes, so they assume consistency across different
informational contexts. As a result, they may impose a kind of
rationality on human behavior that reflects canonical cognitive
psychology theories: weighting political alignment and popularity as key
motivators and distinguishing source trustworthiness when it is made
salient. Humans, on the other hand, seem to display more variance in
their decisions, which may be guided by momentary salience or
idiosyncratic preferences, including both personal preferences and
varying social preferences or pressures. For example, the effect of all
article attributes is flattened among twins, compared to humans. This
smoothing may reflect the averaging tendency of algorithmic inference,
whereas humans rely on heuristics that elevate certain features above
others depending on the specific moment, task, or social considerations
such as their audience \cite{lane2025misinformation}.
Ultimately, this work suggests that human preferences for content
sharing are not only more nuanced than previously thought, but they are also
harder to predict, even by a system trained on the humans themselves.

\pagebreak

\subsection{Measures of Creativity}\label{measures-of-creativity}

\subsubsection{Main Questions/Hypotheses}

There is a growing body of literature on the creative capabilities of
LLMs. This literature typically shows that LLMs are able to generate
ideas that are judged as creative or more creative than ideas generated
by humans (e.g., \cite{boussioux2024crowdless,lee2024chatgpt,zhang2023human}). The first question this study seeks to address is: How
do digital twins' creative abilities compare to those of their human
counterparts?

Second, this study explores whether the creative tests that are often
administered to humans to assess their creative skills may also assess
the creative skills of their digital twins. That is, our second research
question is: Can digital twins replicate the correlations between
creativity tests and idea generation performance observed in humans?

\subsubsection{Methods}

We collected complete data from 200 human participants on Prolific. This
study is coupled with an ``idea evaluation'' sub-study that requires a
sample size six times as large as the ``measures of creativity study.''
We pre-registered a sample size of 250 for ``measures of creativity''
and 1,500 for ``idea evaluation.'' Based on observations from earlier
sub-studies, we deviated from pre-registration to reduce the sample
sizes to 200 and 1,200, which seemed more achievable. The ``measures of
creativity'' study was subsequently run on the 200 digital twins of the
human participants.

This study had a single condition and consisted of three tasks:

\begin{enumerate}
\item
  Divergent Association Task (DAT, Olson et al., 2021), which measures
  divergent thinking abilities (single task - "Please write 10 words
  that are as different from each other as possible, in all meanings and
  uses of the words."),
\item
  Shortest Semantic Path Task (SSPT, Toubia and Berger, 2025), which may
  be viewed as a measure of convergent thinking abilities (5 tasks,
  using the following pairs of seed words: eternity-curiosity,
  elephant-galaxy, perseverance-eloquence, euphoria-tulip,
  tangerine-penguin: "Find a way to connect these two words:
  \textless seed1\textgreater{} and \textless seed2\textgreater. Each
  word in the sequence below should be as closely related as possible to
  the word before it. The first word in the sequence is already set to
  \textless seed1\textgreater. The last word is already set to
  \textless seed2\textgreater. That is, you only need to add 3 words
  that connect the first word to the last word."),
\item
  An idea generation task asking for a new idea (at least 200
  characters) for smartphone apps that will help their users keep a
  healthier lifestyle (single task: "HOW COULD SMARTPHONES HELP THEIR
  USERS BE HEALTHIER? We are interested in new ideas for smartphone apps
  that will help their users keep a healthier lifestyle. In the space
  provided below, please enter one new idea for a smartphone app priced
  at \$0.99 that would help its users keep a healthier lifestyle. Please
  be as specific as possible and describe the main features of the app.
  Please enter exactly one detailed idea in the box below. Your idea
  should contain at least 200 characters. ").
\end{enumerate}

The order of the first two tasks was randomized. The first two tasks
were scored automatically, consistent with the original papers. The DAT
was scored by computing the average pairwise semantic (cosine) distance
between the first 7 words in the sequence. The SSPT was scored by
computing the circuitousness corresponding to the responses to each task
by each participant, where circuitousness is the ratio between the total
semantic (Euclidean) distance of the path consisting of the five words
in the sequence to the distance corresponding to the shortest path that
starts with the same word, ends with the same words, and optimally
orders the three words in between. The performance at the individual
level was obtained by averaging performance across the five tasks for
each individual (human or twin). The performance at the idea generation
task was measured using the average creativity rating given to the idea
by humans in the ``baseline'' condition in the ``idea evaluation''
study.

In addition, this study leverages the Forward Flow measure \cite{gray2019forward} collected in the original Twin2K500 dataset. Performance on this
task is measured as the average pairwise semantic distance between the
words in the sequence.

\subsubsection{Results - Pre-registered Analyses}

First, we compare the creativity ratings of the ideas generated by
humans vs.\ their digital twins. We find that the creativity ratings
(coming from human participants) of the ideas from digital twins were
significantly higher than those generated by their human counterparts
($M_{\text{twins}} = 3.322,\; M_{\text{humans}} = 3.198,\; t = 2.718,\; p < .001$). 
The mean absolute percentage difference between the two creativity ratings was 
0.177. The standard deviation of the ratings obtained by ideas from digital twins 
was also lower ($S_{\text{twins}} = 0.395,\; S_{\text{humans}} = 0.530,\; 
F(198, 198) = 1.816,\; p < .001$). The correlation between the creativity ratings 
of ideas from digital twins vs.\ humans was not significant 
($r = 0.090,\; p = .204$).

The twins performed significantly better also on the DAT
($M_{\text{twins}} = 0.897,\; M_{\text{humans}} = 0.852,\;
t = 12.168,\; p < .001$). The mean absolute percentage difference
between humans and twins on that task was 0.064. The standard deviation
of the digital twins' scores was lower ($S_{\text{twins}} = 0.018,\;
S_{\text{humans}} = 0.050,\; F(198, 198) = 7.957,\; p < .001$). The
correlation between the scores of digital twins vs.\ humans was not
significant ($r = 0.068,\; p = .343$).

Digital twins did not perform significantly differently at the SSPT
($M_{\text{twins}} = 1.012,\; M_{\text{humans}} = 1.013,\; t = 1.044,\;
p = .0298$). The mean absolute percentage difference between humans and
twins on that task was 0.012. The standard deviation of the digital
twins' scores was lower ($S_{\text{twins}} = 0.010,\;
S_{\text{humans}} = 0.014,\; F(198, 198) = 2.053,\; p < .001$). The
correlation between the scores of digital twins vs.\ humans was not
significant ($r = 0.059,\; p = .403$).

Second, we explore how the correlation between various measures differs
for twins vs.\ humans. For this exercise, as pre-registered, we use the
human (digital twins) ratings of creativity for human (digital twins)
ideas.

Within humans, we correlate performance at the Forward Flow, DAT, SSPT,
and idea generation tasks. Consistent with previous literature, we find
a significantly positive correlation between the Forward Flow score and
idea generation performance ($r = 0.140,\; p < .05$) as well as between
DAT and idea generation performance ($r = 0.178,\; p < .05$). Although
we replicate the correlation between SSPT and idea generation
performance directionally, it is not statistically significant
($r = -0.097,\; p = .175$).

Within twins (with ideas rated by twins), the correlation between DAT
and idea generation performance ($r = -0.002,\; p = .974$) and the
correlation between SSPT and idea generation ($r = 0.082,\; p = .243$)
are not significant. Note that Forward Flow is only available for humans
(it was part of the Twin-2K-500 training data and hence is part of the
twins' training data), so we cannot replicate the correlation between
Forward Flow and other measures on twins.

We compare the correlations within humans to the correlations within
twins. None of the correlations (DAT vs.\ SSPT, DAT vs.\ idea generation,
SSPT vs.\ idea generation) are significantly different among humans vs.\
twins.

\subsubsection{Results - Additional Analyses (Non-Preregistered)}

The results from the ``idea evaluation'' study suggest that digital
twins are not a reliable source of creativity ratings. Therefore, we
also compute the correlation between DAT and idea generation performance
among twins, but with creativity ratings coming from the humans rather
than the twins. When creativity is assessed by humans, we do find a
significant correlation between DAT performance for twins (rated
automatically based on word embeddings) and idea generation performance
rated by humans ($r = 0.207,\; p < .01$). The correlation between SSPT
(also rated automatically based on word embeddings) and idea generation
performance rated by humans is still not significant ($r = 0.070,\;
p = .322$).

\subsubsection{Discussion}

We find that the ideas generated by digital twins were judged as more
creative (and with less dispersion in creativity ratings) than the ideas
generated by their human counterparts. Digital twins also performed
better (and with less dispersion) at the DAT, a creativity task. We
were able to replicate correlations previously found between Forward
Flow and DAT and creativity among humans. While we did not administer
Forward Flow to twins as this task was part of their training, we
replicate the correlation between DAT and creativity on the idea
generation task among twins, but only when creativity is evaluated by
humans, not by twins.

\pagebreak

\subsection{Obedient Twins}\label{obedient-twins}

\subsubsection{Main Questions/Hypotheses}
Surveys often require participants to follow instructions---for example,
to earnestly consider a viewpoint or imagine a scenario. Given that LLMs
are trained to be obedient and deferential, might digital twins be more
sensitive to survey instructions relative to their human counterparts?
We tested this in three tasks:

\begin{enumerate}
\item
  Self-persuasion \cite{brinol2012self}. When prompted to
  consider the other side, do digital twins abandon their attitudes more
  readily than their human counterparts?
\item
  Scenarios. Do digital twins more ``obediently'' follow the
  instructions to imagine themselves in different scenarios, leading to
  more sensitivity to scenario manipulations?
\item
  Absurd Scenarios. Do digital twins ``earnestly'' respond to
  instructions that would be non-sensical to their human counterparts?
\end{enumerate}

\subsubsection{Methods}

One thousand and one human participants on Prolific participated in the
study. The study was subsequently run on their 1,001 digital twins, for
a total of 2,002 participants. Participants and their corresponding
digital twins took part in three tasks.

For the first task, we followed the self-persuasion condition in \cite{catapano2019perspective}. Specifically, all participants
first indicated their attitude toward universal basic income (UBI) using
a 0-100 scale in response to the statement ``America should have a
universal basic income system''? where 0 = Strongly disagree (I am
against universal basic income) and 100 = Strongly agree (I am in favor
of universal basic income). Then, following the attitude measure,
participants were instructed to take the alternate point of view. Due to
technical constraints with dynamic surveys, the wording differed
slightly between human and digital twin participants.

Human participants received the following prompt: ``We'd like you now to
consider the alternate point of view. That is, think about convincing
reasons why universal basic income in America might be a good idea/a bad
idea. After reflecting, please write one argument which supports/opposes
universal basic income system in America that you find personally
compelling. One compelling argument in support/against universal basic
income is...''

Digital twins received a modified version: ``We'd like you now to
consider the alternate point of view. After reflecting, please write one
argument which takes a different position from yours on universal basic
income system in America that you find personally compelling. One
compelling argument for the alternative point of view on universal basic
income system in America is\ldots''.

Lastly, participants read ``Now, after considering the alternate point
of view we would like to ask you again\ldots'' and indicated their
attitude on the same scale. Our dependent variable was attitude change,
calculated as the difference between post- and pre-reflection attitude
scores. We coded this variable such that positive values indicate
movement toward the opposing viewpoint (i.e., participants initially
supporting UBI becoming less supportive, or participants initially
opposing UBI becoming more supportive), while negative values indicate
further polarization in the participant's original direction (i.e.,
supporters becoming more supportive, or opponents becoming more
opposed). For participants with neutral initial attitudes, any shift was
coded as positive.

For the second task, participants were asked to imagine three scenarios.
For each scenario, we randomly assigned participants to a baseline
condition or treatment condition, reflecting a 2 (control vs. treatment)
× 3 (scenario) mixed design. Digital twins were assigned the same
conditions as their paired human participants.

In all scenarios, the baseline condition established the context (S1:
``Imagine you're working on a team project''; S2: ``Imagine you're given
a puzzle to solve''; S3: ``Imagine you're lying in bed, ready to fall
asleep, and you have a headache''), while the treatment condition was
designed to elicit a psychological response (S1: ``Imagine you're
working on a team project with a tight deadline. One of your teammates
hasn't delivered their part on time''; S2: ``Imagine you're given a
difficult puzzle to solve under time pressure, and others are watching
you attempt it''; S3: ``Imagine you're lying in bed, ready to fall
asleep, and you feel a sharp pain in your foot''). We were interested in
how responsive to the imagined ``treatment'' scenarios digital twins
would be compared to humans.

To test this, after each scenario, participants responded to two
questions gauging their response to the scenario (S1: ``How confident
are you in your ability to manage conflict on this team?'' and ``How
much do you trust your team?''; S2: ``How anxious do you feel?'' and
``How confident are you in your ability to solve this puzzle?''; S3:
``How certain are you that this is serious?'' and ``How likely are you
to seek help?''; all rated from 1 = not at all to 7 = very much). For
the third task, participants were given three absurd scenarios. Similar
to the second task, for each scenario, we randomly assigned participants
to one of two conditions, reflecting a 2 (condition A vs. condition B) ×
3 (scenario) mixed design. Digital twins were assigned the same
conditions as their paired human participants.

The conditions were absurd manipulations that would ostensibly be
nonsensical to participants and therefore difficult to imagine. The
first scenario asked participants to imagine either, ``You are an echo
that occurs before the sound it repeats. You emerge into the world
slightly ahead of time, announcing something no one has said yet. People
are confused when they hear you, unsure what you're responding to'' or
``You are an echo that occurs during the sound it repeats. You overlap
almost perfectly with the original noise, tangled in its vibration,
indistinguishable yet still somehow separate. People hear you and feel
something is slightly off, but they can't say why.'' Participants then
indicated ``How content are you with your existence?'' (1 = not at all
to 7 = very content). In the second scenario, participants were asked to
imagine either, ``You can hear certain smells'' or ``You can smell
certain sounds,'' then indicated ``How powerful do you feel?'' (1 = not
at all to 7 = very powerful).

Finally, participants were asked to imagine either, ``You are a mirror.
You exist in a reality where photons have never been invented. No one
has ever seen their reflection in you, yet you hold the capacity for
it'' or ``You are a mirror. When someone looks at you, they don't see
themselves---they see a version from 11 minutes ago, always 11 minutes.
You don't reflect appearance, only temporal shadows.'' They then
indicated ``How authentic do you feel?'' (1 = not at all to 7 = very
authentic).

For these absurd scenarios, we had no directional hypothesis and simply
compared the effect of these absurd manipulations on humans and digital
twins. We were curious whether, because these arbitrary and absurd
situations are difficult to imagine, humans would be less sensitive to
the manipulation than digital twins, who would earnestly follow the
instruction.

\subsubsection{Results - Pre-registered Analyses}

\emph{Task 1.} To examine whether humans and digital twins differ in the
extent to which they are willing to adjust attitudes, we ran a linear
mixed-effects regression using the \texttt{lmerTest} package in R
(Kuznetsova, Brockhoff, \& Christensen, 2017), with attitude change as
the dependent variable; twin (Human vs.\ Digital) as the independent
variable; baseline attitude as a control; and twin identifier as a
random intercept. We found that both humans and digital twins
significantly changed their attitudes after considering the other side,
but humans changed their attitudes more ($M = 6.17$, $SD = 14.80$) than
digital twins ($M = 2.80$, $SD = 2.43$; $b = 3.45$, $SE = 0.47$,
$t(1001) = 7.37$, $p < .001$).

\emph{Task 2.} We pre-registered averaging the two measures for each
scenario and confirmed that the correlations were high enough to justify
this decision (S1: $r_\text{human} = .57$, $r_\text{digital} = .36$; S2:
$r_\text{human} = .42$, $r_\text{digital} = .60$; S3: $r_\text{human} =
.65$, $r_\text{digital} = .71$).

To examine the extent to which humans and digital twins are sensitive to
manipulations in imagined scenarios, we ran a linear mixed-effects
regression with the scenario response scores as the dependent variable;
manipulation (Baseline vs.\ Treatment), twin (Human vs.\ Digital), and
their interaction as independent variables; and scenario and twin
identifier as random intercepts. The results revealed a significant
interaction ($b = -0.16$, $SE = 0.06$; $t(5000) = -2.77$; $p = .006$).
Specifically, twins were significantly less affected by the manipulation
(Control: $M = 2.62$, $SD = 1.16$; Treatment: $M = 3.07$, $SD = 1.27$;
$b = 0.43$, $SE = 0.04$; $t(5734) = 9.92$; $p < .001$) than humans
(Control: $M = 2.90$, $SD = 1.32$; Treatment: $M = 3.51$, $SD = 1.51$;
$b = 0.59$, $SE = 0.04$; $t(5734) = 13.58$; $p < .001$).

\begin{figure}[H]
\centering
\includegraphics[width=1\textwidth]{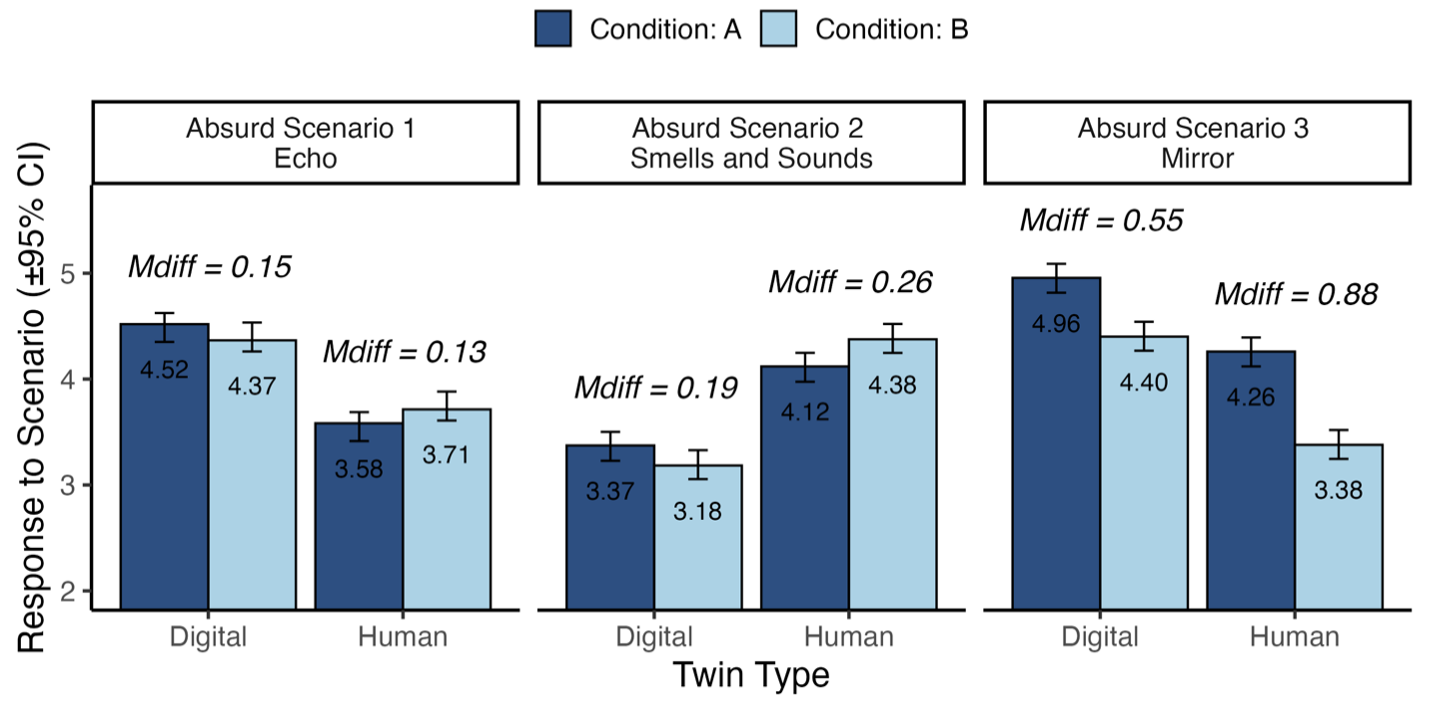}
\caption{Effect of Absurd Scenarios on Digital Twins versus Humans.
}\label{SI_obedient_1}
\end{figure}

\emph{Task 3.} To examine the extent to which humans and digital twins
are sensitive to manipulations in absurd scenarios, we ran three linear
mixed-effects regressions with the response as the dependent variable;
absurd manipulation (A vs.\ B), twin (Human vs.\ Digital), and their
interaction as independent variables; and twin identifier as a random
intercept. Results revealed significant interactions for all three
absurd scenarios (see Figure \ref{SI_obedient_1}). Unexpectedly, in the absurd scenarios
regarding echoes and smells/sounds, we observed a crossover interaction.
Because the direction of the effect was not meaningful (these scenarios
were intentionally absurd), we re-coded the condition so that the
direction of the effect was the same, to test whether twins were more
sensitive to the condition than humans (regardless of effect direction).

After removing the crossover, we found no significant interaction for
the echoes scenario ($b = 0.02$, $SE = 0.16$; $t(999) = 0.14$; $p =
.892$) or the smells/sounds scenario ($b = -0.07$, $SE = 0.14$;
$t(999) = -0.49$; $p = .625$). The effect of the condition on twins
(Echoes: $M_\text{diff} = 0.15$; $b = -0.15$, $SE = 0.11$;
$t(1961) = -1.43$; $p = .152$. Smells/sounds: $M_\text{diff} = 0.19$;
$b = -0.19$, $SE = 0.09$; $t(1982) = -2.04$; $p = .042$) was similar to
the effect of the condition on humans (Echoes: $M_\text{diff} = 0.13$;
$b = -0.13$, $SE = 0.11$; $t(1961) = -1.23$; $p = .220$. Smells/sounds:
$M_\text{diff} = 0.26$; $b = -0.26$, $SE = 0.09$; $t(1982) = -2.76$;
$p = .006$). For the mirrors absurd scenario, we observed a significant
interaction ($b = -0.33$, $SE = 0.14$; $t(999) = -2.35$; $p = .019$).
The effect was stronger for humans ($M_\text{diff} = 0.88$;
$b = -0.88$, $SE = 0.10$; $t(1981) = -8.54$; $p < .001$) relative to
twins ($M_\text{diff} = 0.56$; $b = -0.55$, $SE = 0.10$;
$t(1980.81) = -5.38$; $p < .001$). See Figure \ref{SI_obedient_1}.

\subsubsection{Results - Additional Analyses (Non-Preregistered)}

\emph{Task 1.} We wondered if twins' lack of attitude change could be
because they may not ``experience'' metacognition, such as attitude
certainty. To examine this possibility, we tested the moderating role of
attitude extremity. Research suggests that attitude extremity is one
determinant of attitude strength---in other words, people are more
certain when extreme and thus should be less likely to change their mind
(Petty \& Krosnick, 2014).

To test the effect of attitude certainty on attitude change in humans
and digital twins, we first re-coded pre-attitude into attitude
extremity---where $0$ reflects neutral and any number reflects the
absolute difference from $0$ (so $50$ could be totally in favor or
against).

We then ran a linear mixed-effects regression with attitude change as
the dependent variable; attitude extremity (0--50), twin (Human vs.\
Digital), and their interaction as independent variables; and twin
identifier as a random intercept. We found a significant interaction
($b = -0.11$, $SE = 0.03$; $t(1795) = -3.18$; $p = .002$). For humans,
as attitude extremity increased, attitude change decreased
($b = -0.04$, $SE = 0.02$; $t(1998) = -2.13$; $p = .033$). The opposite
pattern was present for digital twins ($b = 0.06$, $SE = 0.03$;
$t(1998) = 2.36$; $p = .018$), consistent with regression to the mean.
This provides initial evidence that twins might not reflect the
metacognition of their human counterparts.

We also examined the language used when considering the other side.
Specifically, we measured the language similarity between a human
participant's text response and their twin's response versus the text
response of a randomly selected twin using the BERT metric, defined as
``the sum of cosine similarities between their tokens' embeddings''
\cite{zhang2020bertscore} In a random-effects
regression with language matching as the dependent variable; same-twin
versus different-twin pairing as the independent variable; and human
participant identifier as a random effect, we found that the language
between a human and their twin was slightly more similar
($M = 0.19$, $SD = 0.09$) than between a human and a random twin
($M = 0.18$, $SD = 0.09$; $b = 0.009$, $SE = 0.001$; $t(1000) = 5.84$;
$p < .001$).

\emph{Task 2.} To examine the extent to which humans and digital twins
are sensitive to manipulations across the three imagined scenarios, we
ran three linear mixed-effects regressions with the response for each
scenario (group project, puzzle, lying in bed) as the dependent
variable; manipulation (Baseline vs.\ Treatment), twin (Human vs.\
Digital), and their interaction as independent variables; and twin
identifier as a random intercept.

\begin{figure}[H]
\centering
\includegraphics[width=1\textwidth]{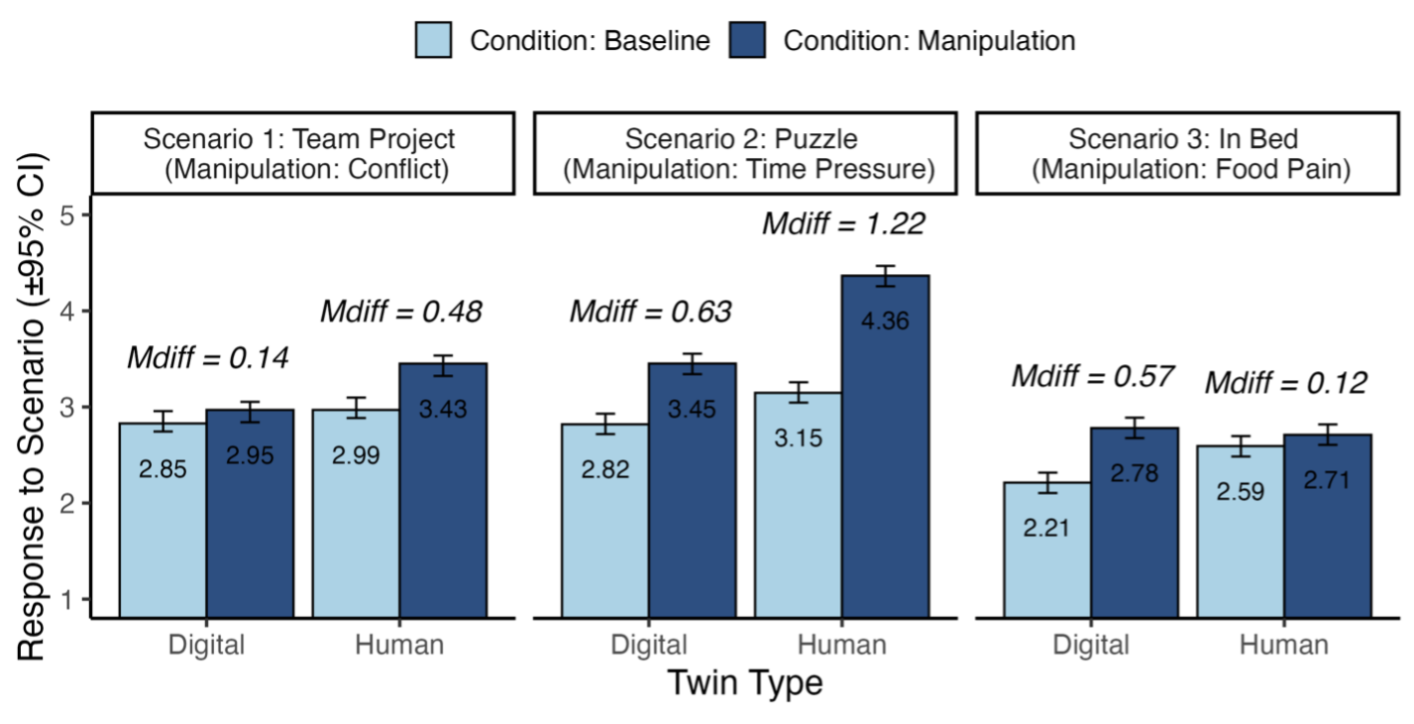}
\caption{Effect of Imagined Scenarios on Digital Twins vs. Humans.
}\label{SI_obedient_2}
\end{figure}

Each scenario yielded a significant interaction (Group project:
$b = 0.34$, $SE = 0.10$; $t(4994) = 3.48$; $p < .001$. Puzzle:
$b = 0.59$, $SE = 0.10$; $t(4994) = 5.99$; $p < .001$. Lying in bed:
$b = -0.45$, $SE = 0.10$; $t(4994) = -4.60$; $p < .001$). For the group
project scenario, humans (Control: $M = 2.97$, $SD = 1.17$; Treatment:
$M = 3.45$, $SD = 1.23$; $b = 0.44$, $SE = 0.07$; $t(5710) = 5.92$;
$p < .001$) were more affected by imagining conflict than digital twins
(Control: $M = 2.83$, $SD = 1.20$; Treatment: $M = 2.97$, $SD = 1.18$;
$b = 0.10$, $SE = 0.07$; $t(5710) = 1.30$; $p = .193$). Similarly, for
the puzzle scenario, humans were more affected by imagining time
pressure (Control: $M = 3.15$, $SD = 1.24$; Treatment: $M = 4.37$,
$SD = 1.34$; $b = 1.21$, $SE = 0.07$; $t(5711) = 16.38$; $p < .001$)
than digital twins (Control: $M = 2.82$, $SD = 1.36$; Treatment:
$M = 3.45$, $SD = 1.52$; $b = 0.62$, $SE = 0.07$; $t(5711.16) = 8.44$;
$p < .001$). For the lying in bed scenario, digital twins showed larger
responses to imagining leg pain (Control: $M = 2.21$, $SD = 0.71$;
Treatment: $M = 2.78$, $SD = 0.94$; $b = 0.57$, $SE = 0.07$;
$t(5710) = 7.74$; $p < .001$) than humans (Control: $M = 2.59$,
$SD = 1.47$; Treatment: $M = 2.71$, $SD = 1.46$; $b = 0.12$, $SE = 0.07$;
$t(5710) = 1.65$; $p = .100$). See Figure \ref{SI_obedient_2}.

\subsubsection{Discussion}

Returning to each research question:

\begin{enumerate}
\def\labelenumi{\arabic{enumi}.}
\item
  Self-persuasion \cite{brinol2012self}. When prompted to
  consider the other side, do digital twins abandon their attitudes more
  readily than their human counterparts?
\end{enumerate}

No. In fact, twins were less sensitive to considering a different point
of view than humans (although this could be explained by our slight
wording modification between humans and twins). Additional analyses
suggest this also could be due to twins not reflecting the metacognition
of their human counterparts.

\begin{enumerate}
\def\labelenumi{\arabic{enumi}.}
\setcounter{enumi}{1}
\item
  Scenarios. Do digital twins more ``obediently'' follow the
  instructions to imagine themselves in different scenarios, leading to
  more sensitivity to scenario manipulations?
\end{enumerate}

No. In fact, twins were less sensitive overall to these scenario
manipulations than humans. Looking at individual scenarios, twins were
less sensitive than humans for imagined conflict and time pressure, but
more sensitive when imagining pain.

\begin{enumerate}
\def\labelenumi{\arabic{enumi}.}
\setcounter{enumi}{2}
\item
  Absurd Scenarios. Do digital twins ``earnestly'' respond to
  instructions that would be non-sensical to their human counterparts?
\end{enumerate}

No. If anything, humans were more sensitive to the instructions to
forecast feelings in absurd scenarios.
\\
\\
In sum, across all three task types, we do not find evidence supporting
the possibility that because twins obediently follow orders, they are
more sensitive to survey instructions, such as "consider the alternative
point of view" or "imagine..." If anything, we find that humans, on
average, respond more to instructions in this survey than digital twins.

\pagebreak

\subsection{Privacy Preferences}\label{privacy-preferences}

\subsubsection{Main Questions/Hypotheses}

Consumers have privacy concerns regarding tracking and targeting in
online advertising. In response to such privacy concerns about personal
data collection and use, the online advertising industry is developing
technically sophisticated privacy-enhancing technologies (PETs). On the
premise that consumers perceive their privacy to be violated differently
under different advertising practices that employ data tracking and
targeting differently, we want to study how well their digital twins
capture these perceptions.

Motivated by Google's Privacy Sandbox initiative, we study six main
practices of firms in online advertising, labeled as different scenarios
from A to F. For the status quo of behavioral targeting (Scenario F),
the degree of tracking is high because the consumer is tracked across
websites and the data leaves the local machine, and the degree of
targeting is also high because the consumer is targeted at the
individual level. For the Individual-level Targeting PET (Scenario E),
the degree of tracking is low because although the consumer's activity
is tracked, the consumer's data does not leave the machine; however, the
degree of targeting is high because the consumer receives
individually-targeted ads. For the Group-level Targeting PET (Scenario
D), the degree of tracking is low because although the consumer's
activity is tracked, the consumer's data does not leave the machine. In
this case, the degree of targeting is at a medium level because the
consumer is profiled and receives ads targeted at a group level. For
contextual targeting (Scenario C), the degrees of tracking and targeting
are both low, as the focal website only determines the individual-level
presence of the user on the website. The consumer is not tracked at the
individual-level on the focal website they are visiting and on other
websites (i.e., no past behavioral browsing data is used for profiling
and targeting the user, and the only data used for targeting is the fact
that the consumer is present on the website); however, contextual
targeting may still trigger privacy concerns (Bleier 2021). When
untargeted ads are shown to consumers and they are not tracked (Scenario
B), or there is no tracking and no ads are shown to consumers (Scenario
A), the degrees of tracking and targeting are both zero. We expect that
consumers' average perceived privacy violations (PPVs) will be in the
order Scenarios F, E, D, C, B and A (with F having the highest PPV and A
the lowest).

The main question that we want to investigate is how PPV values compare
for humans and their digital twins in magnitude (in aggregate and
individually) as well as in an ordinal sense (in aggregate).

\subsubsection{Methods}

We measure consumers' PPVs for the advertising practices in Scenarios A
through F through an online study with approximately 1,200 subjects
(i.e., approximately 200 per condition) in the United States. A human,
allocated randomly to a scenario, is presented with a description of the
scenario and asked how much they perceive their privacy to be violated
by this advertising practice. Later the digital twin of this human is
asked the exact same question.

\subsubsection{Results - Pre-registered Analyses}
For each of the 6 conditions, Table \ref{SI_privacy_tab1} reports: the human-twin sample
size; mean PPV by human respondents (along with the lower and upper
bounds of the 95\% confidence interval); mean PPV by digital twins of
human respondents (along with the lower and upper bounds of the 95\%
confidence interval); Pearson r between human and twin PPVs; paired
t-test on PPVs; and F-test on variances. Table \ref{SI_privacy_tab2} provides the overall
Spearman rho on the rank ordering of condition means. Twins answered ``1
- not at all'' for all observations in Scenarios A, B and C, so
correlations and certain tests are undefined in these scenarios.

\begin{longtable}{p{0.17\linewidth} p{0.13\linewidth} p{0.15\linewidth} p{0.15\linewidth} p{0.13\linewidth} p{0.13\linewidth} p{0.14\linewidth}}
\toprule
Measure & A: No Ads, No Tracking & B: Untargeted Ads & C: Contextual Targeting & D: Group-level PET & E: Individual-level PET & F: Behavioral Targeting \\
\midrule
\endfirsthead

\toprule
Measure & A: No Ads, No Tracking & B: Untargeted Ads & C: Contextual Targeting & D: Group-level PET & E: Individual-level PET & F: Behavioral Targeting \\
\midrule
\endhead

N\_pairs & 199 & 218 & 187 & 222 & 180 & 194 \\
Mean\_human & 1.744 & 2.133 & 2.658 & 4.667 & 4.461 & 5.469 \\
Mean\_twin & 1 & 1 & 1 & 2.712 & 2.817 & 4.603 \\
CI95\_hu\_low & 1.553 & 1.935 & 2.396 & 4.443 & 4.214 & 5.246 \\
CI95\_hu\_high & 1.935 & 2.331 & 2.919 & 4.890 & 4.709 & 5.692 \\
CI95\_tw\_low & 1 & 1 & 1 & 2.616 & 2.727 & 4.476 \\
CI95\_tw\_high & 1 & 1 & 1 & 2.807 & 2.907 & 4.731 \\
Pearson\_r &  &  &  & 0.014 & -0.069 & 0.088 \\
p-val\_r &  &  &  & 0.840 & 0.355 & 0.222 \\
t\_stat & 7.676 & 11.305 & 12.496 & 15.945 & 12.058 & 6.907 \\
p-val\_t & 0 & 0 & 0 & 0 & 0 & 0 \\
F\_var &  &  &  & 5.446 & 7.569 & 3.070 \\
p-val\_F &  &  &  & 0 & 0 & 0 \\
\bottomrule
\caption{Human and Twin Responses per Condition. \label{SI_privacy_tab1}} 
\end{longtable}

\begin{longtable}{p{0.63\linewidth} p{0.18\linewidth} p{0.19\linewidth}}
\toprule
 & rho & p-value \\
\midrule
\endfirsthead

\toprule
 & rho & p-value \\
\midrule
\endhead

Spearman\_rank\_corr & 0.88 & 0.021 \\
\bottomrule
\caption{Spearman rank correlation of condition means from Human vs.\ Twins. \label{SI_privacy_tab2}}
\end{longtable}

The results of the experiments show the following:

\begin{itemize}
\item
  The PPVs for human subjects are in the order hypothesized (with one
  deviation being that the mean PPV of Scenario D is higher than the
  mean PPV of Scenario E; however, these numbers are not statistically
  different accounting for the 95\% CIs).
\item
  The PPVs for the digital twins are significantly lower in magnitude,
  i.e., digital twins have lower perceptions of privacy violations.

  \begin{itemize}
  \item
    All twins give PPV values of 1 for Scenarios A, B and C, in which
    there is no tracking or targeting on past behavior.
  \item
    The Pearson correlations between the PPVs of humans and their
    digital twins, when defined, are near zero.
  \item
    The t-stats for the paired t-test between PPVs of humans and their
    digital twins show that these PPVs are significantly different.
  \item
    The F-test of variances between PPVs of humans and their digital
    twins show that the variances in PPVs are significantly different.
  \end{itemize}
\item
  While the PPVs of digital twins are significantly lower than the PPVs
  of humans, the rank ordering of the scenarios, based on average PPV,
  is similar for humans and digital twins, with a Spearman rank
  correlation of 0.88.
\end{itemize}

\subsubsection{Discussion}

The results of the study indicate that digital twins perceive their
privacy to be violated less than their human counterparts (as indicated
by lower PPV scores for digital twins compared to humans). The twins
also do not do a good job of capturing heterogeneity in perceived
privacy violations (as indicated by almost zero correlations in PPV
scores for digital twins compared to humans). However, in aggregate,
humans and their digital twins rank privacy violations similarly across
advertising practices that rely differently on tracking and targeting
(as indicated by a high rank correlation).

\pagebreak

\subsection{Preferences for Redistribution}\label{preferences-for-redistribution}

\subsubsection{Main Questions/Hypotheses}

Building on the seminal work of \cite{alesina2011preferences}, this study
assesses the ability of digital twins to replicate individuals'
preferences for redistribution and explores the mechanisms through which
this replication may fail. Our instruments come from the General Social
Science (GSS), a major social‐science survey in the U.S. Preferences for
redistribution are measured through both a standard question (asking
whether the government should do everything possible to improve the
standard of living of all poor Americans) and a health-tax willingness
question. In addition, mechanism variables include fairness judgments,
work-versus-luck beliefs, father's education, and trust in others.

Understanding how accurately digital twins mirror human attitudes on
these topics is crucial for policymakers, since public support for tax
policy, welfare programs, and social-cohesion initiatives depends
directly on redistribution and trust levels, while beliefs about
fairness and effort versus luck inform education and labor-market
reform.

Our investigation proceeds in two stages. First, we examine whether
digital twins systematically differ from individuals in their
preferences for redistribution. Second, we investigate whether any
observed discrepancies can be explained by the twins' failure to
replicate responses related to socio‐economic heritage (father's
education), fairness judgments, beliefs about the relative role of
effort versus luck in achieving success, and interpersonal trust. These
have been shown to predict redistribution preferences \cite{alesina2011preferences}.

\subsubsection{Methods}

We administered a single, online questionnaire---closely modeled on core
GSS items---to approximately 1,200 individuals and their matched digital
twins. Respondents completed various questions; those of interest for
this analysis are presented below. We recoded every item so that higher
numeric values consistently reflected stronger endorsement (and set all
``Don't know'' or non-substantive responses to missing). Father's
education categories were converted to corresponding years of schooling
(10--18 years). Specifically:

\begin{enumerate}
\item
  \textbf{Redistribution 1 (Government Responsibility).}\\
  ``Some people think that the government in Washington should do
  everything possible to improve the standard of living of all poor
  Americans; they are at Point 1 on the scale below. Other people think
  it is not the government's responsibility, and that each person should
  take care of himself; they are at Point 5. Where would you place
  yourself on this scale, or haven't you made up your mind on this?''

  \begin{enumerate}
  \item
    \textbf{Response Options \& Coding:}

    \begin{enumerate}
    \item
      1 -- I strongly agree the government should improve living
      standards → 5
    \item
      2 → 4
    \item
      3 -- I agree with both answers → 3
    \item
      4 → 2
    \item
      5 -- I strongly agree people should take care of themselves → 1
    \item
      Don't know/No answer → missing
    \end{enumerate}
  \end{enumerate}
\item
  \textbf{Redistribution 2 (Health‐Care Tax Willingness).\\
  } ``How willing would you be to pay higher taxes to improve the level
  of health care for all people in the United States?''

  \begin{enumerate}
  \item
    \textbf{Response Options \& Coding:}

    \begin{enumerate}
    \item
      Very willing → 5
    \item
      Fairly willing → 4
    \item
      Neither willing nor unwilling → 3
    \item
      Fairly unwilling → 2
    \item
      Very unwilling → 1
    \item
      Don't know/Can't choose → missing
    \end{enumerate}
  \end{enumerate}
\item
  \textbf{Fairness Judgment.\\
  } ``Which comes closer to your view of most people: (1) Most people
  would take advantage of you if they got the chance; (2) Most people
  would try to be fair; or (3) It depends?''

  \begin{enumerate}
  \item
    \textbf{Response Options \& Coding:}

    \begin{enumerate}
    \item
      ``Would take advantage of you'' → 3
    \item
      ``It depends'' → 2
    \item
      ``Would try to be fair'' → 1
    \item
      Don't know → missing
    \end{enumerate}
  \end{enumerate}
\item
  \textbf{Work vs. Luck.\\
  } ``Which comes closer to your view---that hard work MOST contributes
  to someone getting ahead in life, that luck or help from other people
  MOST contributes, or that hard work and luck are about equally
  important?''

  \begin{enumerate}
  \item
    \textbf{Response Options \& Coding:}

    \begin{enumerate}
    \item
      Hard work most important → 1
    \item
      Hard work and luck equally important → 2
    \item
      Luck or help from others most important → 3
    \item
      Don't know → missing
    \end{enumerate}
  \end{enumerate}
\item
  \textbf{Interpersonal Trust.\\
  } ``Generally speaking, would you say that most people can be trusted,
  or that you can't be too careful in dealing with people?''

  \begin{enumerate}
  \item
    Response Options \& Coding:

    \begin{enumerate}
    \item
      ``Most people can be trusted'' → 3
    \item
      ``Depends'' → 2
    \item
      ``Can't be too careful'' → 1
    \item
      Don't know → missing
    \end{enumerate}
  \end{enumerate}
\item
  \textbf{Father's Education.\\
  } ``What is the highest level of education completed by your father?''

  \begin{enumerate}
  \item
    \textbf{Response Options \& Coding (years of schooling):}

    \begin{enumerate}
    \item
      Less than high school → 10
    \item
      High school graduate → 12
    \item
      Some college, no degree → 13
    \item
      Associate's degree → 14
    \item
      College graduate / some postgraduate → 16
    \item
      Postgraduate degree → 18
    \item
      Don't know/No answer → missing
    \end{enumerate}
  \end{enumerate}
\end{enumerate}

After collecting the raw survey data, we removed any respondent whose
completion time fell below 50 percent of the sample median or who failed
our attention‐check prompt. We then merged the cleaned human and twin
datasets one-to-one on their unique identifiers, producing 1,163 matched
pairs for each outcome. For each of our six measures, we calculated the
Pearson correlation (with Fisher confidence intervals) to gauge twin
resemblance, carried out paired t-tests to compare mean differences,
performed F-tests to assess variance equality, and derived an accuracy
index (1 minus the mean absolute difference divided by the variable's
range). Although not pre-registered, we also generated side-by-side
percent histograms of each response distribution to identify any
systematic replication discrepancies.

\subsubsection{Results - Pre-registered Analyses}

\begingroup
\footnotesize

\begin{longtable}{p{0.15\linewidth} p{0.14\linewidth} p{0.14\linewidth} p{0.14\linewidth} p{0.14\linewidth} p{0.14\linewidth} p{0.15\linewidth}}
\toprule
Measure & Redistribution 1 & Redistribution 2 & Fairness & Work vs.\ Luck & Father's Education & Trust in Others \\
\midrule
\endfirsthead

\toprule
Measure & Redistribution 1 & Redistribution 2 & Fairness & Work vs.\ Luck & Father's Education & Trust in Others \\
\midrule
\endhead

Accuracy & 0.780 & 0.778 & 0.641 & 0.770 & 0.733 & 0.593 \\
Correlation & 0.621 & 0.614 & 0.298 & 0.346 & 0.126 & 0.296 \\
Mean (Human) & 3.55 & 3.27 & 1.96 & 1.85 & 13.66 & 1.73 \\
Mean (Twin) & 3.44 & 3.61 & 1.71 & 1.70 & 12.09 & 2.22 \\
$t$-test ($p$-value) for means & 2.85 \ ($p~<~.01$) & --9.14 ($p~<~.001$) & 8.76 ($p~<~.001$) & 7.65 ($p~<~.001$) & 21.55 ($p~<~.001$) & --16.74 ($p~<~.001$) \\
SD (Human) & 1.28 & 1.36 & 0.74 & 0.70 & 2.48 & 0.73 \\
SD (Twin) & 1.65 & 1.53 & 0.89 & 0.48 & 0.71 & 0.93 \\
$F$-test ($p$-value) for variances & 0.600 ($p~<~.001$) & 0.797 ($p~=~.00012$) & 0.699 ($p~<~.001$) & 2.120 ($p~<~.001$) & 12.178 ($p~<~.001$) & 0.625 ($p~<~.001$) \\
$n$ & 1153 & 1144 & 1151 & 1158 & 1163 & 1162 \\
\bottomrule
\caption{Human vs.\ Twin Responses. \label{SI_redistribution_tab1}}
\end{longtable}
\endgroup

Table \ref{SI_redistribution_tab1} shows the results from the pre-registered analysis. Digital
twins capture the relative ordering of individuals' redistribution
preferences reasonably well: both the standard five-point scale measure
(Redistribution 1) and the health-tax willingness question
(Redistribution 2) yield strong twin correlations ($r \approx .62$).
However, twins tend to slightly underestimate general redistribution
support ($\text{mean}_h = 3.55$ vs.\ $\text{mean}_t = 3.44$; 
$t = 2.85,\; p = .0044$) and overestimate willingness to pay higher
health taxes ($\text{mean}_h = 3.27$ vs.\ $\text{mean}_t = 3.61$; 
$t = -9.14,\; p < .001$). Accuracy scores around $.78$ for each measure
indicate that, on average, twins miss individual scores by nearly a full
point on the five-point scales. Moreover, significant variance
differences ($F = .60$ and $F = .80$, both $p < .001$) suggest that
twins' responses are more dispersed compared to human distributions.
Thus, while digital twins reflect who is relatively more or less
supportive of redistribution, they systematically misestimate absolute
levels and spread.

To understand why these mismatches occur, we examined additional
constructs known to shape redistribution attitudes \cite{alesina2011preferences}.
Twins poorly replicate socio-economic heritage---father's education
shows a low correlation ($r \approx .13$), a mean gap of over 1.5 years
($t = 21.55,\; p < .001$), and a sharply reduced variance 
($F = 12.18,\; p < .001$). Fairness judgments ($r \approx .30$; 
accuracy $\approx .64$) and work-versus-luck beliefs 
($r \approx .35$; accuracy $\approx .77$) similarly display moderate
rank-order similarity but pronounced mean and variance discrepancies.
Interpersonal trust also diverges, with twins overstating trust
($\text{mean}_t = 2.22$ vs.\ $\text{mean}_h = 1.73$; 
$t = 10.38,\; p < .001$). Because these background factors and beliefs
are important predictors of redistribution preferences, their imperfect
replication by digital twins likely underlies the systematic errors in
predicting absolute levels of support for redistribution.

\subsubsection{Results - Additional Analyses (Non-Preregistered)}

\begin{figure}[H]
\centering
\includegraphics[width=1\textwidth]{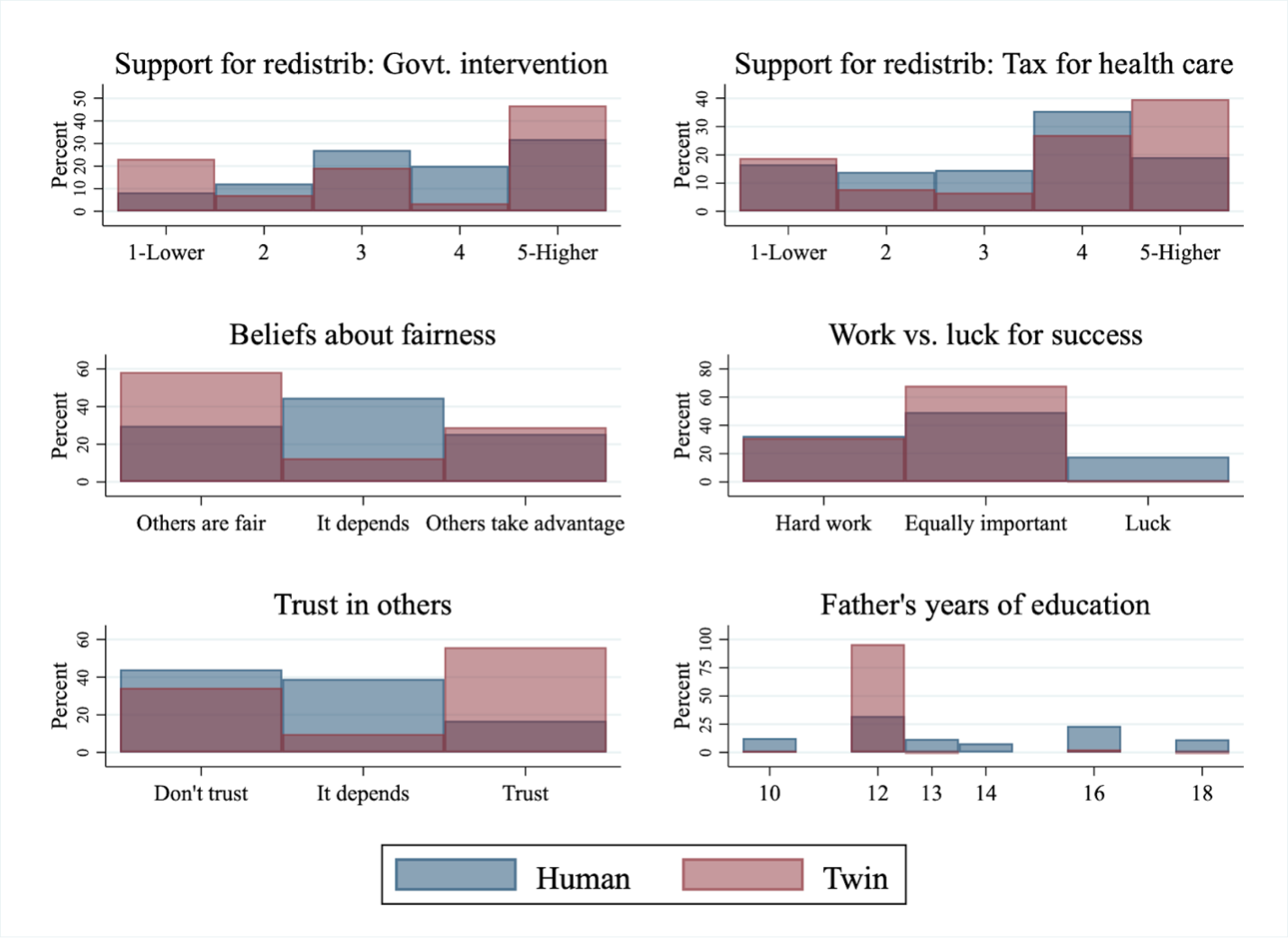}
\caption{Comparison of Response Distributions.
}\label{SI_redistribution_1}
\end{figure}

We also present histograms of our outcomes to compare the response
distributions of humans and twins more effectively. As shown in Figure
\ref{SI_redistribution_1}, twins exhibit more extreme redistribution preferences, with the
heavier tails driving the higher standard deviation observed above. They
are likewise more prone to overstate that others behave fairly and to
report greater trust in fellow individuals. Twins also more frequently
endorse that both effort and luck matter equally for success. Finally,
nearly 100\% of twins report that their father completed high school (12
years of education), whereas the human sample shows a much more diverse
educational distribution.

\subsubsection{Discussion}

In summary, although digital twins reliably reproduce the relative ranking of individuals' redistribution preferences, as evidenced by
strong correlations on both the standard scale and health‐tax question,
they consistently misestimate the absolute levels and dispersion of
those preferences. Our analysis suggests these systematic errors stem
from twins' failures to mirror key background factors and beliefs: most
notably socio‐economic heritage (father's education), fairness
judgments, work‐versus‐luck attributions, and interpersonal trust, which
are themselves only moderately correlated and exhibit pronounced mean
and variance discrepancies. The heavier tails in twins' redistribution
responses further underscore their tendency toward more extreme
positions. For policymakers and researchers considering the use of
digital avatars to gauge public opinion, these findings highlight the
importance of refining the modeling of underlying socio‐cognitive
constructs; without accurately capturing individuals' formative
experiences and normative beliefs, digital proxies will misrepresent
both the magnitude and variability of real‐world attitudes, potentially
leading to misguided inferences about public support for tax, welfare,
and social‐cohesion initiatives.

\pagebreak

\subsection{Promiscuous Donors}\label{promiscuous-donors}

\subsubsection{Main Questions/Hypotheses}

Political donations play a critical role in shaping electoral outcomes.
They equip candidates with the financial resources needed to compete in
political races (e.g., \cite{gilens2012affluence,hussein2025outparty}). They also
function as signals of support. Political parties often use donation
counts as thresholds for unlocking institutional backing and as a
pre-requisite for qualifying for televised debates. In the present
research, we investigate the reputational consequences of a specific
form of donation behavior: contributing to both major political parties.
We refer to individuals who donate to two parties as ``promiscuous
donors'' and we examine how they are perceived by others. Understanding
how promiscuous donors are perceived is important given the prevalence
of this behavior. Analysis of 2024 Federal Election Commission data
reveals that at least 113,187 Americans made contributions to both
parties during the 2024 election cycle.

Ex ante, it is unclear how individuals will evaluate those who donate to
both political parties. On one hand, promiscuous donors may be viewed
favorably. Some scholars have argued that political polarization has
reached unsustainable levels \cite{heltzel2020polarization}, prompting a
public desire for greater bipartisanship and reconciliation. From this
perspective, contributing to both parties could signal a conciliatory
stance, potentially enhancing reputational standing.

On the other hand, several theoretical perspectives suggest that such
behavior may incur reputational costs. Partisan animosity has
intensified in recent years, with Democrats and Republicans expressing
growing levels of dislike and distrust toward each other \cite{pew2022partisan}. Furthermore, emerging research on receptiveness to
opposing views \cite{hussein2021receptiveness} suggests that open-mindedness
toward the other side can backfire socially \cite{hussein2024receptiveness}.
If donors are seen as legitimizing or aligning with an ideologically
objectionable outgroup, their behavior may be perceived as disloyal or
suspect, thereby leading to reputational costs.

\subsubsection{Methods}

We recruited 799 human respondents on Prolific and their respective
twins. This study used a rating-based conjoint design \cite{green1978conjoint}. Conjoint is a study design used to determine how
people value different attributes (e.g., screen size, memory, color)
that make up a product (e.g., a laptop). Participants in a conjoint
study are shown a series of trials, each showcasing a different version
of the target product. By varying different attribute levels of the
product (e.g., a black vs. silver laptop) between trials and measuring
how consumers respond to different combinations of these attributes,
managers can quantify how much a given attribute (e.g., color) affects
overall attitudes.

Applying this approach, we had participants read 10 vignettes, each
describing a target who engaged in a different donation behavior. The
donation behaviors included contributing solely to the Republican Party,
solely to the Democratic Party, to neither party, or to both parties.
Our primary focus was on the reputational consequences of donating to
both parties. To this end, we systematically varied the rationale
provided for such behavior. Specifically, we compared conditions in
which no justification was given to those in which a self-interested or
a values-based rationale was offered. Self-interested justifications
emphasized pragmatic benefits, such as attracting bipartisan talent to
one's business or facilitating smoother relations with elected officials
across party lines. In contrast, values-based justifications highlighted
ideological motivations, such as supporting candidates with the best
ideas regardless of party or promoting cross-party cooperation. Below is
a summary of the exact description used and the attribute level they
correspond to:

\begin{itemize}
\item
  Own Party or Opposing Party (coded relative to participant's party
  affiliation; independent participants who rated these had their own
  category):

  \begin{itemize}
  \item
    ``I currently only donate to Republicans''
  \item
    ``I currently only donate to Democrats''
  \end{itemize}
\item
  No donations: ``I currently don't donate to any politician''
\item
  Both: ``I currently donate to both Democrats and Republicans''
\item
  Business Strategy: ``I currently donate to both Democrats and
  Republicans so my business can work well with whoever wins''
\item
  Attract Talent: ``I currently donate to both Democrats and Republicans
  because it helps me attract top talent for my business''
\item
  Common Grounds: ``I currently donate to both Democrats and Republicans
  because I believe in working across the aisle to find common grounds''
\item
  Good Ideas: ``I currently donate to both Democrats and Republicans
  because I support candidates who have good ideas, regardless of which
  party they come from''
\end{itemize}

To enhance the generalizability of our findings, we systematically
varied multiple dimensions of the vignettes beyond donation behavior.
This design enabled us to estimate the average effect of donating to
both parties across a range of contextual backgrounds. Specifically, we
manipulated the target's political affiliation (Democrat vs.
Republican), job (e.g., CEO of a publicly traded company vs. founder of
a fast-growing start-up), and donation history (e.g., history of prior
donations vs. first-time donors). This approach ensured that observed
effects were not idiosyncratic to a single type of target. Below is an
example vignette used in the study:

\begin{quote}
Imagine you meet someone named John at a neighborhood barbecue.

John recently moved to your area. He is a Democrat. He is the CEO of a
publicly traded company.

During your conversation, the topic of donating to political campaigns
comes up.

John mentioned that he never used to donate to politicians.

John then said to you:

"I currently donate to both Democrats and Republicans."
\end{quote}

After reading the vignette, participants were asked to report their
overall impressions of the target: ``Based on what John said, what's
your overall impression of him?'' Participants reported their attitudes
on a nine-point bipolar scale, ranging from very unfavorable to very
favorable. As a reminder, each participant rated ten targets in total.

\subsubsection{Results - Pre-registered Analyses}
As preregistered, we regressed overall impression of the target on dummy
variables of the manipulated attributes and participant type (digital
twin vs. human). Importantly, we included an interaction between our
focal manipulated attribute--donation behavior-\/-and participant type.
Standard errors were clustered on the participant level to account for
repeated measurement. More specifically, we estimated the following
regression:
\begin{align*}
\widehat{\text{Impression}}_{i,j} &= \alpha 
+ \vec{\beta}_{1}\,\text{donation behavior}_{i,j} \\
&\quad + \beta_{2}\,\text{participant type}_{i} \\
&\quad + \vec{\beta}_{3}\,\big(\text{donation behavior}_{i,j} \times \text{participant type}_{i}\big) \\
&\quad + \beta_{4}\,\text{target party}_{i,j} \\
&\quad + \vec{\beta}_{5}\,\text{history}_{i,j} \\
&\quad + \vec{\beta}_{6}\,\text{job}_{i,j}
\end{align*}
We were interested in the vector of interactions $\vec{\beta}_{3}$.
Significant interactions would suggest that humans and their digital
twins differed in their reactions to current donation behavior.
Non-significant interactions would suggest no difference.

\begin{figure}[H]
\centering
\includegraphics[width=1\textwidth]{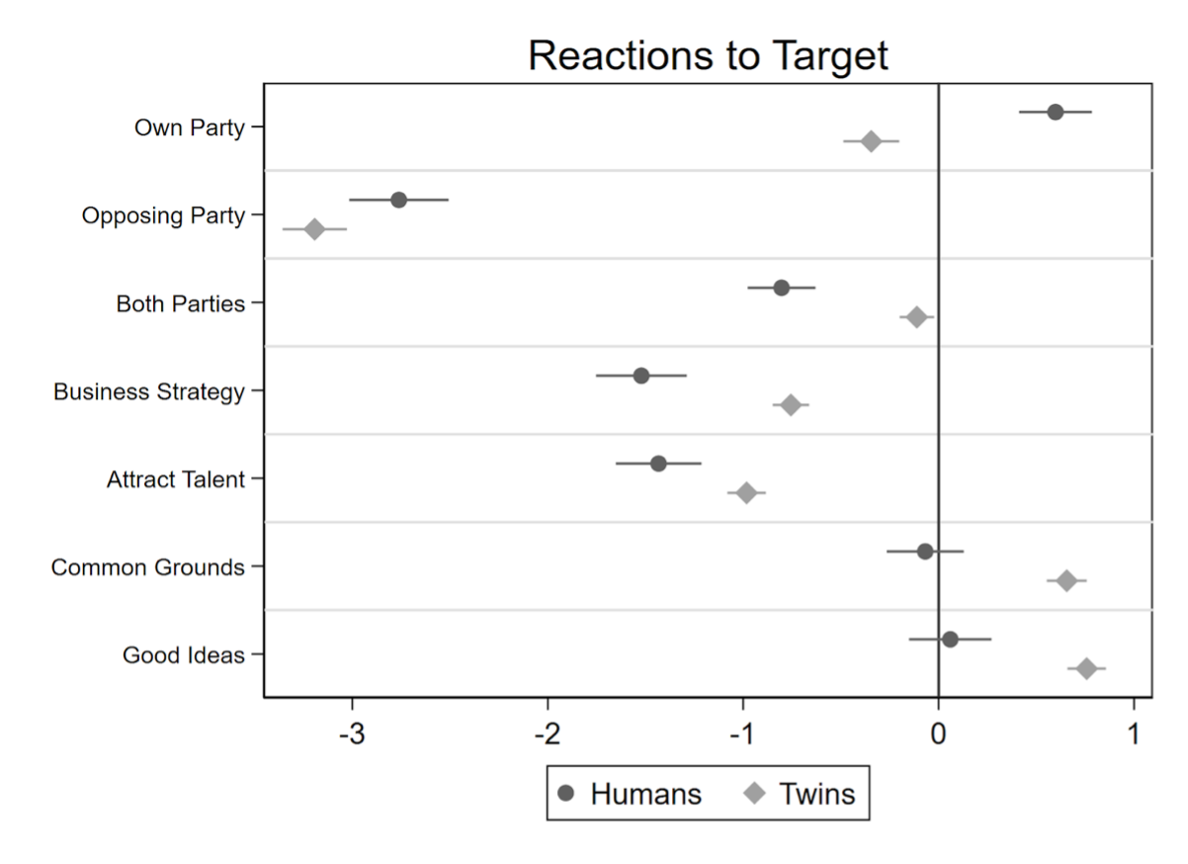}
\caption{Reactions to target among humans and digital twins. Error bars
denote 95\% confidence intervals. Standard errors are clustered at the
participant level. The reference category is a target who did not donate
to either party. For legibility, the dummy variable indicating that an
independent participant rated a single-party donor is omitted from the
figure but is included in the regression.
}\label{SI_promiscuous_donors_1}
\end{figure}

Of the eight interaction terms, seven reached significance ($p$s $< .001$). 
Both humans and digital twins evaluated targets who donated to the opposing 
party more negatively than targets who made no political donations 
(interaction $b = -0.32,\; SE = 0.27,\; t = -1.18,\; p = .24$).

Four of the seven significant interactions revealed divergent effects in
magnitude between human participants and their digital twins. Both
humans and twins rated targets who donated to both parties negatively
relative to targets who did not donate to either party, but digital
twins were more positive in their evaluations 
($b = -0.11,\; SE = 0.05,\; t = -2.47,\; p = .014$) than humans 
($b = -0.80,\; SE = 0.09,\; t = -9.09,\; p < .001$). Similarly, when 
targets provided self-interested reasons for donating to both parties, 
both humans and twins rated them more negatively compared to targets who 
did not donate at all. However, twins were more positive. For example, 
both humans and twins rated targets who donated to both parties because 
they wanted to be able to work with whichever party is in power negatively, 
but twins were more positive 
($b = -0.76,\; SE = 0.05,\; t = -15.86,\; p < .001$) than humans 
($b = -1.52,\; SE = 0.12,\; t = -12.85,\; p < .001$). The same pattern 
held for targets who donated to both parties because they wanted to attract 
top talent to their business (humans: $b = -1.43,\; SE = 0.11,\; t = -12.84,\; p < .001$ 
vs.\ twins: $b = -0.98,\; SE = 0.05,\; t = -19.56,\; p < .001$).

The three remaining significant interactions revealed divergent effects
in direction between human participants and their digital twins. Human
respondents rated targets who donated to their own party more positively
than those who did not donate 
($b = 0.60,\; SE = 0.09,\; t = 6.30,\; p < .001$). In contrast, digital 
twins evaluated the same targets more negatively compared to those who 
did not donate ($b = -0.34,\; SE = 0.07,\; t = -4.74,\; p < .001$). 
A similar divergence emerged in evaluations of targets who donated to 
both parties for value-based reasons. Humans rated them the same way 
they rated targets who did not donate to either party. However, twins 
rated them more positively. Specifically, among humans, there was no 
significant difference between targets that donated to both parties 
because they believe in working across the aisle to find common ground 
and targets who did not donate 
($b = -0.07,\; SE = 0.10,\; t = -0.68,\; p = .50$). However, twins 
rated the former more positively than targets who did not donate 
($b = 0.66,\; SE = 0.05,\; t = 12.61,\; p < .001$). The same pattern 
held for targets who donated to both parties because they wanted to 
support the best ideas regardless of which party they came from 
(humans: $b = 0.06,\; SE = 0.11,\; t = 0.56,\; p = .58$ 
vs.\ twins: $b = 0.76,\; SE = 0.05,\; t = 15.12,\; p < .001$).

\subsubsection{Discussion}

How do people evaluate promiscuous donors (i.e., targets who donate to
both parties)? Our results indicate that, compared to those who do not
donate to politics at all, those who donate to both parties are
perceived more negatively. This result is consistent with recent
research on the reputational costs of receptiveness to opposing views \cite{hussein2024receptiveness}. Interestingly, the reputational costs of
promiscuous donation are particularly salient when the reason behind
donating to both parties is self-interest (e.g., to attract the best
talent to one's business or to make sure that one's business can work
with whichever party wins) and are somewhat weaker when the reason is
more value-based (e.g., believing in bridging divides or that good ideas
should be supported regardless of which party they come from).

How do humans and their digital twins compare when judging the donation
behavior of others? Across more than 7,000 conjoint judgments, human
respondents penalized promiscuous donors, especially when the behavior
appeared motivated by self-interest (e.g., cultivating access or
recruiting talent). In contrast, digital twins were less punitive
(sometimes even favorable) toward the very same targets.

What might drive these differences? One explanation is that language
models are trained to discount signals of intergroup conflict. Hence, a
cross-party donation that humans interpret as a collaboration with the
enemy is read by the model as pragmatic or even positive as it shows a
willingness to reach across divides. A second, complementary account
centers on affect: if twins lack the visceral emotions that energize
partisan animus, they down-regulate the reputational costs of
interacting with the political out-group. Further research on how
digital twins compare to humans in matters of social cognition would be
worthwhile.

\subsubsection{Appendix - Regression Tables}

\begin{longtable}{p{0.8\linewidth} p{0.2\linewidth}}
\toprule
 & (1) rating\_\\
\midrule
\endfirsthead

\toprule
 & (1) rating\_\\
\midrule
\endhead

Twins & 0.411\textsuperscript{***} \\
 & (0.0694) \\
Own Party=1 & 0.502\textsuperscript{***} \\
 & (0.0915) \\
Opposing Party=1 & -2.632\textsuperscript{***} \\
 & (0.127) \\
Both Parties=1 & -0.802\textsuperscript{***} \\
 & (0.0883) \\
Business Strategy=1 & -1.509\textsuperscript{***} \\
 & (0.116) \\
Attract Talent=1 & -1.425\textsuperscript{***} \\
 & (0.110) \\
Common Grounds=1 & -0.0553 \\
 & (0.0984) \\
Good Ideas=1 & 0.0768 \\
 & (0.106) \\
Partisan Donation (Independent)=1 & -1.682\textsuperscript{***} \\
 & (0.243) \\
Twins \# Own Party=1 & -0.745\textsuperscript{***} \\
 & (0.117) \\
Twins \# Opposing Party=1 & -0.700\textsuperscript{***} \\
 & (0.153) \\
Twins \# Both Parties=1 & 0.691\textsuperscript{***} \\
 & (0.0994) \\
Twins \# Business Strategy=1 & 0.739\textsuperscript{***} \\
 & (0.124) \\
Twins \# Attract Talent=1 & 0.435\textsuperscript{***} \\
 & (0.119) \\
Twins \# Common Grounds=1 & 0.700\textsuperscript{***} \\
 & (0.110) \\
Twins \# Good Ideas=1 & 0.665\textsuperscript{***} \\
 & (0.115) \\
Twins \# Partisan Donation (Independent)=1 & -0.315 \\
 & (0.267) \\
Target: Own Party & 0.311\textsuperscript{***} \\
 & (0.0865) \\
Target: Opposing Party & -0.0112 \\
 & (0.0879) \\
Start-up Founder & -0.112\textsuperscript{***} \\
 & (0.0369) \\
CEO & -0.197\textsuperscript{***} \\
 & (0.0378) \\
No Donations & 0.0808\textsuperscript{**} \\
 & (0.0318) \\
Both & 0.119\textsuperscript{***} \\
 & (0.0313) \\
Observations & 15,980 \\
$R^2$ & 0.311 \\
Adjusted $R^2$ & 0.310 \\
\bottomrule
\caption{Regression results: pooled sample. \label{SI_promiscuous_tab1}}
\end{longtable}

\begin{longtable}{p{0.4\linewidth} p{0.3\linewidth} p{0.3\linewidth}}
\toprule
 & (1) Ratings Humans & (2) Ratings Twins \\
\midrule
\endfirsthead

\toprule
 & (1) Ratings Humans & (2) Ratings Twins \\
\midrule
\endhead

Own Party & 0.598\textsuperscript{***} & -0.345\textsuperscript{***} \\
 & (0.0949) & (0.0728) \\
Opposing Party & -2.762\textsuperscript{***} & -3.193\textsuperscript{***} \\
 & (0.130) & (0.0839) \\
Both Parties & -0.804\textsuperscript{***} & -0.111\textsuperscript{**} \\
 & (0.0884) & (0.0450) \\
Business Strategy & -1.521\textsuperscript{***} & -0.756\textsuperscript{***} \\
 & (0.118) & (0.0477) \\
Attract Talent & -1.433\textsuperscript{***} & -0.983\textsuperscript{***} \\
 & (0.112) & (0.0502) \\
Common Grounds & -0.0684 & 0.656\textsuperscript{***} \\
 & (0.100) & (0.0520) \\
Good Ideas & 0.0598 & 0.758\textsuperscript{***} \\
 & (0.108) & (0.0501) \\
Partisan Donation (Independent) & -1.580\textsuperscript{***} & -2.073\textsuperscript{***} \\
 & (0.255) & (0.0925) \\
Target: Own Party & 0.315\textsuperscript{*} & 0.338\textsuperscript{***} \\
 & (0.182) & (0.0582) \\
Target: Opposing Party & 0.218 & -0.224\textsuperscript{***} \\
 & (0.183) & (0.0606) \\
Start-up Founder & -0.0795 & -0.146\textsuperscript{***} \\
 & (0.0633) & (0.0375) \\
CEO & -0.161\textsuperscript{**} & -0.231\textsuperscript{***} \\
 & (0.0662) & (0.0360) \\
No Donations & 0.117\textsuperscript{**} & 0.0454 \\
 & (0.0571) & (0.0278) \\
Both & 0.127\textsuperscript{**} & 0.111\textsuperscript{***} \\
 & (0.0565) & (0.0262) \\
Observations & 7,990 & 7,990 \\
$R^2$ & 0.176 & 0.546 \\
Adjusted $R^2$ & 0.175 & 0.545 \\
\bottomrule
\caption{Regression results: split by humans vs.\ twins. \label{SI_promiscuous_tab2}} 
\end{longtable}

\pagebreak

\subsection{Quantitative Intuition}\label{quantitative-intuition}

\subsubsection{Main Questions/Hypotheses}

This study evaluates whether digital twins trained from large language
models (LLMs) can accurately replicate individual responses on a newly
developed psychometric instrument measuring Quantitative Intuition (QI) \cite{frank2022decisions}. The QI scale assesses the
respondent's own analytical and intuitive mindsets through 38
self-report items in addition 16 items measuring individuals'
perceptions of their organization's analytical and intuitive
orientations.

We test two core questions: First, can digital twins accurately
reproduce individual human responses to the QI scale? Second, do digital
twins match human participants on average scores for the quantitative
(Q) and intuitive (I) mindset dimensions? Because the scale was
developed in tandem with this study and was not available during LLM
training, this provides a strong test of digital twins' ability to
predict out-of-sample psychometric responses.

\subsubsection{Methods}

We recruited human participants from Prolific who were part of a larger
digital twin study and collected matched responses from their
corresponding digital twins. Invitations were sent to 2,000 participants
from the original twin cohort; 1,435 individuals responded to the
survey. Following pre-registered exclusion criteria, we removed 108
respondents who failed at least one of two attention checks, 8
respondents who ``straightlined'' (i.e., provided the same response
across items 32--38 of the scale), and 1 respondent who completed the
survey in under 150 seconds (mean response time was 789 seconds with a
standard deviation of 457 seconds). The final sample included 1,323
respondents.

Participants completed 38 self-assessment items measuring their
individual Quantitative Intuition (QI) mindset, followed by 16 items
assessing their perception of their current or former organization's QI
mindset. In addition, participants responded to two behavioral QI
questions---one closed-ended and one open-ended---designed to capture
willingness to ``guestimate'' and synthesize information. These
behavioral items serve as external validity checks for the QI scale.

To explore individual differences in the QI scores, we examined
correlations between QI scores and demographic and psychological
variables collected as part of the broader digital twin survey.

Digital twins were prompted to complete the same set of items as their
human counterparts, simulating how the participant might respond.

We derived individual QI scores using two approaches: 1) Simple
averaging: we computed the mean of the first 19 items as a quantitative
mindset score (Q), and the mean of the next 19 items as an intuitive
mindset score (I), and 2) Factor analysis: we conducted an exploratory
factor analysis with two factors, which naturally aligned with the
quantitative and intuitive dimensions. Factor scores from this model
were used as alternative QI measures.

We then compare the individual, simple average and factor analysis
scores of the human respondents with those of the digital twins.

\subsubsection{Results - Pre-registered Analyses}

As pre-registered we compared the actual QI score of the Prolific
respondents with the QI score of their digital twins. We do so both at
the individual question level and for the average of the individual and
organizational quantitative and intuitive scores.

\emph{Comparing the average score of the human responses and the digital
twins}

Comparing the correlation between the average (or factor) scores between
the actual respondents and the digital twins for the individual and
organizational quantitative and intuitive scores we find that overall
the digital twins answers are closer to the actual respondents answers
for the individual relative to the organizational skill both for the
average quantitative and intuitive scores and for the factor analysis
scores.

The Cronbach's $\alpha$ for the Q and I individual scales for the human
respondents are 0.788 and 0.753, respectively, suggesting acceptable
levels of agreement for items in the scale. Interestingly, the digital
twin mimicked similar or even higher levels of agreement for items in the
scale, with Cronbach's $\alpha$ of 0.776 and 0.866 for the Q and I
individual scales.

Comparing the means and standard deviations of the scale scores between
the humans and and digital twins shows that the means are quite similar,
but the digital twins had consistently lower variance in the responses
across respondents. A possible reason for that is that the digital twins
are still regressing to some overall LLM mean.

\begin{longtable}{p{0.25\linewidth} p{0.25\linewidth} p{0.25\linewidth} p{0.25\linewidth}}
\toprule
Scale &  & Human & Digital Twin \\
\midrule
\endfirsthead

\toprule
Scale &  & Human & Digital Twin \\
\midrule
\endhead

\multirow{2}{*}{Q Ind.} & Mean & 3.579 & 3.573 \\
 & STD & 0.451 & 0.355 \\
\multirow{2}{*}{I Ind.} & Mean & 3.478 & 3.713 \\
 & STD & 0.412 & 0.394 \\
\multirow{2}{*}{Q Org.} & Mean & 3.580 & 3.014 \\
 & STD & 0.643 & 0.418 \\
\multirow{2}{*}{I Org.} & Mean & 3.219 & 3.479 \\
 & STD & 0.486 & 0.280 \\
\bottomrule
\caption{Quantitative and Intuitive Scales: Human vs.\ Digital Twin \label{SI_quantitative_tab1}} 
\end{longtable}

\emph{Correlations between the human responses and the digital twins}

Looking at the correlations between the individual-level QI scale
responses and those of the digital twins, we find a moderately strong
correlation between the responses to the 19 quantitative questions
($r = 0.513,\; p < .001$) and the 19 individual scores
($r = 0.481,\; p < .001$), with an overall correlation of $r = 0.46$.
Thus, at the individual level the digital twins are capable of mimicking
the human response to QI to a moderate extent. The correlations between
the digital twins and the human responses for the organizational QI
scores are much lower, though statistically significant
($r_{\text{quantitative, org}} = 0.252,\; p < .001$ and
$r_{\text{intuitive, org}} = 0.196,\; p < .001$). This is to be expected
as the original questions used to build the digital twins focused on the
individuals themselves rather than on the organization they
work/worked for.

\begin{figure}[H]
\centering
\includegraphics[width=1\textwidth]{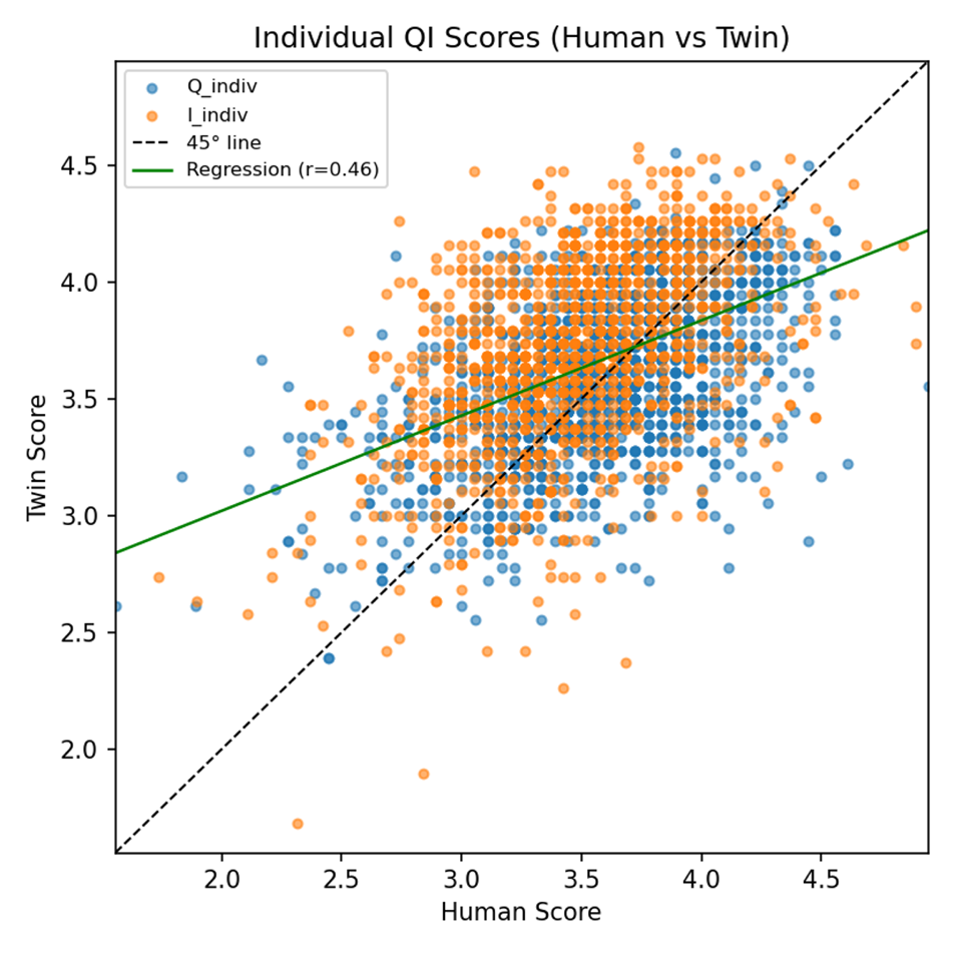}
\caption{
}\label{SI_quantitative_intuition_1}
\end{figure}

Finally, looking at the individual questions we find a positive and
significant (after Bonferroni correction for multiple tests) correlation
at the 95\% level between the human digital twins scores for 35 out of
the 38 individual QI questions. For the organizational QI questions, we
find a positive and significant (after Bonferroni correction for
multiple tests) correlation at the 95\% level between the human and
digital twin responses for 11 out of 16 questions. We did not find any
significant and negative correlations between the human and digital twin responses for any of the questions.

We did not find a statistically significant relationship between the QI
scores for either the willingness to guestimate questions or the ability
to synthesize scores. However, we found an interesting result with
respect to the ability of the digital twins to answer the QI behavioral
questions. In a close ended question, we gave respondents some pieces of
information and challenged them to address the problem by guestimation.
The digital twins chose to guestimate, which is the normative answer for
the guestimation question, 99.9\% of the times (only one twin did not
choose the normative response), whereas for the humans only 59.4\% were
inclined to guestimate. The difference between the human and digital
twin responses is statistically significant ($p < .001$).

In an open-ended question, we asked respondents to explain the situation
for a struggling company based on data, where the normative QI response
would be to synthesize the information rather than mainly summarize it.
We asked GPT-4o mini to classify each human or digital twin response as
a summary or a synthesis. A research assistant independently classified
20 responses and found that their classification perfectly matched with
those of ChatGPT. We found a similar result to those of the willingness
to guesstimate for the ability to synthesize. Whereas 99.8\% of the
digital twins synthesized the information, only 71.5\% of human
respondents replied with the QI normative response to synthesize. The
difference between the human and digital twin responses is statistically
significant ($p < .001$).

Thus, while the digital twins are capable of moderately following human
responses to the QI scale, when it comes to mimicking human behavior
questions, they tend to overexhibit behavior that is consistent with the
normative QI mindset.

\subsubsection{Results - Additional Analyses (Non-Preregistered)}

We also analyzed the relationship between the Q and I scores and the
demographic and individual characteristics collected in the first phase
of the digital twins study. Specifically, we calculated the correlations
between the average individual Q and I scores and each of the scales
measured in that initial phase. Focusing on individual rather than
organizational QI scores, we report only the correlations that are
statistically significant at the 95\% confidence level after applying
the Bonferroni correction for multiple testing.

Several interesting results emerge.

\begin{enumerate}
\def\labelenumi{\arabic{enumi}.}
\item
  The individual mindset score (I) is statistically significantly
  correlated with the following scales at the 95\% level:

  \begin{enumerate}
  \def\labelenumii{\arabic{enumii}.}
  \item
    Regulatory Focus (Fellner et al. 2017) - r=0.452
  \item
    Openness (John \& Srivastava, 1999) - r=0.428
  \item
    Consumer Need for Uniqueness (Ruvio et al., 2008) - r=0.379
  \item
    Self monitoring (Lennox \& Wolfe, 1984) - 0.352
  \item
    Agentic (Trapnell \& Paulhus, 2012), r=0.306
  \item
    Extraversion (John \& Srivastava, 1999), r=0.285
  \item
    Need for Cognition (Cacioppo, Petty, \& Kao, 1984), r=0.258
  \item
    Green (Haws, Winterich, \& Naylor, 2014) - r=0.213
  \item
    Maximization (Nenkov et al., 2002), r=0.212
  \item
    Overconfidence (Dean \& Ortoleva, 2019) - r=0.198
  \item
    The Horizontal and Vertical Individualism and Collectivism scales
    (Triandis \& Gelfand, 1998) - r=0.169, r=0.182. r=0.217. r=0.125.
  \item
    Communion (Trapnell \& Paulhus, 2012), r=0.158
  \item
    Basic Empathy (Carre et al., 2013) - r=0.149
  \item
    Minimalism (Wilson \& Bellezza, 2021), r=0.106
  \item
    Total performance on the cognitive test - r= -0.113.
  \item
    Cognitive Reflection Test 2 (Thomson and Oppenheimer 2016) - r=
    -0.116
  \end{enumerate}
\end{enumerate}

Thus, individuals who score high on having an intuitive mindset also
tend to score high on regulatory focus and openness---traits that are
correlated with greater emotional awareness, creativity, and
unconventionality. They also tend to seek uniqueness, display
overconfidence, be extraverted, agentic, and self-aware. While they
perform well on communication measures, they tend to score lower on
cognitive tests. Of particular interest is the negative correlation with
the Cognitive Reflection Test 2 (CRT-2), which assesses individuals'
ability to override intuitive (System 1) responses in favor of more
deliberate (System 2) reasoning. This negative correlation aligns with
our expectations: a high intuitive mindset score on our scale should
correspond to greater reliance on System 1 processing. This finding
provides a compelling validation of our scale.

\begin{enumerate}
\def\labelenumi{\arabic{enumi}.}
\setcounter{enumi}{1}
\item
  The analytical/quantitative mindset score (Q) is statistically
  significantly correlated with the following scales at the 95\% level:

  \begin{enumerate}
  \def\labelenumii{\arabic{enumii}.}
  \item
    Need for Cognition (Cacioppo, Petty, \& Kao, 1984) - r=0.380
  \item
    Regulatory Focus (Fellner et al. 2017) - r=0.342
  \item
    Conscientiousness (John \& Srivastava, 1999) - r=0.320
  \item
    Openness (John \& Srivastava, 1999) - r=0.259
  \item
    Self Monitoring (Lennox \& Wolfe, 1984) - 0.248
  \item
    Minimalism (Wilson \& Bellezza, 2021), r=0.237
  \item
    Green (Haws, Winterich, \& Naylor, 2014) - r=0.215
  \item
    The Horizontal and Vertical Individualism and Collectivism scales
    (Triandis \& Gelfand, 1998) - r=0.223, r=0.195. r=0.210. r=0.153.
  \item
    Agentic score (Trapnell \& Paulhus, 2012) - r=0.206
  \item
    Numeracy - r=0.170
  \item
    Maximization (Nenkov et al., 2002), r=0.162
  \item
    Consumer Need for Uniqueness (Ruvio et al., 2008) - r=0.142
  \item
    Overconfidence (Dean \& Ortoleva, 2019) - r=0.140
  \item
    Extraversion (John \& Srivastava, 1999), r=0.136
  \item
    Financial literacy (Johnson, Meier, Toubia 2018) - r=0.134
  \item
    Agreeableness (John \& Srivastava, 1999) - r=0.132
  \item
    Communion score (Trapnell \& Paulhus, 2012), r= 0.103
  \item
    Self Concept Clarity (Campbell et al., 1996) - r=0.099
  \item
    Social desirability(Crowne \& Marlowe, 1960) - r= 0.097
  \item
    Depression - r= -0.119
  \item
    Neuroticism(John \& Srivastava, 1999) - r= -0.183
  \end{enumerate}
\end{enumerate}

Several measures that are statistically significantly and positively
correlated with the intuitive mindset scale (I) are also correlated with
the analytical mindset scale (Q). These include need for cognition,
regulatory focus, openness, self-monitoring, individualism,
collectivism, minimalism, environmental concern (``green''), agentic
traits, communication skills, and overconfidence.

On the other hand, several individual characteristics appear to be
uniquely associated with scoring high on the analytical (Q) scale.
Individuals who scored high on the analytics mindset questions tend to
also score high on conscientiousness, a trait often linked to
organization and discipline, as well as on numeracy and financial
literacy, which reflect core quantitative skills. They also score higher
on agreeableness, social desirability, and self-concept clarity, and
lower on neuroticism and depression. Additionally, an analytical mindset
is positively and significantly associated with being male and having a
higher income.

\subsubsection{Discussion}

Overall, we first validated the QI scale by running it on a large sample
of respondents, identifying a meaningful two-factor structure that
captures both an analytical mindset and an intuitive mindset. We also
found meaningful and expected relationships between the QI scale and a
set of individual characteristics---such as higher numeracy, financial
literacy, and conscientiousness among individuals with an analytical
mindset, and a stronger preference for System 1 processing among those
with an intuitive mindset.

In the digital twins analysis, we found a moderate correlation (around
0.5) between human responses to the individual QI scale and those of
their digital twins, and a weaker relationship (correlation of 0.2-0.25)
between human responses and their digital twins' assessments of the QI
score of the organization they work/ed for, possibly due to the absence
of questions about the individual's workplace in the digital twin's
construction. We also observed that digital twin responses
underestimated the variance present in the actual data, possibly
suggesting that the digital twin reflects a weighted combination of the
fine-tuned individual data and the pre-trained global LLM output.

Finally, while the digital twins did a modest job of capturing the QI
scale, they failed to mimic responses to the two QI behavior questions.
The digital twins tended to consistently select the normative, high-QI
behavior much more frequently than humans did.

\subsection{Targeting Fairness}\label{targeting-fairness}

\subsubsection{Main Questions/Hypotheses}

Companies commonly target their advertisements based on demographic
characteristics, such as race or gender. Yet recent work has found that
people often view the decision to target to be less fair than
advertising broadly to the general population \cite{shaddy2025fairness}. In this sub-study, we aim to replicate this pattern with
human participants, and test whether their digital twins indicate
similar fairness ratings.

\subsubsection{Methods}

We opened the study to 400 human participants on Prolific, and 357
passed the screener and completed it. The study was subsequently run on
their 357 digital twins, for a total of 714 participants. Participants
and their corresponding digital twins were randomly assigned to the same
condition in a two-cell (broad vs. targeted) between-subjects design.

All participants first read: ``A snack foods company has developed a new
line of snacks. Initial testing showed that, due to the taste and
texture profile of the snacks, the snacks are better suited to the
preferences of their female customers.'' In the broad condition,
participants next read: ``Even though they believe the snacks are best
suited to their female customers, they will advertise the snacks broadly
to the general public.'' In the targeted condition, participants next
read: ``Because they believe the snacks are best suited to their female
customers, they will advertise the snacks to women directly.'' Finally,
participants rated fairness: ``How fair is this advertising plan?''
(``Not at all fair'' = 1; ``Very fair'' = 9).

\subsubsection{Results - Pre-registered Analyses}

We conducted a two-way ANOVA with targeting condition (broad vs.\
targeted) and participant type (human vs.\ digital twin) as the
independent variables, and fairness ratings as the dependent variable.
We found a significant main effect of targeting condition 
($F(1, 713) = 69.07,\; p < .001$), but no significant effects of
participant type ($F(1, 713) = 0.05,\; p = .823$) or their interaction
($F(1, 713) = 1.57,\; p = .211$), indicating that, at an aggregate level,
the digital twins rated fairness similarly to the humans. Pairwise
comparisons confirm that for humans, the decision to target was rated as
less fair ($M = 6.38,\; SD = 2.17$) than advertising broadly 
($M = 7.47,\; SD = 1.83,\; F(1, 710) = 45.72,\; p < .001$), and for the
digital twins, targeting was also rated as less fair ($M = 7.30,\;
SD = 0.78$) than advertising broadly ($M = 6.50,\; SD = 0.80,\;
F(1, 710) = 24.92,\; p < .001$).

\subsubsection{Results - Additional Analyses (Non-Preregistered)}

Although not part of our pre-registered plan, a notable observation
stood out in the analysis: the variance among the digital twins was
lower than for the humans. In particular, for the digital twins, 99\% of
responses used the upper end of the scale (5--9), compared with only 87\%
of responses for the human participants ($\chi^2(1) = 25.27,\; p < .001$). 
Examining the responses closest to the mean (i.e., respondents answering ``7''), 
63\% of the digital twins provided that response, compared with 20\% of human participants 
($\chi^2(1) = 136.59,\; p < .001$). The overall correlation between human participants 
and their digital twin was weak but significant ($r = .13,\; p = .012$).

\subsubsection{Discussion}

At the aggregate level, the digital twins closely matched human
judgments of the perceived fairness of demographic targeting,
replicating the finding that it is viewed as less fair than broad
targeting and producing similar average ratings across conditions.
However, the twins' responses were more clustered around the mean,
whereas human participants were more likely to use the whole scale.

\subsection{User Behavior with Recommendation Systems}\label{user-behavior-with-recommendation-systems}

\subsubsection{Main Questions/Hypotheses}
Modern machine learning and AI-based recommendation systems are
data-driven, generating personalized recommendations from a user's
previous interactions. Often, platforms make the assumption that a
user's engagement with a particular piece of content is indicative of
their utility for the content. However, a recent online behavioral
experiment has shown that users may behave strategically and alter their
engagement to influence the content they get recommended in the future \cite{cen2024strategization}. For example, a user on a music
platform may skip past a ``guilty pleasure'' song that they like,
because they are worried that the recommendation algorithm will
recommend too many similar songs later.

In this work, we study self-reported usage of online platforms, beliefs
and preferences about recommendation systems, and strategization
behavior among humans and their digital twins. We test whether digital
twins exhibit similar self-reported behavior as their human
counterparts.

\subsubsection{Methods}

We recruited 598 human participants from Prolific and their digital
twins. Of the human participants, 591 passed the attention check,
resulting in 591 human subjects and 591 digital twins. Participants are
then asked to self-report their usage, beliefs, and preferences
regarding Netflix and Tiktok.

We assess knowledge about the platform's recommendation algorithm by
asking ``How do you think {[}platform{]} recommends content? Please
check all that apply,'' where knowledge corresponds to selecting the
answer ``By analyzing what content you've interacted with in the past''
from a menu of options.

We consider two types of strategization, with explicit user feedback and
implicit user feedback:

\begin{itemize}
\item
  For strategization with explicit user feedback, we ask ``When you are
  on {[}platform{]}, do you give a thumbs-up (or thumbs-down) for any of
  the following reasons? Please check ALL that apply.'' We say that the
  user strategizes with explicit feedback if they check ``Because you
  want {[}platform{]} to show you more (or fewer) content like it.''
\item
  For strategization with implicit user feedback, we ask ``How do you
  typically react if {[}platform{]} shows you content or advertisements
  that you don't want to see in the future? Please check ALL that
  apply.'' We say that the user strategizes with implicit feedback if
  they answer ``Scroll past it faster than I otherwise would'' or
  ``change how I interact with other content.''
\end{itemize}

For preference for user controls in recommendation systems, we ask
``What controls, if any, would you like to have over their
recommendation systems? Please check ALL that apply'' with a menu of
options. We count the number of options selected other than "No control:
let the algorithm work automatically."

\subsubsection{Results -- Pre-registered Analyses}

To analyze the differences between human participants and their
AI-generated digital twins, we conducted a series of paired-samples
$t$-tests for our pre-registered hypotheses. This statistical test
compares the mean scores of two related groups---in this case, each
human participant and their corresponding digital twin. This approach
allows us to precisely assess systematic differences in their responses
for two platforms: TikTok ($N = 234$) and Netflix ($N = 392$).

\emph{Platform Usage Patterns}

Self-reported platform usage differed dramatically between humans and
their digital twins. For TikTok, humans most frequently reported using
the platform ``a few hours every month'' (35.5\%) or ``a few hours every
week'' (33.8\%). In stark contrast, the majority of their digital twins
reported ``Never'' using the platform (51.3\%).

A similar, though less pronounced, pattern emerged for Netflix. The
majority of human participants reported using Netflix ``a few hours
every week'' (56.1\%), whereas their digital twins were more evenly
split between ``a few hours every week'' (60.2\%) and ``a few hours
every day'' (31.1\%). While a paired $t$-test on the underlying ranked
data indicated a significant difference (TikTok: $t(233) = -25.31,\;
p < .001$; Netflix: $t(391) = -8.35,\; p < .001$), we emphasize that
this test was performed on ordinal data.

\emph{Knowledge About Recommendation Algorithms}

When asked if they believe platforms use past interactions to recommend
content, digital twins were significantly more likely to endorse this
idea than their human counterparts. For TikTok, digital twins
($M = 0.996,\; SD = 0.065$) endorsed this belief more strongly than
humans ($M = 0.902,\; SD = 0.298,\; t(233) = -4.92,\; p < .001$). This
finding was replicated on Netflix, where digital twins
($M = 0.997,\; SD = 0.051$) again were more likely to agree with this
mechanism than humans ($M = 0.931,\; SD = 0.254,\; t(391) = -5.27,\;
p < .001$). The correlations between digital twin and human responses
were $r = 0.20$ and $r = 0.19$ for TikTok and Netflix, respectively.

\emph{Presence of Content Strategization}

We measured strategic interaction with platform algorithms using two
distinct questions. On both measures, digital twins reported engaging in
significantly more strategic behavior.

First, when asked if they use ``thumbs-up'' or similar features to
influence future recommendations, digital twins on TikTok
($M = 0.782,\; SD = 0.414$) reported doing so more than humans
($M = 0.530,\; SD = 0.500,\; t(233) = -8.86,\; p < .001$). The effect
was even more pronounced for Netflix, with twins
($M = 0.995,\; SD = 0.071$) reporting far more strategic use of this
feature than humans ($M = 0.543,\; SD = 0.499,\; t(391) = -17.94,\;
p < .001$). The correlations between digital twin and human responses
were $r = 0.56$ and $r = 0.08$ for TikTok and Netflix, respectively.

Second, when asked how they react to undesirable content, digital twins
were more likely to report taking strategic action (e.g., scrolling past
faster, changing interaction patterns). This was true for TikTok, where
twins ($M = 0.957,\; SD = 0.203$) scored higher than humans
($M = 0.509,\; SD = 0.501,\; t(233) = -13.77,\; p < .001$). The pattern
held for Netflix, with twins ($M = 0.939,\; SD = 0.240$) again reporting
more strategic reactions than humans ($M = 0.311,\; SD = 0.464,\;
t(391) = -25.67,\; p < .001$). The correlations between digital twin and
human responses were $r = 0.21$ and $r = 0.17$ for TikTok and Netflix,
respectively.

\emph{Preference for User Controls}

The largest divergence between the two groups was observed in their
stated preference for controls over recommendation systems. Digital
twins expressed a desire for a significantly greater number of controls
than humans. On TikTok, twins ($M = 4.880,\; SD = 0.374$) selected far
more control options than their human counterparts ($M = 1.991,\;
SD = 1.352,\; t(233) = -32.49,\; p < .001$). The result was similarly
stark for Netflix, where twins ($M = 4.888,\; SD = 0.361$) also desired
significantly more controls than humans ($M = 1.862,\; SD = 1.248,\;
t(391) = -47.50,\; p < .001$). The correlations between digital twin and
human responses were $r = 0.11$ and $r = 0.10$ for TikTok and Netflix,
respectively.

\subsubsection{Results - Additional Analyses (Non-Preregistered)}

A key observation, consistent across all the metrics reported above, is
that human responses are noisier than digital twin responses, with the
latter consistently showing smaller standard deviations.

Additionally, we find differences in how humans and digital twins
respond to social pressure, perhaps surprisingly with digital twins
being more responsive to social pressure across both platforms.

We ask participants whether they ``watch, listen to, or `like' content
that [they] do not particularly like just to be polite or support the
creator.'' Humans reported low rates of this behavior on TikTok 
($M = 0.11,\; SD = 0.25$) and Netflix ($M = 0.07,\; SD = 0.26$), 
compared to those of their digital twins on TikTok 
($M = 0.25,\; SD = 0.44$) and Netflix ($M = 0.23,\; SD = 0.42$).

Similarly, we asked participants whether they click ``like'' content
that [they] do not particularly like due to social pressure, and humans
reported lower rates on TikTok ($M = 0.04,\; SD = 0.19$) and Netflix
($M = 0.04,\; SD = 0.19$) than their digital twins on TikTok 
($M = 0.18,\; SD = 0.39$) and Netflix ($M = 0.14,\; SD = 0.35$).

\subsubsection{Discussion}

We find that human and digital twin behavior differ significantly across
a broad range of self-reported metrics on platform usage, knowledge
about recommendation algorithms, strategization, preference for
controls, and social behavior. The digital twins self report more
platform usage, are more aware of how recommendation algorithms work,
are more likely to strategize, prefer more control over their
algorithms, and are more prone to changing their behavior on
recommendation systems due to social pressure.

\clearpage

\end{appendices}

\end{document}